%% file: main_RR.tex
\documentclass[11pt]{article}
\input{myPreliminary_submission}
\usepackage{authblk}
\usepackage{array}
\usepackage{placeins}
\usepackage{booktabs}
\newcolumntype{P}[1]{>{\raggedright\arraybackslash}p{#1}}
\newcommand{\showName}{1}

\begin{document}
\title{\bf Self-normalized jump test for high-frequency data}
\date{}
\if1\showName{
	\author[1]{Ting Tin Ma\thanks{{Email:
				\href{mailto:martinmtt@link.cuhk.edu.hk}
				{\nolinkurl{martinmtt@link.cuhk.edu.hk}}}}}
	\author[1]{Kin Wai Chan\thanks{{Email:
				\href{mailto:kinwaichan@cuhk.edu.hk}
				{\nolinkurl{kinwaichan@cuhk.edu.hk}}}}}
	\affil[1]{\textit{Department of Statistics and Data Science, The Chinese University of Hong Kong}.}

	{\let\footnotesize\normalsize\maketitle}
}\else{
	\author[]{}
	\maketitle
}\fi
{\let\small\normalsize
\spacingset{1.25}
\begin{abstract}
Jump detection using high-frequency log-price data is crucial for decision-making in the financial market.
Existing jump detection tests commonly admit the form of a jump estimator divided by a consistent normalizer, resulting in a pivotal limiting distribution in the absence of jumps.
However, simulation studies have shown that this approach can suffer from an inflated type-I error rate.
In this paper, a new approach for constructing jump-robust self-normalizers is proposed.
The self-normalizer is constructed from differences among several multipower variations of observations that are increasingly spaced away from each other.
After proper scaling, it converges to a non-degenerate distribution with the same volatility scale as the jump estimator.
We first use pairwise differencing and then develop an optimal differencing method that uses all the lag-augmented multipower variations.
The optimal differencing scheme that maximizes local asymptotic power within the proposed class of self-normalized tests is shown to be universal for any given order of multipower in our framework.
In the simulation settings considered, the proposed tests exhibit more accurate empirical size than their non-self-normalized counterparts.
We also establish that the proposed test remains asymptotically valid under sufficiently small Gaussian microstructure noise.
\end{abstract}
}%
\noindent%
\vfill
\newpage
\spacingset{1.5}

\section{Introduction}
\label{sec:intro}
\providecommand{\coloneqq}{\mathrel{\mathop:}=}

\subsection{Motivation}

Testing for jumps in a price process is important for financial applications such as volatility forecasting and risk management; see \citet{das2004systemic}, \citet{bollerslev2008risk}, and \citet{haugom2011realized}.
For example, the quadratic variation of the log-price over $[0,1]$ can be decomposed into continuous and discrete components, which are known as integrated variance and the sum of squared price jumps, respectively; see Section~\ref{sec:inference-theta}.
Some studies showed that using these two components as separate covariates to predict future volatility leads to a significant improvement in prediction accuracy compared to just using the realized variance on its own; see \citet{andersen2007roughing,CorsiPirinoReno2010}.

The main component of a popular class of jump-detection tests is constructed by subtracting a jump-robust estimator of integrated variance, such as bipower variation, minimum realized variance, median realized variance, or threshold bipower variation, from realized quadratic variation; see \citet{BarndorffNielsenShephard2004}, \citet{AndersenDobrevSchaumburg2012}, \citet{CorsiPirinoReno2010}, and \citet{PodolskijZiggel2010}.
A comprehensive overview of existing jump detection methods can be found in \citet{maneesoonthorn2020high}.

In order to construct a test with an asymptotically pivotal null distribution, a standardizer is needed to normalize a specific estimator, say $\hat\theta_g$, of the jump variation $\theta$.
To the best of our knowledge, all existing test statistics of this form are constructed by normalizing $\hat\theta_g$ with the square root of a consistent estimator of its asymptotic variance, which is proportional to integrated quarticity.
This approach can lead to a noticeable amount of size inflation even at a high sampling frequency; see \citet[Sect.~4.2]{barndorff2006econometrics}.
The resulting tests can reject a true null more frequently than the nominal type-I error rate, 
which reduces their reliability. %
This motivates the present paper.

One well-known technique in classical statistics to improve the size accuracy of tests is the method of self-normalization; see \citet{student1908probable,zabell2008student,pena2009self,choi2021general}.
Motivated by the self-normalization technique in the field of time series analysis \citep{lobato2001testing,shao2015self,cheng2022snmct}, we propose to construct a self-normalizer (say, $\hat V$) from $M$ equally spaced returns over $[0,1]$.
Instead of consistently estimating the random factor that scales the limit of $M^{1/2}\hat\theta_g$, we construct $\hat V$ such that $M^{1/2}\hat V$ converges weakly to a non-degenerate limit scaled by the same random factor.
If $\hat V$ retains the same first-order scale as $\hat\theta_g$, the common random factor can be cancelled so that the test statistic is approximated by a pivotal reference null distribution.
Consequently, the type-I error probability can be controlled more accurately.
Constructing such a normalizer under in-fill asymptotics is nontrivial.
Self-normalization has been used in a distinct in-fill problem for jump-activity estimation in pure-jump semimartingales \citep{Todorov2015}.
However, to the best of our knowledge, existing self-normalizing techniques are not designed to test for the presence of price jumps in high-frequency data under in-fill asymptotics.
Unlike \citet{PodolskijZiggel2010}, our test does not use i.i.d.\
multipliers or a consistent estimator of integrated quarticity.
We also remark that bootstrap calibration is another approach to 
improve finite-sample type-I error control. 
For example, 
\citet{DovononGoncalvesHounyoMeddahi2019}
propose using conditionally Gaussian bootstrap returns 
using local estimates of integrated volatility.
However, it can be computationally expensive.

\subsection{Background}
\label{sec:background}
We begin by introducing some notation and the data-generating mechanisms.
All processes are defined on a filtered probability space $(\Omega,\mathcal F,(\mathcal F_t)_{t\ge0},\pr)$ whose filtration is right-continuous and contains all $\pr$-null sets.
For a c\`adl\`ag process $X$, write $\Delta X_t:=X_t-X_{t-}$ for $t>0$.
For semimartingales $X$ and $Y$, let $[X,Y]_t$ denote their quadratic covariation.
Whenever $[X,Y]$ is of locally integrable variation, let $\langle X,Y\rangle_t$ denote its predictable compensator, so that $[X,Y]-\langle X,Y\rangle$ is a local martingale.

For a measurable space $S$, a measure $\mathfrak m$ on $\mathbb R^+\times S$, and an integrand $f$ for which the integral is well defined, write $f\star\mathfrak m_t:=\int_{(0,t]\times S}f(\omega,s,z)\,\mathfrak m(\dd s,\dd z)$.
If $\mathfrak p$ is a Poisson random measure with compensator $\mathfrak q$, then $f\star(\mathfrak p-\mathfrak q)_t$ denotes the compensated Poisson stochastic integral, defined as in \citet[Sect.~2.1]{JacodProtter2012}, whenever $(f^2\wedge|f|)\star\mathfrak q$ is locally integrable.
We write $y^*$ for the logarithmic price of an asset and $\sigma_t^2$ for its spot variance.
When an extension of the probability space is used, $\widetilde\E$ denotes expectation on that extension.
Stable convergence relative to $\mathcal F$ means that for every bounded $\mathcal F$-measurable $Y$ and bounded continuous $f$, $\E\{Yf(Z_M)\}\to\widetilde\E\{Yf(Z)\}$.
When $Z_M$ itself is defined on an extension, the expectation on the left is taken on that extension, as in \citet[eqn.~\textup{(2.2.8)}]{JacodProtter2012}.
We denote this convergence by $Z_M\inST Z$.

\begin{assumption}[Price process]\label{ass:price-model}
The log-price $y^*$ and spot volatility $\sigma$ satisfy 
\begin{align}
    y^*(t) &=y^*(0)+\int_0^t\mu_s^y\dd s+\int_0^t\sigma_s\dd W_s
+\{\delta\ind(|\delta|\le1)\}\star(\mathfrak p-\mathfrak q)_t
+\{\delta\ind(|\delta|>1) \}\star\mathfrak p_t,  \label{eq:grigelionis-y}\\
    \sigma_t&=\sigma_0+\int_0^t\widetilde\mu_s^\sigma\dd s
+\int_0^t\widetilde\sigma_s\dd W_s+L_t^\sigma
+\sum_{s\le t}\Delta\sigma_s\ind(|\Delta\sigma_s|>1). \label{eq:volatility-decomposition}
\end{align}
The variables appearing in (\ref{eq:grigelionis-y})--(\ref{eq:volatility-decomposition}) are defined and regularized as follows: 
\begin{enumerate}[label=(\roman*),ref=\theassumption(\roman*)]
\item \label{ass:jump-regularity}\emph{Price dynamics.}
The processes $\mu^y$ and $\sigma$ are progressively measurable
such that $\mu^y$ is locally bounded and $\sigma$ is c\`adl\`ag.
The process $W$ is an $(\mathcal F_t)$-Brownian motion. 
The $(\mathcal F_t)$-Poisson random measure $\mathfrak p$ on $\mathbb R^+\times E$
has compensator %
$\mathfrak q(\dd t,\dd z)=\dd t\,\lambda(\dd z)$, 
where $\lambda$ is a $\sigma$-finite measure on $(E,\mathcal B(E))$ and
$E$ is a Polish space with Borel $\sigma$-field $\mathcal B(E)$.
The jump-size function $\delta:\Omega\times\mathbb R^+\times E\to\mathbb R$ is predictable.  
There exist stopping times $\tau_m\uparrow\infty$ almost surely and deterministic nonnegative measurable functions $\Gamma_m:E\to[0,\infty)$ such that
\begin{align}\label{eqt:bound}
    \int_E\Gamma_m(z)^2\lambda(\dd z)<\infty,
 \qquad |\delta(\omega,t,z)|\wedge1\le\Gamma_m(z)
 \quad\text{when }t\le\tau_m(\omega).
\end{align}

\item \label{ass:volatility-regularity}
\emph{Volatility dynamics.}
The process $\sigma$ is an It\^o semimartingale.
The local martingale $L^\sigma$ is orthogonal to $W$ and $|\Delta L_t^\sigma|\le1$. 
Denote $\langle L^\sigma,L^\sigma\rangle_t=\int_0^t a_s\dd s$.
The counting process $\sum_{s\le t}\ind(|\Delta\sigma_s|>1)$ 
has intensity $\widetilde {a}_s$, so its compensator is $\int_0^t\widetilde a_s\dd s$.
The processes $\widetilde\mu^\sigma,\widetilde\sigma,a,\widetilde a$ are progressively measurable, $\widetilde\mu^\sigma,a,\widetilde a$ are locally bounded, and $\widetilde\sigma$ is c\`adl\`ag or c\`agl\`ad.
Almost surely, 
$\sigma_t\in(0,\infty)$ for $t\in[0,1]$ and $\sigma_{t-}\in(0,\infty)$ for $t\in(0,1]$.

\item \label{ass:effective-drift}
\emph{Effective drift.}
For each $t$ at which this integral converges absolutely, define 
\[
    \mu_t^\delta=\int_E\delta(t,z)\ind\{ |\delta(t,z)|\le1 \}\lambda(\dd z).
\]
Let $\Lambda_t=\lambda\{z\in E:\delta(t,z)\ne0\}$, $A_t=\int_0^t\Lambda_s\dd s$, and $\varrho=\inf\{t\ge0:A_t=\infty\}$, with $\inf\varnothing=\infty$.
Almost surely, $\mu^y-\mu^\delta$ agrees Lebesgue-almost everywhere on $[0,\varrho\wedge1]$ with a real-valued c\`adl\`ag function on the closed interval $[0,\varrho\wedge1]$.
\end{enumerate}
\end{assumption}

The model \eqref{eq:grigelionis-y} and the regularity conditions in item~\ref{ass:jump-regularity} are the one-dimensional specialization of \citet[Eq.~(6.1.1) and Assumption~6.1.1 with $r=2$]{JacodProtter2012}.
In (\ref{eq:grigelionis-y}), 
the first component, $y^*(0)$, initializes the process; 
the second component, $\int_0^t\mu_s^y\dd s$, permits a non-constant mean;
the third component, $\int_0^t\sigma_s\dd W_s$, accounts for time-varying volatility; 
and
the remaining components are jump integrals that allow for sudden jumps, including infinitely many price jumps.
The local bounds (\ref{eqt:bound}) ensure that both jump integrals appearing in (\ref{eq:grigelionis-y})
are well defined.
The representation determines $y^*$ uniquely up to indistinguishability.
Item~\ref{ass:volatility-regularity} permits leverage and jumps in spot volatility.
Item~\ref{ass:effective-drift} concerns the drift of $y^*$ when there is no jump. 
Almost surely on the event that $y^*$ has no jump on $[0,1]$, 
the compensated integral in \eqref{eq:grigelionis-y} equals $-\int_0^t\mu_s^\delta\dd s$ for $t\in[0,1]$, so that $y^*$ has drift $\mu^y-\mu^\delta$ there; see Lemma~\ref{lem:no-jump-reduction} of the supplement.
We emphasize that independence of the drift from $\sigma$ or $W$ is not assumed.

\subsection{Review of inference on \texorpdfstring{$\theta$}{theta}}\label{sec:inference-theta}
To formulate the problem of testing for jumps, 
we define the jump variation and the events
\[
\theta\coloneqq
\sum_{0<s\le1}\{\Delta y^*(s)\}^2,
\qquad
\Omega_0=\{\omega\in\Omega:\theta(\omega)=0\},
\qquad
\Omega_1=\Omega\backslash \Omega_0 = \{\omega\in\Omega:\theta(\omega)>0\}.
\]
In this article, we aim at testing
\[
H_0:\theta=0
\qquad\text{against}\qquad
H_1:\theta>0.
\]
We can equivalently represent them as $H_0: \omega\in\Omega_0$ and $H_1: \omega\in\Omega_1$, respectively.
These hypotheses concern whether the realized path $y^*(\cdot,\omega)$ is continuous on $[0,1]$.

To provide estimation theories for $\theta$, we need the following notation. 
We say that $Z_M\inST Z$ holds in restriction to $\Omega_0$ if, for every bounded $\mathcal F$-measurable $Y$ and bounded continuous $f$, $\E\{\ind(\Omega_0)Yf(Z_M)\}\rightarrow\widetilde\E\{\ind(\Omega_0)Yf(Z)\}$.
This means stable convergence under $\pr(\cdot\mid\Omega_0)$
if $\pr(\Omega_0)>0$.
For $p>0$, let $\mu_p := \E(|Z|^p) = 2^{p/2} \Gamma((p+1)/2)/\pi^{1/2}$, 
where $Z\sim\Normal(0,1)$ and $\Gamma(\cdot)$ is the gamma function. 
Write $\Phi$ for the distribution function of $\Normal(0,1)$ and $z_\alpha$ for its upper $\alpha$ quantile. 
For $d\ge1$, let $I_d\in\R^{d\times d}$ denote the identity matrix and 
$J_d=(1,\ldots,1)^\T\in\R^d$.
Unless otherwise specified, $\|\cdot\|=\|\cdot\|_2$ denotes the Euclidean norm.

The quadratic variation and integrated variance over $[0,1]$ are, respectively, 
\[
\QV\coloneqq [y^*,y^*]_1
\qquad\text{and}\qquad
\IV\coloneqq\int_0^1\sigma_s^2\dd s . 
\]
So,
$
\QV=\IV+\theta<\infty %
$
almost surely. 
High-frequency returns are needed for estimating $\QV$, $\IV$, and $\theta$.
At sampling frequency $M$, observations are made at $i/M$, $i=0,\ldots,M$, and the $i$th return is
\[
y_i:=y^*(i/M)-y^*((i-1)/M)\qquad(i=1,\ldots,M),
\]
i.e., $y_1,\ldots,y_M$ are the evenly sampled returns over the time period $(0,1]$.
For $r>0$, let
\[
v^{[r]}:=\int_0^1|\sigma_t|^r\dd t
=\int_0^1(\sigma_t^2)^{r/2}\dd t.
\]
In particular, $\IV=v^{[2]}$.
Commonly used estimators of $\QV$ and $\IV$ admit the forms of
\begin{align}\label{eqt:def_QV_IV}
    \hat{\QV}:=\sum_{i=1}^M y_i^2
    \qquad\text{and}\qquad
    \hat{\IV}_g:=\sum_{i=1}^{M-b}g(y_i,\ldots,y_{i+b})
\end{align}
for some function $g:\mathbb R^{b+1}\to\mathbb R$ and positive integer $b$ that is implicitly determined by $g$.
This framework covers, e.g., realized bipower variation proposed in \citet{barndorff2004power2,BarndorffNielsenShephard2004}, the realized minimum variation and median variation proposed in \citet{AndersenDobrevSchaumburg2012} by taking $g$ as
\begin{align*}
g_1(x_1,x_2)=\frac{|x_1x_2|}{2/\pi}, \quad
g_2(x_1,x_2)=\frac{\min(|x_1|,|x_2|)^2}{1-2/\pi},\quad
g_3(x_1,x_2,x_3)=\frac{\med(|x_1|,|x_2|,|x_3|)^2}{1+(6-4\sqrt3)/\pi},
\end{align*}
respectively, where $\med$ denotes the median.
Therefore, whenever $\hat{\IV}_g\inP\IV$ holds as $M\to\infty$, where $\inP$ denotes convergence in probability, we know that
\[
    \hat\theta_g:=\hat{\QV}-\hat{\IV}_g \inP \theta.
\]
\begin{assumption}[Structural condition on $g$]
\label{ass:g-structural}
The function $g$ in (\ref{eqt:def_QV_IV}) satisfies (i)--(v):
\begin{enumerate}[label=(\roman*)]
\item (Sign invariance)
\(
g(s_1x_1,\ldots,s_{b+1}x_{b+1})=g(x_1,\ldots,x_{b+1})
\)
for any $s_j\in\{-1,1\}$.
\item (Two-homogeneity)
\(
g(cx_1,\ldots,cx_{b+1})=c^2g(x_1,\ldots,x_{b+1})
\)
for any $c\ge0$.
\item (Normalization) If $Z_1,\ldots,Z_{b+1}\simIID\Normal(0,1)$, then $\E g(Z_1,\ldots,Z_{b+1})=1$.
\item (Regularity)
The function $g$ is continuous on $\mathbb{R}^{b+1}$ and 
continuously differentiable outside 
a nonempty finite union of affine hyperplanes $B_g\subset\mathbb R^{b+1}$.
There are constants $C>0$, $p_g\ge0$, and $w_g\in(0,1]$ such that, 
for every $z\in \mathbb{R}^{b+1}\setminus B_g$,
\begin{align}
 |g(z)|&\le C(1+\|z\|^{p_g}),  
 \label{eq:g-growth-assumption}\\
 \|\nabla g(z)\|&\le C(1+\|z\|^{p_g})
 [1+\operatorname{dist}(z,B_g)^{-(1-w_g)}], 
 \nonumber\\
 \|\nabla g(z+y)-\nabla g(z)\|&\le C\|y\|(1+\|z\|^{p_g})
 [1+\operatorname{dist}(z,B_g)^{-(2-w_g)}] , 
 \label{eq:g-lip-assumption}
\end{align}
where $\|y\|\le1\wedge\operatorname{dist}(z,B_g)/2$
and $\operatorname{dist}(z,B)=\inf_{x\in B}\|z-x\|_2$.
\item (Non-degeneracy) For $(Z_j)_{j\in\mathbb Z}\simIID\Normal(0,1)$,
\begin{equation}
\label{eq:sigma-g-overlap}
\sigma_g^2:=\sum_{h=-b}^b
\Cov\{Z_0^2-g(Z_0,\ldots,Z_b),
Z_h^2-g(Z_h,\ldots,Z_{h+b})\}\in(0,\infty).
\end{equation}
\end{enumerate}
\end{assumption}

The functions $g_1,g_2,g_3$ satisfy Assumption~\ref{ass:g-structural}; see Remark~\ref{rem:concrete-g-verification} of the supplement for verification.
In particular, $\sigma_{g_1}^2=\pi^2/4+\pi-5\approx0.609$, $\sigma_{g_2}^2\approx1.810$, and $\sigma_{g_3}^2\approx0.959$.
As a special case of our Theorem~\ref{thm:joint-stable-CLT-voljump} to be shown in 
Section \ref{subsec:thresholded-statistics}, we have, in restriction to $\Omega_0$, 
\begin{align}\label{eqt:conv_hat_theta_g}
    M^{1/2}\,\hat\theta_g
 \inST 
 (v^{[4]})^{1/2}\sigma_g Z_0,
\end{align}
where $Z_0\sim\Normal(0,1)$ is independent of $\mathcal F$.
So, the conditional variance of the stable limit is 
$\Var\{ (v^{[4]})^{1/2}\sigma_g Z_0\mid\mathcal F \} = \sigma_g^2v^{[4]}.$
Then, the jump test statistic associated with $g$ is
\[
 T_0(g):=\frac{M^{1/2}\,\hat\theta_g}{(\hat v^{[4]})^{1/2}},
 \quad\text{where}\quad
 \hat v^{[4]}:=\frac{M}{\mu_1^4}
 \sum_{i=1}^{M-3}|y_i y_{i+1}y_{i+2}y_{i+3}|.
\]
Under Assumption~\ref{ass:price-model}, $\hat v^{[4]}\inP v^{[4]}$ by \citet[Thm.~8.4.1 and Example~8.4.4]{JacodProtter2012}.
Hence, (\ref{eqt:conv_hat_theta_g}) and (stable) Slutsky lemma 
give $T_0(g)\inST \sigma_gZ_0$ in restriction to $\Omega_0$.
So, conditionally on $\Omega_0$, an asymptotically level-$\alpha$ %
test rejects $H_0$ when $T_0(g)>\sigma_g z_\alpha$.
However, 
the normal approximation to $T_0(g)$ may be inaccurate in finite samples, even at a high sampling frequency; see \citet[Sect.~4.2]{barndorff2006econometrics} for finite-sample over-rejection of the linear bipower-variation test.
This calls for a new test that can control type-I error accurately even with small $M$.

Constructing a jump-robust self-normalizer for high-frequency data is nontrivial.
We contribute by (i) proposing a methodology for constructing a new class of statistics called lag-augmented multipower variations, (ii) extending the self-normalizing procedure to the class of functions $g$ satisfying Assumption~\ref{ass:g-structural}, which includes important examples from the literature, (iii) deriving the optimal differencing scheme for constructing self-normalizers and showing that the resulting statistic is unchanged across the full-rank differencing matrices considered below, and (iv) deriving the limiting distributions of the self-normalized jump tests in closed form.
Under $H_0$, the limiting distribution is Student-$t$, whereas under the local alternative (Section~\ref{sec:pairwise}), it is noncentral Student-$t$ conditionally on $\mathcal F$.
Coordinatewise thresholding is used to obtain consistency under $H_1$.

The remainder of the article is organized as follows.
Section~\ref{sec:lag-augmented} presents the motivation to use lag-augmented multipower variations as the building blocks of self-normalizers and their formal construction.
Section~\ref{sec:differencing} explains how to use the lag-augmented multipower variations optimally to construct the self-normalizer by a difference-based method.
It also discusses the choice of $n$, $\ell$, and $g$.
Section~\ref{sec:microstructure-noise} studies validity and consistency under vanishing Gaussian microstructure noise that may depend on the latent price.
Simulation studies are presented in Section~\ref{sec:simulation-studies} to illustrate the finite-sample size-accuracy and power of tests.
Real-data analysis is carried out in Section~\ref{sec:real-data-application} by applying the self-normalized jump test to detect potential jumps in returns of assets in different markets.

\section{Lag-augmented multipower variations}
\label{sec:lag-augmented}
\label{sec:vol-jump-null}

\subsection{Failure of $n$-power variations as self-normalizers}
\label{sec:failure}
We first define the notion of jump-robust self-normalizers.
\begin{definition}[Jump-robust self-normalizer]
\label{def:jump-robust-self-normalizer}
A statistic $\hat V$ is a jump-robust self-normalizer if, under Assumption~\ref{ass:price-model}: (i) in restriction to $\Omega_0$, for a jointly pivotal non-degenerate pair $(\mathbb A_1,\mathbb A_2)$ independent of $\mathcal F$,
\[
M^{1/2}(\hat V,\hat\theta_g)
\inST 
(v^{[4]})^{1/2}(\mathbb A_1,\mathbb A_2),
\qquad \pr(\mathbb A_1=0)=0,
\]
where the joint law does not depend on $(\mu_t^y)_{t\in[0,1]}$ or $(\sigma_t)_{t\in[0,1]}$; and (ii) $\hat V\inP0$.
\end{definition}
According to Definition~\ref{def:jump-robust-self-normalizer}, a self-normalized statistic for testing $H_0:\theta=0$ is $T=\hat\theta_g/\hat V$, whose limiting null distribution is pivotal, i.e., $T\inST \mathbb A_2/\mathbb A_1$ in restriction to $\Omega_0$.
In addition, $\hat V$ is jump-robust as $\hat V\inP0$ under Assumption~\ref{ass:price-model}.
We attempt to construct building blocks to define self-normalizers.
The standard $n$-power variation as an estimator of $v^{[2]}$ is 
\[
Q^{(n)}=\mu_{2/n}^{-n}\sum_{i=1}^{M-n+1}
\prod_{j=0}^{n-1}|y_{i+j}|^{2/n},
\]
where $N\ge2$ and $1\le n\le N$.
When $n=1,2$, $Q^{(1)}$ and $Q^{(2)}$ reduce to the realized variance and bipower variation, respectively.
For every $n\ge2$, $Q^{(n)}$ has asymptotic mean $v^{[2]}$, and hence its asymptotic distribution depends on $\sigma(\cdot)$.
So, none of $Q^{(n)}$ is a jump-robust self-normalizer.
It is then natural to consider a weighted sum of $Q^{(2)},\ldots,Q^{(N)}$ with weights that sum to zero.
It eliminates the asymptotic mean $v^{[2]}$.
However, the resulting statistic has the complicated covariance structure illustrated in Theorem~\ref{thm:standard-multipower-variation} below.

\begin{theorem}[$n$-power variation]
\label{thm:standard-multipower-variation}
Under Assumption~\ref{ass:price-model}, for fixed $N\ge2$, there is a centered Gaussian vector $\mathcal Z^{(N)}=(\mathcal Z_2^{(N)},\ldots,\mathcal Z_N^{(N)})^\T$, independent of $\mathcal F$, such that 
\[
M^{1/2}\{(Q^{(2)},\ldots,Q^{(N)})^\T-v^{[2]}J_{N-1}\}
\inST 
(v^{[4]})^{1/2}\mathcal Z^{(N)} 
\]
in restriction to $\Omega_0$,
where the covariance matrix $\Sigma_0^{(N)}= (\Sigma_{0}^{(N)}(r,s))_{1\le r,s\le N-1}$ 
of $\mathcal Z^{(N)}$ is given by the following expression: for $2\le r,s\le N$, 
\begin{equation}
\label{eq:standard-multipower-covariance}
\Sigma_0^{(N)}(r-1,s-1)
=(|r-s|+1)\left(\frac{\mu_{2/r+2/s}}{\mu_{2/r}\mu_{2/s}}\right)^{r\wedge s}
+2\sum_{u=1}^{r\wedge s-1}
\left(\frac{\mu_{2/r+2/s}}{\mu_{2/r}\mu_{2/s}}\right)^u+1-r-s. 
\end{equation}
\end{theorem}
From Theorem~\ref{thm:standard-multipower-variation}, the diagonal entries of $\Sigma_0^{(N)}$ take different values.
The off-diagonal entries $\Sigma_0^{(N)}(r-1,s-1)$ change with $r$ and $s$ in a complicated way.
More importantly, the limiting covariance between $\hat\theta_g$ and $Q^{(n)}$ generally depends on $n$, so a weighted sum of $Q^{(2)},\ldots,Q^{(N)}$ whose weights sum to zero need not be asymptotically uncorrelated with $\hat\theta_g$.
In addition, for a fixed value of $n$, this scheme creates only one $Q^{(n)}$ for normalizing $\hat\theta_g$, which may render the test inefficient.
The lag-augmented variations introduced below have the same limiting covariance with $\hat\theta_g$.
Consequently, any weighted sum of these variations whose weights sum to zero is asymptotically uncorrelated with $\hat\theta_g$.

\subsection{Lag augmentation}
\label{subsec:thresholded-statistics}
\label{subsec:null-covariance-contrasts}

We propose a new class of statistics called lag-augmented multipower variations in this section.
It is used to construct self-normalizers in Sections~\ref{sec:pairwise}--\ref{sec:optimal}.
To resolve the problem stated in Section~\ref{sec:failure}, we attempt to derive a new class of $n$-power variations
\[
(Q_1^{(n)},\ldots,Q_b^{(n)},\ldots,Q_{b+\ell-1}^{(n)}).
\]
In particular, we use the last $\ell$ of them, i.e.,
\[
\mathbb Q_\ell^{(n)}(b)
=(Q_b^{(n)},\ldots,Q_{b+\ell-1}^{(n)})^\T,
\]
to construct self-normalizers for $\hat\theta_g$ such that the following conditions are satisfied:
\begin{enumerate}[label=(\roman*)]
\item Each $Q_k^{(n)}$ has the same asymptotic marginal distribution for $k=b,\ldots,b+\ell-1$.
\item The asymptotic joint distribution of $(Q_k^{(n)},\hat\theta_g)$ is the same for each $k=b,\ldots,b+\ell-1$.
\item The asymptotic covariance of $Q_h^{(n)}$ and $Q_k^{(n)}$ depends on $h$ and $k$ only through whether $h=k$.
Also, they are not perfectly correlated.
\end{enumerate}
Our proposed $\mathbb Q_\ell^{(n)}(b)$ is called a lag-augmented $n$-power variation.

Fix integers $\ell\in\{2,3,\ldots\}$, $n\in\{3,4,\ldots\}$, and set $L=b+\ell+1$ and $K=(n-1)L$.
For $k=1,\ldots,b+\ell-1$, define the index set
\[
\mathcal I_k=\{1\}\cup\{k+2+jL:j=0,\ldots,n-2\}.
\]
The $k$-lag-augmented $n$-power variation is
\begin{equation}
\label{eq:Qk-unthresholded}
Q_k^{(n)}=\mu_{2/n}^{-n}\sum_{i=1}^{M-K+1}
\prod_{r\in\mathcal I_k}|y_{i+r-1}|^{2/n}.
\end{equation}
The definition of $Q_k^{(n)}$ depends on $\ell$ through $L=b+\ell+1$.
We suppress this dependence in the notation.
The lags in $Q_k^{(n)}$ are chosen for the following reasons.
In the $i$th summand of $Q_k^{(n)}$, i.e., $|y_i|^{2/n}\prod_{j=0}^{n-2}|y_{i+k+1+jL}|^{2/n}$, the first lag, i.e., between $y_i$ and $y_{i+k+1}$, is set as $k+1$, so the time lag is further augmented by $k$ more lags.
If it is set as $k$ instead, the asymptotic covariance between $Q_b^{(n)}$ and $\hat\theta_g$ would in general be different from that between $Q_k^{(n)}$ and $\hat\theta_g$ for $k>b$.
The remaining lags between $y_{i+k+1+jL}$ and $y_{i+k+1+(j+1)L}$ for $j=0,\ldots,n-3$ are all set as $L=b+\ell+1$.
In particular, for $i,i'\in\{1,\ldots,M-K+1\}$ and $k,k'\in\{b,\ldots,b+\ell-1\}$,
\[
(y_i,y_{i+k+1},\ldots,y_{i+k+1+(n-2)L})
\qquad\text{and}\qquad
(y_{i'},y_{i'+k'+1},\ldots,y_{i'+k'+1+(n-2)L})
\]
can at most have $(n-1)$ common terms when $k\ne k'$.
This spacing makes the limiting covariance matrix compound-symmetric. 

Under $H_1$, Assumption~\ref{ass:g-structural} alone does not ensure that $\hat{\IV}_g$ remains consistent for $v^{[2]}$ without thresholding.
We therefore threshold every return entering a summand, following the threshold multipower variation of \citet[Definition~2.2]{CorsiPirinoReno2010}.
A summand is removed by truncation if any squared return entering it exceeds its threshold.
The truncation procedure guarantees consistency under $H_1$.
In contrast, the realized quadratic variation is intentionally left unthresholded so that it retains jump quadratic variation under $H_1$, while under $H_0$ both $\hat{\QV}$ and $\hat{\IV}_g$ converge to $v^{[2]}$.

Following the threshold approach of \citet{Mancini2009} and \citet{CorsiPirinoReno2010}, we adopt the following thresholding conditions:
\begin{align}
&\Theta(M^{-1})\to0,
\qquad \frac{M^{-1}\log M}{\Theta(M^{-1})}\to0,
\label{eq:Theta-rate}\\
&\Theta(M^{-1})=c_\vartheta^2M^{-2\varpi},
\qquad0<\varpi<\frac12,
\qquad c_\vartheta>0,
\label{eq:Theta-power}\\
&\vartheta_{i,M}=\xi_{i,M}\Theta(M^{-1}),
\label{eq:local-threshold}\\
&0<\underline\xi\le\xi_{i,M}\le\overline\xi<\infty
\quad\text{for all }i,M,\quad\text{a.s.},
\label{eq:xi-bounded}
\end{align}
where $\Theta:(0,\infty)\to(0,\infty)$ is deterministic, $\xi_{i,M}$ are measurable random variables, and $\underline\xi,\overline\xi$ are deterministic constants.
Accordingly, we define the thresholded counterparts by
\begin{equation}
\label{eq:ivg-tr}
\hat{\IV}_g^{\mathrm{tr}}=\sum_{i=1}^{M-b}g(y_i,\ldots,y_{i+b})
\prod_{r=0}^b\ind(y_{i+r}^2\le\vartheta_{i+r,M}),
\qquad
\hat\theta_g^{\mathrm{tr}}=\hat{\QV}-\hat{\IV}_g^{\mathrm{tr}},
\end{equation}
where $\hat{\QV}$ remains unthresholded, and
\begin{equation}
\label{eq:Qk-tr}
Q_k^{(n),\mathrm{tr}}=\mu_{2/n}^{-n}\sum_{i=1}^{M-K+1}
\prod_{r\in\mathcal I_k}\left\{|y_{i+r-1}|^{2/n}
\ind(y_{i+r-1}^2\le\vartheta_{i+r-1,M})\right\},
\end{equation}
and set
\[
\mathbb Q_\ell^{(n),\mathrm{tr}}(b)
=
(Q_b^{(n),\mathrm{tr}},\ldots,
Q_{b+\ell-1}^{(n),\mathrm{tr}})^\T.
\]
On $\Omega_0$, the thresholded and unthresholded statistics agree almost surely for all sufficiently large $M$; see \eqref{eq:null-eventual-equality} in the supplement.
Define
\begin{equation}
\label{eq:Cgn-def}
C_{g,n}=n\mu_{2/n}^{-1}\sum_{q=1}^{b+1}
\E\left\{ (Z_1^2-g(Z_1,\ldots,Z_{b+1}))|Z_q|^{2/n} \right\},
\end{equation}
where $Z_1,\ldots,Z_{b+1}$ are independent standard normal variables.
The constant $C_{g,n}$ is finite by the condition \eqref{eq:g-growth-assumption} and does not depend on $k$.

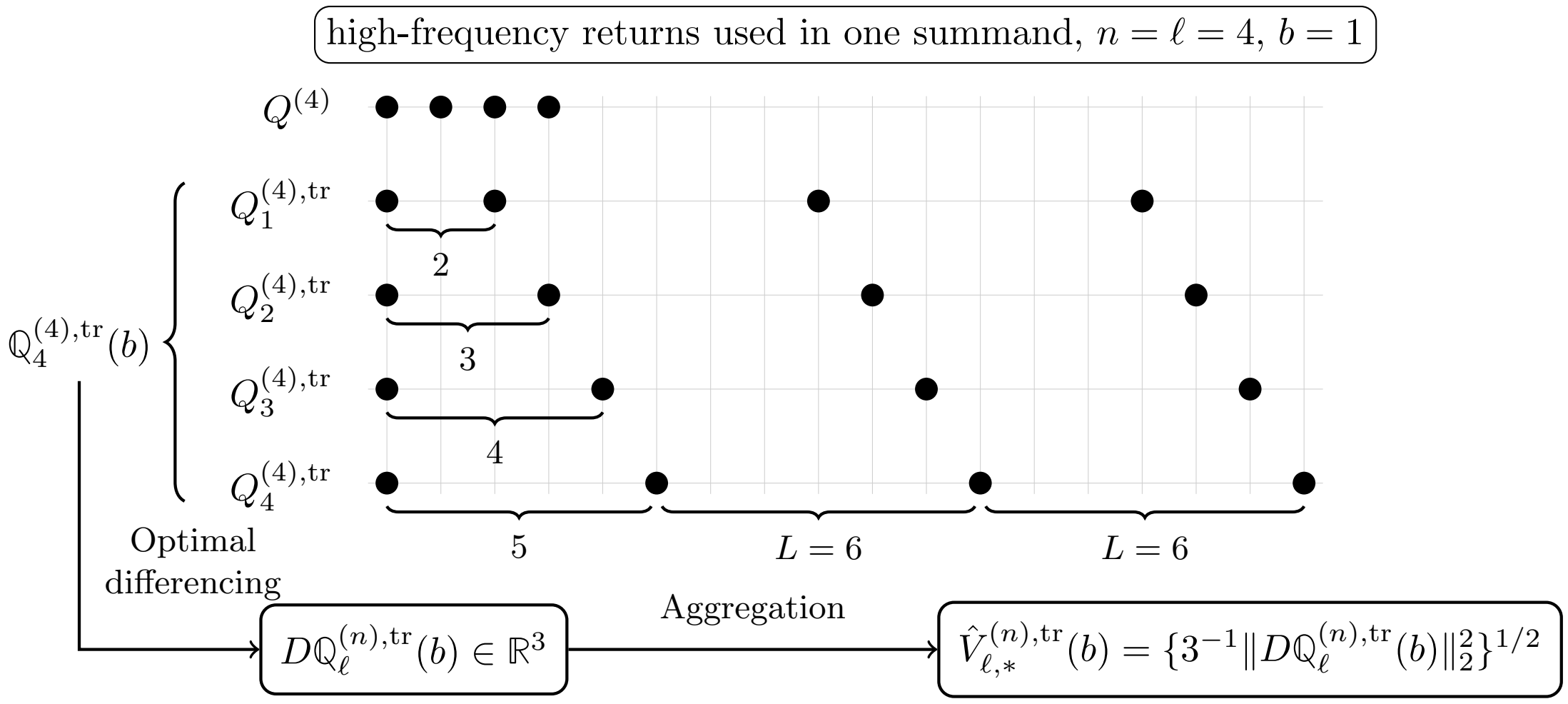
\begin{figure}[t]
\centering
\small
\begin{tikzpicture}[x=.50cm,y=.67cm]
\node[draw,rounded corners,inner sep=3pt] at (8.5,6.20)
{high-frequency returns used in one summand, $n=\ell=4$, $b=1$};
\foreach \x in {0,...,17}\draw[lightgray!75,very thin](\x,-.15)--(\x,5.35);
\foreach \y in {0,1.30,2.60,3.90,5.20}\draw[lightgray!75,very thin](-.35,\y)--(17.35,\y);
\node[anchor=e] at (-.75,5.20){$Q^{(4)}$};
\foreach \x in {0,1,2,3}\fill(\x,5.20)circle(3pt);
\draw[thick,decorate,decoration={brace,amplitude=5pt,mirror}]
(-3.75,4.15)--(-3.75,-.25)
node[midway,left=5pt](qgroup){$\mathbb Q_4^{(4),\mathrm{tr}}(b)$};
\node[anchor=e] at (-.75,3.90){$Q_1^{(4),\mathrm{tr}}$};
\node[anchor=e] at (-.75,2.60){$Q_2^{(4),\mathrm{tr}}$};
\node[anchor=e] at (-.75,1.30){$Q_3^{(4),\mathrm{tr}}$};
\node[anchor=e] at (-.75,0){$Q_4^{(4),\mathrm{tr}}$};
\foreach \x in {0,2,8,14}\fill(\x,3.90)circle(3pt);
\foreach \x in {0,3,9,15}\fill(\x,2.60)circle(3pt);
\foreach \x in {0,4,10,16}\fill(\x,1.30)circle(3pt);
\foreach \x in {0,5,11,17}\fill(\x,0)circle(3pt);
\draw[thick,decorate,decoration={brace,amplitude=3pt,mirror}]
(0,3.58)--(2,3.58)node[midway,below=4pt,font=\footnotesize]{$2$};
\draw[thick,decorate,decoration={brace,amplitude=3pt,mirror}]
(0,2.28)--(3,2.28)node[midway,below=4pt,font=\footnotesize]{$3$};
\draw[thick,decorate,decoration={brace,amplitude=3pt,mirror}]
(0,.98)--(4,.98)node[midway,below=4pt,font=\footnotesize]{$4$};
\draw[thick,decorate,decoration={brace,amplitude=3pt,mirror}]
(0,-.32)--(4.9,-.32)node[midway,below=4pt,font=\footnotesize]{$5$};
\draw[thick,decorate,decoration={brace,amplitude=3pt,mirror}]
(5.1,-.32)--(10.9,-.32)node[midway,below=4pt,font=\footnotesize]{$L=6$};
\draw[thick,decorate,decoration={brace,amplitude=3pt,mirror}]
(11.1,-.32)--(17,-.32)node[midway,below=4pt,font=\footnotesize]{$L=6$};
\node[draw,rounded corners,thick,inner sep=5pt](contrast) at (.5,-2.3)
{$D\mathbb Q_\ell^{(n),\mathrm{tr}}(b)\in\mathbb R^3$};
\draw[->,thick](qgroup.south)--(qgroup.south|-contrast.west)
node[pos=.68,right=3pt,font=\footnotesize]{\shortstack{Optimal\\differencing}}
--(contrast.west);
\node[draw,rounded corners,thick,inner sep=5pt](aggregate) at (16,-2.3)
{$\hat V_{\ell,*}^{(n),\mathrm{tr}}(b)=\{3^{-1}\|D\mathbb Q_\ell^{(n),\mathrm{tr}}(b)\|_2^2\}^{1/2}$};
\draw[->,thick](contrast.east)--(aggregate.west)
node[midway,above=5pt,fill=white,inner sep=1pt,font=\footnotesize]{Aggregation};
\end{tikzpicture}
\caption{Procedure for constructing the self-normalizer when $n=\ell=4$ and $b=1$.
First, we compute $\mathbb Q_\ell^{(n),\mathrm{tr}}(b)$.
The black dots indicate which high-frequency returns are used in one summand of each coordinate.
Second, we apply a differencing matrix $D$ satisfying \eqref{eq:D-conditions}.
Third, we aggregate $D\mathbb Q_\ell^{(n),\mathrm{tr}}(b)$ to form $\hat V_{\ell,*}^{(n),\mathrm{tr}}(b)$.}
\label{fig:proposal-summary}
\end{figure}

Figure~\ref{fig:proposal-summary} visualizes the construction of $\mathbb Q_4^{(4),\mathrm{tr}}(b)$ and our proposed procedure.
The following theorem shows the joint limiting distribution of $\hat\theta_g$ and $\mathbb Q_\ell^{(n)}(b)$.
\FloatBarrier
\begin{theorem}[Lag-augmented $n$-power variation]
\label{thm:joint-stable-CLT-voljump}
Under Assumptions~\ref{ass:price-model} and~\ref{ass:g-structural}, for fixed $n\ge3$ and $\ell\ge2$, there is a centered Gaussian vector $\mathcal Z=(\mathcal Z_0,\ldots,\mathcal Z_\ell)^\T$, independent of $\mathcal F$, such that, in restriction to $\Omega_0$,
\begin{align}\label{eqt:conv_hat_theta_g_Q}
    M^{1/2}\left(\hat\theta_g,
[\mathbb Q_\ell^{(n)}(b)-v^{[2]}J_\ell]^\T\right)^\T
\inST (v^{[4]})^{1/2}\mathcal Z.    
\end{align}
Put $\eta_n=\mu_{4/n}\mu_{2/n}^{-2}$ and $S^{(n)}=2\ind(n\ge4)\sum_{m=2}^{n-2}\eta_n^m$.
For $1\le r,s\le\ell$, define
\begin{equation}
\label{eq:SigmaQ-final}
\Sigma_{\mathbb Q}^{(n)}(r,s)=
\begin{cases}
\eta_n^n+S^{(n)}+2n\eta_n+5-4n,&r=s,\\
\eta_n^{n-1}+S^{(n)}+(2n+1)\eta_n+4(1-n),&r\ne s.
\end{cases}
\end{equation}
Write
$\Sigma_{\mathbb Q}^{(n)}
=(\Sigma_{\mathbb Q}^{(n)}(r,s))_{1\le r,s\le\ell}$.
The covariance matrix of $\mathcal Z$ is
\begin{equation}
\label{eq:Sigma-gQ-block}
\Sigma_{g,\mathbb Q}^{(n)}
=
\begin{pmatrix}
\sigma_g^2&C_{g,n}J_\ell^\T\\
C_{g,n}J_\ell&\Sigma_{\mathbb Q}^{(n)}
\end{pmatrix}.
\end{equation}
If \eqref{eq:Theta-rate}, \eqref{eq:local-threshold}, and the uniform lower bound in \eqref{eq:xi-bounded} hold, 
the result in \eqref{eqt:conv_hat_theta_g_Q} holds when $\hat\theta_g$ and $\mathbb Q_\ell^{(n)}(b)$ are replaced by $\hat\theta_g^{\mathrm{tr}}$ and $\mathbb Q_\ell^{(n),\mathrm{tr}}(b)$, respectively.
\end{theorem}
Write $\mathcal Z_{\mathbb Q}=(\mathcal Z_1,\ldots,\mathcal Z_\ell)^\T$.
From Theorem~\ref{thm:joint-stable-CLT-voljump},
in restriction to $\Omega_0$, the limiting distribution of
$M^{1/2}\{\mathbb Q_\ell^{(n)}(b)-v^{[2]}J_\ell\}$
is multivariate normal conditionally on $\mathcal F$.
Its asymptotic covariance $v^{[4]}\Sigma_{\mathbb Q}^{(n)}(i,j)$ only depends on whether $i=j$ or not.
Each element within $\mathbb Q_\ell^{(n)}(b)$ has the same covariance structure with $\hat\theta_g$.
For example, when $g=g_1$, the common limiting covariance is $C_{g_1,n}v^{[4]}$, where $C_{g_1,n}=2(n+1)-2n\mu_{2/n+1}\mu_{2/n}^{-1}\mu_1^{-1}$ does not depend on $k$; see Remark~\ref{rem:g1-crosscov}.
These features allow us to form difference statistics from $\mathbb Q_\ell^{(n)}(b)$ whose stable limits, after proper scaling, are independent and identically distributed conditionally on $\mathcal F$.
These statistics serve as building blocks for constructing self-normalizers in Sections~\ref{sec:pairwise}--\ref{sec:optimal}.

\section{Difference-based self-normalizers}
\label{sec:differencing}
\label{sec:threshold-extension}

\subsection{Pairwise differencing}
\label{sec:pairwise}
\label{subsec:null-distribution}

As $Q_b^{(n)},\ldots,Q_{b+\ell-1}^{(n)}$ have asymptotic mean $v^{[2]}$, they cannot serve as self-normalizers individually according to Definition~\ref{def:jump-robust-self-normalizer}.
To illustrate the idea, consider the simplest case where $\ell=2$.
Apply differencing to $\mathbb Q_2^{(n)}(b)=(Q_b^{(n)},Q_{b+1}^{(n)})^\T$ to obtain
\[
\hat V_0(b)=|Q_b^{(n)}-Q_{b+1}^{(n)}|,
\quad
\hat V_0^{\mathrm{tr}}(b)=
|Q_b^{(n),\mathrm{tr}}-Q_{b+1}^{(n),\mathrm{tr}}|,
\]
which eliminate the common asymptotic mean $v^{[2]}$.
By Theorem~\ref{thm:joint-stable-CLT-voljump}, in restriction to $\Omega_0$,
\[
M^{1/2}\hat V_0^{\mathrm{tr}}(b)
\inST 
(2\lambda_\perp v^{[4]})^{1/2}|Z|, \quad \text{where} \quad 
\lambda_\perp:=\Sigma_{\mathbb Q}^{(n)}(1,1)-\Sigma_{\mathbb Q}^{(n)}(1,2)
=(\eta_n-1)(\eta_n^{n-1}-1),
\]
and 
$Z\sim\Normal(0,1)$ is independent of $\mathcal F$.
Recall that $\eta_n=\mu_{4/n}\mu_{2/n}^{-2}$. 
So, by Jensen's inequality, $\lambda_\perp >0$.
For a general even integer $\ell\ge2$, we form pairs of consecutive 
$Q_k^{(n)}$s, i.e.,
\begin{align}\label{eqt:mathbb_S_ell_n_b}
\mathbb S_\ell^{(n)}(b)
&=
(Q_b^{(n)}-Q_{b+1}^{(n)},\ldots,
Q_{b+\ell-2}^{(n)}-Q_{b+\ell-1}^{(n)})^\T .
\end{align}
Similarly, the thresholded version is
$\mathbb S_\ell^{(n),\mathrm{tr}}(b)
=(Q_b^{(n),\mathrm{tr}}-Q_{b+1}^{(n),\mathrm{tr}},\ldots,
Q_{b+\ell-2}^{(n),\mathrm{tr}}-Q_{b+\ell-1}^{(n),\mathrm{tr}})^\T.$
Their properties are stated below:

\begin{theorem}[Self-normalizers]
\label{thm:pairwise-self-normalizers}
Under Assumptions~\ref{ass:price-model} and~\ref{ass:g-structural}, for fixed $n\ge3$ and even $\ell\ge2$, suppose that \eqref{eq:Theta-rate}, \eqref{eq:local-threshold}, and the uniform lower bound in \eqref{eq:xi-bounded} hold.
Let $Z=(Z_1,\ldots,Z_{\ell/2})^\T$, where $Z_0,Z_1,\ldots,Z_{\ell/2}\simIID \Normal(0,1)$
are jointly independent of $\mathcal F$.
Then, in restriction to $\Omega_0$,
\[
M^{1/2}
\begin{pmatrix}
\hat\theta_g^{\mathrm{tr}}\\
\mathbb S_\ell^{(n),\mathrm{tr}}(b)
\end{pmatrix}
\inST 
(v^{[4]})^{1/2}
\begin{pmatrix}
\sigma_g Z_0\\
(2\lambda_\perp)^{1/2}Z
\end{pmatrix}.
\]
Replacing $\hat\theta_g^{\mathrm{tr}}$ and $\mathbb S_\ell^{(n),\mathrm{tr}}(b)$ by $\hat\theta_g$ and $\mathbb S_\ell^{(n)}(b)$, respectively, leaves the joint stable limit above unchanged.
Under Assumptions~\ref{ass:price-model} and~\ref{ass:g-structural} and conditions \eqref{eq:Theta-power}--\eqref{eq:xi-bounded}, $\mathbb S_\ell^{(n),\mathrm{tr}}(b)\inP0$.
\end{theorem}

From Theorem~\ref{thm:pairwise-self-normalizers},
the elements of $\mathbb S_\ell^{(n),\mathrm{tr}}(b)$ satisfy the requirements for being jump-robust self-normalizers according to Definition~\ref{def:jump-robust-self-normalizer}.
Moreover, Theorem~\ref{thm:pairwise-self-normalizers} shows that the coordinates of $\mathbb S_\ell^{(n),\mathrm{tr}}(b)$ are asymptotically independent normal random variables conditionally on $\mathcal F$.
So, provided Condition~(i) below holds, the thresholded self-normalizer
\[
\hat V_\ell^{(n),\mathrm{tr}}(b)
=h\!\left(Q_b^{(n),\mathrm{tr}}-Q_{b+1}^{(n),\mathrm{tr}},\ldots,
Q_{b+\ell-2}^{(n),\mathrm{tr}}-Q_{b+\ell-1}^{(n),\mathrm{tr}}\right)
\]
satisfies $M^{1/2}\,\hat V_\ell^{(n),\mathrm{tr}}(b)
\inST 
V_\ell:=h(X_1,\ldots,X_d)$ in restriction to $\Omega_0$ as $M\to\infty$.
Here $h:\mathbb R^d\to[0,\infty)$ is continuous and $d=\ell/2$.
Conditionally on $\mathcal F$, $X_1,\ldots,X_d$ are independent $\Normal(0,2\lambda_\perp v^{[4]})$ random variables.
Consequently, it suffices to choose a suitable $h(\cdot)$ using the following proposed principles:
\begin{enumerate}[label=(\roman*)]
    \item Homogeneity:
    $
    h(cx_1,\ldots,cx_d)
    =
    |c|h(x_1,\ldots,x_d)
    $
    for all $c,x_1,\ldots,x_d\in\mathbb{R}$.

\item Exchangeability:
$h(x_1,\ldots,x_d)=h(x_{\pi(1)},\ldots,x_{\pi(d)})$ for any permutation $\pi$ of $1,\ldots,d$ and for any $x_1,\ldots,x_d\in\mathbb{R}$.

    \item Symmetry:
    $
    h(x_1,\ldots,x_d)
    =
    h(|x_1|,\ldots,|x_d|)
    $
    for all $x_1,\ldots,x_d\in\mathbb{R}$.

    \item Non-negativity:
    $h(x)\geq0$ for all $x\in\mathbb{R}^d$, where equality holds iff $x=(0,\ldots,0)^{\T}$.

    \item Orthogonal invariance:
    $
    h(Ox)=h(x)
    $
    for all $x\in\mathbb{R}^d$ 
    and orthogonal matrix $O\in\mathbb{R}^{d\times d}$.
\end{enumerate}
Condition~(i) is imposed so that $\hat V_\ell^{(n),\mathrm{tr}}(b)$ satisfies Definition~\ref{def:jump-robust-self-normalizer}.
With $c=(2\lambda_\perp v^{[4]})^{1/2}$, homogeneity gives
\(
h(X_1,\ldots,X_d)
=(2\lambda_\perp v^{[4]})^{1/2}h(Z_1,\ldots,Z_d),
\)
where the random variables $Z_j=X_j/(2\lambda_\perp v^{[4]})^{1/2}$ are independent $\Normal(0,1)$, jointly independent of $\mathcal F$.
The factor $(v^{[4]})^{1/2}$ then cancels with that in the stable limit of $M^{1/2}\,\hat\theta_g^{\mathrm{tr}}$ when forming the test statistic.
Condition~(ii) reflects that the components of $\mathbb S_\ell^{(n),\mathrm{tr}}(b)$ have the same marginal limit under $H_0$ and all converge to zero under $H_1$.
Condition~(iii) is important.
As $M^{1/2}\mathbb S_\ell^{(n),\mathrm{tr}}(b)$ converges stably in restriction to $\Omega_0$ to a vector whose coordinates, conditionally on $\mathcal F$, are independent centered normal random variables, the signs of the limiting coordinates are ancillary statistics for $v^{[4]}$.
So, only their magnitudes matter.
Condition~(iv) ensures that the normalizer is positive away from the origin, which ensures that the signs of $\hat\theta_g^{\mathrm{tr}}/\hat V_\ell^{(n),\mathrm{tr}}(b)$ and $\hat\theta_g^{\mathrm{tr}}$ are the same.
A zero denominator instead requires the convention below.
Section~\ref{sec:optimal} discusses the necessity and importance of Condition~(v),
which ensures that the self-normalizer is invariant under orthogonal transformations of the normalized differences.

We now show that conditions (i)--(v) determine $h$ uniquely up to a positive multiplicative constant.
Let $e_1=(1,0,\ldots,0)^\T\in\mathbb R^d$.
For each $x$, there exists an orthogonal matrix $O$ such that $Ox=\|x\|_2e_1$.
Then, the conditions (i) and (v) 
imply that $h(x)=h(\|x\|_2e_1)=\|x\|_2h(e_1)$.
Setting $h((1,\ldots,1)^\T)=1$ as the baseline gives $h(e_1)=d^{-1/2}$.
Therefore,
\begin{equation}\label{eq: h L2}
    h(x)
=
d^{-1/2}\|x\|_2
=
\left(
\frac{1}{d}\sum_{j=1}^d x_j^2
\right)^{1/2}.
\end{equation}

As an alternative point of view, conditionally on $\mathcal F$, $X_1^2/(2\lambda_\perp),\ldots,X_d^2/(2\lambda_\perp)$ are independent $v^{[4]}\chi_1^2$ random variables.
Therefore, the maximum-likelihood estimator of $(v^{[4]})^{1/2}$ based on $X_1,\ldots,X_d$ is
$(2\lambda_\perp)^{-1/2}
\{d^{-1}\sum_{j=1}^dX_j^2\}^{1/2}$, which gives the same choice of $h$.

In view of Definition~\ref{def:jump-robust-self-normalizer}, 
our proposed jump-robust self-normalizer is
\begin{equation}\label{eq: pairwise SN}
\hat V_\ell^{(n),\mathrm{tr}}(b)=
\left\{\frac2\ell\sum_{j=1}^{\ell/2}
(Q_{b+2j-2}^{(n),\mathrm{tr}}-Q_{b+2j-1}^{(n),\mathrm{tr}})^2
\right\}^{1/2},    
\end{equation}
which satisfies
\(
M^{1/2}\hat V_\ell^{(n),\mathrm{tr}}(b)
\inST 
(2\lambda_\perp v^{[4]})^{1/2}
\{(2/\ell)\sum_{j=1}^{\ell/2}Z_j^2\}^{1/2}
\) in restriction to $\Omega_0$.
The self-normalized test based on $\hat\theta_g^{\mathrm{tr}}$ and \eqref{eq: pairwise SN} is
\[
T_\ell^{(n),\mathrm{tr}}(g)=
\frac{(2\lambda_\perp)^{1/2}\hat\theta_g^{\mathrm{tr}}}
{\sigma_g\hat V_\ell^{(n),\mathrm{tr}}(b)}
\]
when $\hat V_\ell^{(n),\mathrm{tr}}(b)>0$.
When $\hat V_\ell^{(n),\mathrm{tr}}(b)=0$, set $T_\ell^{(n),\mathrm{tr}}(g)=+\infty$ if $\hat\theta_g^{\mathrm{tr}}>0$ and $T_\ell^{(n),\mathrm{tr}}(g)=-\infty$ if $\hat\theta_g^{\mathrm{tr}}\le0$.
We reject $H_0$ for large values of $T_\ell^{(n),\mathrm{tr}}(g)$.
Their unthresholded counterparts $\hat V_\ell^{(n)}(b)$ and $T_\ell^{(n)}(g)$ can be defined analogously.
For $a\in\mathbb R$, define the rejection indicator as
\begin{equation}
\label{eq:pairwise-rejection-rule}
\Psi_\ell^{(n),\mathrm{tr}}(a)=
\ind\{T_\ell^{(n),\mathrm{tr}}(g)>a\}.
\end{equation}
Using difference statistics to remove the influence of the mean
when constructing normalizers, 
such as variance estimators, is a popular method in 
nonparametric regression and time series analysis; 
see, e.g., \cite{hall1990} and \cite{chan2020mean,chan2022optimal}.

We consider local alternatives with individual jumps of order $M^{-1/4}$, so that the jump variation is of order $M^{-1/2}$, matching the order of the asymptotic conditional standard deviation $\{\sigma_g^2 v^{[4]}/M\}^{1/2}$ of $\hat\theta_g$ under $H_0$.
Since the jump variation alone does not determine which sampled returns contain jumps or whether thresholding removes them, we study the effects of $\ell$ and $n$ on the power of $T_\ell^{(n),\mathrm{tr}}(g)$ under the local alternative hypothesis $H_1'$, which is characterized by the following assumption:
\begin{assumption}[Local alternative $H_1'$]
\label{ass:abstract-local-alt}
On $(\Omega,\mathcal F,(\mathcal F_t)_{t\ge0},\pr)$, for each $M$, let
\[
y_M^*(t)=y^c(t)+J_M(t),\qquad
J_M(t)=M^{-1/4}\sum_{q=1}^{N_{\mathrm{loc}}}\kappa_q\ind(T_q\le t),
\]
where the continuous component $y^c(\cdot)$ satisfies
\[
y^c(t)=y^c(0)+\int_0^t\mu_s^y\dd s+\int_0^t\sigma_s\dd W_s.
\]
Here $W$ is an $(\mathcal F_t)$-Brownian motion.
The process $\mu^y$ is progressively measurable, c\`adl\`ag, and locally bounded.
The process $\sigma$ satisfies Assumption~\ref{ass:volatility-regularity}.
The integer $N_{\mathrm{loc}}\ge1$ and jump sizes $\kappa_q\neq 0$ are deterministic.
The jump locations $T_q:\Omega\to(0,1)$ are $\mathcal F$-measurable and pairwise distinct almost surely.
In addition, \eqref{eq:Theta-power}--\eqref{eq:xi-bounded} hold with $1/4<\varpi<1/2$.
\end{assumption}
For each $M$, let $y_i=y_M^*(i/M)-y_M^*((i-1)/M)$, $i=1,\ldots,M$, and put $\varphi=\sum_{q=1}^{N_{\mathrm{loc}}}\kappa_q^2$.
Then, under Assumption~\ref{ass:abstract-local-alt}, the jump variation of $y_M^*$ is $\theta=\varphi/M^{1/2}$ for each $M$, with $\varphi>0$.
We aim to test
\[
H_0:\theta=0 \qquad\text{against}\qquad
H_1':\theta=\varphi/M^{1/2}\text{ for some }\varphi>0.
\]
For an integer $\nu\ge1$, $\alpha\in(0,1)$, and $\lambda\in\mathbb R$, let $t_\nu[\lambda]$ denote the law of $(Z+\lambda)/(W_\nu/\nu)^{1/2}$, i.e., 
the noncentral $t$-distribution,
for independent $Z\sim\Normal(0,1)$ and $W_\nu\sim\chi_\nu^2$.
For any $\mathcal F$-measurable random variable $\lambda$, we use $Z_M\inST t_\nu[\lambda]$ to denote stable convergence to $(Z+\lambda)/(W_\nu/\nu)^{1/2}$, where $Z\sim\Normal(0,1)$ and $W_\nu\sim\chi_\nu^2$ are independent of each other and jointly independent of $\mathcal F$.
In addition, we write $t_\nu=t_\nu[0]$.

Let $t_{\nu,\alpha}$ be the $100(1-\alpha)\%$ quantile of the central Student-$t_\nu$ distribution, and define
\[
\beta_{\nu,\alpha}(\lambda)=
\pr\left\{
\frac{Z+\lambda}{(W_\nu/\nu)^{1/2}}>t_{\nu,\alpha}
\right\},
\qquad
\lambda_J=\frac{\varphi}{\sigma_g(v^{[4]})^{1/2}}.
\]

The asymptotic properties of $T_\ell^{(n),\mathrm{tr}}(g)$ are stated in the following theorem.
\begin{theorem}[Properties of $T_\ell^{(n),\mathrm{tr}}(g)$]
\label{thm:pairwise-properties}
Suppose Assumption~\ref{ass:g-structural} holds.
For fixed $n\ge3$, even $\ell\ge2$ and $\alpha_0\in(0,1)$, the following statements hold:

\begin{enumerate}[label=(\roman*)]
\item\label{item:pairwise-properties-1} (Validity) If Assumption~\ref{ass:price-model}, \eqref{eq:Theta-rate}, \eqref{eq:local-threshold}, and the 
uniform lower bound in \eqref{eq:xi-bounded} hold, 
then
\[
T_\ell^{(n),\mathrm{tr}}(g)
\inST t_{\ell/2},
\qquad
T_\ell^{(n)}(g)
\inST t_{\ell/2}
\]
in restriction to $\Omega_0$.
In addition,
\(
\pr\{\Psi_\ell^{(n),\mathrm{tr}}(t_{\ell/2,\alpha_0})=1\mid\Omega_0\}
\to\alpha_0
\) when $\pr(\Omega_0)>0$.

\item\label{item:pairwise-properties-2} (Consistency) If Assumption \ref{ass:price-model} and \eqref{eq:Theta-power}--\eqref{eq:xi-bounded} hold and $\pr(\Omega_1)>0$, then
\[
\pr\{\Psi_\ell^{(n),\mathrm{tr}}(t_{\ell/2,\alpha_0})=1\mid\Omega_1\}
\to1.
\]
\item\label{item:pairwise-properties-3} (Local alternative distribution) If Assumption~\ref{ass:abstract-local-alt} holds,
then 
\[
T_\ell^{(n),\mathrm{tr}}(g)
\inST 
t_{\ell/2}[\lambda_J] .
\]

\item\label{item:pairwise-properties-4} 
(Local power function) If Assumption~\ref{ass:abstract-local-alt} holds,
then the local asymptotic power function of the test $T_\ell^{(n),\mathrm{tr}}(g)$ at size $\alpha_0$ under $H_1'$ is 
\[
\lim_{M\to\infty}
\pr\{\Psi_\ell^{(n),\mathrm{tr}}(t_{\ell/2,\alpha_0})=1\}
=
\E\{\beta_{\ell/2,\alpha_0}(\lambda_J)\}.
\]
\end{enumerate}
\end{theorem}
From \ref{item:pairwise-properties-1}--\ref{item:pairwise-properties-2}, we should reject $H_0$ at size $\alpha_0$ asymptotically 
if $T_\ell^{(n),\mathrm{tr}}(g)>t_{\ell/2,\alpha_0}$.
Theorem~\ref{thm:pairwise-properties} also ensures that the order $n$ does not affect the size and power of the test asymptotically.
From \ref{item:pairwise-properties-2}, the power of the test grows to 1 under $H_1$ as $M\to\infty$.
Those properties guarantee the validity and consistency of $T_\ell^{(n),\mathrm{tr}}(g)$ as a jump detection test.
Since $\lambda_J=\varphi/\{\sigma_g(v^{[4]})^{1/2}\}$ and $\beta_{\ell/2,\alpha_0}(\lambda)$ is increasing in $\lambda$, a smaller $\sigma_g$ gives higher local power.
We postpone the discussion of the choices of $n$, $\ell$, and $g$ to Section~\ref{sec:choice-n-ell-g}.

\subsection{Optimal differencing}
\label{sec:optimal}
By Theorem~\ref{thm:pairwise-properties}, the local asymptotic power of $T_\ell^{(n),\mathrm{tr}}(g)$ is $\E\{\beta_{\ell/2,\alpha_0}(\lambda_J)\}$.
Since $\beta_{\nu,\alpha_0}(\lambda)$ is strictly increasing in $\nu$ for $\lambda>0$ and $0<\alpha_0<1/2$ \citep[Thm.~8]{ghosh1973some}, the power increases with the number $\ell$ of building blocks $Q_k^{(n)}$ involved in the self-normalizer (\ref{eq: pairwise SN}).
We will justify it in Corollaries~\ref{cor:comparison-tests} and~\ref{cor:local-power-n-ell}.
However, the pairwise scheme in (\ref{eqt:mathbb_S_ell_n_b})
can generate at most $\ell/2$ asymptotically independent differences. 
This motivates us to develop a general difference-based approach to 
use more information from 
$Q_b^{(n)},Q_{b+1}^{(n)},\ldots,Q_{b+\ell-1}^{(n)}$.

To remove the common asymptotic mean $v^{[2]}J_\ell$, 
each difference in (\ref{eqt:mathbb_S_ell_n_b}) 
has coefficients that sum to 0.
Since the subspace orthogonal to $J_\ell$ has dimension $\ell-1$, 
at most $\ell-1$ such differences can be linearly independent.
For $1\le r < \ell$, consider the self-normalized statistic
\begin{align}\label{eqt:opt_SN_test_stat_pre}
    \frac{\hat\theta_g^{\mathrm{tr}}}
{\sigma_g\{r^{-1}\|D_r\mathbb Q_\ell^{(n),\mathrm{tr}}(b)\|_2^2\}^{1/2}},    
\end{align}
for some $D_r\in\mathbb R^{r\times\ell}$ satisfying $D_rJ_\ell=0$ and $D_r\Sigma_{\mathbb Q}^{(n)}D_r^\T=I_r$.
By the joint local limit in Theorem~\ref{thm:trunc-local-power}, the test using the critical value $t_{r,\alpha_0}$ 
has local asymptotic power $\E\{\beta_{r,\alpha_0}(\lambda_J)\}$.
Since this power is increasing in $r$ for $\alpha_0\in(0,1/2)$ \citep[Thm.~8]{ghosh1973some}, it is maximized at $r=\ell-1$.
In this sense, using all $\ell-1$ normalized differences is optimal within this class.
Therefore, we set $r=\ell-1$ and let $D\in\mathbb R^{(\ell-1)\times\ell}$ satisfy
\begin{equation}
\label{eq:D-conditions}
DJ_\ell=0,\qquad\operatorname{rank}(D)=\ell-1,
\qquad D\Sigma_{\mathbb Q}^{(n)}D^\T=I_{\ell-1},
\end{equation}
which eliminates the common asymptotic mean $v^{[2]}J_\ell$, 
ensures optimality, and 
standardizes the limiting covariance, respectively.
By Theorem~\ref{thm:joint-stable-CLT-voljump} together with the 
threshold conditions, 
\[
M^{1/2}D\mathbb Q_\ell^{(n),\mathrm{tr}}(b)
\inST 
(v^{[4]})^{1/2}Z,
\qquad Z\sim\Normal(0,I_{\ell-1}),
\]
in restriction to $\Omega_0$, where $Z$ is independent of $\mathcal F$.
This motivates using $D\mathbb Q_\ell^{(n),\mathrm{tr}}(b)$ to construct a self-normalizer.

\begin{theorem}[Optimal differencing]
\label{lem:contrast-space-whitening}
For fixed $n\ge3$ and $\ell\ge2$, the covariance matrix has the form
\(
\Sigma_{\mathbb Q}^{(n)}=\lambda_\perp I_\ell
+\Sigma_{\mathbb Q}^{(n)}(1,2)J_\ell J_\ell^\T.
\)
An explicit choice of $D$ satisfying \eqref{eq:D-conditions} is given by
\begin{equation}
\label{eq:Helmert-contrast}
(D_{\ell,*}^{(n)})_{rj}=\lambda_\perp^{-1/2}
\begin{cases}
\{r(r+1)\}^{-1/2},&j\le r,\\
-r\{r(r+1)\}^{-1/2},&j=r+1,\\
0,&j>r+1,
\end{cases}
\quad r=1,\ldots,\ell-1.
\end{equation}
For every $D$ satisfying \eqref{eq:D-conditions}, we have 
$
D^\T D=\lambda_\perp^{-1}
\left(I_\ell-\ell^{-1}J_\ell J_\ell^\T\right) 
$
and, for each $q\in\mathbb R^\ell$,
\begin{equation}
\label{eq:contrast-norm-invariance}
\|Dq\|_2^2=\lambda_\perp^{-1}\sum_{j=1}^\ell
\left(q_j-\frac1\ell\sum_{m=1}^\ell q_m\right)^2.
\end{equation}
The right-hand side of \eqref{eq:contrast-norm-invariance} does not depend on the particular $D$ satisfying \eqref{eq:D-conditions}.
Any two matrices satisfying \eqref{eq:D-conditions} differ by left multiplication by an orthogonal matrix.
\end{theorem}

Theorem~\ref{lem:contrast-space-whitening} guarantees the existence and provides the explicit form of $D_{\ell,*}^{(n)}$.
In addition, \eqref{eq:contrast-norm-invariance} implies that using $h$ as in \eqref{eq: h L2} makes the self-normalizer invariant to the choice of $D$.
It thus explains why orthogonal invariance is an important criterion for deciding $h$.
Now, 
according to (\ref{eqt:opt_SN_test_stat_pre}), 
the thresholded self-normalizer with any matrix $D$ satisfying \eqref{eq:D-conditions} is 
\begin{align}
\hat V_{\ell,*}^{(n),\mathrm{tr}}(b)
&:=\{(\ell-1)^{-1}
\|D\mathbb Q_\ell^{(n),\mathrm{tr}}(b)\|_2^2\}^{1/2}  \nonumber \\
&=\lambda_\perp^{-1/2}
\left\{\frac{1}{\ell-1}
\sum_{k=b}^{b+\ell-1}
(Q_k^{(n),\mathrm{tr}}-\bar Q)^2\right\}^{1/2},
\quad \text{where}\quad 
\bar Q=\frac1\ell\sum_{k=b}^{b+\ell-1}Q_k^{(n),\mathrm{tr}}. \label{eq:VD-tr}
\end{align}
The second representation of $\hat V_{\ell,*}^{(n),\mathrm{tr}}(b)$ 
in (\ref{eq:VD-tr}) follows from \eqref{eq:contrast-norm-invariance}.
So, $\hat V_{\ell,*}^{(n),\mathrm{tr}}(b)$ 
is simply the sample standard deviation of $Q_b^{(n),\mathrm{tr}},\ldots,Q_{b+\ell-1}^{(n),\mathrm{tr}}$ divided by $\lambda_\perp^{1/2}$.
We then define our main proposed test statistic
\begin{equation}
\label{eq:T-tr}
T_{\ell,*}^{(n),\mathrm{tr}}(g)
=
\frac{\hat\theta_g^{\mathrm{tr}}}
{\sigma_g\hat V_{\ell,*}^{(n),\mathrm{tr}}(b)}, 
\end{equation}
for $\hat V_{\ell,*}^{(n),\mathrm{tr}}(b)>0$. Otherwise, 
set $T_{\ell,*}^{(n),\mathrm{tr}}(g)=+\infty$ if $\hat\theta_g^{\mathrm{tr}}>0$ and $T_{\ell,*}^{(n),\mathrm{tr}}(g)=-\infty$ if $\hat\theta_g^{\mathrm{tr}}\le0$.
For each fixed $a$, define the rejection rule and its $p$-value as
\begin{equation}
\label{eq:cross-multiplied-rejection}
\Psi_{\ell,*}^{(n),\mathrm{tr}}(a)
=\ind\{T_{\ell,*}^{(n),\mathrm{tr}}(g)>a\},
\qquad
p_{\ell,*}^{(n),\mathrm{tr}}(g)=
1-F_{t_{\ell-1}}\bigl(T_{\ell,*}^{(n),\mathrm{tr}}(g)\bigr),
\end{equation}
where $F_{t_{\ell-1}}$ denotes the distribution function of the central Student-$t$ distribution with $\ell-1$ degrees of freedom.
The asymptotic properties of $T_{\ell,*}^{(n),\mathrm{tr}}(g)$ are presented as follows:

\begin{theorem}[Properties of $T_{\ell,*}^{(n),\mathrm{tr}}(g)$]
\label{thm:canonical-properties}
Suppose Assumption~\ref{ass:g-structural} holds.
For fixed $n\ge3$, $\ell\ge2$ and $\alpha_0\in(0,1)$, the following statements hold:
\begin{enumerate}[label=(\roman*)]
\item (Validity) Under Assumption~\ref{ass:price-model}, if \eqref{eq:Theta-rate}, \eqref{eq:local-threshold}, and the uniform lower bound in \eqref{eq:xi-bounded} hold, then
\[
T_{\ell,*}^{(n),\mathrm{tr}}(g)
\inST t_{\ell-1}
\]
in restriction to $\Omega_0$.
 In addition,
\(
\pr\{\Psi_{\ell,*}^{(n),\mathrm{tr}}(t_{\ell-1,\alpha_0})=1\mid\Omega_0\}
\to\alpha_0
\) if $\pr(\Omega_0)>0$.

\item (Consistency) Under Assumption~\ref{ass:price-model} and \eqref{eq:Theta-power}--\eqref{eq:xi-bounded}, if $\pr(\Omega_1)>0$, then
\[
\pr\{\Psi_{\ell,*}^{(n),\mathrm{tr}}(t_{\ell-1,\alpha_0})=1\mid\Omega_1\}
\to1.
\]

\item (Local alternative distribution)
Under Assumption~\ref{ass:abstract-local-alt},
\[
T_{\ell,*}^{(n),\mathrm{tr}}(g)
\inST 
t_{\ell-1}[\lambda_J].
\]

\item (Local power function) 
If Assumption~\ref{ass:abstract-local-alt} holds, then the local asymptotic power function of the test $T_{\ell,*}^{(n),\mathrm{tr}}(g)$ at size $\alpha_0$ under $H_1'$ is 
\[
\lim_{M\to\infty}
\pr\{\Psi_{\ell,*}^{(n),\mathrm{tr}}
(t_{\ell-1,\alpha_0})=1\}
=
\E\{\beta_{\ell-1,\alpha_0}(\lambda_J)\}.
\]
\end{enumerate}
\end{theorem}
From (i)--(ii), we reject $H_0$ at size $\alpha_0$ asymptotically if $T_{\ell,*}^{(n),\mathrm{tr}}(g)>t_{\ell-1,\alpha_0}$.
The local power of the test $T_{\ell,*}^{(n),\mathrm{tr}}(g)$ 
also increases with $\ell$, decreases with $\sigma_g^2$, but does not depend on $n$. 
The following corollary compares the power of $\{T_{\ell}^{(n),\mathrm{tr}}(g)\}_{\ell\in 2\mathbb N,n\ge3}$ and $\{T_{\ell,*}^{(n),\mathrm{tr}}(g)\}_{\ell\ge2,n\ge3}$.
\begin{corollary}[Comparison of tests]
\label{cor:comparison-tests}
Suppose Assumption~\ref{ass:g-structural} holds.
Let $n\ge3$ and $\ell\ge1$.
\begin{enumerate}
\item Under Assumption~\ref{ass:price-model}, if \eqref{eq:Theta-rate}, \eqref{eq:local-threshold}, and the uniform lower bound in \eqref{eq:xi-bounded} hold, then, in restriction to $\Omega_0$,
$T_{2\ell}^{(n),\mathrm{tr}}(g)
\inST t_\ell$
and 
$T_{\ell+1,*}^{(n),\mathrm{tr}}(g)
\inST t_\ell$.

\item Under Assumption~\ref{ass:abstract-local-alt}, for every $\alpha_0\in(0,1)$,
\[
\lim_{M\to\infty}
\pr\{\Psi_{2\ell}^{(n),\mathrm{tr}}(t_{\ell,\alpha_0})=1\}
=
\lim_{M\to\infty}
\pr\{\Psi_{\ell+1,*}^{(n),\mathrm{tr}}(t_{\ell,\alpha_0})=1\}
=
\E\{\beta_{\ell,\alpha_0}(\lambda_J)\}.
\]
\end{enumerate}
\end{corollary}
Corollary~\ref{cor:comparison-tests} implies that $T_{2\ell}^{(n),\mathrm{tr}}(g)$ and $T_{\ell+1,*}^{(n),\mathrm{tr}}(g)$ have the same local asymptotic power.
However, 
$T_{2\ell}^{(n),\mathrm{tr}}(g)$ needs $2\ell$ lag-augmented multipower variations 
to achieve the same local asymptotic power as $T_{\ell+1,*}^{(n),\mathrm{tr}}(g)$, which needs only $\ell+1$.
When $M$ is not large, this difference is important because increasing 
$\ell$ 
reduces the number of available summands.

\subsection{Choice of \texorpdfstring{$n$, $\ell$, and $g$}{n, ell, and g}}
\label{sec:choice-n-ell-g}
Notice that
$\sigma_{g_1}^2\approx0.609<\sigma_{g_3}^2\approx0.959<
\sigma_{g_2}^2\approx1.810$.
In addition, recall that $\lambda_J=\varphi/\{\sigma_g(v^{[4]})^{1/2}\}$ and $\beta_{\ell-1,\alpha_0}(\lambda)$ is increasing in $\lambda$.
Hence, for fixed $(\ell,\alpha_0)$, a smaller $\sigma_g$ yields higher first-order local asymptotic power.
So, $g_1$ is recommended among these three functions.
The following corollary describes the effect of $n$ and $\ell$ on the local power function.

\begin{corollary}[Effects of $n$ and $\ell$]
\label{cor:local-power-n-ell}
Under Assumptions~\ref{ass:g-structural} and~\ref{ass:abstract-local-alt}, for fixed $b$, $g$, and $\alpha_0\in(0,1/2)$, we have 
$
\pr\{\Psi_{\ell,*}^{(n),\mathrm{tr}}(t_{\ell-1,\alpha_0})=1\}
\to
\E\{\beta_{\ell-1,\alpha_0}(\lambda_J)\},
$
where neither $\lambda_J$ nor $\beta_{\ell-1,\alpha_0}(\lambda)$ depends on $n\ge3$, and $\beta_{\ell,\alpha_0}(\lambda)>
\beta_{\ell-1,\alpha_0}(\lambda)$ for every $\lambda>0$ and $\ell\ge2$.
\end{corollary}

First, we discuss $n$. 
Corollary~\ref{cor:local-power-n-ell} shows that first-order local asymptotic power is independent of fixed $n\ge3$.
By \eqref{eq:Qk-tr}, each $Q_k^{(n),\mathrm{tr}}$ contains $M-K+1$ summands, where $K=(n-1)(b+\ell+1)$.
For fixed $M$, $b$, and $\ell$, increasing $n$ reduces the number of available summands.
Because it does so without changing first-order local asymptotic power, we recommend $n=3$, the smallest value covered by our theory, as a rule of thumb.

Second, the choice of $\ell$ is different.
Corollary~\ref{cor:local-power-n-ell} implies that local asymptotic power increases with $\ell$, so power alone favors a larger $\ell$.
A larger $\ell$ also increases $K$ and, for finite $M$, reduces the number of available summands.
We therefore assess how close the finite-$\ell$ power function is to its limit as the degrees of freedom $\ell-1$ tend to infinity.
Fix $\alpha_0\in(0,1/2)$.
The limiting power function is defined as
\[
\beta_{\infty,\alpha_0}(\lambda)
:=\lim_{\nu\to\infty}\beta_{\nu,\alpha_0}(\lambda)
=\Phi(\lambda-z_{\alpha_0}).
\]
For any $g$ satisfying Assumption~\ref{ass:g-structural}, define
\(
T_0^{\mathrm{tr}}(g):=
M^{1/2}\hat\theta_g^{\mathrm{tr}}/(\hat v^{[4]})^{1/2}.
\)
Under $H_1'$, Theorem~\ref{thm:trunc-conventional-local-power} gives
\(
T_0^{\mathrm{tr}}(g)/\sigma_g
\inST Z_0+\lambda_J.
\)
Thus, the upper-tail test with critical value $\sigma_g z_{\alpha_0}$ has conditional limiting power $\beta_{\infty,\alpha_0}(\lambda_J)$.
This motivates us to propose the following criterion for choosing $\ell$ in practice.
For $p\in(0,1)$, define
\begin{align}
R_{\nu,\alpha_0}&=
\inf_{\lambda\ge0}\frac{\beta_{\nu,\alpha_0}(\lambda)}
{\beta_{\infty,\alpha_0}(\lambda)},\qquad
\ell^*(\alpha_0,p)=
\min\{\ell\ge2:R_{\ell-1,\alpha_0}\ge p\}.
\label{eq:ell-wc-definition}
\end{align}
Choosing $\ell=\ell^*(\alpha_0,p)$ retains at least $100p\%$ of the local asymptotic  power of the corresponding thresholded non-self-normalized test $T_0^{\mathrm{tr}}(g)$ under $H_1'$.

\begin{corollary}[Properties of $\ell^*$]
\label{cor:local-power-retention}
Fix $\alpha_0\in(0,1/2)$.
For any $\nu\in\mathbb{Z}^+$, 
we have 
(i) $\beta_{\nu,\alpha_0}(\lambda)<\beta_{\infty,\alpha_0}(\lambda)$ for all $\lambda>0$, 
(ii) the infimum defining $R_{\nu,\alpha_0}$ is attained at some $\lambda\in(0,\infty)$, (iii)
\[
R_{\nu,\alpha_0}<R_{\nu+1,\alpha_0}<1,
\qquad
R_{\nu,\alpha_0}\to1
\quad\text{as }\nu\to\infty,
\]
and (iv) $\ell^*(\alpha_0,p)$ is finite for every $p\in(0,1)$ and nondecreasing in $p$.
\end{corollary}
By Corollary \ref{cor:local-power-retention}, choosing $\ell=\ell^*(\alpha_0,p)$ ensures that
$\beta_{\ell-1,\alpha_0}(\lambda)
\ge p\,\beta_{\infty,\alpha_0}(\lambda)$ for every $\lambda\ge0$.
Moreover, since $R_{\nu,\alpha_0}$ depends only on $(\nu,\alpha_0)$, the choice of $\ell^*$ does not require knowledge of the jump magnitudes or volatility path.
Table~\ref{tab:local-power-retention} reports $R_{\ell-1,\alpha_0}$ at the displayed significance levels and can be used to choose $\ell^*(\alpha_0,p)$.
For example, at $\alpha_0=0.1$ and $p=0.9$, $R_{4,0.1}\approx.892$ and $R_{5,0.1}\approx.914$, so the criterion gives $\ell^*(\alpha_0,p)=6$.

\begin{table}[!ht]
\centering
\setlength{\tabcolsep}{3pt}\renewcommand{\arraystretch}{1.05}
\begin{tabular}{@{}c*{14}{c}@{}}\hline
$\alpha_0\backslash\ell$&2&3&4&5&6&7&8&9&10&11&12&13&14&15\\\hline
.005&.053&.143&.251&.349&.429&.495&.548&.592&.628&.659&.685&.707&.727&.744\\
.01&.099&.235&.367&.470&.548&.607&.653&.690&.720&.745&.765&.783&.798&.812\\
.025&.218&.419&.559&.649&.710&.754&.786&.811&.831&.847&.860&.871&.881&.889\\
.05&.378&.600&.716&.781&.823&.851&.872&.888&.900&.910&.918&.925&.931&.935\\
.10&.609&.786&.856&.892&.914&.928&.939&.947&.953&.957&.961&.965&.967&.970\\\hline
\end{tabular}
\caption{Values of $R_{\ell-1,\alpha_0}$, rounded to three decimal places.} 
\label{tab:local-power-retention}
\end{table}

Although $\ell^*(\alpha_0,p)<\infty$ for all fixed choices of $(\alpha_0,p)$, a large $\ell$ reduces the number of available summands at finite $M$.
So, in practice, we recommend choosing $\ell$ as
\begin{equation}
\label{eq:ell-M-definition}
\ell_M^*(\alpha_0,p)=\min\left\{\ell^*(\alpha_0,p),\,\max\{2,\lfloor M^{1/3}\rfloor\}\right\}.
\end{equation}
As $\ell^*(\alpha_0,p)<\infty$, we have $\ell_M^*(\alpha_0,p)=\ell^*(\alpha_0,p)$ for all sufficiently large $M$ and hence we can still retain at least $100p\%$ of the limiting local power as $M\to\infty$.
When the cap is active and $M\ge8$, 
we have $\ell_M^*(\alpha_0, p) = \lfloor M^{1/3}\rfloor$ so as to ensure a significant proportion of summands is available for the evaluation of the self-normalizer.
Such a choice of $\ell_M^*(\alpha_0, p)$ is examined via simulation; 
see Section~\ref{sec:supp-ell-M} of the supplement.
Algorithm~\ref{algo:OJT_computation} summarizes the implementation of the proposed self-normalized jump test.

\begin{algorithm}[t]
\caption{Self-normalized jump test for $H_0$ against $H_1$.}
\label{algo:OJT_computation}

\textbf{Input}:\\
(i) $y_1,\ldots,y_M$ ($M\geq K$) -- the series of high-frequency returns.\\
(ii) $\alpha_0\in(0,1/2)$ -- the nominal type-I error rate, e.g., $\alpha_0=5\%$.\\
(iii) $g:\R^{b+1}\to\R$ -- the function satisfying Assumption~\ref{ass:g-structural}. Default: $g=g_1$.\\
(iv) $n\ge3$ -- the order of lag-augmented
$n$-power variation in \eqref{eq:Qk-unthresholded}.
Default: $n=3$.\\
(v) $\ell\ge2$ -- the number of statistics used by the self-normalizer
in \eqref{eq:VD-tr}. 
Default:
$\ell=\ell_M^*(\alpha_0, 90\%)$ from \eqref{eq:ell-M-definition}, or  
set $\ell=6$ as a less recommended quick choice.  
\\
(vi) $\vartheta_{i,M}$, $i=1,\ldots,M$ -- the squared-return thresholds satisfying \eqref{eq:Theta-rate}--\eqref{eq:xi-bounded}.\\

\Begin{
(1) Compute
$\lambda_\perp
=\Sigma_{\mathbb Q}^{(n)}(1,1)
-\Sigma_{\mathbb Q}^{(n)}(1,2)$
from \eqref{eq:SigmaQ-final}.\\

(2) Compute
$\mathbb Q_\ell^{(n),\mathrm{tr}}(b)$
from \eqref{eq:Qk-tr} and
$\hat V_{\ell,*}^{(n),\mathrm{tr}}(b)$
from \eqref{eq:VD-tr}.\\

(3) Compute
$\hat\theta_g^{\mathrm{tr}}$
from \eqref{eq:ivg-tr} and
$\sigma_g$ from \eqref{eq:sigma-g-overlap}.\\

(4) Compute
$T_{\ell,*}^{(n),\mathrm{tr}}(g)$
from \eqref{eq:T-tr}. \\

(5) Compute
$\Psi=\Psi_{\ell,*}^{(n),\mathrm{tr}}
(t_{\ell-1,\alpha_0})$
from \eqref{eq:cross-multiplied-rejection}.\\
}

\textbf{return}
$\Psi$ -- the test result. If $\Psi=1$, then $H_0$ is rejected.
\end{algorithm}

\section{Extension under microstructure noise}
\label{sec:microstructure-noise}

At relatively high sampling frequencies, transaction prices can deviate from the latent efficient price because of bid--ask bounce and price rounding; see \citet{JiangOomen2008} and \citet{KalninaLinton2008}.
For related asymptotic results on realized volatility and bipower variation under vanishing microstructure noise, see \citet{Kurisu2018}.
For each fixed $M$, let $(\varepsilon_{i,M})_{i=0}^M$ be random variables on an extension of $(\Omega,\mathcal F,(\mathcal F_t)_{t\ge0},\pr)$, and let $\omega_M\ge0$ be deterministic.
Consider
\[
\begin{aligned}
Y_{i/M}^{\mathrm{obs}}
&=y^*(i/M)+\varepsilon_{i,M},
\quad i=0,\ldots,M,\\
\widetilde y_i
&=Y_{i/M}^{\mathrm{obs}}-Y_{(i-1)/M}^{\mathrm{obs}}
=y_i+\varepsilon_{i,M}-\varepsilon_{i-1,M},
\quad i=1,\ldots,M.
\end{aligned}
\]
Throughout this section, we evaluate the existing statistics with $y_i$ replaced by $\widetilde y_i$ and retain their original notation.
The law of the noise is specified by the following assumption for validity and in Theorem~\ref{thm:microstructure-noise}(ii) for consistency.

\begin{assumption}[Microstructure noise]
\label{ass:microstructure-noise-null}
Suppose Assumption~\ref{ass:price-model} holds.
In addition, $\sigma$ is continuous and $\sigma_t>0$ for every $t\in[0,1]$ almost surely.
Let $(\mathcal G_t)_{t\ge0}$ be a sub-filtration of $(\mathcal F_t)_{t\ge0}$ that is right-continuous and contains all $\pr$-null sets, with $W$ independent of $\mathcal G_1$ and $\sigma_0$ measurable with respect to $\mathcal G_0$.
In \eqref{eq:volatility-decomposition}, $\widetilde\sigma\equiv0$, $L^\sigma$ is a continuous $(\mathcal G_t)$-local martingale, and the processes $\widetilde\mu^\sigma$ and $a$ are $(\mathcal G_t)$-progressively measurable.
For each $M$, $\rho_M\in[-1,1]$ is deterministic, and 
$G_{0,M},\ldots,G_{M,M}\simIID \Normal(0,1)$ defined on an extension of $(\Omega,\mathcal F,(\mathcal F_t)_{t\ge0},\pr)$, independent of $\mathcal F$.
The noise satisfies $\varepsilon_{0,M}=\omega_MG_{0,M}$ and
\[
\varepsilon_{i,M}=\omega_M\{\rho_MZ_{i,M}+(1-\rho_M^2)^{1/2}G_{i,M}\},
\qquad
Z_{i,M}=\frac{\int_{(i-1)/M}^{i/M}\sigma_s\dd W_s}
{\{\int_{(i-1)/M}^{i/M}\sigma_s^2\dd s\}^{1/2}},
\qquad i=1,\ldots,M.
\]
\end{assumption}

Assumption~\ref{ass:microstructure-noise-null} requires the volatility process to be continuous and independent of \(W\), and hence excludes the leverage effect and volatility jumps.
The sub-filtration $(\mathcal G_t)$ ensures that, conditional on $\mathcal G_1$, the diffusion increments over disjoint intervals are independent centered Gaussian variables.
Consequently, $Z_{1,M},\ldots,Z_{M,M}$ are independent standard normal variables, independent of $\mathcal G_1$; see Lemma~\ref{lem:noise-conditional-law}(i) of the supplement.
Hence $\varepsilon_{0,M},\ldots,\varepsilon_{M,M}$ are i.i.d.\
$\Normal(0,\omega_M^2)$, and, if $\omega_M>0$, $\operatorname{Corr}(\varepsilon_{i,M},Z_{i,M}\mid\mathcal G_1)=\rho_M$ for $i=1,\ldots,M$.
When $\rho_M\omega_M\ne0$, the noise depends on the latent price through $Z_{i,M}$.

\begin{theorem}[Microstructure noise]
\label{thm:microstructure-noise}
Take $g=g_1$ and $b=1$.
Fix $n\ge3$ and $\ell\ge2$, and suppose $M^{5/4}\omega_M^2\to c\in[0,\infty)$ as $M\to\infty$ and \eqref{eq:Theta-power}--\eqref{eq:xi-bounded} hold.
\begin{enumerate}[label=(\roman*)]
\item (Validity) Under Assumption~\ref{ass:microstructure-noise-null}, suppose $M^{3/4}\rho_M\omega_M\to\iota\in\mathbb R$ as $M\to\infty$.
Then, in restriction to $\Omega_0$,
\[
T_{\ell,*}^{(n),\mathrm{tr}}(g_1)
\inST 
t_{\ell-1}[\lambda_{c,\iota}], 
\quad\text{where} \quad 
\lambda_{c,\iota}
=-\frac{\int_0^1(c/\sigma_t+\iota)^2\dd t}
{2\sigma_{g_1}(v^{[4]})^{1/2}}.
\]
For every $\alpha_0\in(0,1)$, if $\pr(\Omega_0)>0$, then
\[
\pr\{\Psi_{\ell,*}^{(n),\mathrm{tr}}(t_{\ell-1,\alpha_0})=1\mid\Omega_0\}
\longrightarrow\E\{\beta_{\ell-1,\alpha_0}(\lambda_{c,\iota})\mid\Omega_0\}
\le\alpha_0
\qquad\text{as }M\to\infty.
\]
\item (Consistency) Under Assumption~\ref{ass:price-model}, suppose $\varepsilon_{i,M}\sim\Normal(0,\omega_M^2)$ for $i=0,\ldots,M$.
Then, if $\pr(\Omega_1)>0$, for every $\alpha_0\in(0,1)$,
\(
\pr\{\Psi_{\ell,*}^{(n),\mathrm{tr}}(t_{\ell-1,\alpha_0})=1\mid\Omega_1\}
\to1.
\)
\end{enumerate}
\end{theorem}
From Theorem \ref{thm:microstructure-noise}, in restriction to $\Omega_0$, the proposed test statistic has conditional limiting law $t_{\ell-1}[\lambda_{c,\iota}]$ given $\mathcal{F}$.
The corresponding test is asymptotically level-$\alpha_0$ since $\lambda_{c,\iota}\le 0$, and it is asymptotically exact when $c=\iota=0$.

\section{Simulation studies}
\label{sec:simulation-studies}

In this section, we examine the finite-sample size accuracy of $T_{\ell,*}^{(n),\mathrm{tr}}(g_1)$ under different volatility models.
We also assess its empirical power through distinct settings of alternatives, and compare its size-adjusted power with existing tests.
In particular, its size robustness to microstructure noise is studied in Section~\ref{subsec:simulation-microstructure-noise}.

\subsection{Simulation setup}
Under $H_0$, the price process is generated through
$y^*(t)=y^*(0)+\int_0^t\sigma_s\dd W_s$
for $t\in[0,1]$.
For the stochastic-volatility scenarios, the volatility follows a log-Ornstein--Uhlenbeck (log-OU) model:
\begin{equation}
\label{eq:simulation-log-volatility}
\dd(\log\sigma_t)
=
5\{\log(0.20)-\log\sigma_{t-}\}\dd t
+0.30\,\dd B_t +\dd J_t^\sigma,
\qquad \sigma_0=0.20.
\end{equation}
Here $B$ is a standard Brownian motion.
Let $J_t^\sigma=\sum_{j=1}^{N_t^\sigma}\zeta_j^\sigma$, where $N^\sigma$ is a Poisson process with intensity 2 and $(\zeta_j^\sigma)_{j\ge1}$ are independent and identically distributed with $\pr(\zeta_j^\sigma=0.20)=\pr(\zeta_j^\sigma=-0.20)=1/2$.
The processes $W,B,N^\sigma$ and $(\zeta_j^\sigma)_{j\ge1}$ are mutually independent.
We assess the empirical size of our proposal through three scenarios: (i) constant volatility $\sigma_t\equiv0.20$, (ii) \eqref{eq:simulation-log-volatility} with $J^\sigma\equiv0$, and (iii) \eqref{eq:simulation-log-volatility} with $J^\sigma$.

Let $M=288$, $\Delta_{\mathrm{fine}}=(10M)^{-1}$, and $t_q=q\Delta_{\mathrm{fine}}$, $q=0,\ldots,10M$.
Let $(U_q)_{q=0}^{10M-1}$ and $(Z_q)_{q=0}^{10M-1}$ be two independent i.i.d. $\Normal(0,1)$ sequences, jointly independent of $N^\sigma$ and $(\zeta_j^\sigma)_{j\ge1}$.
We approximate the stochastic differential equation \eqref{eq:simulation-log-volatility} and the price process on $(t_q)$ through the Euler--Maruyama scheme:
\[
\begin{aligned}
\log\sigma_{t_{q+1}}
&=
\log\sigma_{t_q}
+5\{\log(0.20)-\log\sigma_{t_q}\}\Delta_{\mathrm{fine}}
+0.30\sqrt{\Delta_{\mathrm{fine}}}U_q
+(J_{t_{q+1}}^\sigma-J_{t_q}^\sigma),\\
y^*(t_{q+1})
&=y^*(t_q)+
\sigma_{t_q}\sqrt{\Delta_{\mathrm{fine}}}Z_q,
\end{aligned}
\]
for $q=0,\ldots,10M-1$.
For our proposed test $T_{\ell,*}^{(n),\mathrm{tr}}(g_1)$, take $n=3$ and, unless otherwise specified, $\ell=\ell_M^*(\alpha_0,0.9)$ from \eqref{eq:ell-M-definition}.
\citet[Sects.~5.3 and~6]{AitSahaliaJacod2009} truncate absolute returns at a level of the form $c_\vartheta M^{-\varpi}$ and recommend $\varpi$ close to $1/2$, such as $\varpi=0.47$, with $c_\vartheta$ between three and five times an average value of the volatility.
We take $\varpi=0.47$ and $c_\vartheta=5\sigma_{\mathrm{ref}}$, where $\sigma_{\mathrm{ref}}$ is a reference volatility level chosen below.

We adapt this choice by estimating the volatility level from the returns entering the test and clipping it to deterministic bounds, which gives the truncation threshold
\begin{equation}
\label{eq:simulation-threshold-choice}
u_M=5\bar\sigma_M M^{-0.47},
\qquad
\bar\sigma_M=\min\{10\sigma_{\mathrm{ref}},\max(\sigma_{\mathrm{ref}}/10,\hat\sigma_M)\},
\qquad
\hat\sigma_M=(\hat{\IV}_{g_1})^{1/2},
\end{equation}
where $\hat{\IV}_{g_1}$ is the unthresholded realized bipower variation and $\sigma_{\mathrm{ref}}=0.20$.
Because $\hat\sigma_M$ scales with the volatility of the returns, $u_M$ adapts to the volatility level of each price path.
The squared-return threshold $\vartheta_{i,M}=u_M^2=\xi_{i,M}\Theta(M^{-1})$ satisfies \eqref{eq:Theta-rate}--\eqref{eq:xi-bounded} with $c_\vartheta=5\sigma_{\mathrm{ref}}$, $\varpi=0.47$, 
$\Theta(M^{-1})=(5\sigma_{\mathrm{ref}})^2M^{-0.94}$,
and 
$\xi_{i,M}=(\bar\sigma_M / \sigma_{\mathrm{ref}})^2\in[0.01,100]$.
For $M=288$ and all values of $\alpha_0$ we considered in the following experiments, $\ell_M^*(\alpha_0,0.9)=6$, so the statistic does not change as $\alpha_0$ varies.
Implementation details for all competing procedures are given in Section~\ref{sec:supp-simulation-implementations} of the supplement.

\subsection{Finite-sample size and power}
\begin{figure}[!tbp]
\centering
\includegraphics[width=\textwidth]{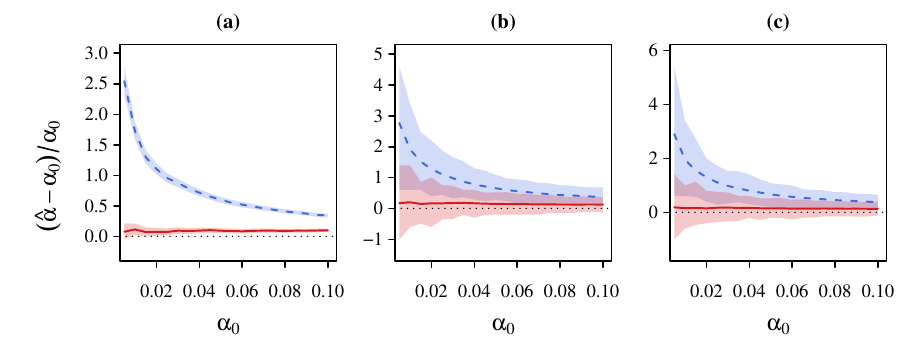}\vspace{-0.4cm}
\caption{Relative size error for the proposed test and $T_0^{\mathrm{tr}}(g_1)$.
Panels (a)--(c) use constant volatility, continuous log-OU volatility, and log-OU volatility with compound-Poisson jumps, respectively.
Red solid and blue dashed curves denote the proposed test and $T_0^{\mathrm{tr}}(g_1)$.
Shading gives an exact 95\% interval %
in panel (a) and the 10th to 90th percentiles across volatility paths in panels (b)--(c).
The curves in panels (b)--(c) are the corresponding means.}
\label{fig:sim-clean-size-standard}
\end{figure}

First, we compare the empirical size accuracy of $T_{\ell,*}^{(n),\mathrm{tr}}(g_1)$ with $T_0^{\mathrm{tr}}(g_1)$, the thresholded counterpart of the conventional test $T_0(g_1)$ \citep[Sect.~2.2, eqn.~(10)]{barndorff2006econometrics}.
On $\Omega_0$, $T_0^{\mathrm{tr}}(g_1)$ and $T_0(g_1)$ agree almost surely for all sufficiently large $M$.
For scenario (i) $\sigma_t\equiv0.20$, we generate 50,000 independent price paths.
For each of scenarios (ii) and (iii), we generate 200 independent volatility paths and 250 conditionally independent price paths per volatility path, again giving 50,000 paths.
For each price-jump alternative and each value of the jump-amplitude parameter $a$, we generate 200 independent volatility paths and 100 conditionally independent price paths per volatility path, giving 20,000 paths.
For every size or power comparison, all statistics are evaluated on the same simulated returns.
We consider nominal sizes $\alpha_0=0.005,\allowbreak 0.010,\allowbreak\ldots,\allowbreak 0.100$.
Writing $\widehat\alpha$ for the empirical rejection frequency at nominal size $\alpha_0$, we report the relative size error $(\widehat\alpha-\alpha_0)/\alpha_0$.
Figure~\ref{fig:sim-clean-size-standard} shows that the relative size error of $T_{\ell,*}^{(n),\mathrm{tr}}(g_1)$ is closer to zero than that of $T_0^{\mathrm{tr}}(g_1)$ in all three scenarios.
Similar comparisons for $g_2$ and $g_3$ are reported in Figures~\ref{fig:supp-size-g2} and~\ref{fig:supp-size-g3} of the supplement.

Then, we restrict to scenario (iii) and compare our proposal with the competing tests $T_{\mathrm{AJ}}$ of \citet{AitSahaliaJacod2009}, $T_{\mathrm{JO}}$ of \citet{JiangOomen2008}, $T_{\mathrm{CPR}}$ of \citet{CorsiPirinoReno2010}, and $T_{\mathrm{LM}}$ of \citet{LeeMykland2008}.
\begin{figure}[!tbp]
\centering
\includegraphics[width=\textwidth]{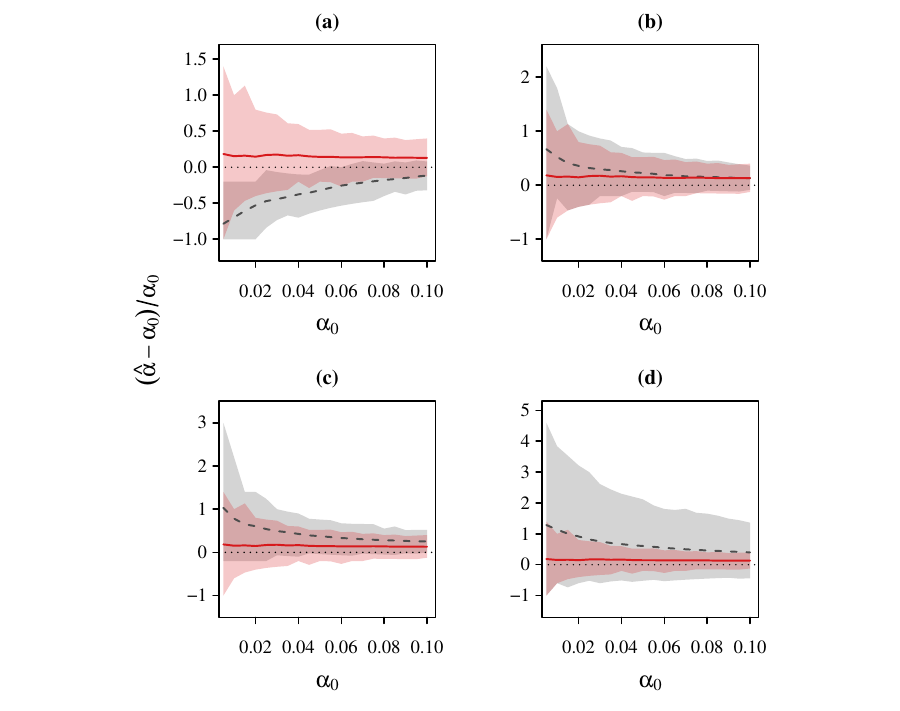}\vspace{-0.4cm}
\caption{Relative size error under log-OU volatility with compound-Poisson jumps.
Panels (a)--(d) compare the proposed test with $T_{\mathrm{AJ}}$, $T_{\mathrm{JO}}$, $T_{\mathrm{CPR}}$, and $T_{\mathrm{LM}}$, respectively.
Red solid curves denote the proposed test and gray dashed curves the competing test.
Shading gives the 10th to 90th percentiles across volatility paths.
The curves are the corresponding means.}
\label{fig:sim-clean-size-competitors}
\end{figure}
Figure~\ref{fig:sim-clean-size-competitors} shows that the mean relative size error of $T_{\ell,*}^{(n),\mathrm{tr}}(g_1)$ is generally closer to zero than that of each competing test over the nominal sizes considered.
For example, when $\alpha_0=0.01$, the proposed test has the smallest absolute size distortion.
Its type-I error is 1.15\%, compared with 0.30\%, 1.52\%, 1.78\%, and 2.14\% for $T_{\mathrm{AJ}}$, $T_{\mathrm{JO}}$, $T_{\mathrm{CPR}}$, and $T_{\mathrm{LM}}$, respectively.

\begin{figure}[!tbp]
\centering
\includegraphics[width=\textwidth]{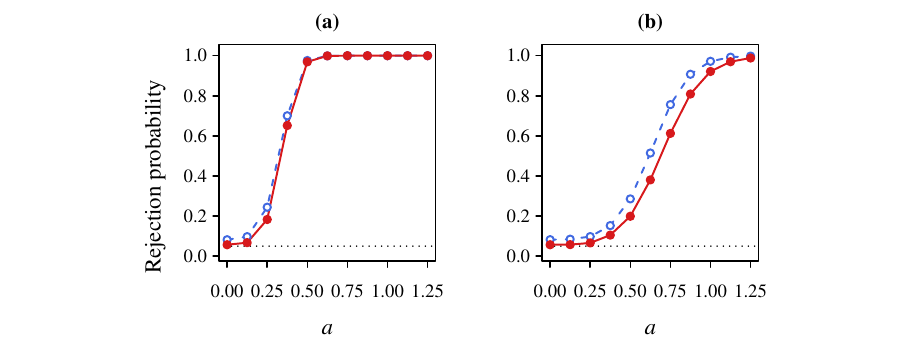}
\vspace{-1.2cm}
\caption{Raw rejection probabilities at the nominal size $\alpha_0=0.05$.
Here the proposed self-normalized test statistic is $T_{6,*}^{(3),\mathrm{tr}}(g_1)$~(\SNProposedCaptionKey), and the conventional thresholded test statistic is $T_0^{\mathrm{tr}}(g_1)$~(\SNConventionalCaptionKey).
Panels (a) and (b) use the rare/large and frequent/small price-jump alternatives, respectively.
The dotted horizontal line marks 0.05.
Both tests use the same simulated paths and threshold~\eqref{eq:simulation-threshold-choice}.}
\label{fig:sim-clean-power}
\label{fig:sim-power-proposed-vs-T0-5pct}
\end{figure}

\begin{figure}[!tbp]
\centering
\includegraphics[width=\textwidth]{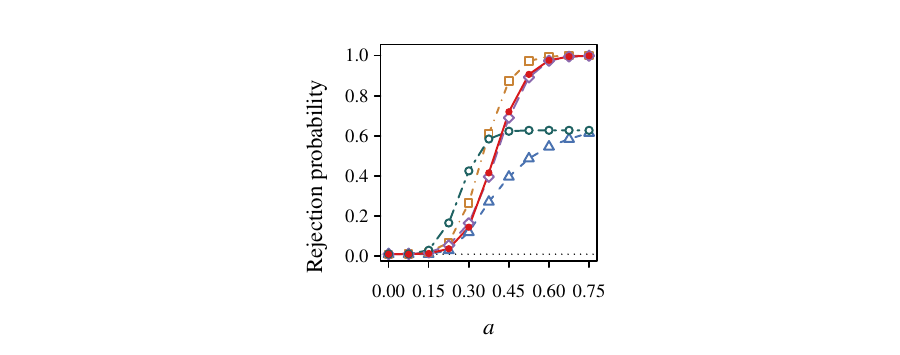}\vspace{-0.4cm}
\caption{Size-adjusted power at the 1\% nominal size under the rare/large price-jump alternative.
It compares
the proposed test~(\SNProposedCaptionKey) with $T_{\mathrm{AJ}}$~(\SNAJCaptionKey), $T_{\mathrm{JO}}$~(\SNJOCaptionKey), $T_{\mathrm{CPR}}$~(\SNCPRCaptionKey), and $T_{\mathrm{LM}}$~(\SNLMCaptionKey).
}
\label{fig:sim-clean-power-competitors-adjusted}
\end{figure}

We consider \eqref{eq:simulation-log-volatility} with $J^\sigma$.
Under the rare/large alternative, consider the jump location $\tau\sim\operatorname{Unif}[0.15,0.85]$ and $\Delta y_\tau^*=0.20aS$.
For the frequent/small alternative, assume jump locations $\tau_1,\ldots,\tau_{20}\simIID\operatorname{Unif}[0.15,0.85]$ and $\Delta y_{\tau_j}^*=0.20aS_j/\sqrt{20}$, $j=1,\ldots,20$.
Here $S,S_1,\ldots,S_{20}$ are independent Rademacher variables.
In both cases, the price-jump times and signs are mutually independent and jointly independent of $(U_q)$, $(Z_q)$, $N^\sigma$, and $(\zeta_j^\sigma)_{j\ge1}$.
Both cases satisfy $\sum_j(\Delta y_{\tau_j}^*)^2=0.20^2a^2$.
Figures~\ref{fig:sim-clean-power} and~\ref{fig:sim-clean-power-competitors-adjusted} evaluate empirical rejection probability on the grids $a\in\{0,0.125,\ldots,1.25\}$ and $a\in\{0,0.075,\ldots,0.75\}$, respectively.

Figure~\ref{fig:sim-clean-power} compares the rejection probabilities of the two tests at the 5\% nominal size.
In both cases, $T_0^{\mathrm{tr}}(g_1)$ has higher raw rejection probabilities than $T_{6,*}^{(3),\mathrm{tr}}(g_1)$, but it also rejects more frequently when $a=0$.
The self-normalized test shows more accurate empirical size in the settings considered.
While comparing across panels, we found that the frequent but small jumps (Panel (b)) are harder to detect than rare but large jumps (Panel (a)).

We next compare size-adjusted power with the other competing tests at the 1\% nominal size.
In Figure~\ref{fig:sim-clean-power-competitors-adjusted}, the size-adjusted power of the proposed test is slightly lower than that of $T_{\mathrm{JO}}$, whereas its relative power against $T_{\mathrm{CPR}}$, $T_{\mathrm{AJ}}$, and $T_{\mathrm{LM}}$ varies with jump magnitude.
For $T_{\mathrm{LM}}$, the absence of a presample restricts testing to returns $y_{120},\ldots,y_M$ ($M=288$) and its plotted size-adjusted power levels off at around 0.65; see Section~\ref{sec:supp-simulation-implementations} of the supplement.
On the other hand, Figures~\ref{fig:sim-clean-size-standard} and~\ref{fig:sim-clean-size-competitors} demonstrate that our proposed test exhibits more accurate empirical size in the settings considered.
The size-adjusted comparison when jumps are frequent but small and the raw rejection probabilities for both alternatives are reported in Section~\ref{sec:supp-simulation-implementations} of the supplement.

\subsection{Empirical performance under microstructure noise}
\label{subsec:simulation-microstructure-noise}

We consider \eqref{eq:simulation-log-volatility} with $J^\sigma\equiv0$ and no price jumps.
We examine the size accuracy of our proposal through three scenarios: (i) $\varepsilon_{i,M}\equiv0$, (ii) $\varepsilon_{i,M}\sim\Normal(0,\omega_M^2)$, and (iii) $\varepsilon_{i,M}\sim(\omega_M^2/2)^{1/2}t_4$, where $\omega_M^2=0.05M^{-3/2}$.
Motivated by \citet[Eq.~(3)]{KalninaLinton2008}, we construct $\varepsilon$ with dependence on $y^*$ governed by $\rho\in\{-0.25,0,0.25\}$ in cases~(ii) and~(iii).

For $i=1,\ldots,M$, let
$Z_{i,M}=y_i/
\{\sum_{q=10(i-1)}^{10i-1}\sigma_{t_q}^2
\Delta_{\mathrm{fine}}\}^{1/2}$.
Let $(G_{i,M})_{i=0}^M$ and $(H_{i,M})_{i=0}^M$ be independent i.i.d. sequences with $G_{i,M}\sim\Normal(0,1)$ and $H_{i,M}\sim\chi_4^2$, independent of $(U_q)$ and $(Z_q)$.
Set $s_{i,M}=1$ for case~(ii) and $s_{i,M}=(2/H_{i,M})^{1/2}$ for case~(iii), $i=0,\ldots,M$.
We define $\varepsilon_{0,M}=\omega_M G_{0,M}s_{0,M}$ and 
\[
\varepsilon_{i,M}
=\omega_M\{\rho Z_{i,M}+(1-\rho^2)^{1/2}G_{i,M}\}s_{i,M},
\qquad i=1,\ldots,M.
\]
For $i=1,\ldots,M$, $\operatorname{Corr}(\varepsilon_{i,M},Z_{i,M})=\rho$ in case~(ii) and $(\sqrt{\pi}/2)\rho$ in case~(iii).
When $\rho=0$, $(\varepsilon_{i,M})_{i=0}^M$ is independent of $(y^*(t))_{t\in[0,1]}$, whereas $\rho\ne0$ makes $\varepsilon_{i,M}$ dependent on $y_i$.
The variables $(\varepsilon_{i,M})_{i=0}^M$ remain independent across $i$.
The construction in case~(ii) follows the Euler-scheme counterpart of Assumption~\ref{ass:microstructure-noise-null}.  

Since $M^{5/4}\omega_M^2=0.05M^{-1/4}\to0$ and $M^{3/4}\rho\omega_M=0.05^{1/2}\rho$, the noise specification corresponds to $c=0$ and $\iota=0.05^{1/2}\rho$ in Theorem~\ref{thm:microstructure-noise}.
Table~\ref{tab:microstructure-noise-simulation} reports the type-I error at $\alpha_0=0.01$ for $\ell=2,\ldots,6$.
We set $M=288$ throughout the simulation, with 200 volatility paths and 250 conditional price paths for each volatility path.
Cases~(i) and (ii) are covered by the scope of Theorem~\ref{thm:microstructure-noise}, while case~(iii) is not.
It is included to examine finite-sample robustness to heavy-tailed microstructure noise.

\begin{table}[!ht]
\centering
\setlength{\tabcolsep}{7pt}\renewcommand{\arraystretch}{1.05}
\begin{tabular}{@{}lccccccc@{}}\hline
Noise distribution & $\rho$ & $T_0^{\mathrm{tr}}(g_1)$ &
\multicolumn{5}{c}{$T_{\ell,*}^{(3),\mathrm{tr}}(g_1)$}\\
\cline{4-8}
& & & $\ell=2$ & $\ell=3$ & $\ell=4$ & $\ell=5$ & $\ell=6$\\\hline
(i) $\varepsilon_{i,M}\equiv0$ & --- & 2.83\% & 1.08\% & 1.07\% & 1.05\% & 1.12\% & 1.14\%\\
(ii) $\varepsilon_{i,M}\sim\Normal(0,\omega_M^2)$ & $-0.25$ & 2.89\% & 1.10\% & 0.99\% & 1.05\% & 1.13\% & 1.21\%\\
& $0$ & 2.56\% & 1.09\% & 0.90\% & 0.94\% & 1.01\% & 1.06\%\\
& $+0.25$ & 2.09\% & 0.85\% & 0.81\% & 0.78\% & 0.83\% & 0.89\%\\
(iii) $\varepsilon_{i,M}\sim(\omega_M^2/2)^{1/2}t_4$ & $-0.25$ & 2.92\% & 1.04\% & 1.12\% & 1.25\% & 1.35\% & 1.36\%\\
& $0$ & 2.57\% & 0.97\% & 1.06\% & 1.11\% & 1.16\% & 1.16\%\\
& $+0.25$ & 2.24\% & 0.91\% & 0.92\% & 0.93\% & 1.04\% & 1.00\%\\\hline
\end{tabular}
\caption{Finite-sample type-I error with no noise, Gaussian noise, and scaled $t_4$ noise.
The nominal size is $\alpha_0=1\%$.}
\label{tab:microstructure-noise-simulation}
\end{table}

Table~\ref{tab:microstructure-noise-simulation} shows that $T_0^{\mathrm{tr}}(g_1)$ over-rejects in every case, whereas all self-normalized tests have rejection rates closer to the nominal 1\% level.

\section{Real-data application}
\label{sec:real-data-application}

In this section, we compare our proposed self-normalized jump test with the conventional jump test based on a thresholded jump estimator, using five-minute returns of the Standard and Poor's 500 (S\&P 500) index, the USD/JPY exchange rate, and the price of gold.
All three datasets were retrieved from Bloomberg and cover 2014 to 2023.
For the S\&P 500, we retain $N=2{,}495$ trading days, each containing $M_d=81$ or $82$ five-minute intervals.
For USD/JPY, we retain $N=2{,}064$ days, each containing $M_d=288$ five-minute intervals.
For gold, we retain $N=2{,}497$ trading days, each containing $M_d=259$ five-minute intervals.
For each market and day $d$, the returns are denoted by $y_{i,d}$ for $i=1,\ldots,M_d$.

We consider the tests $T\in\{T_0^{\mathrm{tr}}(g_1),T_{\ell_{M_d}^*,*}^{(3),\mathrm{tr}}(g_1)\}$, where $n=3$ and $\ell_{M_d}^*=\ell_{M_d}^*(\alpha_0,p)$ is defined by \eqref{eq:ell-M-definition} with $M=M_d$.
We test the hypothesis at the nominal size $\alpha_0=0.1$ with $p=0.9$.
For the S\&P 500, all trading days have $M_d=81$ or $82$ available 5-minute returns, so $\ell_{M_d}^* = 4<6=\ell^*$ and the finite-$M$ adjustment in \eqref{eq:ell-M-definition} limits the reduction in available summands.
Therefore, unlike $\ell^*$, the capped choice is not guaranteed to retain at least $100p\%$ of the local asymptotic power of $T_0^{\mathrm{tr}}(g_1)$ while $\ell_{M_d}^*<\ell^*$.
For the USD/JPY and gold data, the same choice of $(\alpha_0,p)$ gives $\ell_{M_d}^*=\ell^*=6$.

We compute the $p$-values of $T_0^{\mathrm{tr}}(g_1)/\sigma_{g_1}$ from the standard normal distribution and those of the proposed test from the Student-$t_3$ distribution for the S\&P 500 and the Student-$t_5$ distribution for USD/JPY and gold.
For both tests, we apply the truncation threshold \eqref{eq:simulation-threshold-choice} to each day's returns $y_{1,d},\ldots,y_{M_d,d}$, with $M=M_d$ and $\sigma_{\mathrm{ref}}=0.01$, allowing it to adapt to volatility differences across trading days.
Spurious jump detections arising from multiple testing are discussed by \citet{BajgrowiczScailletTreccani2016}.
To account for the multiple-testing problem across trading days, we consider the procedures of \citet{BenjaminiHochberg1995} and \citet{BenjaminiYekutieli2001}, denoted by BH and BY, respectively.
The BH and BY procedures aim to control the false discovery rate (FDR).
The BH procedure controls the FDR under independence or certain forms of positive dependence, whereas the more conservative BY procedure allows arbitrary dependence among $p$-values.

For each market and test $T$, let $p_d(T)$ denote its $p$-value.
For $a\in\{0,\mathrm{BH},\mathrm{BY}\}$, denote the adjusted $p$-value by $p_d^{(a)}(T)$, where $p_d^{(0)}(T)=p_d(T)$ represents the unadjusted $p$-value.
Define the set of detected trading days and the number of detections as
\[
I^{(a)}(T)=\{d:p_d^{(a)}(T)<\alpha_0\},
\qquad
N_J^{(a)}(T)=|I^{(a)}(T)|.
\]
Each multiple-testing adjustment is applied across all $N$ retained days within each market, separately for each test statistic.
Since volatility differs across trading days, we standardize the estimated jump component by the median squared return of the corresponding day.
Accordingly, we define
\[
\rho_{\mathrm{BH}}(T)=
\med_{d\in I^{(\mathrm{BH})}(T)}
\left|
\frac{\hat\theta_{g_1,d}^{\mathrm{tr}}}
{\mathop{\med}_{1\le i\le M_d}(y_{i,d}^2)}
\right|,
\]
where $\hat\theta_{g_1,d}^{\mathrm{tr}}$ is the value of $\hat\theta_{g_1}^{\mathrm{tr}}$ evaluated on $(y_{1,d},\ldots,y_{M_d,d})$.
The proportions $N_J^{(a)}(T)/N$ and the values of $\rho_{\mathrm{BH}}(T)$ are reported in Table~\ref{tab:real-data-application}.

\begin{table}[!t]
\centering
\begin{tabular}{@{}lcccccc@{}}\hline
& \multicolumn{2}{c}{S\&P 500 ($N=2{,}495$)}
& \multicolumn{2}{c}{USD/JPY ($N=2{,}064$)}
& \multicolumn{2}{c}{Gold ($N=2{,}497$)}\\
\cmidrule(r){2-3}\cmidrule(lr){4-5}\cmidrule(l){6-7}
Measure$\backslash\, T$ & $T_0^{\mathrm{tr}}(g_1)$ & $T_{\ell_{M_d}^*,*}^{(3),\mathrm{tr}}(g_1)$
& $T_0^{\mathrm{tr}}(g_1)$ & $T_{\ell_{M_d}^*,*}^{(3),\mathrm{tr}}(g_1)$
& $T_0^{\mathrm{tr}}(g_1)$ & $T_{\ell_{M_d}^*,*}^{(3),\mathrm{tr}}(g_1)$\\\hline
$N_J^{(0)}(T)/N$ & 47.6\% & 38.8\% & 71.3\% & 64.5\% & 57.5\% & 53.9\%\\
$N_J^{(\mathrm{BH})}(T)/N$ & 39.4\% & 20.4\% & 67.2\% & 56.8\% & 51.3\% & 45.4\%\\
$N_J^{(\mathrm{BY})}(T)/N$ & 28.4\% & 7.9\% & 51.1\% & 28.0\% & 37.4\% & 25.2\%\\\hline
$\rho_{\mathrm{BH}}(T)$ & $82.13$ & $164.81$ & $223.32$ & $263.69$ & $264.54$ & $304.38$\\\hline
\end{tabular}
\caption{Percentages of detected trading days and median standardized absolute estimates of jump variation for the S\&P 500, USD/JPY, and gold.}
\label{tab:real-data-application}
\end{table}

\begin{figure}[!tbp]
\centering
\includegraphics[width=\textwidth]{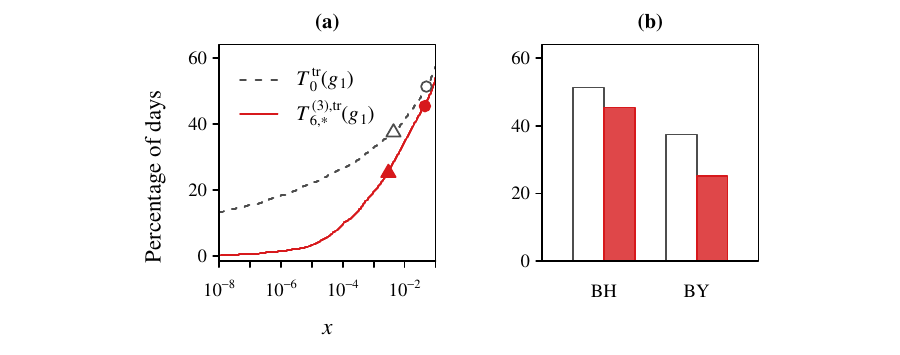}
\vspace{-1.2cm}
\caption{Distribution of daily jump-test $p$-values and rejection rates for gold ($N=2{,}497$).
(a) Percentage of days with raw $p$-values at most $x$.
The symbols $\bigcirc$ and $\triangle$ indicate the largest rejected raw $p$-values under BH and BY, respectively.
(b) Percentage of days rejected under each procedure.
White and red bars refer to $T_0^{\mathrm{tr}}(g_1)$ and $T_{6,*}^{(3),\mathrm{tr}}(g_1)$, respectively.}
\label{fig:gold-pvalues}
\end{figure}

We find that $T_0^{\mathrm{tr}}(g_1)$ detects a higher percentage of potential jump days than $T_{\ell_{M_d}^*,*}^{(3),\mathrm{tr}}(g_1)$ under all three testing procedures in all three markets.
For example, after applying the BH adjustment to the S\&P 500, $T_0^{\mathrm{tr}}(g_1)$ detects 39.4\% of days, compared with 20.4\% for the proposed test.
On the other hand, $\rho_{\mathrm{BH}}(T_{\ell_{M_d}^*,*}^{(3),\mathrm{tr}}(g_1))>\rho_{\mathrm{BH}}(T_0^{\mathrm{tr}}(g_1))$ in all three markets.
Across the tests and procedures considered, USD/JPY and gold generally have higher proportions of detected trading days than the S\&P 500.
This is consistent with \citet{LahayeLaurentNeely2011}, who report more jump days for JPY/USD than for S\&P 500 futures, and \citet{Jurdi2020}, who find more intraday jumps in a gold ETF than in an S\&P 500 ETF.
For the gold market, Figure~\ref{fig:gold-pvalues} shows that $T_0^{\mathrm{tr}}(g_1)$ yields more extremely small $p$-values.
This is consistent with the larger difference in detected-day proportions under BY (12.25 percentage points) than under BH (5.93 percentage points).
Overall, the empirical results suggest that the proposed self-normalized test provides a more selective identification of potential jump days in high-frequency returns.

\section{Discussion}
\label{sec:discussion}

This paper develops a general framework for constructing a class of jump detection tests for high-frequency data based on self-normalization.
The procedure is easy to implement as in Algorithm \ref{algo:OJT_computation} and does not require complicated hyperparameter tuning.
The proposed optimal differencing approach can efficiently boost the power of the test while preserving satisfactory size accuracy.
It also indicates that the principle of self-normalization shows great potential in handling high-frequency return data.

One of the possible directions for future research is to investigate whether pre-averaging can be combined with our proposed self-normalization framework to handle non-vanishing microstructure noise.
Indeed, pre-averaging has been developed for volatility and jump inference under microstructure noise \citep{JacodLiMyklandPodolskijVetter2009,AitSahaliaJacodLi2012}.
Such a generalization would require verifying whether the lag-augmented multipower variations retain equal asymptotic covariances with $\hat\theta_g$ and a compound-symmetric covariance structure after pre-averaging.
In addition, more general dependence between $y^*$ and $\varepsilon$ than that specified in Assumption~\ref{ass:microstructure-noise-null} remains to be studied.

\clearpage
\input{supp_RR}
\FloatBarrier

\vspace{0.2cm}
\noindent\textbf{Competing interests:} The authors declare none.

\noindent\textbf{Funding:} 
This research was partially supported by grants
GRF-14307922 and 14307925 provided by the Research Grants Council of HKSAR.
{\spacingset{1.5}
\interlinepenalty=10000
\bibliographystyle{apacite}
\bibliography{myRef}
}

\end{document}

%% file: myPreliminary_submission.tex
\usepackage[OT1]{fontenc}
\usepackage{amsmath,amssymb,amsthm}
\usepackage{bbold}
\usepackage[cal=boondox]{mathalfa}

\usepackage{graphicx}
\usepackage{tikz}
\usetikzlibrary{decorations.pathreplacing}
\usepackage{enumitem}
\usepackage[lined,boxed,commentsnumbered,ruled,linesnumbered]{algorithm2e}
\usepackage[natbibapa,nodoi]{apacite}
\makeatletter
\renewcommand{\NAT@sort}{0}
\makeatother

\newif\ifshowbiburls
\showbiburlsfalse
\ifshowbiburls\else
  
\fi
\usepackage{xcolor}

\definecolor{myred}{rgb}{0.7,0.1,0.1}
\definecolor{myblue}{rgb}{0.1,0.1,0.7}

\usepackage[colorlinks]{hyperref}
\hypersetup{citecolor=myblue,linkcolor=myred,urlcolor=myblue}
\makeatletter
\newif\ifinsupplement
\let\SN@internalhyperlink\hyper@link
\def\SN@linktype{link}
\def\hyper@link#1#2#3{%
  \begingroup
  \edef\SN@linkstyle{#1}%
  \ifx\SN@linkstyle\SN@linktype
    \edef\SN@checktarget{\noexpand\in@{.supp.}{#2}}%
    \SN@checktarget
    \ifinsupplement
      \ifin@\else\def\SN@linkstyle{file}\fi
    \else
      \ifin@\def\SN@linkstyle{file}\fi
    \fi
  \fi
  \edef\SN@next{\endgroup\noexpand\SN@internalhyperlink{\SN@linkstyle}}%
  \SN@next{#2}{#3}%
}
\makeatother
\providecommand{\doi}[1]{#1}
\renewcommand{\doi}[1]{%
  \begingroup
  \urlstyle{APACsame}%
  \href{#1}{\textcolor{black}{\nolinkurl{#1}}}%
  \endgroup}

\usepackage[letterpaper,margin=1.25in]{geometry}

\newcommand{\spacingset}[1]{%
  \renewcommand{\baselinestretch}{#1}\small\normalsize}
\spacingset{1.5}

\newtheorem{lemma}{Lemma}
\newtheorem{corollary}{Corollary}
\newtheorem{proposition}{Proposition}
\newtheorem{assumption}{Assumption}
\newtheorem{remark}{Remark}
\newtheorem{definition}{Definition}
\newtheorem{theorem}{Theorem}

\newcommand{\E}{\mathsf{E}}
\newcommand{\pr}{\mathsf{P}}
\newcommand{\Cov}{\mathop{\mathsf{Cov}}}
\newcommand{\Var}{\mathop{\mathsf{Var}}}
\newcommand{\inP}{\overset{\textnormal{pr}}{\rightarrow}}
\newcommand{\R}{\mathbb{R}}
\newcommand{\med}{\mathop{\mathrm{med}}\limits}
\newcommand{\simIID}{\overset{\mathrm{i.i.d.}}{\sim}}
\newcommand{\T}{\textsc{T}}
\newcommand{\dd}{\textnormal{d}}
\newcommand{\Normal}{\textnormal{N}}
\newcommand{\ind}{\mathbb{1}}
\def\IV{\textsc{iv}}
\def\QV{\textsc{qv}}
\newcommand{\inST}{\xrightarrow{\mathrm{st}}}

\definecolor{SNcaptionred}{HTML}{D7191C}
\definecolor{SNcaptionblue}{HTML}{4169E1}
\DeclareRobustCommand{\SNProposedCaptionKey}{%
  \textcolor{SNcaptionred}{red}\nobreak\hspace{0.3em}%
  \tikz[baseline=-0.5ex,x=1em,y=1em]{%
    \draw[SNcaptionred,line width=0.65pt,line cap=round] (0,0)--(2.4,0);
    \fill[SNcaptionred] (1.2,0) circle[radius=0.15em];
  }%
}
\DeclareRobustCommand{\SNConventionalCaptionKey}{%
  \textcolor{SNcaptionblue}{blue}\nobreak\hspace{0.3em}%
  \tikz[baseline=-0.5ex,x=1em,y=1em]{%
    \draw[SNcaptionblue,line width=0.65pt,line cap=round,
      dash pattern=on 1.95pt off 3.25pt] (0,0)--(1.05,0);
    \draw[SNcaptionblue,line width=0.65pt,line cap=round,
      dash pattern=on 1.95pt off 3.25pt] (1.35,0)--(2.4,0);
    \draw[SNcaptionblue,line width=0.65pt,fill=none]
      (1.2,0) circle[radius=0.15em];
  }%
}

\definecolor{SNcaptionAJ}{HTML}{4A72B0}
\definecolor{SNcaptionJO}{HTML}{C98336}
\definecolor{SNcaptionCPR}{HTML}{8B68AD}
\definecolor{SNcaptionLM}{HTML}{1F6262}
\DeclareRobustCommand{\SNAJCaptionKey}{%
  \textcolor{SNcaptionAJ}{blue}\nobreak\hspace{0.3em}%
  \tikz[baseline=-0.5ex,x=1em,y=1em]{%
    \draw[SNcaptionAJ,line width=0.65pt,line cap=round,
      dash pattern=on 1.95pt off 3.25pt] (0,0)--(0.998,0);
    \draw[SNcaptionAJ,line width=0.65pt,line cap=round,
      dash pattern=on 1.95pt off 3.25pt] (1.402,0)--(2.4,0);
    \draw[SNcaptionAJ,line width=0.65pt,line join=round]
      (1.2,0.233)--(1.402,-0.117)--(0.998,-0.117)--cycle;
  }%
}
\DeclareRobustCommand{\SNJOCaptionKey}{%
  \textcolor{SNcaptionJO}{orange}\nobreak\hspace{0.3em}%
  \tikz[baseline=-0.5ex,x=1em,y=1em]{%
    \draw[SNcaptionJO,line width=0.65pt,line cap=round,
      dash pattern=on 0pt off 2.6pt on 1.95pt off 2.6pt] (0,0)--(1.05,0);
    \draw[SNcaptionJO,line width=0.65pt,line cap=round,
      dash pattern=on 0pt off 2.6pt on 1.95pt off 2.6pt] (1.35,0)--(2.4,0);
    \draw[SNcaptionJO,line width=0.65pt] (1.05,-0.15) rectangle (1.35,0.15);
  }%
}
\DeclareRobustCommand{\SNCPRCaptionKey}{%
  \textcolor{SNcaptionCPR}{purple}\nobreak\hspace{0.3em}%
  \tikz[baseline=-0.5ex,x=1em,y=1em]{%
    \draw[SNcaptionCPR,line width=0.65pt,line cap=round,
      dash pattern=on 3.9pt off 2.6pt] (0,0)--(0.988,0);
    \draw[SNcaptionCPR,line width=0.65pt,line cap=round,
      dash pattern=on 3.9pt off 2.6pt] (1.412,0)--(2.4,0);
    \draw[SNcaptionCPR,line width=0.65pt,line join=round]
      (0.988,0)--(1.2,0.212)--(1.412,0)--(1.2,-0.212)--cycle;
  }%
}
\DeclareRobustCommand{\SNLMCaptionKey}{%
  \textcolor{SNcaptionLM}{teal}\nobreak\hspace{0.3em}%
  \tikz[baseline=-0.5ex,x=1em,y=1em]{%
    \draw[SNcaptionLM,line width=0.65pt,line cap=round,
      dash pattern=on 0.65pt off 1.95pt on 3.25pt off 1.95pt] (0,0)--(1.05,0);
    \draw[SNcaptionLM,line width=0.65pt,line cap=round,
      dash pattern=on 0.65pt off 1.95pt on 3.25pt off 1.95pt] (1.35,0)--(2.4,0);
    \draw[SNcaptionLM,line width=0.65pt] (1.2,0) circle[radius=0.15em];
  }%
}

\definecolor{SNcaptiongray}{HTML}{4D4D4D}
\DeclareRobustCommand{\SNEllMCaptionKey}{%
  \textcolor{SNcaptionred}{red}\nobreak\hspace{0.3em}%
  \tikz[baseline=-0.5ex,x=1em,y=1em]{%
    \draw[SNcaptionred,line width=0.65pt,line cap=round] (0,0)--(2.4,0);
    \fill[SNcaptionred] (1.2,0) circle[radius=0.17em];
  }%
}
\DeclareRobustCommand{\SNEllSixCaptionKey}{%
  \textcolor{SNcaptionblue}{blue}\nobreak\hspace{0.3em}%
  \tikz[baseline=-0.5ex,x=1em,y=1em]{%
    \draw[SNcaptionblue,line width=0.65pt,line cap=round,
      dash pattern=on 1.95pt off 2.6pt] (0,0)--(0.95,0);
    \draw[SNcaptionblue,line width=0.65pt,line cap=round,
      dash pattern=on 1.95pt off 2.6pt] (1.45,0)--(2.4,0);
    \draw[SNcaptionblue,line width=0.55pt,line join=round]
      (1.03,-0.12)--(1.37,-0.12)--(1.2,0.18)--cycle;
  }%
}
\DeclareRobustCommand{\SNEllStarCaptionKey}{%
  \textcolor{SNcaptiongray}{gray}\nobreak\hspace{0.3em}%
  \tikz[baseline=-0.5ex,x=1em,y=1em]{%
    \draw[SNcaptiongray,line width=0.65pt,line cap=round,
      dash pattern=on 2.6pt off 1.3pt on 0.4pt off 1.3pt] (0,0)--(0.95,0);
    \draw[SNcaptiongray,line width=0.65pt,line cap=round,
      dash pattern=on 2.6pt off 1.3pt on 0.4pt off 1.3pt] (1.45,0)--(2.4,0);
    \draw[SNcaptiongray,line width=0.55pt] (1.05,-0.15) rectangle (1.35,0.15);
  }%
}
\DeclareRobustCommand{\SNTzeroCaptionKey}{%
  black\nobreak\hspace{0.3em}%
  \tikz[baseline=-0.5ex,x=1em,y=1em]{%
    \draw[black,line width=0.65pt,line cap=round,
      dash pattern=on 3.6pt off 2pt] (0,0)--(0.95,0);
    \draw[black,line width=0.65pt,line cap=round,
      dash pattern=on 3.6pt off 2pt] (1.45,0)--(2.4,0);
    \draw[black,line width=0.55pt,line join=miter]
      (1.01,0)--(1.2,0.19)--(1.39,0)--(1.2,-0.19)--cycle;
  }%
}
\DeclareRobustCommand{\SNEllMPointKey}{%
  \textcolor{SNcaptionred}{red}\nobreak\hspace{0.3em}%
  \tikz[baseline=-0.5ex,x=1em,y=1em]{\fill[SNcaptionred] (0,0) circle[radius=0.17em];}%
}
\DeclareRobustCommand{\SNEllSixPointKey}{%
  \textcolor{SNcaptionblue}{blue}\nobreak\hspace{0.3em}%
  \tikz[baseline=-0.5ex,x=1em,y=1em]{%
    \draw[SNcaptionblue,line width=0.55pt,line join=round]
      (-0.17,-0.12)--(0.17,-0.12)--(0,0.18)--cycle;}%
}
\DeclareRobustCommand{\SNEllStarPointKey}{%
  \textcolor{SNcaptiongray}{gray}\nobreak\hspace{0.3em}%
  \tikz[baseline=-0.5ex,x=1em,y=1em]{%
    \draw[SNcaptiongray,line width=0.55pt] (-0.15,-0.15) rectangle (0.15,0.15);}%
}

%% file: supp_RR.tex
\insupplementtrue
\setcounter{section}{0}
\setcounter{equation}{0}
\setcounter{theorem}{0}
\setcounter{assumption}{0}
\setcounter{figure}{0}
\setcounter{table}{0}

\renewcommand{\thesection}{S\arabic{section}}
\renewcommand{\theequation}{S\arabic{equation}}
\renewcommand{\thetheorem}{S\arabic{theorem}}
\renewcommand{\thelemma}{S\arabic{lemma}}
\renewcommand{\theproposition}{S\arabic{proposition}}
\renewcommand{\thecorollary}{S\arabic{corollary}}
\renewcommand{\theassumption}{S\arabic{assumption}}
\renewcommand{\theremark}{S\arabic{remark}}
\renewcommand{\thedefinition}{S\arabic{definition}}
\renewcommand{\thefigure}{S\arabic{figure}}
\renewcommand{\thetable}{S\arabic{table}}

\renewcommand*{\theHsection}{supp.\arabic{section}}
\renewcommand*{\theHequation}{supp.\arabic{equation}}
\renewcommand*{\theHtheorem}{supp.\arabic{theorem}}
\renewcommand*{\theHlemma}{supp.\arabic{lemma}}
\renewcommand*{\theHproposition}{supp.\arabic{proposition}}
\renewcommand*{\theHassumption}{supp.\arabic{assumption}}
\renewcommand*{\theHremark}{supp.\arabic{remark}}
\renewcommand*{\theHfigure}{supp.\arabic{figure}}
\renewcommand*{\theHtable}{supp.\arabic{table}}

\pdfbookmark[0]{Supplementary Material}{supplement}
\section*{Supplementary Material}

This supplement contains additional calculations and all proofs omitted from the main text.
Numbers with the prefix~S refer to the supplement.
Plain numbers refer to the main text, whose notation, sampling design, statistics, assumptions, and stable-convergence conventions are used throughout, as defined in the main text.

Section~\ref{sec:supp-ell-M} motivates the choice $\ell=\ell_M^*(\alpha_0,p)$ in \eqref{eq:ell-M-definition} and reports the numerical evidence for it.
Section~\ref{sec:supp-simulation-implementations} presents implementation details and additional Monte Carlo results.
Section~\ref{sec:supp-unthresholded} proves all theoretical results.

\section{A practical choice of \texorpdfstring{$\ell$}{l} for finite \texorpdfstring{$M$}{M}}
\label{sec:supp-ell-M}

The criterion \eqref{eq:ell-wc-definition} determines $\ell^*(\alpha_0,p)$ by comparing asymptotic local power, and thus $\ell^*(\alpha_0,p)$ can be chosen independently of $M$.
However, there are only $N_M(\ell)=M-(n-1)(b+\ell+1)+1$ summands available for evaluating each of $Q_k^{(n),\mathrm{tr}}$ in \eqref{eq:Qk-tr}.
For example, when $n=3$ and $b=1$, the proportion of retained summands is $N_M(\ell)/M=1-(2\ell+3)/M$.
In this case, if $M$ is relatively small, e.g., $M=81$, then $\ell^*(0.01,0.9)=29$ leaves only 20 summands for evaluating the self-normalizer.
This will harm the finite-sample performance of the test.
Therefore, we truncate $\ell^*(\alpha_0,p)$ at $M^{1/3}$ to define
\[
\ell_M^*(\alpha_0,p)=\min\left\{\ell^*(\alpha_0,p),\,\max\{2,\lfloor M^{1/3}\rfloor\}\right\}.
\]
Under the same scenario, $\ell_{81}^*(0.01,0.9)=4$ and 70 summands are left for the evaluation of the self-normalizer, which is much better than when $\ell=29$.

For $n=3$, $b=1$, and $M\ge8$ (so that $M^{1/3}\geq2$), we have $\ell_M^*(\alpha_0,p)\le M^{1/3}$ and hence the proportion of retained summands satisfies
\[
1-2M^{-2/3}-\frac{3}{M}
\le \frac{N_M(\ell_M^*)}{M}
\le 1
\]
whatever $\alpha_0$ and $p$.
Thus, $\ell_M^*(\alpha_0,p)$ may grow with $M$, while the proportion of retained summands converges to one.
A similar phenomenon can be observed if we use $M^q$ instead of $M^{1/3}$ regardless of the value of $q\in(0,1)$.
We pick $q=1/3$ since it gives the most satisfactory balance between type-I error and power under various simulation setups we have examined.

We have $\ell_M^*(\alpha_0,p)=\ell^*(\alpha_0,p)$ when $M\ge\ell^*(\alpha_0,p)^3$, so the two choices coincide for all sufficiently large $M$.
Therefore, both results in Theorem~\ref{thm:canonical-properties} and Corollary~\ref{cor:local-power-retention} hold when we adopt $\ell=\ell_M^*(\alpha_0,p)$, since they apply to the eventually identical procedure with the fixed value $\ell=\ell^*(\alpha_0,p)$.
In particular, at least $100p\%$ of the local asymptotic power of the non-self-normalized test is retained.
This means that it can automatically lower the value of $\ell$ to guarantee that a sufficient proportion of summands is retained for evaluating the self-normalizer.
However, no such power retention is guaranteed for finite $M$ while the cap binds, that is, while $\ell_M^*(\alpha_0,p)<\ell^*(\alpha_0,p)$.

\begin{figure}[!tbp]
\centering
\includegraphics[width=\textwidth]{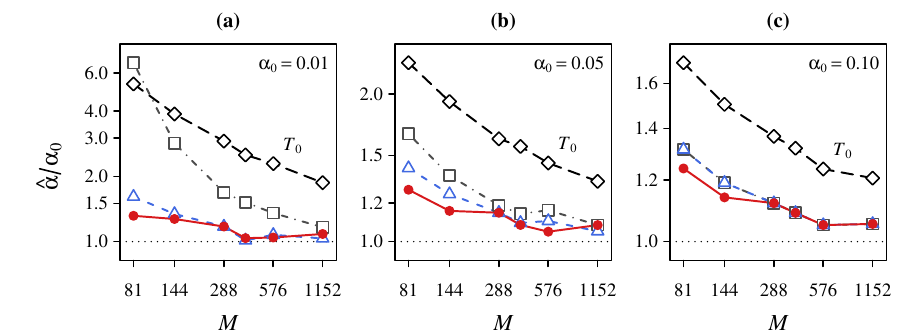}
\caption{Ratio $\widehat\alpha/\alpha_0$ of empirical to nominal size against $M$ under scenario~(iii) for $T_{\ell,*}^{(3),\mathrm{tr}}(g_1)$ with 
$\ell=\ell_M^*(\alpha_0,0.9)$~(\SNEllMCaptionKey), $\ell=6$~(\SNEllSixCaptionKey) and $\ell=\ell^*(\alpha_0,0.9)$~(\SNEllStarCaptionKey), and for $T_0^{\mathrm{tr}}(g_1)$~(\SNTzeroCaptionKey).
The value of $\ell^*(\alpha_0,p)$ is 29, 10 and 6 in panels (a)--(c), respectively.
Both axes are displayed on a logarithmic scale.
The horizontal dotted line represents $\widehat\alpha/\alpha_0=1$.}
\label{fig:supp-ell-M-size}
\end{figure}

\begin{figure}[!tbp]
\centering
\includegraphics[width=\textwidth]{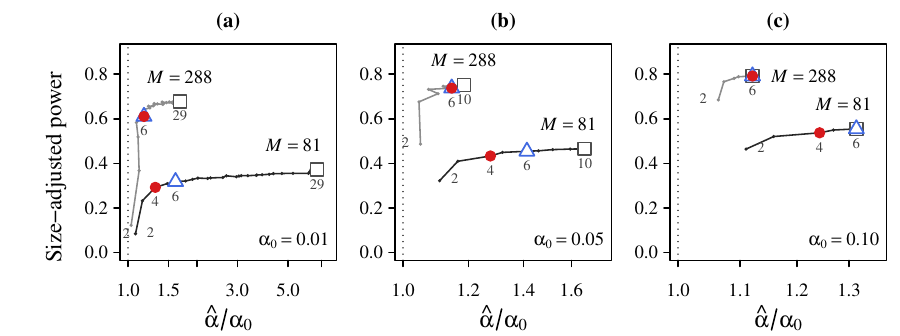}
\caption{The plot of size-adjusted power against $\widehat\alpha/\alpha_0$ under scenario~(iii) for $T_{\ell,*}^{(3),\mathrm{tr}}(g_1)$, with $\ell=2,\ldots,\ell^*(\alpha_0,0.9)$ joined in increasing order at $M=81$ (dark path) and $M=288$ (light path).
The numbers below selected points indicate the value of $\ell$.
The marked points correspond to $\ell=\ell_M^*(\alpha_0,0.9)$~(\SNEllMPointKey), $\ell=6$~(\SNEllSixPointKey), and $\ell=\ell^*(\alpha_0,0.9)$~(\SNEllStarPointKey).
The horizontal axes are displayed on a logarithmic scale.
The vertical dotted line marks $\widehat\alpha/\alpha_0=1$.
Monte Carlo standard errors are below 0.14 for the ratio and 0.004 for power.}
\label{fig:supp-ell-M-tradeoff}
\end{figure}

Figures~\ref{fig:supp-ell-M-size}--\ref{fig:supp-ell-M-tradeoff} and Table~\ref{tab:supp-ell-M} follow the same simulation designs as in Section~\ref{sec:simulation-studies} with $n=3$, $b=1$, $g=g_1$ and the threshold \eqref{eq:simulation-threshold-choice}.
We vary $M$ from 81 to 1152 and the jump sizes under the alternative are set to be proportional to $M^{-1/4}$ as in Assumption~\ref{ass:abstract-local-alt}.
The amplitude parameter is $a_M(c)=c(288/M)^{1/4}$, with $c\in\{0.25,0.375,0.5,0.625,0.75\}$ for the rare/large alternative and $c\in\{0.5,0.75,1,1.25,1.5\}$ for the frequent/small alternative.
For each nominal size $\alpha_0$, size-adjusted power is computed using the empirical $(1-\alpha_0)$ null quantile from independent simulations under scenario~(iii), and the reported power is the equally weighted average over the ten combinations of jump alternative and nonzero amplitude.
Monte Carlo standard errors are computed from rejection rates across independent volatility paths, treating the conditional price paths associated with each volatility path as one cluster.

\begin{table}[!tbp]
\centering
\small
\setlength{\tabcolsep}{3.5pt}
\begin{tabular}{@{}llcccccccccc@{}}\hline
& & \multicolumn{3}{c}{$\ell=\ell_M^*(\alpha_0,0.9)$} & & \multicolumn{2}{c}{$\ell=6$} & &
\multicolumn{3}{c}{$\ell=\ell^*(\alpha_0,0.9)$}\\
\cline{3-5}\cline{7-8}\cline{10-12}
$M$ & $\alpha_0$ & $\ell$ & size (\%) & power & & size (\%) & power & &
$\ell$ & size (\%) & power\\\hline
81 & 0.01 & 4 & 1.23--1.33 & 0.292 & & 1.55--1.65 & 0.318 & & 29 & 6.21--6.69 & 0.371\\
 & 0.05 & 4 & 6.21--6.38 & 0.433 & & 6.92--7.07 & 0.454 & & 10 & 7.83--8.30 & 0.465\\
 & 0.10 & 4 & 12.15--12.43 & 0.537 & & 13.06--13.15 & 0.554 & & 6 & 13.06--13.15 & 0.554\\[3pt]
288 & 0.01 & 6 & 1.08--1.17 & 0.610 & & 1.08--1.17 & 0.610 & & 29 & 1.68--1.73 & 0.678\\
 & 0.05 & 6 & 5.47--5.76 & 0.738 & & 5.47--5.76 & 0.738 & & 10 & 5.62--5.93 & 0.751\\
 & 0.10 & 6 & 10.63--11.26 & 0.792 & & 10.63--11.26 & 0.792 & & 6 & 10.63--11.26 & 0.792\\\hline
\end{tabular}
\caption{Empirical size and size-adjusted power of $T_{\ell,*}^{(3),\mathrm{tr}}(g_1)$ for three choices of $\ell$, with ${p=0.9}$.
The size column indicates the range of the empirical rejection probabilities over scenarios (i)--(iii).
The power column indicates rejection probability under scenario~(iii) averaged over the rare/large and frequent/small alternatives with five different nonzero values of jump variation each.
Monte Carlo standard errors are below $0.2\%$ (size) and $0.004$ (power).}
\label{tab:supp-ell-M}
\end{table}

At extremely small $M$ with large $p$, the self-normalized test using $\ell^*(\alpha_0,p)$ may over-reject even more than the conventional test.
Adopting $\ell_M^*(\alpha_0,p)$ can sharply reduce the false-positive error.
For example, when $M=81$ and $\alpha_0=0.01$, the ratio $\widehat\alpha/\alpha_0$ falls from 6.7 to 1.3, compared with 5.3 for $T_0^{\mathrm{tr}}(g_1)$ (Figure~\ref{fig:supp-ell-M-size}).
Across all three panels in Figure~\ref{fig:supp-ell-M-size}, using $\ell_M^*(\alpha_0,p)$ gives empirical size control that is at least as accurate as that obtained with $\ell=6$ in most cases.

Table~\ref{tab:supp-ell-M} shows that when $\lfloor M^{1/3}\rfloor<\ell^*(\alpha_0,p)$, 
using $\ell_M^*(\alpha_0,p)$ gives more accurate size control than using $\ell^*(\alpha_0,p)$ in all three scenarios.
As a tradeoff, it gives lower size-adjusted power. 

Figure~\ref{fig:supp-ell-M-tradeoff} provides further empirical evidence to support  adopting $\ell_M^*(\alpha_0,p)$ in practice.
Informally, the closer a point is to $(1,1)$, the better the test performs.
This is because $(1,1)$ corresponds to perfect size accuracy and power equal to one.
The test adopting $\ell=\ell_M^*(\alpha_0,p)$ performs satisfactorily, as it is relatively close to this ideal point in the plot.
In particular, for $M=81$, reducing $\ell$ to 2 costs substantial size-adjusted power, while values well above $\ell_M^*(\alpha_0,p)$ add modest power with much worse empirical size control.
A similar phenomenon appears across all three panels.

We also examined additional simulation settings involving larger fluctuations in spot volatility and an intraday pattern with higher volatility near the beginning and end of the observation period.
Although these results are not reported here, they generally support the proposed finite-$M$ adjustment in improving empirical size accuracy. 
\FloatBarrier

\section{Implementation details and additional Monte Carlo results}
\label{sec:supp-simulation-implementations}

As in Section~\ref{sec:simulation-studies}, the proposed test uses $\ell=\ell_M^*(\alpha_0,0.9)$ from \eqref{eq:ell-M-definition} unless otherwise specified.
For $M=288$ and the nominal sizes considered, we have $\ell_M^*(\alpha_0,0.9)=6$.

For detailed formulations of the competing tests, we refer readers to the original articles.
Here, we summarize selected tuning parameters, chosen following the recommendations in the respective references.

The test $T_0^{\mathrm{tr}}(g_1)$ uses the same thresholded jump estimator as the proposed test and the quarticity estimator of \citet[Sect.~2.2]{barndorff2006econometrics}.
For $T_{\mathrm{AJ}}$ \citep{AitSahaliaJacod2009}, we set $p=4$ and $k=2$ (in the notation of their article), adopt the threshold $0.80M^{-0.47}$ in its variance estimator, and reverse the sign of the statistic so that rejection of $H_0$ is based on large values.
For $T_{\mathrm{CPR}}$ \citep{CorsiPirinoReno2010}, we use thresholds set at three estimated local standard deviations and a Gaussian-kernel local variance estimator with bandwidth 25.
For $T_{\mathrm{LM}}$ \citep{LeeMykland2008}, we use a local volatility estimation window of $\operatorname{round}(4M^{0.6})$ returns (120 at $M=288$).
Since the simulated returns cover only a single observation period, no presample is available for $T_{\mathrm{LM}}$.
Consequently, testing is restricted to $y_{120},\ldots,y_M$, which cover approximately 62.4\% of the possible jump times in $[0.15,0.85]$ under the rare/large-jump alternative.

\begin{figure}[t]
\centering
\includegraphics[width=\textwidth]{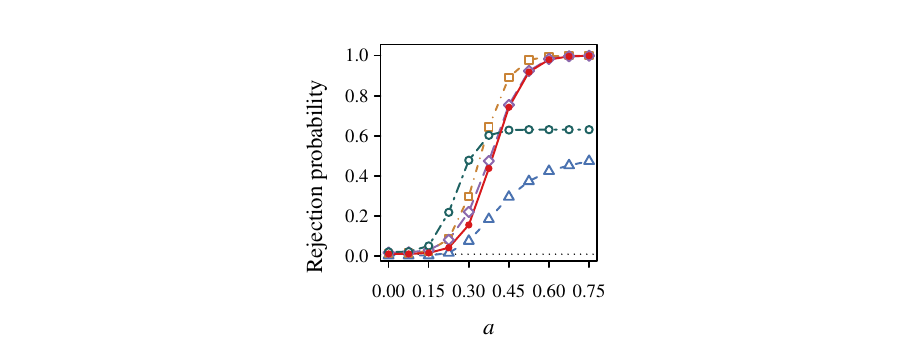}\vspace{-0.4cm}
\caption{Nominal rejection probabilities at the 1\% size under the rare/large alternative.
All other settings and plotting conventions are as in Figure~\ref{fig:sim-clean-power-competitors-adjusted}.}
\label{fig:supp-clean-power-rare-large-nominal}
\end{figure}

Note that $T_{\mathrm{JO}}$ is a two-sided test, whereas all other competing tests are one-sided.
Therefore, we use $|T_{\mathrm{JO}}|$ in the following comparisons so that rejection is based on an upper-tail critical value.
In Figures~\ref{fig:sim-clean-power-competitors-adjusted} and~\ref{fig:supp-clean-power-rare-large-nominal}, we compare the size-adjusted and nominal power, respectively, of the proposed test and the competing tests.
Here, size-adjusted power refers to the empirical probability, under the alternative, that a test statistic exceeds its empirical 99\% quantile under the null.

\begin{figure}[t]
\centering
\includegraphics[width=\textwidth]{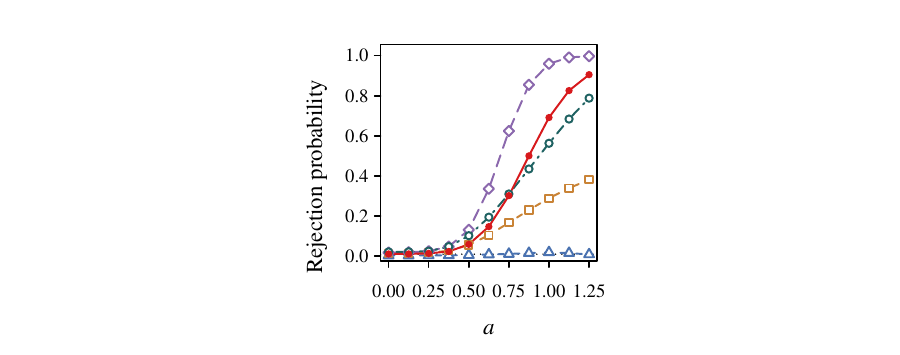}\vspace{-0.4cm}
\caption{Nominal rejection probabilities at the 1\% size under the frequent/small alternative.
The jump setting and grid are as in Figure~\ref{fig:sim-clean-power}(b), while all other settings and plotting conventions are as in Figure~\ref{fig:sim-clean-power-competitors-adjusted}.}
\label{fig:supp-clean-power-frequent-small-nominal}
\end{figure}

\begin{figure}[t]
\centering
\includegraphics[width=\textwidth]{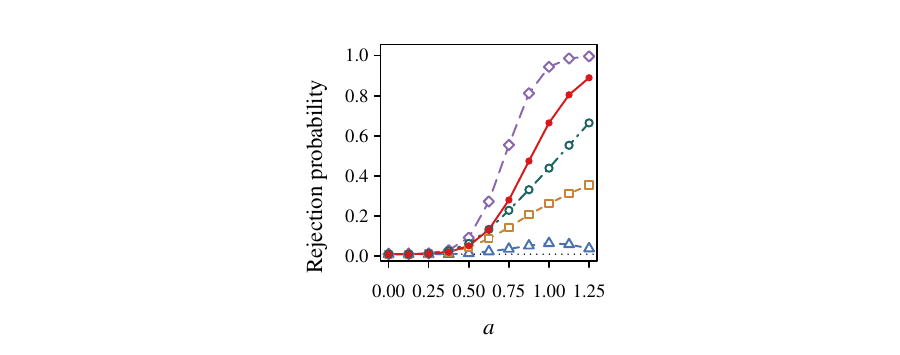}\vspace{-0.7cm}
\caption{Size-adjusted rejection probabilities at the 1\% size when jumps are frequent but small.
The jump setting and all plotting conventions are as in Figure~\ref{fig:supp-clean-power-frequent-small-nominal}.}
\label{fig:supp-clean-power-frequent-small-adjusted}
\end{figure}

Figures~\ref{fig:supp-clean-power-rare-large-nominal}, \ref{fig:supp-clean-power-frequent-small-nominal}, and~\ref{fig:supp-clean-power-frequent-small-adjusted} present the empirical power curves of the proposed test and the competing tests.
The size-adjusted results allow comparisons that account for differences in empirical size across tests.

\begin{figure}[!tbp]
\centering
\includegraphics[width=\textwidth]{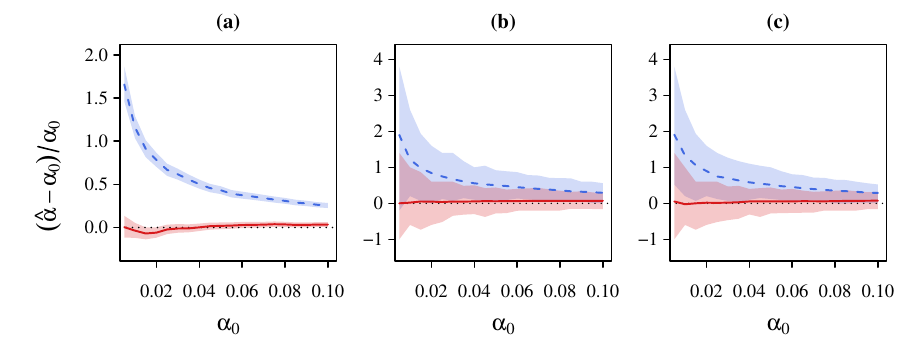}\vspace{-0.7cm}
\caption{Relative size error for $T_{6,*}^{(3),\mathrm{tr}}(g_2)$ (red) and $T_0^{\mathrm{tr}}(g_2)$ (blue).
Panels and shading conventions are as in Figure~\ref{fig:sim-clean-size-standard}.}
\label{fig:supp-size-g2}
\end{figure}

\begin{figure}[!tbp]
\centering
\includegraphics[width=\textwidth]{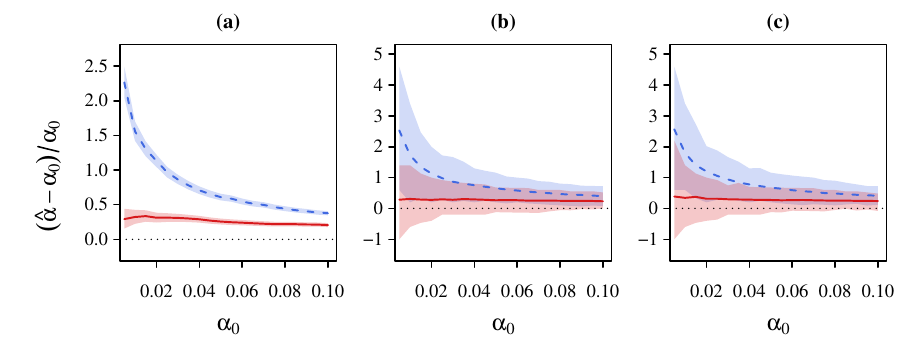}
\caption{Relative size error for $T_{6,*}^{(3),\mathrm{tr}}(g_3)$ (red) and $T_0^{\mathrm{tr}}(g_3)$ (blue).
Panels and shading conventions are as in Figure~\ref{fig:sim-clean-size-standard}.}
\label{fig:supp-size-g3}
\end{figure}
Figures~\ref{fig:supp-size-g2} and~\ref{fig:supp-size-g3} repeat the size comparison in Figure~\ref{fig:sim-clean-size-standard} with $g_2$ and $g_3$, respectively, using the corresponding value of $b$.
For both $g_2$ and $g_3$, the self-normalized test has relative size errors substantially closer to zero than its conventional counterpart across the three volatility scenarios.
The gap in empirical size accuracy is particularly pronounced for small values of $\alpha_0.$

\begin{figure}[!tbp]
\centering
\includegraphics[width=\textwidth]{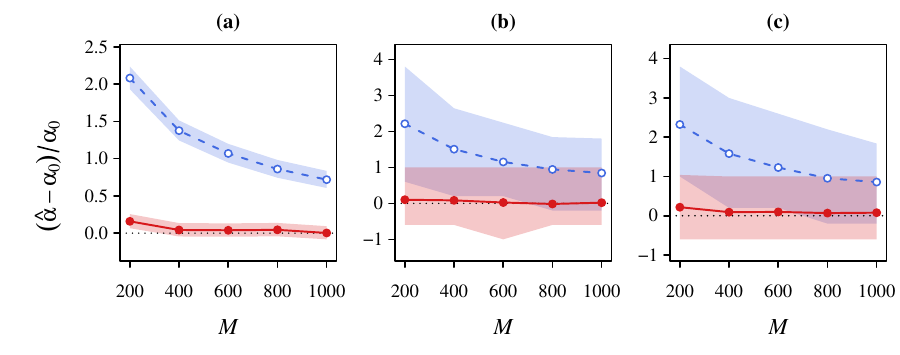}
\caption{Relative size error at the 1\% nominal size against $M$ for $T_{6,*}^{(3),\mathrm{tr}}(g_1)$~(\SNProposedCaptionKey) and $T_0^{\mathrm{tr}}(g_1)$~(\SNConventionalCaptionKey).
Points mark the simulated values of $M$.
Panels, curves, and shading conventions are as in Figure~\ref{fig:sim-clean-size-standard}.}
\label{fig:supp-size-by-M}
\end{figure}

To examine how sampling frequency affects empirical size accuracy, we extend the comparison in Figure~\ref{fig:sim-clean-size-standard} by fixing $\alpha_0=0.01$ and $M\in\{200,\allowbreak400,\allowbreak600,\allowbreak800,\allowbreak1000\}$, with $\ell=6$ fixed and all other settings unchanged.
Figure~\ref{fig:supp-size-by-M} shows that $T_{6,*}^{(3),\mathrm{tr}}(g_1)$ has relative size errors closer to zero than $T_0^{\mathrm{tr}}(g_1)$ across all three volatility scenarios.
For all tests considered, their empirical size accuracy improves as $M$ grows.
For all values of $M$ considered, our proposed test shows more accurate empirical size than $T_0^{\mathrm{tr}}(g_1)$.

\begin{figure}[!tbp]
\centering
\includegraphics[width=\textwidth]{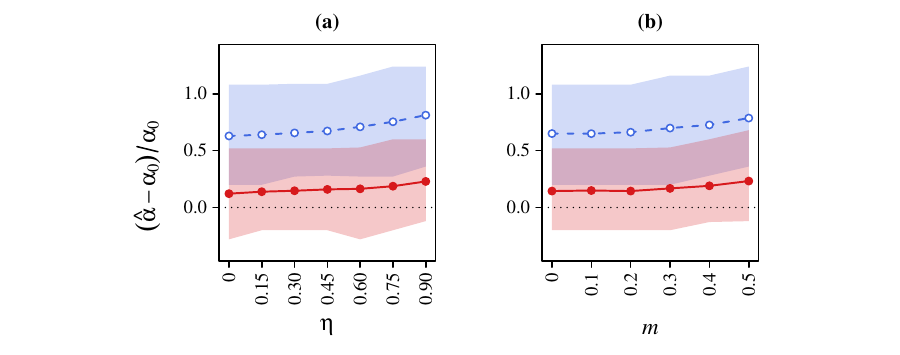}
\caption{Relative size error at the 5\% nominal size for $T_{6,*}^{(3),\mathrm{tr}}(g_1)$~(\SNProposedCaptionKey) and $T_0^{\mathrm{tr}}(g_1)$~(\SNConventionalCaptionKey) against (a) the volatility-of-volatility coefficient $\eta$ and (b) the absolute log-volatility jump magnitude $m$.
Points mark the simulated parameter values.
Curves and shading conventions are as in Figure~\ref{fig:sim-clean-size-standard}(b)--(c).}
\label{fig:supp-volatility-sensitivity}
\end{figure}

We next vary the volatility parameters in \eqref{eq:simulation-log-volatility} for the size comparison in Figure~\ref{fig:sim-clean-size-standard}, with $\alpha_0=0.05$.
In panel~(a), we take $J^\sigma\equiv0$ and replace the coefficient $0.30$ of $\dd B_t$ by
$\eta\in\{0,\allowbreak0.15,\allowbreak0.30,\allowbreak0.45,
\allowbreak0.60,\allowbreak0.75,\allowbreak0.90\}$.
In panel~(b), we keep $\eta=0.30$ and replace the jump sizes $\pm0.20$ of $J^\sigma$ by $\pm m$ with $m\in\{0,0.1,0.2,0.3,0.4,0.5\}$.
Figure~\ref{fig:supp-volatility-sensitivity} shows smaller size distortion for the proposed test throughout the displayed grids.
Both procedures over-reject, with size errors generally increasing as $\eta$ or $m$ increases.

\FloatBarrier

We then investigate the effect of microstructure-noise intensity on empirical size.
We consider a generalization of case~(ii) of Table~\ref{tab:microstructure-noise-simulation}.
All other configurations remain unchanged.
Table~\ref{tab:supp-noise-intensity} shows that the self-normalized tests have empirical size closer to the nominal size $\alpha_0=0.01$ than $T_0^{\mathrm{tr}}(g_1)$.
\begin{table}[!ht]
\centering
\setlength{\tabcolsep}{7pt}\renewcommand{\arraystretch}{1.05}
\begin{tabular}{@{}lccccccc@{}}\hline
$c_\omega$ & $\rho$ & $T_0^{\mathrm{tr}}(g_1)$ &
\multicolumn{5}{c}{$T_{\ell,*}^{(3),\mathrm{tr}}(g_1)$}\\
\cline{4-8}
& & & $\ell=2$ & $\ell=3$ & $\ell=4$ & $\ell=5$ & $\ell=6$\\\hline
$0$ & --- & 2.83\% & 1.08\% & 1.07\% & 1.05\% & 1.12\% & 1.14\%\\
$0.025$ & $-0.25$ & 2.86\% & 0.99\% & 1.05\% & 0.99\% & 1.07\% & 1.21\%\\
 & $0$ & 2.71\% & 1.09\% & 1.01\% & 0.99\% & 1.03\% & 1.16\%\\
 & $+0.25$ & 2.45\% & 0.96\% & 0.89\% & 0.91\% & 0.95\% & 1.09\%\\
$0.05$ & $-0.25$ & 2.89\% & 1.10\% & 0.99\% & 1.05\% & 1.13\% & 1.21\%\\
 & $0$ & 2.56\% & 1.09\% & 0.90\% & 0.94\% & 1.01\% & 1.06\%\\
 & $+0.25$ & 2.09\% & 0.85\% & 0.81\% & 0.78\% & 0.83\% & 0.89\%\\
$0.075$ & $-0.25$ & 2.69\% & 1.00\% & 0.95\% & 1.00\% & 1.10\% & 1.12\%\\
 & $0$ & 2.26\% & 0.92\% & 0.89\% & 0.91\% & 0.92\% & 0.95\%\\
 & $+0.25$ & 1.69\% & 0.82\% & 0.71\% & 0.66\% & 0.72\% & 0.73\%\\\hline
\end{tabular}
\caption{Finite-sample type-I error under Gaussian microstructure noise with variance  $\omega_M^2=c_\omega M^{-3/2}$ with $c_\omega\in\{0,0.025,0.05,0.075\}$ and $\rho\in\{-0.25,0,0.25\}$ for $c_\omega>0$.}
\label{tab:supp-noise-intensity}
\end{table}

\FloatBarrier

\section{Theory and proofs}
\label{sec:supp-unthresholded}

The preliminary calculations in Section~\ref{subsec:supp-preliminaries} derive technical results to simplify the proofs of joint stable limits in Section~\ref{subsec:supp-stable-limits}.
That section derives the joint stable limits and uses the compound-symmetric covariance structure in \eqref{eq:SigmaQ-final} to construct differenced statistics.
Section~\ref{sec:supp-threshold} proves the asymptotic properties of the proposed tests and compares their local power.
Section~\ref{subsec:supp-microstructure-noise} treats microstructure noise.

\subsection{Preliminaries}
\label{subsec:supp-preliminaries}

Under the following assumption, the unthresholded joint stable limit in Theorem~\ref{thm:joint-stable-CLT-voljump} and the joint stable limit in Theorem~\ref{thm:standard-multipower-variation} follow from Theorem~11.2.1(b) of \citet{JacodProtter2012} once its hypotheses are verified for a suitable $F$.
That theorem requires a continuous price process.

\begin{assumption}[Continuous price process]
\label{ass:continuous-price}
The log-price process is continuous and admits
\[
y^*(t)=y^*(0)+\int_0^t\mu_s^y\dd s+\int_0^t\sigma_s\dd W_s,
\qquad t\in[0,1].
\]
Here $W$ is an $(\mathcal F_t)$-Brownian motion.
The process $\mu^y$ is progressively measurable, c\`adl\`ag, and locally bounded.
The process $\sigma$ satisfies Assumption~\ref{ass:volatility-regularity}.
\end{assumption}

Under Assumption~\ref{ass:continuous-price}, Assumption~\ref{ass:price-model} holds with $\delta\equiv0$, and $\Omega_0=\Omega$.

Recall that $K=(n-1)L$.
Under Assumption~\ref{ass:continuous-price}, suppose $F:\R^K\to\R^{1+\ell}$ satisfies the function conditions of Theorem~11.2.1(b) of \citet{JacodProtter2012}.
Then, after proper centering and scaling, that theorem yields a central limit theorem for
\[M^{-1}\sum_{i=1}^{M-K+1}F(M^{1/2}y_i,\ldots,M^{1/2}y_{i+K-1}).\]
  Conditionally on $\mathcal F$, the stable limit is centered Gaussian, and the covariance of its $i$-th and $j$-th coordinates is $\int_0^1R_{\sigma_s^2}(F_i,F_j)\dd s$ \citep[eqn.~(11.2.11)]{JacodProtter2012}, where $R_a$ is defined in \eqref{eq:Ra-explicit-definition}.
  We will show that $(Q_k^{(n)},\hat\theta_g)$ also admits this form with a specific choice of $F$ up to a negligible error.

As a first step, Proposition~\ref{prop:JP-common-hypotheses} gives the stable central limit theorem for this choice of $F$ by verifying the regularity conditions.
Lemma~\ref{lem:boundary-negligible} shows that $\hat\theta_g-\widetilde\theta_{M,g}=O_P(M^{-1})$, which is negligible at the $M^{1/2}$ scale of the stable central limit theorem.
Then, Lemma~\ref{lem:cross-kernel} rewrites $R_1$ as a finite covariance sum, which is a useful representation for evaluating the explicit forms of $\sigma_g^2$, $C_{g,n}$, and $\Sigma_{\mathbb Q}^{(n)}$.

Proposition~\ref{prop:JP-common-hypotheses} is established under Assumption~\ref{ass:continuous-price} and therefore does not directly yield a stable limit in restriction to $\Omega_0$ under Assumption~\ref{ass:price-model}.
Although $y^*$ is continuous on $\Omega_0$, conditioning on this event need not preserve the Brownian property of $W$.
Instead, Lemma~\ref{lem:no-jump-reduction} bypasses this issue by constructing, for each $q\ge1$, a continuous process $y^{(q)}$ on the original filtered probability space that satisfies Assumption~\ref{ass:continuous-price}.
In particular, it implies that
\[
\pr\left(
\Omega_0\cap
\left\{\sup_{t\in[0,1]}
|y^{(q)}(t)-y^*(t)|>0\right\}
\right)
\le\frac{2}{q}\longrightarrow0.
\]
Thus, $y^{(q)}$ agrees with $y^*$ on $\Omega_0$ outside an event whose probability tends to zero.
We can hence apply the continuous-price CLT to $y^{(q)}$ under $\pr$, where $W$ remains Brownian.
We can then use Lemma~\ref{lem:no-jump-reduction} to obtain the corresponding stable limit for $y^*$ in restriction to $\Omega_0$, provided the stable limit is the same for every $q$.

\subsubsection{Representation of the statistics}

Recall that $\mathbb Q_\ell^{(n)}(b)
=(Q_b^{(n)},\ldots,Q_{b+\ell-1}^{(n)})^\T
$.
We can rewrite $Q_k^{(n)}$, for $k=b,\ldots,b+\ell-1$, as
\begin{equation*}
Q_k^{(n)}
=
\frac1M\sum_{i=1}^{M-K+1}
 F_{Q,k}(M^{1/2}\,y_i,\dots,M^{1/2}\,y_{i+K-1}),
\qquad
F_{Q,k}(x):=\mu_{2/n}^{-n}\prod_{r\in \mathcal I_k}|x_r|^{2/n}.
\end{equation*}
Take $F_\theta$ as
\begin{equation*}
F_\theta(x_1,\ldots,x_K)=x_1^2-g(x_1,\ldots,x_{b+1}).
\end{equation*}
We then define
\[
\widetilde\theta_{M,g}
:=
M^{-1}\sum_{i=1}^{M-K+1}
 F_\theta(M^{1/2}\,y_i,\dots,M^{1/2}\,y_{i+K-1}).
\]
It differs from \(\hat\theta_g=\hat{\QV}-\hat{\IV}_g\) only at the right boundary:
\begin{equation}
\label{eq:theta-tail}
\hat{\theta}_g-\widetilde\theta_{M,g}
=
\frac1M\sum_{i=M-K+2}^{M-b}
 \{(M^{1/2}\,y_i)^2-g(M^{1/2}\,y_i,\dots,M^{1/2}\,y_{i+b})\}
+
\frac1M\sum_{i=M-b+1}^{M}(M^{1/2}\,y_i)^2.
\end{equation}
We will show that this difference is of size $O_P(M^{-1})$ and hence negligible.

\subsubsection{Stable central limit theorem for \texorpdfstring{$F$}{F}}

Define
\[
F:=(F_\theta,F_{Q,b},\dots,F_{Q,b+\ell-1}):\R^K\to\R^{1+\ell}.
\]

For an integer \(k\ge2\) and \(a\ge0\), let \(U_1,U_2,\ldots\simIID \Normal(0,a)\).
Put
\[
 \mathcal U_{k-1}:=\sigma(U_1,\ldots,U_{k-1}),
 \qquad
 \mathcal U_k:=\sigma(U_1,\ldots,U_k).
\]
For Borel-measurable \(H,G:\R^k\to\R\) such that \(H(U_1,\ldots,U_k)\) and \(G(U_1,\ldots,U_k)\) are square-integrable, set
\[
 A_j^H:=H(U_{k-j},\ldots,U_{2k-j-1}),
 \qquad
 B_j^G:=G(U_{k-j},\ldots,U_{2k-j-1}),
 \qquad 0\le j<k,
\]
and define
\begin{align}
\label{eq:Ra-explicit-definition}
R_a(H,G)
&:=\sum_{j,j'=0}^{k-1}
 \E\!\left\{
  \left[\E(A_j^H\mid\mathcal U_k)
       -\E(A_j^H\mid\mathcal U_{k-1})\right]
  \E(B_{j'}^G\mid\mathcal U_k)
 \right\}.
\end{align}
By the tower property,
\[
\E\!\left\{
\E(A_j^H\mid\mathcal U_{k-1})
\E(B_{j'}^G\mid\mathcal U_k)
\right\}
=
\E\!\left\{
\E(A_j^H\mid\mathcal U_{k-1})
\E(B_{j'}^G\mid\mathcal U_{k-1})
\right\}.
\]
Hence \eqref{eq:Ra-explicit-definition} agrees with \citet[eqn.~\textup{(11.2.4)}]{JacodProtter2012}.
For each $j,j'$, conditional-expectation contraction and Cauchy--Schwarz give
\[
\E\!\left|
\{\E(A_j^H\mid\mathcal U_k)-\E(A_j^H\mid\mathcal U_{k-1})\}
\E(B_{j'}^G\mid\mathcal U_k)
\right|
\le 2\{\E|A_j^H|^2\}^{1/2}\{\E|B_{j'}^G|^2\}^{1/2}<\infty.
\]

\begin{proposition}[Stable CLT for $F$]
\label{prop:JP-common-hypotheses}
Under Assumptions~\ref{ass:continuous-price} and~\ref{ass:g-structural}, as $M\to\infty$,
\[
\begin{aligned}
&M^{1/2}\left[
 M^{-1}\sum_{i=1}^{M-K+1}
 F(M^{1/2}y_i,\ldots,M^{1/2}y_{i+K-1})\right.\\
&\hspace{3em}\left.
 -\int_0^1\widetilde\E\left\{
 F(\sqrt{\sigma_s^2}U_1,\ldots,\sqrt{\sigma_s^2}U_K)
 \mid\mathcal F\right\}\dd s
 \right]
 \inST \mathcal V(F,y^*),
\end{aligned}
\]
where on an extension of the underlying probability space, $U_1,\ldots,U_K\simIID\Normal(0,1)$ are independent of $\mathcal F$.
Conditionally on $\mathcal F$, $\mathcal V(F,y^*)$ is a mean-zero multivariate normal random variable with
\[
\widetilde\E\left\{
\mathcal V(F_i,y^*)\mathcal V(F_j,y^*)\mid\mathcal F
\right\}
=\int_0^1R_{\sigma_s^2}(F_i,F_j)\dd s,
\qquad i,j=1,\ldots,\ell+1.
\]
Here $R_a$ is defined by \eqref{eq:Ra-explicit-definition} with $k=K$.
\end{proposition}

\begin{proof}
The stated stable convergence follows from Theorem~11.2.1(b) of \citet{JacodProtter2012} once its hypotheses are verified.
Since the theorem is formulated on $\mathbb R_+$, extend $\mu_t^y=\mu_1^y$ and $\sigma_t=\sigma_1$ for $t>1$, and continue $y^*$ by its representation in Assumption~\ref{ass:continuous-price}.
The volatility dynamics in \eqref{eq:volatility-decomposition} are stopped at time~1, with the coefficient extensions preserving their stated regularity conditions.
The extended processes satisfy Assumptions~(H) and~(K) of \citet{JacodProtter2012} by Assumption~\ref{ass:continuous-price} and \eqref{eq:volatility-decomposition}.
In the univariate setting, the strict positivity of $\sigma_t^2$ and $\sigma_{t-}^2$ implies Assumption~(K') of \citet{JacodProtter2012}.

By Assumption~\ref{ass:g-structural}(i), $F_\theta(-x)=F_\theta(x)$.
Also, we have $F_{Q,k}(-x)=F_{Q,k}(x)$.
Thus $F(-x)=F(x)$ and hence $F$ satisfies the global-evenness condition~(11.2.5) of \citet{JacodProtter2012}.

For convenience, we state conditions~(11.2.6)--(11.2.9) of \citet{JacodProtter2012} specialized to $F:\R^K\to\R^{1+\ell}$ with $q'=0$, using our notation.
It remains to show that \eqref{eq:JP-regularity-1}--\eqref{eq:JP-regularity-4} hold:

There exists a nonempty finite union \(B\subset\mathbb R^K\) of affine hyperplanes such that, for \(j=1,\ldots,\ell+1\),
\begin{equation}
F_j \text{ is continuous on } \mathbb R^K
\quad\text{and}\quad
F_j\in C^1(B^c).
\label{eq:JP-regularity-1}
\end{equation}
Moreover, for some \(C>0\), \(p\ge0\), and \(w\in(0,1]\),
\begin{equation}
\|F(z)\|
\le
C(1+\|z\|^p),
\qquad z\in\mathbb R^K.
\label{eq:JP-regularity-2}
\end{equation}
For \(j=1,\ldots,\ell+1\),
\begin{equation}
\|\nabla F_j(z)\|
\le
C(1+\|z\|^p)
\left\{
1+\operatorname{dist}(z,B)^{-(1-w)}
\right\},
\qquad z\in B^c.
\label{eq:JP-regularity-3}
\end{equation}
Finally, for \(j=1,\ldots,\ell+1\),
\begin{equation}
\|\nabla F_j(z+y)-\nabla F_j(z)\|
\le
C\|y\|
\left\{
1+\operatorname{dist}(z,B)^{-(2-w)}
\right\}
(1+\|z\|^p),
\label{eq:JP-regularity-4}
\end{equation}
whenever $z\in B^c$ and $\|y\|\le1\wedge\operatorname{dist}(z,B)/2$.

Indeed, they can be proved by taking
\[
\begin{aligned}
B:=\widetilde B_g\cup\bigcup_{k=b}^{b+\ell-1}
     \bigcup_{r\in\mathcal I_k}\{x:x_r=0\},
     \qquad
\widetilde B_g&:=\{x\in\mathbb R^K:(x_1,\ldots,x_{b+1})\in B_g\},
\end{aligned}
\]
with $\bar p=2\vee p_g$ and $\bar w=w_g\wedge(2/n)$.
Recall from Assumption~\ref{ass:g-structural}(iv) that $|g(z)|\le C(1+\|z\|^{p_g})$ for $z\notin B_g$.
Since $B_g$ is a finite union of affine hyperplanes, $B_g^c$ is a dense set.
Therefore, for any $z\in B_g$, we can pick a sequence $z_m\in B_g^c$ such that $z_m\to z$.
Then the continuity of $g$ and of the norm together gives
\[
|g(z)|=\lim_m|g(z_m)|\le \lim_mC(1+\|z_m\|^{p_g}) \le C(1+\|z\|^{p_g}).
\]
Therefore, this bound holds over all of $\mathbb R^{b+1}$.

We first notice that \(F_\theta(x)=x_1^2-g(x_1,\dots,x_{b+1})\) inherits continuity, \(C^1\)-regularity off \(\widetilde B_g\), and conditions~\textup{(11.2.7)--(11.2.9)} of \citet{JacodProtter2012} from Assumption~\ref{ass:g-structural}(iv), after replacing \(p_g\) by \(2\vee p_g\).

From $F_{Q,k}(x)=\mu_{2/n}^{-n}\prod_{r\in\mathcal I_k}|x_r|^{2/n}$, the coordinates indexed by $\mathcal I_k$ have exponent $2/n$, and the other coordinates have exponent zero.
Applying \citet[Example~11.2.2]{JacodProtter2012} to the function of the $n$ coordinates indexed by $\mathcal I_k$, on which $F_{Q,k}$ depends, gives continuity, \(C^1\)-regularity away from \(B_{Q,k}:=\cup_{r\in \mathcal I_k}\{x:x_r=0\}\), and \citet[conditions~\textup{(11.2.7)--(11.2.9)}]{JacodProtter2012} with \(p=2\) and \(w=2/n\).

Notice that the choices of $(p,w)$ used in the argument above need not be the same for $F_\theta$ and $F_{Q,k}$.
We next show that the same argument holds for both functions when $p=\bar p$ and $w=\bar w$, so that $F$ satisfies \eqref{eq:JP-regularity-2}--\eqref{eq:JP-regularity-4}.

The finite union \(B=\widetilde B_g\cup_kB_{Q,k}\) is a nonempty finite union of affine hyperplanes, and every component is \(C^1\) on \(B^c\).
Write \(F_1=F_\theta\) and \(F_r=F_{Q,b+r-2}\), \(r=2,\dots,1+\ell\), for the components of \(F\).
Set \((p_1,w_1,B_1)=(2\vee p_g,w_g,\widetilde B_g)\) and \((p_r,w_r,B_r)=(2,2/n,B_{Q,b+r-2})\), \(r=2,\ldots,1+\ell\).
Thus $B=\bigcup_{r=1}^{1+\ell}B_r$, $\bar p=\max_r p_r$, and $\bar w=\min_r w_r$.
It remains to verify \eqref{eq:JP-regularity-2}--\eqref{eq:JP-regularity-4} with $B,\bar p,\bar w$.
For every $r$, we have $p_r\le\bar p$, $B_r\subset B$, and $\bar w\le w_r$.
In particular, $\operatorname{dist}(x,B)\le\operatorname{dist}(x,B_r)$.

For each $r$, $F_r$ satisfies conditions \eqref{eq:JP-regularity-2}--\eqref{eq:JP-regularity-4} with $(B_r,p_r,w_r)$ and a constant $C_r$.
For $r=1,\ldots,\ell+1$ and $j=1,2$, we have
\begin{equation}
\label{eq:JP-component-bounds}
\begin{aligned}
1+\|x\|^{p_r}
&\le2(1+\|x\|^{\bar p}),
&&x\in\mathbb R^K,\\
1+\operatorname{dist}(x,B_r)^{-(j-w_r)}
&\le2\bigl\{1+\operatorname{dist}(x,B)^{-(j-\bar w)}\bigr\},
&&x\in B^c.
\end{aligned}
\end{equation}
The first inequality follows from $p_r\le\bar p$.
Notice that $0\le j-w_r\le j-\bar w$ implies $t^{-(j-w_r)}\le1+t^{-(j-\bar w)}$ for $t>0$.
The second inequality follows by applying this with $t=\operatorname{dist}(x,B)>0$ and using the fact that $\operatorname{dist}(x,B_r)\ge\operatorname{dist}(x,B)$.

By \eqref{eq:JP-component-bounds}, $\|F(x)\|\le\sum_r|F_r(x)|\le2\sum_rC_r(1+\|x\|^{\bar p})$ for $x\in\mathbb R^K$.
For $x\in B^c\subset B_r^c$, \eqref{eq:JP-component-bounds} with $j=1$ gives
\[
\begin{aligned}
\|\nabla F_r(x)\|
&\le C_r(1+\|x\|^{p_r})\{1+\operatorname{dist}(x,B_r)^{-(1-w_r)}\}\\
&\le4C_r(1+\|x\|^{\bar p})\{1+\operatorname{dist}(x,B)^{-(1-\bar w)}\}.
\end{aligned}
\]
Since $B_r\subset B$, we have
\[
\|y\|\le 1\wedge\frac{\operatorname{dist}(x,B)}2
\quad\Longrightarrow\quad
\|y\|\le 1\wedge\frac{\operatorname{dist}(x,B_r)}2.
\]

Recall that $F_r$ satisfies \eqref{eq:JP-regularity-4} with $p=p_r$, $w=w_r$, and $B_r$.
For $x\in B^c$ and $\|y\|\le1\wedge\operatorname{dist}(x,B)/2$, using \eqref{eq:JP-component-bounds} with $j=2$ gives
\[
\begin{aligned}
\|\nabla F_r(x+y)-\nabla F_r(x)\|
&\le C_r\|y\|\{1+\operatorname{dist}(x,B_r)^{-(2-w_r)}\}(1+\|x\|^{p_r})\\
&\le4C_r\|y\|\{1+\operatorname{dist}(x,B)^{-(2-\bar w)}\}(1+\|x\|^{\bar p}).
\end{aligned}
\]
Hence \eqref{eq:JP-regularity-2}--\eqref{eq:JP-regularity-4} hold with $B$, $\bar p$, $\bar w$, and $C=4\sum_rC_r$.

Each $F_r$ is continuous, and $B_r\subset B$ implies $B^c\subset B_r^c$, so its $C^1$-regularity on $B_r^c$ gives $C^1$-regularity on $B^c$.
Thus \eqref{eq:JP-regularity-1} also holds.
Therefore, all conditions of Theorem~11.2.1(b) of \citet{JacodProtter2012} hold, which completes the proof of this proposition.
\end{proof}

\subsubsection{Boundary terms and the covariance quantity \texorpdfstring{$R_a$}{R\_a}}

A Borel-measurable function \(G:\R^k\to\R\) is said to be of polynomial growth if there exist \(C>0\) and \(p\ge0\) such that \(|G(x)|\le C(1+\|x\|^p)\) for all \(x\in\R^k\).
Also, for a process \(X\), denote \(\Delta_i^MX:=X_{i/M}-X_{(i-1)/M}\).

\begin{lemma}[Negligibility of boundary terms]
\label{lem:boundary-negligible}
Under Assumption~\ref{ass:continuous-price}, let $R\ge1$ be fixed.
For $r=1,\ldots,R$, let $G_r:\R^{k_r}\to\R$ be Borel-measurable and of polynomial growth.
For each \(M\), let \(\mathcal B_{r,M}\subseteq\{1,\ldots,M-k_r+1\}\) be deterministic and satisfy \(\sup_M|\mathcal B_{r,M}|<\infty\), \(r=1,\ldots,R\).
Then
\[
\max_{1\le r\le R}\left|
 M^{-1}\sum_{i\in\mathcal B_{r,M}}
 G_r(M^{1/2}\,y_i,\dots,M^{1/2}\,y_{i+k_r-1})\right|=O_P(M^{-1}).
\]
\end{lemma}

\begin{proof}
For \(q\ge1\), let
\[
\tau_q:=\inf\{t\ge0:|\mu_t^y|+|\sigma_t|>q\}
\]
and set
\[
y_t^{*,q}:=y_0^*
+\int_0^t\ind(s<\tau_q)\mu_s^y\dd s
+\int_0^t\ind(s<\tau_q)\sigma_s\dd W_s.
\]
By construction, the two integrands are bounded by \(q\), and \(y^{*,q}=y^*\) on \([0,1]\) whenever \(\tau_q>1\).
Since the paths of \(\mu^y\) and \(\sigma\) are locally bounded, \(\pr(\tau_q\le1)\to0\) as \(q\to\infty\).

For every \(i=1,\ldots,M\),
\[
\Delta_i^M y^{*,q}
=
\int_{(i-1)/M}^{i/M}\ind(s<\tau_q)\mu_s^y\,\dd s
+
\int_{(i-1)/M}^{i/M}\ind(s<\tau_q)\sigma_s\,\dd W_s.
\]
Since \(|\mu_s^y|\le q\) and \(|\sigma_s|\le q\) for \(s<\tau_q\),
\(M^{1/2}|\int_{(i-1)/M}^{i/M}\ind(s<\tau_q)\mu_s^y\,\dd s|
\le qM^{-1/2}\).
For every fixed \(p>0\), the Burkholder--Davis--Gundy inequality gives
\[
\begin{aligned}
\E\left|
M^{1/2}\int_{(i-1)/M}^{i/M}\ind(s<\tau_q)\sigma_s\,\dd W_s
\right|^p
\le C_pM^{p/2}\E\left(
\int_{(i-1)/M}^{i/M}\ind(s<\tau_q)\sigma_s^2\,\dd s
\right)^{p/2}
\le C_pq^p.
\end{aligned}
\]
As \(|a+b|^p\le 2^{\max\{p-1,0\}}(|a|^p+|b|^p)\) for any \(p>0\), combining the bounds for those two integrals above yields
\begin{equation}
\label{eq:boundary-increment-moment}
\sup_{M\ge\max_{1\le r\le R}k_r}\sup_{1\le i\le M}
\E|M^{1/2}\Delta_i^M y^{*,q}|^p\le C_{p,q}<\infty,
\end{equation}
where \(C_{p,q}\) does not depend on \(M\) or \(i\).

For each \(r\), as $G_r$ is of polynomial growth, there exist \(A_r>0\) and \(p_r\ge0\) such that \(|G_r(x)|\le A_r(1+\|x\|^{p_r})\) for all \(x\in\R^{k_r}\).
For fixed \(k_r\),
\(\|(x_1,\ldots,x_{k_r})\|^{p_r}
\le C(1+\sum_{j=1}^{k_r}|x_j|^{p_r})\) when \(p_r>0\).
Thus it follows from \eqref{eq:boundary-increment-moment} that for each fixed \(q\),
\[
\sup_{M\ge\max_{1\le s\le R}k_s}\sup_{1\le i\le M-k_r+1}
\E\left|G_r(M^{1/2}\Delta_i^M y^{*,q},\ldots,
M^{1/2}\Delta_{i+k_r-1}^M y^{*,q})\right|<\infty.
\]
For \(p_r=0\), it follows directly from \(|G_r(x)|\le 2A_r\).

By the triangle inequality, for every \(r=1,\ldots,R\),
\[
\begin{aligned}
&\sup_{M\ge\max_{1\le s\le R}k_s}
\E\left|\sum_{i\in\mathcal B_{r,M}}
G_r(M^{1/2}\Delta_i^M y^{*,q},\ldots,
M^{1/2}\Delta_{i+k_r-1}^M y^{*,q})\right|\\
&\qquad\le \sup_M|\mathcal B_{r,M}|
\sup_{\substack{M\ge\max_{1\le s\le R}k_s\\1\le i\le M-k_r+1}}
\E\left|G_r(M^{1/2}\Delta_i^M y^{*,q},\ldots,
M^{1/2}\Delta_{i+k_r-1}^M y^{*,q})\right|<\infty.
\end{aligned}
\]
Since \(\max_{1\le r\le R}|x_r|\le\sum_{r=1}^R|x_r|\) and \(R\) is a finite positive integer, we have
\begin{equation}
\label{eq:boundary-stopped-sum-moment}
\sup_{M\ge\max_{1\le s\le R}k_s}
\E\left[\max_{1\le r\le R}\left|
\sum_{i\in\mathcal B_{r,M}}
G_r(M^{1/2}\Delta_i^M y^{*,q},\ldots,
M^{1/2}\Delta_{i+k_r-1}^M y^{*,q})\right|\right]<\infty.
\end{equation}
For any \(\varepsilon>0\), choose \(q\) such that \(\pr(\tau_q\le1)<\varepsilon/2\).
For this fixed \(q\), \eqref{eq:boundary-stopped-sum-moment} and Markov's inequality imply that we can pick a constant \(C<\infty\) such that
\[
\pr\!\left\{
\max_{1\le r\le R}\left|
\sum_{i\in\mathcal B_{r,M}}
G_r(M^{1/2}\Delta_i^M y^{*,q},\ldots,
M^{1/2}\Delta_{i+k_r-1}^M y^{*,q})\right|>C
\right\}<\varepsilon/2
\]
for every \(M\ge\max_{1\le r\le R}k_r\).
Notice that on \(\{\tau_q>1\}\), \(y^{*,q}=y^*\) on \([0,1]\), and hence \(\Delta_i^M y^{*,q}=y_i\) for every \(i=1,\ldots,M\).
It follows that
\[
\begin{aligned}
&\pr\!\left\{
\max_{1\le r\le R}\left|
\sum_{i\in\mathcal B_{r,M}}
G_r(M^{1/2}y_i,\ldots,M^{1/2}y_{i+k_r-1})\right|>C
\right\}\\
&\qquad\le\pr(\tau_q\le1)
+\pr\!\left\{
\max_{1\le r\le R}\left|
\sum_{i\in\mathcal B_{r,M}}
G_r(M^{1/2}\Delta_i^M y^{*,q},\ldots,
M^{1/2}\Delta_{i+k_r-1}^M y^{*,q})\right|>C
\right\}
<\varepsilon.
\end{aligned}
\]
Thus we can conclude
\[
\max_{1\le r\le R}\left|
\sum_{i\in\mathcal B_{r,M}}
G_r(M^{1/2}y_i,\ldots,M^{1/2}y_{i+k_r-1})\right|=O_P(1),
\]
and multiplying both sides by \(M^{-1}\) completes the proof.
\end{proof}

Recall the boundary error of approximating \(\hat\theta_g\) through \(\widetilde\theta_{M,g}\).
An immediate consequence of Lemma~\ref{lem:boundary-negligible} is that this approximation error vanishes.
In particular, under Assumptions~\ref{ass:continuous-price} and \ref{ass:g-structural} we have \(M^{1/2}(\hat\theta_g-\widetilde\theta_{M,g})\inP0\).

It remains to evaluate $R_a(H,G)$ for the components of $F$, which identifies $\sigma_g^2$, $C_{g,n}$, and $\Sigma_{\mathbb Q}^{(n)}$ for the joint stable limits in Section~\ref{subsec:supp-stable-limits}.

\begin{lemma}[Scaling of $R_a$]
\label{lem:Ra-scaling}
For an integer \(k\ge2\), if \(H,G:\R^k\to\R\) are Borel measurable, have polynomial growth, and are two-homogeneous, then
\[
R_a(H,G)=a^2R_1(H,G),\qquad a\ge0.
\]
\end{lemma}

\begin{proof}
For \(a>0\), we can write $U_j$ in \eqref{eq:Ra-explicit-definition} as \(U_j=a^{1/2}Z_j\), where \(Z_1,Z_2,\ldots\simIID \Normal(0,1)\).
For \(a\ne0\), \(\mathcal U_m=\sigma(U_1,\ldots,U_m)=\sigma(Z_1,\ldots,Z_m)\) for \(m=k-1,k\).
Two-homogeneity gives
\[
\begin{aligned}
&\E(A_j^H\mid\mathcal U_k)-\E(A_j^H\mid\mathcal U_{k-1})\\
&\qquad=a\left\{
\E\{H(Z_{k-j},\ldots,Z_{2k-j-1})\mid\mathcal U_k\}
-\E\{H(Z_{k-j},\ldots,Z_{2k-j-1})\mid\mathcal U_{k-1}\}
\right\},\\
&\E(B_{j'}^G\mid\mathcal U_k)
=a\E\{G(Z_{k-j'},\ldots,Z_{2k-j'-1})\mid\mathcal U_k\}.
\end{aligned}
\]
Thus every term in \eqref{eq:Ra-explicit-definition} is \(a^2\) times its counterpart in \(R_1(H,G)\).
For \(a=0\), the statement becomes $0=0$ and holds trivially.
\end{proof}

Lemma \ref{lem:Ra-scaling} implies that it suffices to evaluate $R_1(H,G)$ in order to evaluate $R_a(H,G)$, while the following lemma gives the equivalent representation of $R_1(H,G)$.
For sigma-fields $\mathcal A_1$ and $\mathcal A_2$, $\mathcal A_1\vee\mathcal A_2$ denotes the smallest sigma-field containing both.

\begin{lemma}[Representation of $R_1(H,G)$]
\label{lem:cross-kernel}
Let \(k\ge2\) be an integer, let \(H,G:\R^k\to\R\) be Borel measurable, let \((Z_r)_{r\in\mathbb Z}\simIID \Normal(0,1)\), and suppose\newline
$
\E\{H(Z_0,\dots,Z_{k-1})^2\},
\E\{G(Z_0,\dots,Z_{k-1})^2\}<\infty.
$
Set
\[
\zeta_r^H:=H(Z_r,\dots,Z_{r+k-1}),\qquad
\zeta_r^G:=G(Z_r,\dots,Z_{r+k-1}).
\]
For \(X\in L^2\), define the projection operator \(\mathcal P_t\) by
\[
\mathcal P_tX:=
\E\{X\mid\sigma(Z_u:u\le t)\}
-\E\{X\mid\sigma(Z_u:u\le t-1)\}.
\]
Then
\begin{align}
R_1(H,G)
&=
\sum_{j,j'=0}^{k-1}
 \E\{(\mathcal P_0\zeta_{-j}^H)
       (\mathcal P_0\zeta_{-j'}^G)\},
\label{eq:cross-kernel-proj}\\
&=\sum_{h\in\mathbb Z}\Cov(\zeta_0^H,\zeta_h^G).
\label{eq:cross-kernel-covsum}
\end{align}
The covariance sum has at most \(2k-1\) nonzero terms.
Both \eqref{eq:cross-kernel-proj} and \eqref{eq:cross-kernel-covsum} are symmetric in \((H,G)\).
\end{lemma}

\begin{proof}
We first show \eqref{eq:cross-kernel-proj}.
For $a=1$, \eqref{eq:Ra-explicit-definition} gives
\begin{align*}
R_1(H,G)
&=\sum_{j,j'=0}^{k-1}
 \E\!\left\{
  \left[\E(A_j^H\mid\mathcal U_k)
       -\E(A_j^H\mid\mathcal U_{k-1})\right]
  \E(B_{j'}^G\mid\mathcal U_k)
 \right\}.
\end{align*}
Setting $U_m=Z_{m-k}$ ($m\ge1$), we can write
\[
A_j^H=\zeta_{-j}^H,\quad B_{j'}^G=\zeta_{-j'}^G,
\qquad
\mathcal U_{k-1}=\sigma(Z_{-k+1},\ldots,Z_{-1}),\quad
\mathcal U_k=\sigma(Z_{-k+1},\ldots,Z_0).
\]
Notice that for $0\le j,j'<k$, $\sigma(Z_u:u\le-k)$ is independent of $\sigma(\zeta_{-j}^H,\zeta_{-j'}^G)\vee\mathcal U_k$.
Moreover, by definition,
\[
\begin{aligned}
\mathcal U_k\vee\sigma(Z_u:u\le-k)=\sigma(Z_u:u\le0),\qquad
\mathcal U_{k-1}\vee\sigma(Z_u:u\le-k)=\sigma(Z_u:u\le-1).
\end{aligned}
\]
Therefore, we can write
\[
\begin{aligned}
\E(A_j^H\mid\mathcal U_k)
=\E(\zeta_{-j}^H\mid\mathcal U_k)
=\E(\zeta_{-j}^H\mid\mathcal U_k \vee \sigma(Z_u:u\le-k))
=\E\{\zeta_{-j}^H\mid\sigma(Z_u:u\le0)\}
\end{aligned}
\]
and similarly $\E(A_j^H\mid\mathcal U_{k-1})
=\E\{\zeta_{-j}^H\mid\sigma(Z_u:u\le-1)\}.$
It follows that
\[
\E(A_j^H\mid\mathcal U_k)-\E(A_j^H\mid\mathcal U_{k-1})
=\mathcal P_0\zeta_{-j}^H.
\]
Applying the same argument to $B_{j'}^G$ yields
\[
\begin{aligned}
\E(B_{j'}^G\mid\mathcal U_k)
=\E\{\zeta_{-j'}^G\mid\sigma(Z_u:u\le0)\}
=\mathcal P_0\zeta_{-j'}^G
+\E\{\zeta_{-j'}^G\mid\sigma(Z_u:u\le-1)\}.
\end{aligned}
\]
The second term \(\E\{\zeta_{-j'}^G\mid\sigma(Z_u:u\le-1)\}\) is measurable with respect to $\sigma(Z_u:u\le-1)$, and
\begin{align*}
\E\!\left\{
\mathcal P_0\zeta_{-j}^H
\,\middle|\,
\sigma(Z_u:u\le-1)
\right\}
&=
\E\!\left[
\E\{\zeta_{-j}^H\mid\sigma(Z_u:u\le0)\}
\,\middle|\,
\sigma(Z_u:u\le-1)
\right]
\\
&\quad-
\E\!\left[
\E\{\zeta_{-j}^H\mid\sigma(Z_u:u\le-1)\}
\,\middle|\,
\sigma(Z_u:u\le-1)
\right]
\\
&=
\E\{\zeta_{-j}^H\mid\sigma(Z_u:u\le-1)\}
-
\E\{\zeta_{-j}^H\mid\sigma(Z_u:u\le-1)\}
=0.
\end{align*}
Therefore, we can write
\begin{align*}
R_1(H,G)
&=\sum_{j,j'=0}^{k-1}
 \E\!\left\{
  (\mathcal P_0\zeta_{-j}^H)
  \left[\mathcal P_0\zeta_{-j'}^G
+\E\{\zeta_{-j'}^G\mid\sigma(Z_u:u\le-1)\}\right]
 \right\}\\
& =
 \sum_{j,j'=0}^{k-1}
 \E\{(\mathcal P_0\zeta_{-j}^H)
       (\mathcal P_0\zeta_{-j'}^G)\},
\end{align*}
which proves \eqref{eq:cross-kernel-proj}.
We next prove \eqref{eq:cross-kernel-covsum}.
Because \(\zeta_0^H\) is measurable with respect to \(\sigma(Z_u:u\le k-1)\) and is independent of \(\sigma(Z_u:u\le-1)\), telescoping the projections gives
\begin{align*}
 \zeta_0^H-\E\zeta_0^H
 =\sum_{s=0}^{k-1}\mathcal P_s\zeta_0^H
 \qquad\text{and}\qquad
 \zeta_h^G-\E\zeta_h^G
 =\sum_{t=h}^{h+k-1}\mathcal P_t\zeta_h^G.
\end{align*}
For \(s<t\), \(\mathcal P_s\zeta_0^H\) is measurable with respect to \(\sigma(Z_u:u\le t-1)\), while
\(
\E(\mathcal P_t\zeta_h^G\mid\sigma(Z_u:u\le t-1))=0.
\)
For \(s>t\), \(\mathcal P_t\zeta_h^G\) is measurable with respect to \(\sigma(Z_u:u\le s-1)\), while \(\E(\mathcal P_s\zeta_0^H\mid\sigma(Z_u:u\le s-1))=0\).
Thus projections at distinct times are orthogonal, and
\begin{align*}
\Cov(\zeta_0^H,\zeta_h^G)
&=
\sum_{s=0}^{k-1}\sum_{t=h}^{h+k-1}
\E\!\left[
(\mathcal P_s\zeta_0^H)
(\mathcal P_t\zeta_h^G)
\right]
\\
&=
\sum_{\substack{0\le s\le k-1\\ h\le s\le h+k-1}}
\E\!\left[
(\mathcal P_s\zeta_0^H)
(\mathcal P_s\zeta_h^G)
\right]
\\
&=
\sum_{s=0}^{k-1}
\E\!\left[
(\mathcal P_s\zeta_0^H)
(\mathcal P_s\zeta_h^G)
\right].
\end{align*}
The last equality holds since \(\mathcal P_s\zeta_h^G=0\) whenever $s\not\in \{h,\ldots,h+k-1\}$.
Since $(Z_r)_{r\in\mathbb Z}$ is stationary, the block processes $(\zeta_r^H,\zeta_r^G)$ are jointly stationary.
Shifting all indices by $-s$, together with the corresponding conditioning sigma-fields in the definition of $\mathcal P_s$, gives
\[
(\mathcal P_s\zeta_0^H,\mathcal P_s\zeta_h^G)
\overset{d}{=}
(\mathcal P_0\zeta_{-s}^H,\mathcal P_0\zeta_{h-s}^G),
\]
and hence
\[
\E\{(\mathcal P_s\zeta_0^H)(\mathcal P_s\zeta_h^G)\}
=
\E\{(\mathcal P_0\zeta_{-s}^H)(\mathcal P_0\zeta_{h-s}^G)\}.
\]

If \(|h|\ge k\), \(\zeta_0^H\) and \(\zeta_h^G\) involve disjoint sets of Gaussian random variables, so their covariance is zero.
The covariance sum therefore has at most \(2k-1\) nonzero terms, and hence rearrangements of the double sum are legitimate.
For each \(s\), \(\mathcal P_0\zeta_{h-s}^G\) can be nonzero only when \(h-s\in\{-k+1,\ldots,0\}\), equivalently \(j'=s-h\in\{0,\ldots,k-1\}\).
Reindexing by \(j'=s-h\) gives
\begin{align*}
 \sum_{h\in\mathbb Z}\Cov(\zeta_0^H,\zeta_h^G)
 &=\sum_{s=0}^{k-1}\sum_{h\in\mathbb Z}
   \E\!\left[(\mathcal P_0\zeta_{-s}^H)
              (\mathcal P_0\zeta_{h-s}^G)\right]=\sum_{s=0}^{k-1}\sum_{j'=0}^{k-1}
   \E\!\left[(\mathcal P_0\zeta_{-s}^H)
              (\mathcal P_0\zeta_{-j'}^G)\right].
\end{align*}
The final expression is exactly the right-hand side of \eqref{eq:cross-kernel-proj}, proving \eqref{eq:cross-kernel-covsum}.

Finally, the right-hand side of \eqref{eq:cross-kernel-proj} can be written as
\(
 \E\![
  \{\sum_{j=0}^{k-1}\mathcal P_0\zeta_{-j}^H\}
  \{\sum_{j'=0}^{k-1}\mathcal P_0\zeta_{-j'}^G\}
 ].
\)
It is symmetric in \((H,G)\).
\end{proof}
\subsubsection{Evaluation of \texorpdfstring{$\sigma_g^2$, $C_{g,n}$, and $\Sigma_{\mathbb Q}^{(n)}$}{sigma\_g\textasciicircum 2, C\_g,n, and Sigma\_Q\textasciicircum(n)}}

\begin{proposition}[Evaluation of $\sigma_g^2$, $C_{g,n}$, and $\Sigma_{\mathbb Q}^{(n)}$]
\label{prop:covariance-constants}
Under Assumption~\ref{ass:g-structural}, for every \(a\ge0\) and \(k,k'\in\{b,\ldots,b+\ell-1\}\):
\begin{enumerate}[label=(\roman*)]
\item \(R_a(F_\theta,F_\theta)=a^2\sigma_g^2\), with \(\sigma_g^2\) exactly as in \eqref{eq:sigma-g-overlap}.
\item \(R_a(F_\theta,F_{Q,k})=a^2C_{g,n}\), where \(C_{g,n}\) is defined in \eqref{eq:Cgn-def}.
\item \(R_a(F_{Q,k},F_{Q,k'})
=a^2\Sigma_{\mathbb Q}^{(n)}(k-b+1,k'-b+1)\), with the entries as in \eqref{eq:SigmaQ-final}.
\end{enumerate}
Integrating (i)--(iii) with $a=\sigma_s^2$ over $s\in[0,1]$ replaces $a^2$ by $v^{[4]}$.
\end{proposition}

\begin{proof}
Since $F_\theta$ and $F_{Q,k}$ have polynomial growth, they satisfy the square-integrability requirement of Lemma~\ref{lem:cross-kernel}.
We will use Lemma~\ref{lem:cross-kernel} and count the common indices in the relevant summands.
Then, Lemma~\ref{lem:Ra-scaling} gives $R_a(H,G)=a^2R_1(H,G)$, and $\int_0^1(\sigma_s^2)^2\dd s=v^{[4]}$ gives the integrated covariance.

Let \((Z_r)_{r\in\mathbb Z}\simIID \Normal(0,1)\).
Set
\[
\zeta_r^\theta:=Z_r^2-g(Z_r,\dots,Z_{r+b}),\qquad
\zeta_{k,r}^Q
 :=\mu_{2/n}^{-n}|Z_r|^{2/n}
   \prod_{m=0}^{n-2}|Z_{r+k+1+mL}|^{2/n}.
\]
By Assumption~\ref{ass:g-structural}(iii), we have \(\E\zeta_r^\theta=0\).
Since $F_\theta$ depends only on coordinates $1,\ldots,b+1$, $\Cov(\zeta_0^\theta,\zeta_h^\theta)=0$ for $|h|>b$.
We can write
\[
R_1(F_\theta,F_\theta)
 =\sum_{h=-b}^{b}\Cov(\zeta_0^\theta,\zeta_h^\theta)
 =\sigma_g^2.
\]

We then show that item (ii) holds.
For $R_1(F_\theta,F_{Q,k})$, distinct elements of \(\mathcal I_k\) are more than \(b\) apart.
Hence \(\{0,\dots,b\}\) and \(h+\mathcal I_k-1\) have at most one common element, and such an overlap happens if and only if $h$ satisfies
\[
h=q-r+1,\qquad\text{ for some } q\in\{0,\dots,b\},\quad r\in \mathcal I_k.
\]
Indeed, if two pairs $(q,r)$ and $(q',r')$ give the same shift $h$, then $q-r=q'-r'$, so $|r-r'|=|q-q'|\le b$.
Since distinct elements of $\mathcal I_k$ are more than $b$ apart, we must have $r=r'$, and hence $q=q'$.
Thus there are exactly \(n(b+1)\) shifts \(h\) for which the coordinate sets \(\{0,\dots,b\}\) and \(h+\mathcal I_k-1\) overlap in exactly one coordinate, and there are no other overlaps.
At the shift associated with \((q,r)\), independence of the remaining \(n-1\) \(Q\)-coordinates and their normalized Gaussian moments give
\[
\Cov(\zeta_0^\theta,\zeta_{k,h}^Q)
 =\mu_{2/n}^{-1}\E\{\zeta_0^\theta|Z_q|^{2/n}\}.
\]
Since each $q=0,\ldots,b$ contributes $n$ identical covariance terms, relabelling \(Z_0,\dots,Z_b\) yields
\[
R_1(F_\theta,F_{Q,k})
=n\mu_{2/n}^{-1}\sum_{q=1}^{b+1}
 \E\!\left\{
 \left[Z_1^2-g(Z_1,\dots,Z_{b+1})\right]|Z_q|^{2/n}\right\}
=C_{g,n}.
\]
For every \(q=0,\ldots,b\), Cauchy--Schwarz gives
\[
\left|\E\{\zeta_0^\theta|Z_q|^{2/n}\}\right|
\le \E\{|\zeta_0^\theta||Z_q|^{2/n}\}
\le (\E|\zeta_0^\theta|^2)^{1/2}\mu_{4/n}^{1/2}<\infty.
\]

It remains to evaluate \(R_1(F_{Q,k},F_{Q,k'})\).
Write
\[
m_h(k,k'):=|\mathcal I_k\cap(\mathcal I_{k'}+h)|.
\]
The $n$ indices in $\mathcal I_k$ are distinct and lie in $\{1,\ldots,K\}$.
For $m=m_h(k,k')\ge1$, the two products $\zeta_{k,0}^Q$ and $\zeta_{k',h}^Q$ share exactly $m$ Gaussian variables.
Each shared variable contributes a factor of $\eta_n=\mu_{4/n}/\mu_{2/n}^2$ to the expectation of their product, while each variable appearing in only one product contributes a factor of one.
By independence, $\E(\zeta_{k,0}^Q\zeta_{k',h}^Q)=\eta_n^m$.
Since $\E\zeta_{k,0}^Q=\E\zeta_{k',h}^Q=1$, we have
\[
\Cov(\zeta_{k,0}^Q,\zeta_{k',h}^Q)
=
\eta_n^{m_h(k,k')}-1.
\]
When $m_h(k,k')=0$, then they are independent and the identity above also holds.

Since $\mathcal I_k=\{1\}\cup\{k+2+jL:0\le j\le n-2\}$, shifts $h=k-k'+qL$ with $|q|\le n-2$ produce $n-1-|q|$ matches among the indices spaced $L$ apart.
In addition, the index $1$ matches itself when $h=0$ and matches one of the remaining indices for another $2(n-1)$ shifts.
Counting these cases gives the following number of shifts $h$ for each positive overlap count $m_h(k,k')$:
\[
\begin{array}{|cc|cc|}
\hline
\multicolumn{2}{|c|}{k=k'} &
\multicolumn{2}{c|}{k\ne k'}\\ \hline
\text{overlap count }m & \text{number of shifts }h &
\text{overlap count }m & \text{number of shifts }h\\ \hline
n & 1 & n-1 & 1\\
2,\ldots,n-2 & 2 & 2,\ldots,n-2 & 2\\
1 & 2n & 1 & 2n+1 \\ \hline
\end{array}
\]

Summing \(\eta_n^m-1\) over the table gives
\[
R_1(F_{Q,k},F_{Q,k'})
=
\begin{cases}
\displaystyle
\eta_n^n+2\sum_{m=2}^{n-2}\eta_n^m+2n\eta_n+(5-4n),
 &k=k',\\[4pt]
\displaystyle
\eta_n^{n-1}+2\sum_{m=2}^{n-2}\eta_n^m+(2n+1)\eta_n+4(1-n),
 &k\ne k',
\end{cases}
\]
with $\sum_{m=2}^{n-2}\eta_n^m=0$ when $n=3$, exactly as in \eqref{eq:SigmaQ-final}.
\end{proof}

\begin{remark}[Verification for $g_1,g_2,g_3$]
\label{rem:concrete-g-verification}
\emergencystretch=2em
All three are invariant under coordinatewise sign changes and are two-homogeneous.
Let $b_1=b_2=1$, $b_3=2$, and $H_{j,i}:=Z_i^2-g_j(Z_i,\ldots,Z_{i+b_j})$.
For $g_1$, $\E g_1(Z_0,Z_1)=\mu_1^{-2}(\E|Z|)^2=1$.
The normalizers in \citet[eqn.~\textnormal{(4)}]{AndersenDobrevSchaumburg2012} give
\[
 \E\min(|Z_0|,|Z_1|)^2=\frac{\pi-2}{\pi},\qquad
 \E\med(|Z_0|,|Z_1|,|Z_2|)^2
 =\frac{\pi+6-4\sqrt3}{\pi},
\]
and hence $\E g_2=\E g_3=1$.

For Assumption~\ref{ass:g-structural}(iv), one may take
\[
B_{g_1}=\{x:x_1=0\}\cup\{x:x_2=0\},
\qquad
B_{g_2}=\{x:x_1=x_2\}\cup\{x:x_1=-x_2\},
\]
and
\[
B_{g_3}=\bigcup_{1\le i<j\le3}
\left(\{x:x_i=x_j\}\cup\{x:x_i=-x_j\}\right).
\]
For each \(j=1,2,3\), \(g_j\) is quadratic and \(\nabla g_j\) is linear on each connected component of \(B_{g_j}^c\).
There are finitely many such components, giving \(|g_j(z)|\le C(1+\|z\|^2)\) for all \(z\) and \(\|\nabla g_j(z)\|\le C(1+\|z\|^2)\) for \(z\notin B_{g_j}\).
If \(z\notin B_{g_j}\) and \(\|y\|\le\operatorname{dist}(z,B_{g_j})/2\), the segment from \(z\) to \(z+y\) avoids \(B_{g_j}\), so both points remain in the same connected component of \(B_{g_j}^c\).
Linearity of the gradient there gives \(\|\nabla g_j(z+y)-\nabla g_j(z)\|\le C\|y\|\).
These inequalities imply \eqref{eq:g-growth-assumption}--\eqref{eq:g-lip-assumption} with the sufficient choice \(p_g=2\) and \(w_g=1\).

To verify Assumption~\ref{ass:g-structural}(v), put \(h_{j,r}(z):=\E(H_{j,0}\mid Z_r=z)\).
Stationarity and independence of disjoint Gaussian coordinate sets give, for \(N>b_j\),
\[
N^{-1}\Var\!\left(\sum_{i=1}^NH_{j,i}\right)
=\sum_{h=-b_j}^{b_j}
\left(1-\frac{|h|}{N}\right)\Cov(H_{j,0},H_{j,h})
\to\sigma_{g_j}^2
\]
as \(N\to\infty\), by \eqref{eq:sigma-g-overlap}.
For every \(m=b_j+1,\ldots,N\), the terms containing \(Z_m\) correspond exactly to \(i=m-r\), \(r=0,\ldots,b_j\).
All other terms are independent of \(Z_m\) and have mean zero.
Consequently,
\[
\E\!\left(\sum_{i=1}^NH_{j,i}\,\middle|\,Z_m\right)
=\sum_{r=0}^{b_j}h_{j,r}(Z_m).
\]
The right-hand sides for distinct \(m\) are centered functions of independent \(Z_m\)'s and hence mutually orthogonal.
The first-order Hoeffding projection therefore gives
\[
\Var\!\left(\sum_{i=1}^NH_{j,i}\right)
\ge\sum_{m=b_j+1}^N
\Var\!\left\{\sum_{r=0}^{b_j}h_{j,r}(Z_m)\right\}
=(N-b_j)\Var\!\left\{\sum_{r=0}^{b_j}h_{j,r}(Z)\right\}.
\]
Dividing by \(N\), letting \(N\to\infty\), and using \eqref{eq:sigma-g-overlap} yields
\[
\sigma_{g_j}^2\ge\Var\!\left\{\sum_{r=0}^{b_j}h_{j,r}(Z)\right\}.
\]
The functions $\sum_{r=0}^{b_j}h_{j,r}(z)$, $j=1,2,3$, are respectively
\begin{align*}
 &z^2+1-2\mu_1^{-1}|z|,\\
 &z^2+1-2\frac{\pi}{\pi-2}\E\min(z^2,Z^2),\\
 &z^2+2-3\frac{\pi}{\pi+6-4\sqrt3}
       \E\med(z^2,Z_1^2,Z_2^2).
\end{align*}
For \(j=2,3\), $\E\min(z^2,Z^2)\to1$ and $\E\med(z^2,Z_1^2,Z_2^2)\to\E\max(Z_1^2,Z_2^2)<\infty$ as $|z|\to\infty$.
The functions \(\sum_{r=0}^{b_j}h_{j,r}(z)\) are continuous.
This is immediate for \(j=1\), and for \(j=2,3\) it follows by dominated convergence, using \(\min(z^2,Z^2)\le Z^2\) and \(\med(z^2,Z_1^2,Z_2^2)\le\max(Z_1^2,Z_2^2)\).
For each \(j=1,2,3\), \(\sum_{r=0}^{b_j}h_{j,r}(z)\to+\infty\) as \(|z|\to\infty\), so it is nonconstant.
Since \(Z\sim\Normal(0,1)\) has support \(\mathbb R\), continuity and nonconstancy imply
\[
\Var\!\left\{\sum_{r=0}^{b_j}h_{j,r}(Z)\right\}>0,
\]
and therefore $\sigma_{g_j}^2>0$ for $j=1,2,3$.
\end{remark}

\begin{remark}[Example: \(g=g_1\)]
\label{rem:g1-crosscov}
For \(g_1(x_1,x_2)=\mu_1^{-2}|x_1x_2|\), independence gives
\[
\E\{(Z_1^2-g_1)|Z_1|^{2/n}\}
 =\mu_{2/n+2}-\mu_1^{-1}\mu_{2/n+1},\qquad
\E\{(Z_1^2-g_1)|Z_2|^{2/n}\}
 =\mu_{2/n}-\mu_1^{-1}\mu_{2/n+1}.
\]
Substitution in \eqref{eq:Cgn-def} using \(\mu_{2/n+2}=(2/n+1)\mu_{2/n}\) yields
\[
C_{g_1,n}
=2(n+1)-2n\mu_{2/n+1}\mu_{2/n}^{-1}\mu_1^{-1}.
\]
\end{remark}

\subsubsection{Continuous processes that agree with \texorpdfstring{$y^*$}{y*} on \texorpdfstring{$\Omega_0$}{Omega\_0}}
\label{subsec:supp-no-jump-reduction}

Under Assumption~\ref{ass:price-model}, the following lemma reduces stable limits in restriction to $\Omega_0$ to stable limits under Assumption~\ref{ass:continuous-price}.

\begin{lemma}[Continuous processes that agree with $y^*$ on $\Omega_0$]
\label{lem:no-jump-reduction}
Suppose Assumption~\ref{ass:price-model} holds.
\begin{enumerate}[label=(\roman*)]
    \item For $q\ge1$, there are a bounded, progressively measurable, c\`adl\`ag process $b^{(q)}$ and an event $E_q\in\mathcal F$ with $E_q\subset\Omega_0$ such that $\pr(\Omega_0\setminus E_q)\to0$ as $q\to\infty$ and
\[
y^{(q)}(t)=y^*(0)+\int_0^tb_s^{(q)}\dd s+\int_0^t\sigma_s\dd W_s,
\qquad t\ge0,
\]
satisfies $y^{(q)}(t)=y^*(t)$ for every $t\in[0,1]$, almost surely on $E_q$.
In particular, $y^{(q)}$ satisfies Assumption~\ref{ass:continuous-price} with $b^{(q)}$ in place of $\mu^y$.

\item Let $U_M=\Phi_M\{(y^*(i/M))_{0\le i\le M},\zeta_M\}$, where $\Phi_M$ is measurable with values in $\mathbb R^d$ or $[-\infty,\infty]^d$ and $\zeta_M$ is a random vector possibly defined on an extension of the probability space.
Let $U_M^{(q)}$ be defined in the same way with $y^{(q)}$ in place of $y^*$.
If $U_M^{(q)}\inST U$ for every $q\ge1$, then $U_M\inST U$ in restriction to $\Omega_0$.
\end{enumerate}

\end{lemma}

\begin{proof}
We follow the continuous-path localization argument in the proof of \citet[Thm.~3.5.1(b), pp.~95--96]{JacodProtter2012} and its use for restricted stable convergence in \citet[Thm.~11.4.4(b), p.~335]{JacodProtter2012}.
We verify separately that the continuous approximating processes have progressively measurable, c\`adl\`ag drifts, as required by Assumption~\ref{ass:continuous-price}.

We first show that both $\varrho_q=\inf\{t\ge0:A_t\ge q\}$ and $S=\inf\{t>0: \Delta y^*(t)\neq0\}$ are stopping times.
For $q\ge1$, set $\varrho_q=\inf\{t\ge0:A_t\ge q\}$ and $R_q=\varrho_q\wedge1$.
Since $\Lambda$ is nonnegative and predictable, $A_t=\int_0^t\Lambda_s\dd s$ is adapted, nondecreasing, and left-continuous, with values in $[0,\infty]$ and $A_0=0$.
Thus $A$ is progressively measurable.
Define
\(
D_q:=\{(t,\omega)\in[0,\infty)\times\Omega:A_t(\omega)\ge q\}.
\)
Since $A$ is progressively measurable,
$D_q\cap([0,T]\times\Omega)\in
\mathcal B([0,T])\otimes\mathcal F_T$ for every $T\ge0$.
Moreover,
\[
\inf\{t\ge0:(t,\omega)\in D_q\}
=\inf\{t\ge0:A_t(\omega)\ge q\}
=\varrho_q(\omega).
\]
It follows from the d\'ebut theorem that $\varrho_q$ is a stopping time.
Similarly, since $y^*$ is adapted and c\`adl\`ag, both $y^*$ and its left-limit process $y^*_{-}$ are progressively measurable.
The process $\Delta y^*$ is progressively measurable, and so is the set
\(
D^J:=\{(t,\omega)\in[0,\infty)\times\Omega:
t>0,\ \Delta y^*(t,\omega)\ne0\}.
\)
It then follows from the d\'ebut theorem that $S=\inf\{t\ge0:(t,\omega)\in D^J\}=\inf\{t>0: \Delta y^*(t)\neq0\}$ is also a stopping time.

Notice that for a stopping time $T$, $\ind(s\le T)$ is predictable as a function of $s$.
Since $\delta$ is predictable, the compensation formula in \citet[eqn.~\textup{(3.5.11)}]{JacodProtter2012} for the integrand $\ind(\delta(s,z)\ne0)\ind(s\le T)$ yields
\begin{equation}\label{eq: compensation}
\E\{\ind(\delta\ne0)\star\mathfrak p_T\}
=\E\{\ind(\delta\ne0)\star\mathfrak q_T\}
=\E\int_0^T\Lambda_s\dd s
=\E A_T.    
\end{equation}

We first consider $T=R_q=\varrho_q\wedge1$.
As $A_t<q$ for $t<\varrho_q$, left-continuity implies $A_{R_q}\le q$.
It then follows from \eqref{eq: compensation} that 
\begin{equation}\label{eq: Rq-bound}
    \E\{\ind(\delta\ne0)\star\mathfrak p_{R_q}\}=\E A_{R_q}\le q.
\end{equation}

We then handle $T=S\wedge1$.
The Poisson random measure $\mathfrak p$ has no simultaneous atoms \citep[p.~34]{JacodProtter2012}.
Since the compensator $\mathfrak q(\dd t,\dd z)=\dd t\,\lambda(\dd z)$ has no time atoms, \citet[eqn.~\textup{(2.1.17)} and p.~35]{JacodProtter2012} applied to \eqref{eq:grigelionis-y} gives $\Delta y^*(t)=\delta(t,z)$ when $\mathfrak p$ has an atom $(t,z)$ and $\Delta y^*(t)=0$ otherwise.
Consequently, $\ind(\delta\ne0)\star\mathfrak p_t$ counts the nonzero jumps of $y^*$ in $(0,t]$.
By construction of $S$, there is at most one nonzero jump in $(0,S\wedge1]$.
Therefore, it follows from \eqref{eq: compensation} that 
\[\E\{\ind(\delta\ne0)\star\mathfrak p_{S\wedge1}\}
=\E A_{S\wedge1}\le1.\]

We next identify the effective drift of the stopped price process.
Define $V_q(s,z)=\delta(s,z)\ind(|\delta(s,z)|\le1)\ind(s\le R_q)$.
Notice that $|\delta|\ind(|\delta|\le1)\le\ind(\delta\ne0)$ and it follows from \eqref{eq: Rq-bound} that $\ind(\delta\ne0)\star\mathfrak p_{R_q}<\infty$ almost surely.
Therefore,
\begin{align*}
|V_q|\star\mathfrak p_1
&=\int_{(0,1]\times E}
|\delta(s,z)|\ind(|\delta(s,z)|\le1)
\ind(s\le R_q)\mathfrak p(\dd s,\dd z)\\
&\le\int_{(0,R_q]\times E}
\ind(\delta(s,z)\ne0)\mathfrak p(\dd s,\dd z)\\
&=\ind(\delta\ne0)\star\mathfrak p_{R_q}
<\infty\qquad\text{a.s.}
\end{align*}
Since $|V_q|\star\mathfrak q_1\le q<\infty$, the integral defining $\mu_s^\delta$ is absolutely convergent and hence $\mu_s^\delta$ is well defined for almost every $s\in[0,R_q]$ (in the Lebesgue sense).
Moreover, we have
\begin{align*}
|V_q|\star\mathfrak q_1
&=\int_0^1\int_E
|\delta(s,z)|\ind(|\delta(s,z)|\le1)
\ind(s\le R_q)\lambda(\dd z)\dd s\\
&\le\int_0^{R_q}\int_E
\ind(\delta(s,z)\ne0)\lambda(\dd z)\dd s\\
&=\int_0^{R_q}\Lambda_s\dd s
=A_{R_q}\le q.
\end{align*}
It follows from the definition of $\mu^\delta$ that \(
\int_0^{R_q}|\mu_s^\delta|\dd s
\le |V_q|\star\mathfrak q_1
\le q.
\)

By the stopping property of stochastic integrals with respect to the compensated Poisson random measure $\mathfrak p-\mathfrak q$, we have
\(
V_q\star(\mathfrak p-\mathfrak q)_t
=(\delta\ind(|\delta|\le1))
\star(\mathfrak p-\mathfrak q)_{t\wedge R_q},
\)
for $t\geq0$.
Since $|V_q|\star\mathfrak p_1$ and $|V_q|\star\mathfrak q_1$ are finite almost surely, the compensated stochastic integral coincides with the difference of the corresponding pathwise integrals; see \citet[pp.~31--32 and~35]{JacodProtter2012}.
Thus, for $t\le R_q$,
\[
\begin{aligned}
(\delta\ind(|\delta|\le1))\star(\mathfrak p-\mathfrak q)_t
=V_q\star\mathfrak p_t-V_q\star\mathfrak q_t
=\sum_{0<s\le t}\Delta y^*(s)
  \ind(|\Delta y^*(s)|\le1)-\int_0^t\mu_s^\delta\dd s.
\end{aligned}
\]
Similarly, we have
\(
(\delta\ind(|\delta|>1))\star\mathfrak p_t
=\sum_{0<s\le t}\Delta y^*(s)
\ind(|\Delta y^*(s)|>1),
\)
over $t\le R_q$ almost surely.
Substituting both terms into \eqref{eq:grigelionis-y} gives, almost surely,
\begin{equation}
\label{eq:stopped-price}
y^*(t)=y^*(0)+\int_0^t(\mu_s^y-\mu_s^\delta)\dd s
+\int_0^t\sigma_s\dd W_s
+\sum_{0<s\le t}\Delta y^*(s),
\qquad t\in[0,R_q],
\end{equation}
where the aggregated jump component $\sum_{0<s\le t}\Delta y^*(s)$ is finite almost surely.

As the representation \eqref{eq:stopped-price} is valid over $t\le R_q$, it suffices to investigate how $R_q$ behaves under $\Omega_0$.
On $\Omega_0$, we have $S\ge1$ and therefore $A_1=A_{S\wedge1}$.
Since $\varrho_q<1$ implies $A_1\ge q$, Markov's inequality yields
\begin{equation}
\label{eq:no-jump-intensity}
\begin{aligned}
\pr(\Omega_0\cap\{\varrho_q<1\})
&\le\pr(\Omega_0\cap\{A_1\ge q\})\\
&\le q^{-1}\E\{\ind(\Omega_0)A_1\}
 =q^{-1}\E\{\ind(\Omega_0)A_{S\wedge1}\}\\
&\le q^{-1}\E A_{S\wedge1}\le q^{-1}.
\end{aligned}
\end{equation}
It follows that $R_q=\varrho_q\wedge1=1$ on $\Omega_0$, except on an event of probability at most $q^{-1}$.

The integral defining $\mu_s^\delta$ is meaningful where it converges absolutely, and the argument above shows that this holds for Lebesgue-almost every $s\in[0,R_q]$.
However, $\mu^\delta$ defined in this way may not be progressively measurable.
We show that without loss of generality, we can extend the definition by setting $\mu_s^\delta=0$ whenever
\[
\int_E|\delta(s,z)|\ind(|\delta(s,z)|\le1)
\lambda(\dd z)=\infty.
\]
Since $\delta$ is predictable, Tonelli's theorem implies that the integrals of the positive and negative parts of the integrand, as well as the set where the integral converges absolutely, are predictable.
Hence the modified $\mu^\delta$ is predictable, so $\mu^y-\mu^\delta$ is progressively measurable.
Indeed, \eqref{eq:stopped-price} remains valid after this modification of $\mu^\delta$.
For every $t<\varrho$,
\[
\int_0^t\int_E|\delta(s,z)|\ind(|\delta(s,z)|\le1)
\lambda(\dd z)\dd s\le A_t<\infty.
\]
Thus the convention affects only a Lebesgue-null set of times on $[0,\varrho\wedge1]$, leaving \eqref{eq:stopped-price} and the condition in Assumption~\ref{ass:effective-drift} unchanged.

We then construct a bounded, progressively measurable, c\`adl\`ag drift satisfying Assumption~\ref{ass:continuous-price}.
Although the effective drift is progressively measurable, Assumption~\ref{ass:effective-drift} only guarantees a pathwise c\`adl\`ag representative, whose adaptedness is not immediate.
The construction proceeds through the following steps:
\[
\boxed{r^{(q)}}
\xrightarrow{\text{c\`adl\`ag rep.}}
\boxed{h^{(q)}}
\xrightarrow{\text{averages \eqref{eq: c_r connection}}}
\boxed{c^{(q)}}
\xrightarrow{\text{clipping \eqref{eq: clipping}}}
\boxed{b^{(q)}}.
\]
Each of $r^{(q)}$, $h^{(q)}$, and $c^{(q)}$ serves as an intermediate process for constructing the bounded drift process $b^{(q)}$.
Their properties are summarized below:
\begin{center}
\small
\renewcommand{\arraystretch}{1.25}
\begin{tabular}{|lccc|}
\hline
Process
& {Progressively measurable}
& C\`adl\`ag
& {Uniformly bounded}\\[3pt]
\hline
$r^{(q)}$ & Yes & Not necessarily & Not necessarily\\
$h^{(q)}$ & To be verified & Yes & Not necessarily\\
$c^{(q)}$ & Yes & Yes & Not necessarily\\
$b^{(q)}$ & Yes & Yes & Yes\\
\hline
\end{tabular}
\end{center}
For $h^{(q)}$, progressive measurability is not immediate from its pathwise construction.
It will be established through the construction of $c^{(q)}$.

We now construct $b^{(q)}$.
Let $\Omega'$ be the event on which $\mu^y-\mu^\delta$ agrees Lebesgue-almost everywhere on $[0,\varrho\wedge1]$ with a real-valued c\`adl\`ag function.
By Assumption~\ref{ass:effective-drift}, we have $\pr(\Omega')=1$.
For fixed $q$, define
\[
r_s^{(q)}
=(\mu_s^y-\mu_s^\delta)\ind(s<R_q),
\qquad s\ge0,
\]
and set $r^{(q)}\equiv0$ on $\Omega\setminus\Omega'$.
Since $\mu^y-\mu^\delta$ is progressively measurable, $R_q$ is a stopping time, and $\mathcal F_0$ contains all $\pr$-null sets, $r^{(q)}$ is progressively measurable. 

The fact that $\mu^y-\mu^\delta$ agrees with a c\`adl\`ag function almost everywhere does not imply that it is itself c\`adl\`ag.
Therefore, for each sample path in $\Omega'$, define $h^{(q)}$ as the c\`adl\`ag representative of $\mu^y-\mu^\delta$ on $[0,R_q)$ and zero on $[R_q,\infty)$.
Then $h^{(q)}$ is c\`adl\`ag and agrees with $r^{(q)}$ Lebesgue-almost everywhere on $[0,\infty)$.
Since $R_q\le1$ and $h^{(q)}$ is c\`adl\`ag, it is also bounded and integrable on $[0,\infty)$.

Although $h^{(q)}$ is c\`adl\`ag, its adaptedness has not yet been established.
To bypass this problem, consider
\begin{equation}\label{eq: c_r connection}
c_t^{(q)}
=\limsup_{j\to\infty}
j\int_t^{t+1/j}r_s^{(q)}\dd s,
\qquad t\ge0.
\end{equation}
For every $t\ge0$ and every path in $\Omega'$, the almost-everywhere equivalence of $r^{(q)}$ and $h^{(q)}$ gives
\[
\begin{aligned}
\left|
j\int_t^{t+1/j}r_s^{(q)}\dd s
-h^{(q)}(t)\right|
&=
\left|j\int_t^{t+1/j}
\{h^{(q)}(s)-h^{(q)}(t)\}\dd s\right|\\
&\le
\sup_{t\le s\le t+1/j}
|h^{(q)}(s)-h^{(q)}(t)|
\longrightarrow0
\end{aligned}
\]
by right-continuity.
Hence $c_t^{(q)}=h^{(q)}(t)$ for every $t$ on $\Omega'$.
Since $c^{(q)}$ is zero on $\Omega\setminus\Omega'$, it follows that $c^{(q)}$ has c\`adl\`ag paths.
It remains to verify that $c^{(q)}$ is adapted.
For fixed $t$ and $j$, progressive measurability of $r^{(q)}$ implies that $j\int_t^{t+1/j}r_s^{(q)}\dd s$ is $\mathcal F_{t+1/j}$-measurable.
It follows that for every $j_0\ge1$, $c_t^{(q)}$ is $\mathcal F_{t+1/j_0}$-measurable.
By right-continuity of the filtration, we have
\(
\bigcap_{j_0\ge1}\mathcal F_{t+1/j_0}
=\mathcal F_t.
\)
Thus $c_t^{(q)}$ is $\mathcal F_t$-measurable for each $t$ and hence $c^{(q)}$ is adapted.
As it is also c\`adl\`ag, $c^{(q)}$ is progressively measurable.

Since $c^{(q)}$ is c\`adl\`ag and vanishes after $R_q\le1$, $\sup_{t\le1}|c_t^{(q)}|$ is measurable and finite almost surely.
Therefore, we can pick a deterministic $C_q<\infty$ such that
\(
\pr(\sup_{t\le1}|c_t^{(q)}|>C_q)\le q^{-1},
\)
and define
\begin{equation}\label{eq: clipping}
    b^{(q)}=(-C_q)\vee(c^{(q)}\wedge C_q).
\end{equation}
It follows immediately that $b^{(q)}$ is bounded, progressively measurable, and c\`adl\`ag, while $b^{(q)}=c^{(q)}$ on $[0,1]$ whenever the supremum above is at most $C_q$.
Set
\[
E_q=\Omega_0\cap\Omega'\cap\{\varrho_q\ge1\}
\cap\left\{\sup_{t\le1}|c_t^{(q)}|\le C_q\right\}.
\]
On $E_q$, we have $R_q=1$ and $b_t^{(q)}=c_t^{(q)}=h^{(q)}(t)$ for $t\in[0,1]$.
Since $h^{(q)}$ agrees almost everywhere with $\mu^y-\mu^\delta$ for $t\in[0,R_q)$, for every $t\in[0,1]$,
\(
\int_0^tb_s^{(q)}\dd s
=\int_0^t(\mu_s^y-\mu_s^\delta)\dd s.
\)
The aggregated jump component $\sum_{0<s\le t}\Delta y^*(s)$ in \eqref{eq:stopped-price} also vanishes on $E_q$ since $E_q\subseteq\Omega_0$.
Therefore,
\[
y^{(q)}(t)=y^*(0)+\int_0^tb_s^{(q)}\dd s
+\int_0^t\sigma_s\dd W_s,
\qquad t\ge0,
\]
agrees with $y^*(t)$ for every $t\in[0,1]$, almost surely on $E_q$.
In addition, the drift $b^{(q)}$ and volatility $\sigma$ satisfy Assumption~\ref{ass:continuous-price}.
It then follows from \eqref{eq:no-jump-intensity} that
\[
\begin{aligned}
\pr(\Omega_0\setminus E_q)
\le\pr(\Omega_0\cap\{\varrho_q<1\})
 +\pr\left(\sup_{t\le1}|c_t^{(q)}|>C_q\right)
\le2q^{-1}\to0.
\end{aligned}
\]
This proves the first part of the lemma.

We prove the second statement of the lemma.
Recall that $U_M$ and $U_M^{(q)}$ use the same measurable map $\Phi_M$ and $\zeta_M$.
Their sampled price paths agree on $E_q$, so $U_M=U_M^{(q)}$ almost surely on $E_q$ for every $M$.
Let $Y$ be bounded and $\mathcal F$-measurable, and let $f$ be a bounded continuous real-valued function on $\mathbb R^d$ or $[-\infty,\infty]^d$.
Since $E_q\subseteq\Omega_0$ and $U_M=U_M^{(q)}$ almost surely on $E_q$, we have
\begin{align*}
&\left|\E\{\ind(\Omega_0)Yf(U_M)\}
-\widetilde\E\{\ind(\Omega_0)Yf(U)\}\right|\\
&\le
\left|\E\{\ind(E_q)Yf(U_M^{(q)})\}
-\widetilde\E\{\ind(E_q)Yf(U)\}\right|\\
&\qquad+
\left|\E\{\ind(\Omega_0\setminus E_q)Yf(U_M)\}\right|+
\left|\widetilde\E\{\ind(\Omega_0\setminus E_q)Yf(U)\}\right|\\
&\le
\left|\E\{\ind(E_q)Yf(U_M^{(q)})\}
-\widetilde\E\{\ind(E_q)Yf(U)\}\right|+
2\sup|Y|\sup|f|\,\pr(\Omega_0\setminus E_q).
\end{align*}

For each fixed $q$, the first term goes to zero as $M\to\infty$ since $U_M^{(q)}\inST U$ and $\ind(E_q)Y$ is $\mathcal F$-measurable.
It follows that
\[
\begin{aligned}
\limsup_{M\to\infty}
&\left|\E\{\ind(\Omega_0)Yf(U_M)\}
 -\widetilde\E\{\ind(\Omega_0)Yf(U)\}\right|\\
&\le2\sup|Y|\sup|f|\,\pr(\Omega_0\setminus E_q)
\le4q^{-1}\sup|Y|\sup|f|.
\end{aligned}
\]
Letting $q\to\infty$ yields $U_M\inST U$ in restriction to $\Omega_0$.
\end{proof}

\subsection{Joint stable limits and differencing}
\label{subsec:supp-stable-limits}
The unthresholded counterparts $\hat V_{\ell,*}^{(n)}(b)$ and $T_{\ell,*}^{(n)}(g)$ are defined analogously by replacing $\mathbb Q_\ell^{(n),\mathrm{tr}}(b)$ and $\hat\theta_g^{\mathrm{tr}}$ by $\mathbb Q_\ell^{(n)}(b)$ and $\hat\theta_g$, respectively.

\subsubsection{Standard \texorpdfstring{$n$}{n}-power variations}
\label{subsec:supp-standard-mpv}

\begin{proof}[Proof of Theorem~\ref{thm:standard-multipower-variation}]
Suppose first that Assumption~\ref{ass:continuous-price} holds.
We first write $Q^{(2)},\ldots,Q^{(N)}$ with the same summation range and control the omitted boundary terms.
We then apply Theorem~11.2.1(b) and count the shared indices to calculate the covariance matrix.
Finally, we write the limit as $(v^{[4]})^{1/2}\mathcal Z^{(N)}$ and use its deterministic conditional covariance matrix to verify that $\mathcal Z^{(N)}$ is independent of $\mathcal F$.

In this proof, for \(n=2,\dots,N\), define
\[
F_n(x_1,\dots,x_N):=\mu_{2/n}^{-n}\prod_{j=1}^n|x_j|^{2/n},
\qquad F:=(F_2,\dots,F_N),
\]
and
\[
\widetilde Q^{(n)}
:=M^{-1}\sum_{i=1}^{M-N+1}
F_n(M^{1/2}y_i,\dots,M^{1/2}y_{i+N-1}).
\]
Two-homogeneity gives
\[
M^{-1}F_n(M^{1/2}y_i,\ldots,M^{1/2}y_{i+N-1})
=
\mu_{2/n}^{-n}\prod_{j=0}^{n-1}|y_{i+j}|^{2/n}.
\]
Thus $\widetilde Q^{(n)}$ is the standard $n$-power sum truncated at \(M-N+1\).
Consequently \(Q^{(n)}-\widetilde Q^{(n)}\) consists of exactly \(N-n\) boundary summands (none for \(n=N\)).
Regard $F_n$ as functions $\R^n\to\R$ by omitting their unused last $N-n$ arguments for $n=2,\ldots,N$.
Applying Lemma~\ref{lem:boundary-negligible} to these functions simultaneously gives
\begin{equation}
\label{eq:standard-multipower-boundary}
\max_{2\le n\le N}|Q^{(n)}-\widetilde Q^{(n)}|=O_P(M^{-1}),
\qquad
\max_{2\le n\le N}M^{1/2}
 |Q^{(n)}-\widetilde Q^{(n)}|\inP0.
\end{equation}

For each $n=2,\ldots,N$, $F_n$ has the same product form as $F_{Q,k}$ in the proof of Proposition~\ref{prop:JP-common-hypotheses}.
By \citet[Example~11.2.2]{JacodProtter2012}, $F_n$ satisfies conditions~(11.2.6)--(11.2.9) with
\[
B_n=\bigcup_{j=1}^n\{x\in\R^N:x_j=0\},
\qquad p=2,\qquad w=2/n.
\]
Since $B_n\subseteq B_N$ and $2/n\ge2/N$, the argument using \eqref{eq:JP-component-bounds} in the proof of Proposition~\ref{prop:JP-common-hypotheses} establishes these conditions for $F$ with $B=B_N$, $p=2$, and $w=2/N$.
Moreover, $F(-x)=F(x)$, verifying condition~(11.2.5).
The process hypotheses were also verified in that proof.

Since $F_n$ is two-homogeneous and $\E F_n(U_1,\ldots,U_N)=1$ for $U_1,\ldots,U_N\simIID\Normal(0,1)$, the centering term for the stable central limit theorem is $v^{[2]}J_{N-1}$.
Thus Theorem~11.2.1(b) in \citet{JacodProtter2012} implies that
\[
M^{1/2}\left[
(\widetilde Q^{(2)},\ldots,\widetilde Q^{(N)})^\T
-v^{[2]}J_{N-1}
\right]
\inST 
\mathcal V(F,y^*),
\]
where by Lemma~\ref{lem:Ra-scaling}, $\mathcal V(F,y^*)$ is centered Gaussian conditionally on $\mathcal F$ with covariance matrix $v^{[4]}(R_1(F_r,F_s))_{r,s=2}^N$.
It suffices to evaluate $(R_1(F_r,F_s))_{r,s=2}^N$.
For a two-sided i.i.d.\ standard-normal sequence \((Z_h)_{h\in\mathbb Z}\), put
\[
\zeta_h^{(n)}
=\mu_{2/n}^{-n}\prod_{j=0}^{n-1}|Z_{h+j}|^{2/n}.
\]
By Lemma~\ref{lem:cross-kernel}, \(R_1(F_r,F_s)=\sum_h\Cov(\zeta_0^{(r)},\zeta_h^{(s)})\).
Let
\(
u_h(r,s):=|\{0,\dots,r-1\}\cap\{h,\dots,h+s-1\}|.
\)
Then independence of the \((Z_h)_{h\in\mathbb Z}\) gives, for \(u_h\ge1\),
\[
\Cov\{\zeta_0^{(r)},\zeta_h^{(s)}\}
=\left(\frac{\mu_{2/r+2/s}}{\mu_{2/r}\mu_{2/s}}\right)^{u_h}-1,
\]
and the covariance is zero when \(u_h=0\).
Writing \(m=r\wedge s\), overlap count \(m\) occurs at \(|r-s|+1\) shifts, while each size \(u=1,\dots,m-1\) occurs at the two shifts \(h=u-s\) and \(h=r-u\).
We have
\begin{align*}
R_1(F_r,F_s)
&=
\sum_{h\in\mathbb Z}
\Cov\{\zeta_0^{(r)},\zeta_h^{(s)}\}
\\
&=
\sum_{h:u_h(r,s)\ge1}
\left[
\left(
\frac{\mu_{2/r+2/s}}{\mu_{2/r}\mu_{2/s}}
\right)^{u_h(r,s)}
-1
\right]
\\
&=
(|r-s|+1)
\left(
\frac{\mu_{2/r+2/s}}{\mu_{2/r}\mu_{2/s}}
\right)^m
+
2\sum_{u=1}^{m-1}
\left(
\frac{\mu_{2/r+2/s}}{\mu_{2/r}\mu_{2/s}}
\right)^u
+
1-r-s.
\end{align*}
Therefore, \eqref{eq:standard-multipower-covariance} holds. 

Assumption~\ref{ass:volatility-regularity} implies that \(0<v^{[4]}<\infty\) almost surely.
Then, we define
\(
\mathcal Z^{(N)}
:=(v^{[4]})^{-1/2}\mathcal V(F,y^*)
\) on the same extension.
The covariance matrix \(\Sigma_0^{(N)}\) is deterministic, so the conditional distribution of \(\mathcal Z^{(N)}\) given \(\mathcal F\) is the \(\Normal_{N-1}(0,\Sigma_0^{(N)})\) distribution for every realization.
Hence \(\mathcal Z^{(N)}\) is independent of \(\mathcal F\).
Combining this representation with \eqref{eq:standard-multipower-boundary} and stable Slutsky proves the theorem under Assumption~\ref{ass:continuous-price}.

Under Assumption~\ref{ass:price-model}, let $y^{(q)}$ be the processes in Lemma~\ref{lem:no-jump-reduction}.
The vector in the theorem is a function of $(y^*(i/M))_{0\le i\le M}$ and $v^{[2]}$, and $y^{(q)}$ has the volatility $\sigma$ of $y^*$.
Since $y^{(q)}$ satisfies Assumption~\ref{ass:continuous-price}, the vector computed from $y^{(q)}$ converges stably to $(v^{[4]})^{1/2}\mathcal Z^{(N)}$, where $\mathcal Z^{(N)}$ can be taken the same for every $q$ because it is independent of $\mathcal F$ with the $\Normal_{N-1}(0,\Sigma_0^{(N)})$ distribution.
Lemma~\ref{lem:no-jump-reduction} therefore gives the stated convergence in restriction to $\Omega_0$.
\end{proof}
\subsubsection{Lag-augmented \texorpdfstring{$n$}{n}-power variations and self-normalizers}
\label{subsec:supp-joint-clt}

We split the proof of Theorem~\ref{thm:joint-stable-CLT-voljump} into two parts.
We first show that the results hold for $(\hat\theta_g,\mathbb Q_\ell^{(n)}(b))$, where
$\mathbb Q_\ell^{(n)}(b)
= (Q_b^{(n)},\ldots,Q_{b+\ell-1}^{(n)})^\T$.
We then justify that the same result holds for the thresholded version $(\hat\theta_g^{\mathrm{tr}},\mathbb Q_\ell^{(n),\mathrm{tr}}(b))$.

\begin{proof}[Proof of Theorem~\ref{thm:joint-stable-CLT-voljump} for unthresholded statistics]
Suppose first that Assumption~\ref{ass:continuous-price} holds.
Set \(F=(F_\theta,F_{Q,b},\dots,F_{Q,b+\ell-1})\).
With this $F$, the block average in Proposition~\ref{prop:JP-common-hypotheses} equals $(\widetilde\theta_{M,g},\{\mathbb Q_\ell^{(n)}(b)\}^\T)^\T$.
Proposition~\ref{prop:JP-common-hypotheses} gives, stably relative to $\mathcal F$,
\[
\begin{aligned}
&M^{1/2}\left\{
\bigl(\widetilde\theta_{M,g},\{\mathbb Q_\ell^{(n)}(b)\}^\T\bigr)^\T
-\int_0^1\widetilde\E\{F(\sqrt{\sigma_s^2}U_1,\ldots,\sqrt{\sigma_s^2}U_K)
\mid\mathcal F\}\dd s
\right\}\inST \mathcal V(F,y^*),
\end{aligned}
\]
where, on an extension of the underlying probability space, \(U_1,\ldots,U_K\simIID\Normal(0,1)\) are independent of \(\mathcal F\), and the conditional covariance of coordinates \(i,j\) is \(\int_0^1R_{\sigma_s^2}(F_i,F_j)\dd s\).
By Lemma~\ref{lem:Ra-scaling} and Proposition~\ref{prop:covariance-constants}, the conditional covariance matrix is $v^{[4]}\Sigma_{g,\mathbb Q}^{(n)}$.

For \(X_1,\ldots,X_K\sim\Normal(0,a)\) independently, we have
$
\E\{F_\theta(X_1,\ldots,X_K)\}=0$ \newline and $
\E\{F_{Q,k}(X_1,\ldots,X_K)\}=a.
$
Since \(\sigma_s^2\) is \(\mathcal F\)-measurable and \(U_1,\ldots,U_K\) are independent of \(\mathcal F\), we can take \(a=\sigma_s^2\) and hence
\(\int_0^1\widetilde\E\{F(\sqrt{\sigma_s^2}U_1,\ldots,\sqrt{\sigma_s^2}U_K)
\mid\mathcal F\}\dd s=(0,v^{[2]},\dots,v^{[2]})^\T\).

Since $v^{[4]}>0$ almost surely, we define on the same extension
\[
(\mathcal Z_0,\ldots,\mathcal Z_\ell)^\T
:=(v^{[4]})^{-1/2}\mathcal V(F,y^*).
\]
Conditionally on $\mathcal F$, this vector is centered Gaussian with covariance matrix $\Sigma_{g,\mathbb Q}^{(n)}$.
Since $\Sigma_{g,\mathbb Q}^{(n)}$ is non-random, $(\mathcal Z_0,\ldots,\mathcal Z_\ell)^\T$ is independent of $\mathcal F$.
Finally, \eqref{eq:theta-tail} and Lemma~\ref{lem:boundary-negligible} give $M^{1/2}(\hat\theta_g-\widetilde\theta_{M,g})\inP0$, so the stated stable convergence follows by stable Slutsky.

Under Assumption~\ref{ass:price-model}, the argument in the last paragraph of the proof of Theorem~\ref{thm:standard-multipower-variation}, with $(v^{[4]})^{1/2}\mathcal Z$ in place of $(v^{[4]})^{1/2}\mathcal Z^{(N)}$, gives the stated convergence in restriction to $\Omega_0$.
\end{proof}

\begin{proof}[Proof of Theorem~\ref{thm:joint-stable-CLT-voljump} for thresholded statistics]
Let
\[
\mathcal A_M=
\left\{\max_{1\le i\le M}y_i^2
\le\underline\xi\Theta(M^{-1})\right\}.
\]
We first show that $\mathcal A_M$ occurs eventually almost surely under Assumption~\ref{ass:continuous-price}.
It follows from \eqref{eq:Theta-rate} that
\[
\underline\xi\Theta(M^{-1})\to0,
\qquad
\frac{M^{-1}\log M}{\underline\xi\Theta(M^{-1})}\to0.
\]
Therefore, the threshold requirements of \citet[Theorem~1]{Mancini2009} are satisfied.
Since $y^*$ is continuous under Assumption~\ref{ass:continuous-price}, the condition on nonzero jump sizes holds automatically.
Since $\mu^y$ and $\sigma$ are c\`adl\`ag, they are bounded on $[0,1]$ almost surely.
Therefore, the following bounds hold with probability one:
\[
\begin{aligned}
\max_{i\le M}\left|\int_{(i-1)/M}^{i/M}\mu_s^y\dd s\right|
&\le M^{-1}\sup_{s\le1}|\mu_s^y|
=o\!\left[\left(M^{-1}\log M\right)^{1/2}\right],\\
\max_{i\le M}M\int_{(i-1)/M}^{i/M}\sigma_s^2\dd s
&\le\sup_{s\le1}|\sigma_s|^2<\infty.
\end{aligned}
\]
By \citet[Remark~(iii)]{Mancini2009}, these two bounds verify conditions~(1)--(2) of \citet[Theorem~1]{Mancini2009}.
Together with the threshold conditions and the continuity of $y^*$, \citet[Theorem~1]{Mancini2009} then implies that $\mathcal A_M$ occurs eventually almost surely.

Under Assumption~\ref{ass:price-model}, let $y^{(q)}$ and $E_q$ be as in Lemma~\ref{lem:no-jump-reduction}, and let $\mathcal A_M^{(q)}$ be defined as $\mathcal A_M$ with $y_i$ replaced by the returns of $y^{(q)}$.
Since $y^{(q)}$ satisfies Assumption~\ref{ass:continuous-price}, each $\mathcal A_M^{(q)}$ occurs eventually almost surely, and $\mathcal A_M=\mathcal A_M^{(q)}$ almost surely on $E_q$.
Since $E_q\subset\Omega_0$ and $\pr(\Omega_0\setminus\bigcup_{q\ge1}E_q)=0$, $\mathcal A_M$ occurs eventually almost surely on $\Omega_0$.

We then handle a general choice of $\vartheta_{i,M}$.
By \eqref{eq:xi-bounded}, $\vartheta_{i,M}\ge\underline\xi\Theta(M^{-1})$ for all $i,M$, almost surely.
Therefore, given that $\mathcal A_M$ holds, all threshold indicators in \eqref{eq:ivg-tr} and \eqref{eq:Qk-tr} equal one, and
\begin{equation}
\label{eq:null-eventual-equality}
\hat{\IV}_g^{\mathrm{tr}}=\hat{\IV}_g,
\qquad
\hat\theta_g^{\mathrm{tr}}=\hat\theta_g,
\qquad
Q_k^{(n),\mathrm{tr}}=Q_k^{(n)}
\quad(k=b,\ldots,b+\ell-1)
\end{equation}
eventually almost surely on $\Omega_0$.
Therefore, by dominated convergence, replacing all unthresholded quantities in the previous proof with their thresholded counterparts leaves the stable limit in restriction to $\Omega_0$ unchanged.
This completes the proof.
\end{proof}

\begin{proof}[Proof of Theorem~\ref{thm:pairwise-self-normalizers}]
Let $A\in\mathbb R^{(\ell/2)\times\ell}$ have row $j$ equal to $e_{2j-1}^\T-e_{2j}^\T$, where $e_1,\ldots,e_\ell$ are the standard basis vectors of $\R^\ell$.
Then
\[
\mathbb S_\ell^{(n)}(b)=A\mathbb Q_\ell^{(n)}(b),
\qquad
\mathbb S_\ell^{(n),\mathrm{tr}}(b)
=A\mathbb Q_\ell^{(n),\mathrm{tr}}(b).
\]
Also, we have $
AJ_\ell=0,
$ and $
AA^\T=2I_{\ell/2}.
$
Since $\Var(|Z|^{2/n})>0$, we have $\eta_n=\mu_{4/n}/\mu_{2/n}^2>1$ and$
\lambda_\perp
=(\eta_n-1)(\eta_n^{n-1}-1)>0.
$
Moreover, \eqref{eq:SigmaQ-final} enables us to rewrite $\Sigma_{\mathbb Q}^{(n)}$ as 
$$\Sigma_{\mathbb Q}^{(n)}=\lambda_\perp I_\ell
+\Sigma_{\mathbb Q}^{(n)}(1,2)J_\ell J_\ell^\T.$$
Therefore,
\begin{align}\label{eq: A diff}
A\Sigma_{\mathbb Q}^{(n)}A^\T
=\lambda_\perp AA^\T+\Sigma_{\mathbb Q}^{(n)}(1,2)(AJ_\ell)(AJ_\ell)^\T
=2\lambda_\perp I_{\ell/2},
\quad
\Cov(A\mathcal Z_{\mathbb Q},\mathcal Z_0)=C_{g,n}AJ_\ell=0.
\end{align}
Denote $\mathcal Z_{\mathbb Q}=(\mathcal Z_1,\ldots,\mathcal Z_\ell)^\T$.
Then applying $\operatorname{diag}(1,A)$ to the stable convergence stated in Theorem~\ref{thm:joint-stable-CLT-voljump} implies that, in restriction to $\Omega_0$,
\[
M^{1/2}
\begin{pmatrix}
\hat\theta_g^{\mathrm{tr}}\\
\mathbb S_\ell^{(n),\mathrm{tr}}(b)
\end{pmatrix}
\inST 
(v^{[4]})^{1/2}
\begin{pmatrix}
\sigma_g Z_0\\
(2\lambda_\perp)^{1/2}Z
\end{pmatrix},
\]
where it follows from \eqref{eq: A diff} that $Z_0=\mathcal Z_0/\sigma_g$ is independent of $Z=(2\lambda_\perp)^{-1/2}A\mathcal Z_{\mathbb Q}$, and they are jointly independent of $\mathcal F$.
Replacing $\hat\theta_g^{\mathrm{tr}}$ and $\mathbb S_\ell^{(n),\mathrm{tr}}(b)$ by $\hat\theta_g$ and $\mathbb S_\ell^{(n)}(b)$, respectively, yields the same result.

Finally, Proposition~\ref{prop:tr-Q-robust} gives $\mathbb Q_\ell^{(n),\mathrm{tr}}(b)-v^{[2]}J_\ell\inP0$.
Multiplication by $A$ proves the last assertion.
\end{proof}

\subsubsection{The differencing matrix}

\begin{proof}[Proof of Theorem~\ref{lem:contrast-space-whitening}]
Recall that
\(
\Sigma_{\mathbb Q}^{(n)}
=\lambda_\perp I_\ell
+\Sigma_{\mathbb Q}^{(n)}(1,2)J_\ell J_\ell^\T.
\)
As the rows of \(\lambda_\perp^{1/2}D_{\ell,*}^{(n)}\) in \eqref{eq:Helmert-contrast} are orthonormal and each row sums to zero, we have
\[
D_{\ell,*}^{(n)}J_\ell=0,\qquad
\operatorname{rank}(D_{\ell,*}^{(n)})=\ell-1,
\qquad
D_{\ell,*}^{(n)}
\Sigma_{\mathbb Q}^{(n)}
(D_{\ell,*}^{(n)})^\T
=
\lambda_\perp
D_{\ell,*}^{(n)}
(D_{\ell,*}^{(n)})^\T
=
I_{\ell-1}.
\]
Thus $D_{\ell,*}^{(n)}$ satisfies \eqref{eq:D-conditions}.
For any $D$ satisfying \eqref{eq:D-conditions}, let $B=\lambda_\perp^{1/2}D$ and $u=\ell^{-1/2}J_\ell$.
Then $BB^\T=I_{\ell-1}$, $Bu=0$, and $u^\T u=1$.
Thus the square matrix
\[
U=\begin{pmatrix}B\\u^\T\end{pmatrix}
\]
has orthonormal rows and hence satisfies $U^\T U=I_\ell$.
Consequently,
\[
I_\ell
=U^\T U
=B^\T B+uu^\T
=\lambda_\perp D^\T D
+\ell^{-1}J_\ell J_\ell^\T
\]
and hence
\(D^\T D=\lambda_\perp^{-1}
(I_\ell-\ell^{-1}J_\ell J_\ell^\T)\). 

For any $q\in\mathbb R^\ell$, writing $\bar q=\ell^{-1}J_\ell^\T q$, we obtain
\[
\begin{aligned}
\|Dq\|_2^2
=q^\T D^\T Dq
=\lambda_\perp^{-1}
\left\{\sum_{j=1}^\ell q_j^2
-\ell^{-1}(J_\ell^\T q)^2\right\}
=\lambda_\perp^{-1}
\sum_{j=1}^\ell(q_j-\bar q)^2,
\end{aligned}
\]
which proves \eqref{eq:contrast-norm-invariance}.

For any matrices $D_1,D_2$ satisfying \eqref{eq:D-conditions}, let $P=I_\ell-\ell^{-1}J_\ell J_\ell^\T$ and $O=\lambda_\perp D_2D_1^\T$.
Since $D_i^\T D_i=\lambda_\perp^{-1}P$, $D_iP=D_i$, and $\lambda_\perp D_iD_i^\T=I_{\ell-1}$, we have
\[
OD_1=D_2P=D_2,
\qquad
OO^\T=\lambda_\perp D_2PD_2^\T=I_{\ell-1}.
\]
Thus $O$ is orthogonal, which completes the proof.
\end{proof}
\subsection{Asymptotic properties of the proposed tests}
\label{sec:supp-threshold}

\subsubsection{\texorpdfstring{Student-$t$ limits under $H_0$}
{Student-t limits under H0}}
\label{subsec:supp-null-t}

\begin{theorem}[Student-$t$ limit for a differencing matrix]
\label{thm:general-t-limit}
Under Assumptions~\ref{ass:price-model} and~\ref{ass:g-structural}, in restriction to $\Omega_0$,
\[
T_{\ell,*}^{(n)}(g)\inST t_{\ell-1}.
\]
\end{theorem}

\begin{proof}
\emergencystretch=1em
Take any $D$ satisfying \eqref{eq:D-conditions}.
Let \(Z_0=\mathcal Z_0/\sigma_g\) and \(Z=D\mathcal Z_{\mathbb Q}\in\R^{\ell-1}\).
Because \(DJ_\ell=0\), Theorem~\ref{thm:joint-stable-CLT-voljump} gives, in restriction to $\Omega_0$,
\[
M^{1/2}\bigl(\hat\theta_g,
[D\mathbb Q_\ell^{(n)}(b)]^\T\bigr)^\T
\inST 
(v^{[4]})^{1/2}\bigl(\sigma_gZ_0,Z^\T\bigr)^\T.
\]
It follows from \eqref{eq:Sigma-gQ-block} and \eqref{eq:D-conditions} that $\Var(Z_0)=1$, $\Cov(Z)=I_{\ell-1}$, and $\Cov(Z,Z_0)=\sigma_g^{-1}C_{g,n}DJ_\ell=0$.

Therefore \(Z_0,Z_1,\dots,Z_{\ell-1}\) are independent standard normals and are independent of \(\mathcal F\).
By the stable continuous mapping theorem, we have, in restriction to $\Omega_0$,
\[
M^{1/2}\,\hat V_{\ell,*}^{(n)}(b)
\inST 
(v^{[4]})^{1/2}\{\|Z\|_2^2/(\ell-1)\}^{1/2}.
\]
The limiting random variable is strictly positive almost surely since \((v^{[4]})^{1/2}>0\) and \(\|Z\|_2^2\sim\chi_{\ell-1}^2\).
Applying the Portmanteau theorem for the closed set \(\{0\}\) gives \(\pr(\hat V_{\ell,*}^{(n)}(b)=0,\ \Omega_0)\to0\).
Consider the mapping
\[
(x_0,x)\longmapsto
\frac{x_0}{\sigma_g\{\|x\|_2^2/(\ell-1)\}^{1/2}},
\qquad x\ne0,
\]
with value zero when $x=0$.
The set of discontinuities of this map is contained in $\{x=0\}$, which has probability zero under the limiting distribution.
Evaluated at $M^{1/2}(\hat\theta_g,[D\mathbb Q_\ell^{(n)}(b)]^\T)^\T$, the map differs from $T_{\ell,*}^{(n)}(g)$ only on $\{\hat V_{\ell,*}^{(n)}(b)=0\}$, whose intersection with $\Omega_0$ has probability tending to zero.
Stable continuous mapping therefore gives, in restriction to $\Omega_0$ and with $T_{\ell,*}^{(n)}(g)$ regarded as a random variable in the extended real line,
\[
T_{\ell,*}^{(n)}(g)
\inST 
\frac{Z_0}{\{\|Z\|_2^2/(\ell-1)\}^{1/2}}
\sim t_{\ell-1}.
\]
The factor $(v^{[4]})^{1/2}$ cancels between numerator and denominator and hence the limiting random variable is independent of $\mathcal F$.
\end{proof}

\begin{theorem}[Thresholded statistics under $H_0$]
\label{thm:null-transfer-truncation}
Suppose Assumptions~\ref{ass:price-model} and~\ref{ass:g-structural}, \eqref{eq:Theta-rate}, \eqref{eq:local-threshold}, and the uniform lower bound in \eqref{eq:xi-bounded} hold.
Then, in restriction to $\Omega_0$,
\[
T_{\ell,*}^{(n),\mathrm{tr}}(g)
\inST t_{\ell-1}.
\]
\end{theorem}

\begin{proof}[Proof of Theorem~\ref{thm:null-transfer-truncation}]
By \eqref{eq:null-eventual-equality}, $T_{\ell,*}^{(n),\mathrm{tr}}(g)=T_{\ell,*}^{(n)}(g)$ eventually almost surely on $\Omega_0$.
The stated stable convergence follows from Theorem~\ref{thm:general-t-limit} by dominated convergence.
\end{proof}

\subsubsection{\texorpdfstring{Consistency under $H_1$}
{Consistency under H1}}
\label{subsec:supp-fixed-alt}
We use the following consequence of \citet[Thm.~9.2.1]{JacodProtter2012}.
Suppose that $y^*$ satisfies Assumption~\ref{ass:jump-regularity}, and let $F:\mathbb R^m\to\mathbb R$ be continuous and two-homogeneous, where $m\ge1$ is a fixed integer.
Suppose further that, for some $C>0$,
\[
|F(x_1,\ldots,x_m)|
\le C\prod_{j=1}^m(1+|x_j|^2).
\]
Then, for any deterministic $v_M=cM^{-\varpi}$ with $c>0$ and $0<\varpi<1/2$,
\[
\begin{aligned}
&\sum_{i=1}^{M-m+1}
F(y_i,\ldots,y_{i+m-1})
\prod_{j=0}^{m-1}\ind(|y_{i+j}|\le v_M)\\
&\qquad\inP
\int_0^1\widetilde\E\left\{
F(\sqrt{\sigma_s^2}U_1,\ldots,\sqrt{\sigma_s^2}U_m)
\mid\mathcal F\right\}\dd s,
\end{aligned}
\]
where, on an extension of the underlying probability space, $U_1,\ldots,U_m\simIID\Normal(0,1)$ are independent of $\mathcal F$.

\begin{proposition}[Consistency of $\hat{\IV}_g^{\mathrm{tr}}$ under $H_1$]
\label{prop:tr-iv-robust}
Assume Assumptions~\ref{ass:g-structural} and~\ref{ass:jump-regularity} and \eqref{eq:Theta-power}--\eqref{eq:xi-bounded} hold.
Then
\[
\hat{\IV}_g^{\mathrm{tr}} \inP v^{[2]} = \int_0^1 \sigma_s^2\dd s.
\]
Consequently,
\[
\hat\theta_g^{\mathrm{tr}} = \hat{\QV}-\hat{\IV}_g^{\mathrm{tr}}
\inP [y^*,y^*]_1-v^{[2]} = \theta.
\]
\end{proposition}

\begin{proof}[Proof of Proposition~\ref{prop:tr-iv-robust}]
Define
$
v_M^-:=\underline\xi^{1/2}c_\vartheta M^{-\varpi},
$ and $
v_M^+:=\overline\xi^{1/2}c_\vartheta M^{-\varpi}.$
By \eqref{eq:xi-bounded}, the following inequalities hold almost surely for all $i,M$:
\[
(v_M^-)^2\le \vartheta_{i,M}\le (v_M^+)^2.
\]

For $F\in\{g,|g|\}$, define
\[
\hat{\IV}_{F,M}^{\pm}
:=
\sum_{i=1}^{M-b}
F(y_i,\dots,y_{i+b})
\prod_{r=0}^{b}\ind(y_{i+r}^2\le (v_M^{\pm})^2).
\]
Recall that $g$ and $|g|$ are continuous and two-homogeneous.
Therefore, we have $g(0)=0$ and $C:=\sup_{\|u\|_2=1}|g(u)|<\infty$.
It implies that for both $F=g$ and $F=|g|$,
\begin{equation}
\label{eq:g-quadratic-bound}
|F(x)|=|g(x)|\le C\|x\|_2^2
\le C\sum_{j=1}^{b+1}|x_j|^2
\le C\prod_{j=1}^{b+1}(1+|x_j|^2).
\end{equation}
We then apply the consequence of \citet[Thm.~9.2.1]{JacodProtter2012} with $m=b+1$ and $v_M=v_M^-$ or $v_M=v_M^+$ to yield the following convergence:
\[
\hat{\IV}_{F,M}^{\pm}\inP
\int_0^1\widetilde\E\{F(\sqrt{\sigma_s^2}U_1,\dots,\sqrt{\sigma_s^2}U_{b+1})
\mid\mathcal F\}\dd s,
\qquad F\in\{g,|g|\}.
\]
Therefore, for each $F\in\{g,|g|\}$, $\hat{\IV}_{F,M}^+$ and $\hat{\IV}_{F,M}^-$ converge to the same probability limit.

We next handle $\hat{\IV}_g^{\mathrm{tr}}$.
First, define
\[
\begin{aligned}
A_i^-:=\prod_{r=0}^{b}\ind\{y_{i+r}^2\le(v_M^-)^2\},
\quad
A_i^{\mathrm{tr}}:=\prod_{r=0}^{b}
\ind\{y_{i+r}^2\le\vartheta_{i+r,M}\},
\quad
A_i^+:=\prod_{r=0}^{b}\ind\{y_{i+r}^2\le(v_M^+)^2\}.
\end{aligned}
\]
By \eqref{eq:xi-bounded}, 
$
0\le A_i^-\le A_i^{\mathrm{tr}}\le A_i^+
\le1
$
holds for all $i,M$ almost surely.
Therefore, we have
\[
\left|\hat{\IV}_g^{\mathrm{tr}}-\hat{\IV}_{g,M}^-\right|
\le \sum_{i=1}^{M-b}|g(y_i,\dots,y_{i+b})|
       (A_i^{\mathrm{tr}}-A_i^-)
\le \hat{\IV}_{|g|,M}^+-\hat{\IV}_{|g|,M}^-
\inP0.
\]
It follows from the squeeze theorem and the convergence of $\hat{\IV}_{g,M}^-$ that
\[
\hat{\IV}_g^{\mathrm{tr}}\inP
\int_0^1\widetilde\E\{g(\sqrt{\sigma_s^2}U_1,\dots,\sqrt{\sigma_s^2}U_{b+1})
\mid\mathcal F\}\dd s.
\]
Since $\sigma_s^2$ is $\mathcal F$-measurable and $U_1,\dots,U_{b+1}$ are independent of $\mathcal F$, Assumption~\ref{ass:g-structural}(iii) and two-homogeneity give, when $\sigma_s^2>0$,
\[
\widetilde\E\{g(\sqrt{\sigma_s^2}U_1,\dots,\sqrt{\sigma_s^2}U_{b+1})
\mid\mathcal F\}
=\sigma_s^2\widetilde\E g(U_1,\dots,U_{b+1})
=\sigma_s^2.
\]
At $\sigma_s^2=0$ the same identity follows from $g(0)=0$.
Therefore
\[
\hat{\IV}_g^{\mathrm{tr}}\inP \int_0^1 \sigma_s^2\dd s=v^{[2]}.
\]
Applying \citet[Thm.~3.3.1\textnormal{(B)}]{JacodProtter2012} with $f(x)=x^2$ gives
\[
\hat{\QV}\inP
\int_0^1 \sigma_s^2\dd s+\sum_{0<s\le1}[\Delta y^*(s)]^2
=[y^*,y^*]_1.
\]
Therefore, we have
\[
\hat\theta_g^{\mathrm{tr}}
=\hat{\QV}-\hat{\IV}_g^{\mathrm{tr}}
\inP [y^*,y^*]_1-v^{[2]}=\theta.
\]
\end{proof}

\begin{proposition}[Consistency of $\mathbb Q_\ell^{(n),\mathrm{tr}}(b)$ under $H_1$]
\label{prop:tr-Q-robust}
Assume Assumption~\ref{ass:jump-regularity} and \eqref{eq:Theta-power}--\eqref{eq:xi-bounded} hold.
Then
\[
\mathbb Q_\ell^{(n),\mathrm{tr}}(b)-v^{[2]}J_\ell\inP0.
\]
\end{proposition}

\begin{proof}[Proof of Proposition~\ref{prop:tr-Q-robust}]
Fix $k\in\{b,\dots,b+\ell-1\}$ and the threshold $v_M^-$ defined in the proof of Proposition~\ref{prop:tr-iv-robust}.
Since $|\mathcal I_k|=n$, $F_{Q,k}$ is continuous and two-homogeneous.
The arithmetic--geometric mean inequality gives
\begin{equation}
\label{eq:FQ-quadratic-bound}
0\le F_{Q,k}(x)
\le \frac{\mu_{2/n}^{-n}}{n}\sum_{r\in \mathcal I_k}|x_r|^2
\le C\|x\|_2^2,
\end{equation}
which implies the growth condition in the stated consequence of \citet[Thm.~9.2.1]{JacodProtter2012}.
Define
\[
\overline Q_{k,M}^{(n),-}
:=
\sum_{i=1}^{M-K+1}
F_{Q,k}(y_i,\dots,y_{i+K-1})
\prod_{r=1}^{K}\ind(|y_{i+r-1}|\le v_M^-).
\]
It follows from \citet[Thm.~9.2.1]{JacodProtter2012} with $m=K$, $F=F_{Q,k}$, and $v_M=v_M^-$ that
\[
\overline Q_{k,M}^{(n),-}
\inP\int_0^1\widetilde\E\{F_{Q,k}(\sqrt{\sigma_s^2}U_1,\dots,\sqrt{\sigma_s^2}U_K)
\mid\mathcal F\}\dd s.
\]
Notice that we can write
\(
Q_k^{(n)}
=M^{-1}\sum_{i=1}^{M-K+1}
F_{Q,k}(M^{1/2}\,y_i,\dots,M^{1/2}\,y_{i+K-1}).
\)
Under Assumption~\ref{ass:jump-regularity}, $y^*$ satisfies Assumption~\textup{(H)} of \citet{JacodProtter2012}.
Condition~\textnormal{(8.4.2)} of \citet{JacodProtter2012} holds because the function
\(
u\longmapsto\mu_{2/n}^{-n/K}(1+u^2)^{-(n-1)/n}
\)
is continuous and vanishes at infinity, and
\[
F_{Q,k}(x)
\le \prod_{j=1}^{K}
\mu_{2/n}^{-n/K}(1+|x_j|^2)^{-(n-1)/n}(1+|x_j|^2).
\]
Then, \citet[Thm.~8.4.1 and Example~8.4.4]{JacodProtter2012} imply that
\[
Q_k^{(n)}\inP\int_0^1\widetilde\E\{F_{Q,k}(\sqrt{\sigma_s^2}U_1,\dots,
\sqrt{\sigma_s^2}U_K)\mid\mathcal F\}\dd s.
\]
It follows from the construction that $\vartheta_{j,M}^{1/2}\ge v_M^-$ almost surely for all $j,M$ and hence we have
\[
0\le \overline Q_{k,M}^{(n),-}
\le Q_k^{(n),\mathrm{tr}}
\le Q_k^{(n)},
\]
where $\overline Q_{k,M}^{(n),-}$ and $Q_k^{(n)}$ converge to the same limit.
To identify the common limit, note that if $X_1,\dots,X_K\simIID\Normal(0,a)$ with $a\ge0$, then 
\[
\E\{F_{Q,k}(X_1,\dots,X_K)\}
=\mu_{2/n}^{-n}\left\{a^{1/n}\mu_{2/n}\right\}^{n}
=a.
\]
As \(\sigma_s^2\) is \(\mathcal F\)-measurable and \(U_1,\dots,U_K\) are independent of \(\mathcal F\), taking \(a=\sigma_s^2\) shows that the common limit is \(\int_0^1\sigma_s^2\dd s\).
By the squeeze theorem,
\[
Q_k^{(n),\mathrm{tr}}\inP\int_0^1\sigma_s^2\dd s=v^{[2]}.
\]
Since $\ell$ is fixed, the componentwise convergences imply
\(
\mathbb Q_\ell^{(n),\mathrm{tr}}(b)-v^{[2]}J_\ell \inP 0.
\)
\end{proof}

\begin{theorem}[Consistency under $H_1$]
\label{thm:thresholded-consistency}
Fix $a\in\mathbb R$.
Under Assumptions~\ref{ass:g-structural} and~\ref{ass:jump-regularity}, with \eqref{eq:Theta-power}--\eqref{eq:xi-bounded},
\[
\pr\{\Psi_{\ell,*}^{(n),\mathrm{tr}}(a)=1,\ \Omega_1\}
\to\pr(\Omega_1).
\]
In particular, if $\pr(\Omega_1)>0$, then
$\pr\{\Psi_{\ell,*}^{(n),\mathrm{tr}}(t_{\ell-1,\alpha_0})=1\mid\Omega_1\}
\to1$ for every $\alpha_0\in(0,1)$.
\end{theorem}

\begin{proof}[Proof of Theorem~\ref{thm:thresholded-consistency}]
Under Assumption~\ref{ass:jump-regularity}, Proposition~\ref{prop:tr-iv-robust} gives $\hat\theta_g^{\mathrm{tr}}\inP\theta$.
Proposition~\ref{prop:tr-Q-robust} and \eqref{eq:VD-tr} give $\hat V_{\ell,*}^{(n),\mathrm{tr}}(b)\inP0$, while the structural assumption gives $0<\sigma_g<\infty$.
Hence, for every fixed $a\in\R$,
\[
 \hat\theta_g^{\mathrm{tr}}
 -a\sigma_g\hat V_{\ell,*}^{(n),\mathrm{tr}}(b)\inP\theta.
\]
Since $\sigma_g>0$, \eqref{eq:cross-multiplied-rejection} and the definition \eqref{eq:T-tr}, including the values $\pm\infty$ of $T_{\ell,*}^{(n),\mathrm{tr}}(g)$ when $\hat V_{\ell,*}^{(n),\mathrm{tr}}(b)=0$, give the equivalent form
$\Psi_{\ell,*}^{(n),\mathrm{tr}}(a)
=\ind\{\hat\theta_g^{\mathrm{tr}}
>a\sigma_g\hat V_{\ell,*}^{(n),\mathrm{tr}}(b)\}$.
Hence, for every $\eta>0$,
\[
\{\Psi_{\ell,*}^{(n),\mathrm{tr}}(a)=0\}
=\{\hat\theta_g^{\mathrm{tr}}
 -a\sigma_g\hat V_{\ell,*}^{(n),\mathrm{tr}}(b)\le0\}
\subset
\{|\hat\theta_g^{\mathrm{tr}}
 -a\sigma_g\hat V_{\ell,*}^{(n),\mathrm{tr}}(b)-\theta|>\eta\}
\cup\{\theta\le\eta\}.
\]
Intersecting the inclusion above with $\Omega_1=\{\theta>0\}$ gives
\[
\limsup_{M\to\infty}
\pr\{\Psi_{\ell,*}^{(n),\mathrm{tr}}(a)=0,\ \Omega_1\}
\le\pr(0<\theta\le\eta).
\]
Letting $\eta\downarrow0$ proves the first assertion, and dividing by $\pr(\Omega_1)>0$ with $a=t_{\ell-1,\alpha_0}$ proves the second.
\end{proof}

\subsubsection{\texorpdfstring{Local distribution and power under $H_1'$}
{Local distribution and power under H1'}}
\label{subsec:supp-local-alt}

Throughout Section \ref{subsec:supp-local-alt}, we work under Assumption~\ref{ass:abstract-local-alt}.
For $a_M>0$, the notation $X_M=O_{\mathrm{a.s.}}(a_M)$ means $\limsup_{M\to\infty}|X_M|/a_M<\infty$ almost surely, and $X_M=o_{\mathrm{a.s.}}(a_M)$ means $X_M/a_M\to0$ almost surely.

For \(i=1,\ldots,M\), define the increments of \(y^c\) by
\[
 y_i^c:=y^c(i/M)-y^c((i-1)/M).
\]
Write $m_M^c:=\max_{1\le i\le M}|y_i^c|$.
Since \(y^c\) satisfies Assumption~\ref{ass:continuous-price} and \eqref{eq:Theta-power} implies \eqref{eq:Theta-rate}, the argument establishing the eventual occurrence of $\mathcal A_M$ in the proof of Theorem~\ref{thm:joint-stable-CLT-voljump} for thresholded statistics also applies to \(y^c\).
It follows that
\begin{equation}
\label{eq:local-continuous-threshold-bound}
(m_M^c)^2\le\underline\xi\Theta(M^{-1})
=\underline\xi c_\vartheta^2M^{-2\varpi}
\qquad\text{eventually almost surely}.
\end{equation}
In particular, $m_M^c=O_{\mathrm{a.s.}}(M^{-\varpi})
=o_{\mathrm{a.s.}}(M^{-1/4})$ because $\varpi>1/4$.
Also, recall that $T_q$ indicates the jump location and hence
\[
\Delta J_M(T_q)=M^{-1/4}\kappa_q,
\qquad
\sum_{q=1}^{N_{\mathrm{loc}}}[\Delta J_M(T_q)]^2
=M^{-1/2}\sum_{q=1}^{N_{\mathrm{loc}}}\kappa_q^2
=\frac{\varphi}{M^{1/2}}.
\]

\begin{theorem}[Local distribution and power]
\label{thm:trunc-local-power}
Suppose Assumptions~\ref{ass:g-structural} and \ref{ass:abstract-local-alt} hold, and take $D$ satisfying \eqref{eq:D-conditions}.
Then
\[
M^{1/2}\left(
\hat\theta_g^{\mathrm{tr}},
\left[\mathbb Q_\ell^{(n),\mathrm{tr}}(b)-v^{[2]}J_\ell\right]^\T
\right)^\T
\inST 
(\varphi,0,\ldots,0)^\T+(v^{[4]})^{1/2}\mathcal Z,
\]
where $\mathcal Z$ is the Gaussian vector in Theorem~\ref{thm:joint-stable-CLT-voljump}.
Consequently,
\[
T_{\ell,*}^{(n),\mathrm{tr}}(g)\inST 
\frac{Z_0+\lambda_J}{\{\|Z\|_2^2/(\ell-1)\}^{1/2}},
\qquad
\lambda_J=\frac{\varphi}{\sigma_g(v^{[4]})^{1/2}},
\]
where $Z_0=\mathcal Z_0/\sigma_g$ and $Z=D\mathcal Z_{\mathbb Q}=(Z_1,\ldots,Z_{\ell-1})^\T$.
The random variables $Z_0,Z_1,\ldots,Z_{\ell-1}$ are independent standard normals and are independent of $\mathcal F$.
Conditionally on $\mathcal F$, the limit has law $t_{\ell-1}[\lambda_J]$.
For every $\alpha_0\in(0,1)$,
\[
\pr\{\Psi_{\ell,*}^{(n),\mathrm{tr}}(t_{\ell-1,\alpha_0})=1\}
\to\E\{\beta_{\ell-1,\alpha_0}(\lambda_J)\}.
\]
\end{theorem}

\begin{proof}[Proof of Theorem~\ref{thm:trunc-local-power}]
In this proof, we will show that thresholding removes jump-affected returns, so the integrated-variance and multipower statistics are asymptotically equivalent to those computed from $y^c$ after being scaled by $M^{1/2}$.
In contrast, realized quadratic variation retains the squared local jumps, producing the shift $\varphi$.
Combining these results with the joint stable limit for $y^c$ yields the conditional noncentral Student-$t$ law and the limiting rejection probability.

Let $\hat{\QV}_M^c$, $\hat{\IV}_{g,M}^c$, $\hat\theta_{g,M}^c$, and $Q_{k,M}^{(n),c}$ denote the statistics obtained by replacing $y_i$ with $y_i^c$ in the definitions of $\hat{\QV}$, $\hat{\IV}_g$, $\hat\theta_g$, and $Q_k^{(n)}$, respectively.
As \(y^c\) satisfies Assumption~\ref{ass:continuous-price}, Theorem~\ref{thm:joint-stable-CLT-voljump} applies to \(y^c\) with $\Omega_0=\Omega$ and implies that
\begin{equation}
\label{eq:local-proof-companion-CLT}
M^{1/2}
\bigl(
\hat{\theta}_{g,M}^c,\;
Q_{b,M}^{(n),c}-v^{[2]},\ldots,
Q_{b+\ell-1,M}^{(n),c}-v^{[2]}
\bigr)^\T
\inST 
(v^{[4]})^{1/2}\mathcal Z.
\end{equation}

Since $T_q\in(0,1)$, we have $T_q\in((i_q(M)-1)/M,i_q(M)/M]$ with a unique index $i_q(M):=\lceil MT_q\rceil$ for $q=1,\ldots,N_{\mathrm{loc}}$.
It then follows from the definition of $J_M$ that
\[
\begin{aligned}
y_i-y_i^c
=M^{-1/4}\sum_{q=1}^{N_{\mathrm{loc}}}\kappa_q\ind\{i=i_q(M)\}
\end{aligned}
\]
for $i=1,\ldots,M$.
Since there are only finitely many jumps and $T_q\in(0,1)$ are pairwise distinct almost surely, their associated return indices $i_q(M)$ are also pairwise distinct for all sufficiently large $M$ almost surely.
Therefore, $y_{i_q(M)}=y_{i_q(M)}^c+M^{-1/4}\kappa_q$ for all $q$, eventually almost surely.
By \eqref{eq:local-continuous-threshold-bound}, we have
\(
M^{1/4}m_M^c
=O_{\mathrm{a.s.}}(M^{1/4-\varpi})\to0.
\)
Let $\kappa_*:=\min_{1\le q\le N_{\mathrm{loc}}}|\kappa_q|>0$ and $\mathcal B_M:=\{i_q(M):q=1,\ldots,N_{\mathrm{loc}}\}$.
Then uniformly over $i\in\mathcal B_M$, 
\[
|y_{i}|
\ge
\kappa_*M^{-1/4}-m_M^c
\ge
\frac{\kappa_*}{2}M^{-1/4}
\]
eventually almost surely.
For $\varpi>\frac14$, we have
\(
\max_{1\le i\le M}\vartheta_{i,M}^{1/2}
\le \overline\xi^{1/2}c_\vartheta M^{-\varpi}
=o(M^{-1/4})
\).
Hence, for large $M$, the squared returns $y_i^2$ impacted by jumps will exceed the threshold, i.e.,
\[
\ind(y_{i_q(M)}^2\le\vartheta_{i_q(M),M})=0,
\qquad q=1,\dots,N_{\mathrm{loc}},
\]
eventually almost surely.
On the other hand, for returns not impacted by jumps, \eqref{eq:local-continuous-threshold-bound} gives
\(
(y_i^c)^2\le(m_M^c)^2
\le\underline\xi\Theta(M^{-1})\le\vartheta_{i,M}.
\)
Therefore, uniformly over \(i\notin\mathcal B_M\), we have
\[
y_i=y_i^c,
\qquad
\ind(y_i^2\le\vartheta_{i,M})=1,
\]
eventually almost surely, i.e., $y_i^2$ will not exceed the threshold for large $M$.
We can thus represent the thresholding condition as
\begin{equation}
\label{eq:local-eventual-indicators}
\ind\{y_i^2\le\vartheta_{i,M}\}
=
\ind\{i\notin\mathcal B_M\},
\qquad i=1,\ldots,M,
\end{equation}
eventually almost surely.
In particular, those summands omitted from $\hat{\IV}_g$ due to thresholding can be indexed by
\[
\mathcal A_M^{\IV}
:=\{i\in\{1,\dots,M-b\}:
\{i,\dots,i+b\}\cap\mathcal B_M\ne\varnothing\}.
\]
Similarly, those omitted from $Q_k^{(n)}$ are indexed by
\[
\mathcal A_{k,M}^{Q}
:=\{i\in\{1,\dots,M-K+1\}:
\{i+r-1:r\in\mathcal I_k\}\cap\mathcal B_M\ne\varnothing\}
\]
for $k\in\{b,\dots,b+\ell-1\}$. 

By \eqref{eq:local-eventual-indicators}, the following equations hold eventually almost surely:
\begin{align}
\hat{\IV}_g^{\mathrm{tr}}-\hat{\IV}_{g,M}^c
&=-\sum_{i\in\mathcal A_M^{\IV}}g(y_i^c,\dots,y_{i+b}^c),
\label{eq:local-iv-deleted-set}\\
Q_k^{(n),\mathrm{tr}}-Q_{k,M}^{(n),c}
&=-\sum_{i\in\mathcal A_{k,M}^{Q}}
F_{Q,k}(y_i^c,\dots,y_{i+K-1}^c).
\label{eq:local-Q-deleted-set}
\end{align}
By \eqref{eq:g-quadratic-bound} and \eqref{eq:FQ-quadratic-bound}, after enlarging a constant $C$ independent of $M$, $|g(x)|\le C\|x\|_2^2$ and $0\le F_{Q,k}(x)\le C\|x\|_2^2$.
Each return with index in $\mathcal B_M$ appears in at most $b+1$ summands of $\hat{\IV}_g$ and $n$ summands of $Q_k^{(n)}$.
Hence $|\mathcal A_M^{\IV}|\le N_{\mathrm{loc}}(b+1)$ and $|\mathcal A_{k,M}^{Q}|\le N_{\mathrm{loc}}n\le N_{\mathrm{loc}}K$.
It then follows from \eqref{eq:local-continuous-threshold-bound}, \eqref{eq:local-iv-deleted-set}, and \eqref{eq:local-Q-deleted-set} that
\[
\left|\hat{\IV}_g^{\mathrm{tr}}-\hat{\IV}_{g,M}^c\right|
+\max_{b\le k\le b+\ell-1}
\left|Q_k^{(n),\mathrm{tr}}-Q_{k,M}^{(n),c}\right|
=O_{\mathrm{a.s.}}(M^{-2\varpi}).
\]
Since $M^{1/2}\,M^{-2\varpi}=M^{1/2-2\varpi}\to0$ for $\varpi>1/4$, we obtain
\begin{equation}
\label{eq:local-proof-threshold-vs-cont}
M^{1/2}\left|\hat{\IV}_g^{\mathrm{tr}}-\hat{\IV}_{g,M}^c\right|
+M^{1/2}\max_{b\le k\le b+\ell-1}
\left|Q_k^{(n),\mathrm{tr}}-Q_{k,M}^{(n),c}\right|
\to0
\qquad\text{a.s.}
\end{equation}

We next investigate the effect of those jumps on the realized quadratic variations.
For all sufficiently large $M$, as $i_q(M)$ are pairwise distinct, direct algebra gives
\[
\hat{\QV}-\hat{\QV}_M^c
=
\sum_{q=1}^{N_{\mathrm{loc}}}\left(2M^{-1/4}\kappa_q y_{i_q(M)}^c + M^{-1/2}\kappa_q^2\right).
\]
As
\(
2M^{1/4}\sum_{q=1}^{N_{\mathrm{loc}}}|\kappa_q|m_M^c
=O_{\mathrm{a.s.}}(M^{1/4-\varpi})\to0
\), we have
\begin{equation}
\label{eq:local-proof-qv-shift}
M^{1/2}\left(\hat{\QV}-\hat{\QV}_M^c\right)
=
\varphi+o_{\mathrm{a.s.}}(1).
\end{equation}
By definition,
\(
\hat\theta_g^{\mathrm{tr}}-\hat\theta_{g,M}^c
=
(\hat{\QV}-\hat{\QV}_M^c)
-
(\hat{\IV}_g^{\mathrm{tr}}-\hat{\IV}_{g,M}^c).
\)
Combining \eqref{eq:local-proof-threshold-vs-cont} with \eqref{eq:local-proof-qv-shift} gives
\begin{equation}
\label{eq:local-proof-theta-shift}
M^{1/2}\left(\hat\theta_g^{\mathrm{tr}}-\hat\theta_{g,M}^c\right)
=
\varphi+o_{\mathrm{a.s.}}(1).
\end{equation}
Equations~\eqref{eq:local-proof-threshold-vs-cont} and \eqref{eq:local-proof-theta-shift} give
\[
M^{1/2}\left[
\begin{pmatrix}
\hat\theta_g^{\mathrm{tr}}\\
\mathbb Q_\ell^{(n),\mathrm{tr}}(b)-v^{[2]}J_\ell
\end{pmatrix}
-
\begin{pmatrix}
\hat\theta_{g,M}^c\\
Q_{b,M}^{(n),c}-v^{[2]}\\
\vdots\\
Q_{b+\ell-1,M}^{(n),c}-v^{[2]}
\end{pmatrix}
\right]
\longrightarrow_{\mathrm{a.s.}}
(\varphi,0,\ldots,0)^\T.
\]
The joint stable convergence stated in this theorem then follows from \eqref{eq:local-proof-companion-CLT} and stable Slutsky relative to $\mathcal F$.
Dependence between $T_q$ and $y^c$ is allowed throughout our argument.

Recall $Z_0=\mathcal Z_0/\sigma_g$ and $Z=D\mathcal Z_{\mathbb Q}\in\R^{\ell-1}$.
As established in the proof of Theorem~\ref{thm:general-t-limit}, $Z_0,Z_1,\ldots,Z_{\ell-1}$ are independent standard normal variables and are independent of $\mathcal F$.
Applying $\operatorname{diag}(1,D)$ to the joint limit and using $DJ_\ell=0$ gives
\begin{equation}
\label{eq:local-projected-stable-limit}
M^{1/2}\bigl(
\hat\theta_g^{\mathrm{tr}},
[D\mathbb Q_\ell^{(n),\mathrm{tr}}(b)]^\T
\bigr)^\T
\inST 
\bigl(\varphi+(v^{[4]})^{1/2}\sigma_gZ_0,
(v^{[4]})^{1/2}Z^\T\bigr)^\T.
\end{equation}
The stable continuous mapping theorem for the Euclidean norm gives
\[
M^{1/2}\,\hat V_{\ell,*}^{(n),\mathrm{tr}}(b)
\inST 
(v^{[4]})^{1/2}
\left\{\frac{\|Z\|_2^2}{\ell-1}\right\}^{1/2}.
\]
The limit is strictly positive almost surely because $(v^{[4]})^{1/2}>0$ almost surely and $\|Z\|_2^2\sim\chi_{\ell-1}^2$.
Hence, by the same Portmanteau argument as in the proof of Theorem~\ref{thm:general-t-limit},
\begin{equation}
\label{eq:local-zero-event-vanishes}
\pr\{\hat V_{\ell,*}^{(n),\mathrm{tr}}(b)=0\}\to0.
\end{equation}

Applying the continuous-mapping argument in the proof of Theorem~\ref{thm:general-t-limit} to \eqref{eq:local-projected-stable-limit}, with \eqref{eq:local-zero-event-vanishes}, gives, on the extended real line,
\[
T_{\ell,*}^{(n),\mathrm{tr}}(g)
\inST 
\mathcal T_J
:=
\frac{\varphi+(v^{[4]})^{1/2}\sigma_gZ_0}
{\sigma_g(v^{[4]})^{1/2}\{\|Z\|_2^2/(\ell-1)\}^{1/2}}
=
\frac{Z_0+\lambda_J}{\{\|Z\|_2^2/(\ell-1)\}^{1/2}}.
\]
Conditionally on $\mathcal F$, this has law $t_{\ell-1}[\lambda_J]$.

Since $\lambda_J$ is $\mathcal F$-measurable, $Z_0$ is independent of $(\mathcal F,Z)$, and $\|Z\|_2>0$ almost surely, the conditional distribution of $\mathcal T_J$ given $(\mathcal F,Z)$ is Gaussian with strictly positive variance.
Therefore, $\pr(\mathcal T_J=x)=0$ for every $x\in\R$, so the distribution of $\mathcal T_J$ is continuous at $t_{\ell-1,\alpha_0}$.
Weak convergence plus the definition of $\beta$ gives
\[
\lim_{M\to\infty}
\pr\left\{T_{\ell,*}^{(n),\mathrm{tr}}(g)>t_{\ell-1,\alpha_0}\right\}
=
\pr\{\mathcal T_J>t_{\ell-1,\alpha_0}\}
=\E\left[\beta_{\ell-1,\alpha_0}(\lambda_J)\right].
\]
This completes the proof.
\end{proof}

\subsubsection{The pairwise and optimal-differencing tests}

\begin{proof}[Proof of Theorem~\ref{thm:pairwise-properties}]
\textit{(Validity)} Theorem~\ref{thm:pairwise-self-normalizers} and continuous mapping give, in restriction to $\Omega_0$,
\[
M^{1/2}\,\hat V_\ell^{(n),\mathrm{tr}}(b)
\inST 
(2\lambda_\perp v^{[4]})^{1/2}
\left\{\frac2\ell\sum_{j=1}^{\ell/2}Z_j^2\right\}^{1/2}.
\]
Since the limiting variable is strictly positive almost surely, the Portmanteau theorem gives $\pr\{\hat V_\ell^{(n),\mathrm{tr}}(b)=0,\ \Omega_0\}\to0$.
The joint convergence in Theorem~\ref{thm:pairwise-self-normalizers}, stable continuous mapping and $\pr\{\hat V_\ell^{(n),\mathrm{tr}}(b)=0,\ \Omega_0\}\to0$ yield, as in the proof of Theorem~\ref{thm:general-t-limit} and in restriction to $\Omega_0$,
\[
T_\ell^{(n),\mathrm{tr}}(g)
\inST 
\frac{Z_0}{\{(2/\ell)\sum_{j=1}^{\ell/2}Z_j^2\}^{1/2}}
\sim t_{\ell/2}.
\]
The unthresholded statement of the same theorem gives the same limit for $T_\ell^{(n)}(g)$ as well.
Since $\Psi_\ell^{(n),\mathrm{tr}}(a)
=\ind\{T_\ell^{(n),\mathrm{tr}}(g)>a\}$,
the convergence above and continuity of the $t_{\ell/2}$ distribution imply that
\(
\pr\{\Psi_\ell^{(n),\mathrm{tr}}(t_{\ell/2,\alpha_0})=1,\ \Omega_0\}
\to\alpha_0\pr(\Omega_0)
\)
for $\alpha_0\in(0,1)$.
Dividing by $\pr(\Omega_0)>0$ gives the stated conditional limit.\\
\textit{(Consistency)} Proposition~\ref{prop:tr-iv-robust} gives $\hat\theta_g^{\mathrm{tr}}\inP\theta$, while Theorem~\ref{thm:pairwise-self-normalizers} gives $\mathbb S_\ell^{(n),\mathrm{tr}}(b)\inP0$ and hence $\hat V_\ell^{(n),\mathrm{tr}}(b)\inP0$ by \eqref{eq: pairwise SN}.
It follows that for every fixed $a$, we have
\[
(2\lambda_\perp)^{1/2}\hat\theta_g^{\mathrm{tr}}
-a\sigma_g\hat V_\ell^{(n),\mathrm{tr}}(b)
\inP(2\lambda_\perp)^{1/2}\theta.
\]
Since $\lambda_\perp,\sigma_g>0$, \eqref{eq:pairwise-rejection-rule} gives
\[
\Psi_\ell^{(n),\mathrm{tr}}(a)
=\ind\{(2\lambda_\perp)^{1/2}\hat\theta_g^{\mathrm{tr}}
-a\sigma_g\hat V_\ell^{(n),\mathrm{tr}}(b)>0\}.
\]
The argument in the proof of Theorem~\ref{thm:thresholded-consistency}, applied to this event, gives
\(
\pr\{\Psi_\ell^{(n),\mathrm{tr}}(a)=1,\ \Omega_1\}
\to\pr(\Omega_1),
\)
and dividing by $\pr(\Omega_1)>0$ with $a=t_{\ell/2,\alpha_0}$ gives the stated conditional limit.\\
\textit{(Local alternative distribution)} Let $A$ be the matrix defined in the proof of Theorem~\ref{thm:pairwise-self-normalizers}.
Under $H_1'$, applying $\operatorname{diag}(1,A)$ to the joint limit in Theorem~\ref{thm:trunc-local-power} for
$\bigl(\hat\theta_g^{\mathrm{tr}},
[\mathbb Q_\ell^{(n),\mathrm{tr}}(b)-v^{[2]}J_\ell]^\T\bigr)^\T$, and using $AJ_\ell=0$, gives
\[
M^{1/2}
\begin{pmatrix}
\hat\theta_g^{\mathrm{tr}}\\
\mathbb S_\ell^{(n),\mathrm{tr}}(b)
\end{pmatrix}
\inST 
(\varphi,0,\ldots,0)^\T
+(v^{[4]})^{1/2}
\begin{pmatrix}\sigma_g Z_0\\(2\lambda_\perp)^{1/2}Z\end{pmatrix},
\]
where $Z_0=\mathcal Z_0/\sigma_g$ and
$Z=(2\lambda_\perp)^{-1/2}A\mathcal Z_{\mathbb Q}
=(Z_1,\ldots,Z_{\ell/2})^\T$.
The proof of Theorem~\ref{thm:pairwise-self-normalizers} shows that $Z_0,Z_1,\ldots,Z_{\ell/2}$ are independent standard normal variables, jointly independent of $\mathcal F$.
The limit of $M^{1/2}\,\hat V_\ell^{(n),\mathrm{tr}}(b)$ is positive almost surely and hence the Portmanteau theorem implies that $\pr\{\hat V_\ell^{(n),\mathrm{tr}}(b)=0\}\to0$.
Therefore, stable continuous mapping gives
\[
T_\ell^{(n),\mathrm{tr}}(g)
\inST 
\frac{Z_0+\lambda_J}
{\{(2/\ell)\sum_{j=1}^{\ell/2}Z_j^2\}^{1/2}}
\]
as in the proof of Theorem~\ref{thm:general-t-limit}.
It follows that the limiting random variable has law $t_{\ell/2}[\lambda_J]$ conditionally on $\mathcal F$.\\
\textit{(Local power function)} As the conditional limiting distribution is continuous and $\Psi_\ell^{(n),\mathrm{tr}}(a)=\ind\{T_\ell^{(n),\mathrm{tr}}(g)>a\}$ for every $a\in\mathbb R$, taking $a=t_{\ell/2,\alpha_0}$ gives the limiting rejection probability $\E\{\beta_{\ell/2,\alpha_0}(\lambda_J)\}$.
\end{proof}

\begin{proof}[Proof of Theorem~\ref{thm:canonical-properties}]
Theorem~\ref{lem:contrast-space-whitening} shows that $D_{\ell,*}^{(n)}$ satisfies \eqref{eq:D-conditions}.
For every fixed $q\in\mathbb R^\ell$, equation~\eqref{eq:contrast-norm-invariance} shows that $\|Dq\|_2$ has the same value for every $D$ satisfying \eqref{eq:D-conditions}.
Validity follows from Theorem~\ref{thm:null-transfer-truncation}, consistency from Theorem~\ref{thm:thresholded-consistency}, and the local distribution and power from Theorem~\ref{thm:trunc-local-power}, with $D=D_{\ell,*}^{(n)}$.
Since $\Psi_{\ell,*}^{(n),\mathrm{tr}}(a)=\ind\{T_{\ell,*}^{(n),\mathrm{tr}}(g)>a\}$ for every $a\in\mathbb R$, continuity of the $t_{\ell-1}$ law gives
$\pr\{\Psi_{\ell,*}^{(n),\mathrm{tr}}(t_{\ell-1,\alpha_0})=1,\ \Omega_0\}
\to\alpha_0\pr(\Omega_0)$
for every $\alpha_0\in(0,1)$, and dividing by $\pr(\Omega_0)>0$ gives the stated conditional limit.
This completes the proof.
\end{proof}

\begin{proof}[Proof of Corollary~\ref{cor:comparison-tests}]
Apply Theorem~\ref{thm:pairwise-properties} with $2\ell$ in place of $\ell$ and Theorem~\ref{thm:canonical-properties} with $\ell+1$ in place of $\ell$.
\end{proof}

\subsubsection{Local power comparisons}
\label{subsec:supp-local-power}

\begin{theorem}[Local distribution and power of $T_0^{\mathrm{tr}}(g)$]
\label{thm:trunc-conventional-local-power}
Under Assumptions~\ref{ass:g-structural} and \ref{ass:abstract-local-alt}, we have $\hat v^{[4]}\inP v^{[4]}$ and
\[
\frac{T_0^{\mathrm{tr}}(g)}{\sigma_g}
\inST Z_0+\lambda_J,
\]
where $Z_0\sim\Normal(0,1)$ is independent of $\mathcal F$ and $\lambda_J=\varphi/\{\sigma_g(v^{[4]})^{1/2}\}$.
Conditionally on $\mathcal F$, the limiting distribution is $\Normal(\lambda_J,1)$.
Moreover, for every $\alpha_0\in(0,1)$,
\[
\pr\{T_0^{\mathrm{tr}}(g)>\sigma_g z_{\alpha_0}\}
\to\E\{\beta_{\infty,\alpha_0}(\lambda_J)\},
\]
where $\beta_{\infty,\alpha_0}(\lambda)
=\Phi(\lambda-z_{\alpha_0})$.
\end{theorem}

\begin{proof}
Define the quarticity estimator computed from $y^c$ by
\(
\hat v^{[4]c}
:=\frac{M}{\mu_1^4}
\sum_{i=1}^{M-3}\prod_{j=0}^3|y_{i+j}^c|.
\)
Since $y^c$ satisfies Assumption~\ref{ass:continuous-price}, it also satisfies Assumption~(H), as verified in the proof of Proposition~\ref{prop:JP-common-hypotheses}.
Theorem~8.4.1 and Example~8.4.4 of \citet{JacodProtter2012} give $\hat v^{[4]c}\inP v^{[4]}$.

Since the finitely many jump times $T_q$ are pairwise distinct almost surely, their associated return indices $i_q(M):=\lceil MT_q\rceil$ are eventually separated by more than three.
Thus, each product of four consecutive returns contains at most one jump-affected return.
Recall that \(
y_i-y_i^c
=M^{-1/4}\sum_{q=1}^{N_{\mathrm{loc}}}\kappa_q\ind\{i=i_q(M)\}.
\) Then for any product containing the $q$th jump, the reverse-triangle inequality gives
\[
\left|
\prod_{j=0}^3|y_{i+j}|
-
\prod_{j=0}^3|y_{i+j}^c|
\right|
\le M^{-1/4}|\kappa_q|(m_M^c)^3.
\]
As each jump-affected return appears in at most four such products, summing over them for each $q$ and using \eqref{eq:local-continuous-threshold-bound} gives
\[
\begin{aligned}
|\hat v^{[4]}-\hat v^{[4]c}|
&\le \frac{4M^{3/4}}{\mu_1^4}
\Big(\textstyle\sum_q|\kappa_q|\Big)(m_M^c)^3
=O_{\mathrm{a.s.}}(M^{3/4-3\varpi})\to0.
\end{aligned}
\]
Since $\hat v^{[4]c}-v^{[4]}\inP 0$, we conclude that $\hat v^{[4]}\inP v^{[4]}$.

Theorem~\ref{thm:trunc-local-power} and stable Slutsky give
\[
\frac{T_0^{\mathrm{tr}}(g)}{\sigma_g}
\inST 
\frac{\varphi+(v^{[4]})^{1/2}\sigma_g Z_0}
{\sigma_g(v^{[4]})^{1/2}}
=Z_0+\lambda_J.
\]
Conditionally on $\mathcal F$, the limiting random variable is normally distributed with mean $\lambda_J$ and unit variance.
Its distribution function is thus continuous and we have
\[
\pr\{T_0^{\mathrm{tr}}(g)>\sigma_g z_{\alpha_0}\}
\to\E\{\beta_{\infty,\alpha_0}(\lambda_J)\}.
\]
\end{proof}

\begin{proof}[Proof of Corollary~\ref{cor:local-power-n-ell}]
For every fixed $n\ge3$ and $\ell\ge2$, Theorem~\ref{thm:trunc-local-power} gives the limiting rejection probability
\(
 \E\{\beta_{\ell-1,\alpha_0}(\lambda_J)\}
\).
As the formula for $\lambda_J$ contains no $n$, $\E\{\beta_{\ell-1,\alpha_0}(\lambda_J)\}$ is the same for all $n\ge3$.
 If $\ell'>\ell$, then applying \citet[Thm.~8]{ghosh1973some} with degrees of freedom $\ell-1$ and $\ell'-1$ and $\alpha_0\in(0,1/2)$ gives
\(
 \beta_{\ell'-1,\alpha_0}(\lambda_J)
 >
 \beta_{\ell-1,\alpha_0}(\lambda_J)
\)
almost surely.
Both random variables are bounded by one, so taking expectations preserves the strict inequality.
\end{proof}

\begin{proof}[Proof of Corollary~\ref{cor:local-power-retention}]
Fix $\alpha_0\in(0,1/2)$.
As $Z$ is independent of $W_\nu$, we have
\[
\begin{aligned}
\beta_{\nu,\alpha_0}(\lambda)
=\E\left[
\pr\left\{
Z>t_{\nu,\alpha_0}(W_\nu/\nu)^{1/2}-\lambda
\,\middle|\,W_\nu
\right\}
\right]
=\E\left\{
\Phi\left(\lambda-t_{\nu,\alpha_0}(W_\nu/\nu)^{1/2}\right)
\right\}.
\end{aligned}
\]
Since $t_{\nu,\alpha_0}\to z_{\alpha_0}$ and $(W_\nu/\nu)^{1/2}\inP1$ as $\nu\to\infty$, boundedness and continuity of $\Phi$ give, for any fixed $\lambda\ge0$,
\begin{equation}
\label{eq:supp-retention-pointwise-limit}
\beta_{\nu,\alpha_0}(\lambda)
\to\Phi(\lambda-z_{\alpha_0})
=\beta_{\infty,\alpha_0}(\lambda)
\quad\text{as }\nu\to\infty.
\end{equation}
By \citet[Thm.~8]{ghosh1973some}, $\beta_{\nu,\alpha_0}(\lambda)$ is strictly increasing in $\nu$ for every $\lambda>0$.
Together with \eqref{eq:supp-retention-pointwise-limit}, this implies
$\beta_{\nu,\alpha_0}(\lambda)
<\beta_{\infty,\alpha_0}(\lambda)$.

For fixed $\nu$, let
$q_\nu(\lambda):=\beta_{\nu,\alpha_0}(\lambda)/
\beta_{\infty,\alpha_0}(\lambda)$.
The conditional representation above and dominated convergence show that $\beta_{\nu,\alpha_0}(\lambda)$ is continuous in $\lambda$.
Since $\beta_{\infty,\alpha_0}(\lambda)\ge\alpha_0>0$ for $\lambda\ge0$, $q_\nu$ is continuous on $[0,\infty)$.
By definition, we have $q_\nu(0)=1$.
For every fixed $\nu$, both power functions converge to one as $\lambda\to\infty$, so $q_\nu(\lambda)\to1$ as $\lambda\to\infty$.
Since $q_\nu(\lambda)<1$ for every $\lambda>0$, its minimum is attained at a finite positive $\lambda$.
In particular, $R_{\nu,\alpha_0}<1$.

Let $\lambda>0$ be a minimizer of $q_{\nu+1}$.
By \citet[Thm.~8]{ghosh1973some},
\[
R_{\nu+1,\alpha_0}
=q_{\nu+1}(\lambda)
>q_\nu(\lambda)
\ge R_{\nu,\alpha_0}.
\]

To show that $R_{\nu,\alpha_0}\to1$ as $\nu\to\infty$, we establish uniform convergence of $q_\nu$ to one on $[0,\infty)$.
Since $W_\nu\sim\chi_\nu^2$, we have $\E(W_\nu/\nu)=1$ and $\Var(W_\nu/\nu)=2/\nu$.
Thus, by $|\sqrt{x}-1|\le|x-1|$ for $x\ge0$ and Cauchy--Schwarz,
\[
\begin{aligned}
\E\left|(W_\nu/\nu)^{1/2}-1\right|
\le\E|W_\nu/\nu-1|
\le\{\Var(W_\nu/\nu)\}^{1/2}
=(2/\nu)^{1/2}.
\end{aligned}
\]
For every $\lambda\ge0$, using
$\beta_{\infty,\alpha_0}(\lambda)
=\Phi(\lambda-z_{\alpha_0})\ge\alpha_0$
and the Lipschitz bound $|\Phi(x)-\Phi(y)|\le|x-y|/\sqrt{2\pi}$, we obtain
\[
\begin{aligned}
|q_\nu(\lambda)-1|
&=\frac{
\left|\E\left\{
\Phi\left(\lambda-t_{\nu,\alpha_0}
(W_\nu/\nu)^{1/2}\right)
-\Phi(\lambda-z_{\alpha_0})
\right\}\right|}
{\Phi(\lambda-z_{\alpha_0})}\\
&\le\frac{1}{\alpha_0\sqrt{2\pi}}
\E\left|t_{\nu,\alpha_0}
(W_\nu/\nu)^{1/2}-z_{\alpha_0}\right|\\
&=\frac{1}{\alpha_0\sqrt{2\pi}}
\E\left|t_{\nu,\alpha_0}
\left\{(W_\nu/\nu)^{1/2}-1\right\}
+(t_{\nu,\alpha_0}-z_{\alpha_0})\right|\\
&\le\frac{
t_{\nu,\alpha_0}
\E\left|(W_\nu/\nu)^{1/2}-1\right|
+|t_{\nu,\alpha_0}-z_{\alpha_0}|}
{\alpha_0\sqrt{2\pi}}\\
&\le\frac{
|t_{\nu,\alpha_0}-z_{\alpha_0}|
+t_{\nu,\alpha_0}(2/\nu)^{1/2}}
{\alpha_0\sqrt{2\pi}}.
\end{aligned}
\]
The upper bound is independent of $\lambda$ and converges to zero as $\nu\to\infty$, since $t_{\nu,\alpha_0}\to z_{\alpha_0}$.
Therefore, $\sup_{\lambda\ge0}|q_\nu(\lambda)-1|\to0$ as $\nu\to\infty$, which implies $R_{\nu,\alpha_0}\to1$ as $\nu\to\infty$.

By \eqref{eq:ell-wc-definition}, this convergence ensures that $\ell^*(\alpha_0,p)$ is finite for every $p\in(0,1)$.
Moreover, for $0<p_1\le p_2<1$, the set defining $\ell^*(\alpha_0,p_2)$ is contained in that defining $\ell^*(\alpha_0,p_1)$.
Hence $\ell^*(\alpha_0,p_1)\le\ell^*(\alpha_0,p_2)$.
\end{proof}

\subsection{Microstructure noise}
\label{subsec:supp-microstructure-noise}
Throughout this subsection, $g=g_1$ and $b=1$, so $L=\ell+2$, $K=(n-1)(\ell+2)$, and $k=1,\ldots,\ell$.
Brackets $[z]$ indicate evaluation at a return sequence $z=(z_1,\ldots,z_M)$, including for $\hat V_{\ell,*}^{(n),\mathrm{tr}}(1)[z]$.
For the vector $\mathbb Q_\ell^{(n)}$ we write $\mathbb Q_\ell^{(n)}(1;z)$.

We begin by outlining the proof of Theorem~\ref{thm:microstructure-noise}.
Under $H_0$, although microstructure noise vanishes as $M\to\infty$, its effect on the test statistics can remain non-negligible.
To quantify this effect, we compare $\hat\theta_{g_1}[\widetilde y]$ and $\mathbb Q_\ell^{(n)}(1;\widetilde y)$ with their noise-free counterparts $\hat\theta_{g_1}[y]$ and $\mathbb Q_\ell^{(n)}(1;y)$.
Lemma~\ref{lem:noise-conditional-law} shows that, conditional on $\mathcal G_1$, the increments of the diffusion component of the latent price and the microstructure-noise increments are jointly Gaussian, and derives their covariances.
Lemma~\ref{lem:noise-drift} shows that removing the price drift changes $\hat\theta_{g_1}[\widetilde y]-\hat\theta_{g_1}[y]$ and $Q_k^{(n)}[\widetilde y]-Q_k^{(n)}[y]$ by only $o_P(M^{-1/2})$.
Lemma~\ref{lem:noise-effect} then identifies the explicit form of the nonpositive shift in $\hat\theta_{g_1}$ due to microstructure noise and shows that the leading noise-induced changes in $Q_k^{(n)}$ are common across $k$ and eliminated by $D$ because $DJ_\ell=0$.
Together with the thresholding argument, these results establish, under the conditions of Theorem~\ref{thm:microstructure-noise}(i) and Assumption~\ref{ass:continuous-price},
\[
\begin{aligned}
M^{1/2}\bigl\{
\hat\theta_{g_1}^{\mathrm{tr}}[\widetilde y]
-\hat\theta_{g_1}[y]
\bigr\}
&\inP
-\frac12\int_0^1
\Bigl(\frac c{\sigma_t}+\iota\Bigr)^2\dd t,\\
M^{1/2}D\bigl\{
\mathbb Q_\ell^{(n),\mathrm{tr}}(1;\widetilde y)
-\mathbb Q_\ell^{(n)}(1;y)
\bigr\}
&\inP0.
\end{aligned}
\]
Combining these comparisons with the joint stable limit of the noise-free statistics proves Theorem~\ref{thm:microstructure-noise}(i) under Assumption~\ref{ass:continuous-price}, and Lemma~\ref{lem:no-jump-reduction} extends it to $\Omega_0$ under Assumption~\ref{ass:microstructure-noise-null}.
For Theorem~\ref{thm:microstructure-noise}(ii), we separately show that sufficiently small noise preserves the consistency of the thresholded statistics under $H_1$.

By two-homogeneity, $\hat{\QV}[z]$, $\hat{\IV}_{g_1}[z]$, and $Q_k^{(n)}[z]$ can each be written in the form
\begin{equation}
\label{eq:noise-sum-form}
\begin{gathered}
\frac1M\sum_{i=1}^{M-d+1}
H(M^{1/2}z_i,\ldots,M^{1/2}z_{i+d-1}),\quad
(H,d)=(v\mapsto v^2,1),\ (g_1,2),\ (F_{Q,k},K),
\end{gathered}
\end{equation}
respectively.

Put $x_M=M\omega_M^2$, $\gamma_M=\rho_Mx_M^{1/2}$, and $I_{i,M}=((i-1)/M,i/M]$.
Under Assumption~\ref{ass:continuous-price}, for $i=1,\ldots,M$, we can decompose $\widetilde y_i$ as
\[
\begin{aligned}
M^{1/2}\widetilde y_i
=M^{1/2}\int_{I_{i,M}}\mu_s^y\dd s
+
\overbrace{
\underbrace{M^{1/2}\int_{I_{i,M}}\sigma_s\dd W_s}_{=:\, X_{i,M}^0}
+
\underbrace{M^{1/2}(\varepsilon_{i,M}-\varepsilon_{i-1,M})}
_{=:\, e_{i,M}^{\varepsilon}}
}^{=:\, X_{i,M}}.
\end{aligned}
\]
The corresponding noise-free return satisfies
$M^{1/2}y_i=X_{i,M}^0+
M^{1/2}\int_{I_{i,M}}\mu_s^y\dd s$.
Also, define $q_{i,M}^2=M\int_{I_{i,M}}\sigma_s^2\dd s$ and write $X^0=(X_{i,M}^0)_{i=1}^M$ and $X=(X_{i,M})_{i=1}^M$.
Therefore, $M^{-1/2}X$ and $M^{-1/2}X^0$ can be regarded as the drift-free counterparts of $\widetilde y$ and $y$, respectively.

\begin{assumption}[Localization bound]
\label{ass:noise-bounded-coefficients}
For some fixed constant $m\ge2$, almost surely,
\[
m^{-1}\le\sigma_t\le m,\quad|\mu_t^y|\le m,\quad
|\widetilde\mu_t^\sigma|+|a_t|\le m
\qquad\text{for all }t\in[0,1].
\]
\end{assumption}

Assumption~\ref{ass:noise-bounded-coefficients} will only be imposed in the technical lemmas, i.e., Lemma~\ref{lem:noise-conditional-law}(iii) and Lemmas~\ref{lem:noise-drift}--\ref{lem:noise-effect}.
This assumption is not required to justify Theorem~\ref{thm:microstructure-noise}.
In the proof of part~(i), we will relax this assumption through a localization argument.
Throughout these lemmas, the values of the constant $C$ may change from line to line.
It will only depend on $m$, $n$, and $\ell$, but not on $M$ or $i$.

\begin{lemma}[Conditional Gaussian structure of drift-free returns]
\label{lem:noise-conditional-law}
Suppose Assumption~\ref{ass:microstructure-noise-null} holds.
\begin{enumerate}[label=(\roman*)]
\item The path $\{\sigma_t\}_{t\in[0,1]}$ is $\mathcal G_1$-measurable, and $Z_{1,M},\ldots,Z_{M,M},G_{0,M},\ldots,G_{M,M}$ are mutually independent standard normal variables, jointly independent of $\mathcal G_1$.

\item Conditionally on $\mathcal G_1$,
$(X_{1,M}^0,\ldots,X_{M,M}^0,e_{1,M}^{\varepsilon},\ldots,
e_{M,M}^{\varepsilon})$ is centered Gaussian, $X_{i,M}^0$ are independent with variances $q_{i,M}^2$, and $\widetilde\E\{(e_{i,M}^{\varepsilon})^2\mid\mathcal G_1\}=2x_M$.
For $1\le i\le j\le M$,
\begin{equation}
\label{eq:noise-covariance}
\widetilde\E(X_{i,M}X_{j,M}\mid\mathcal G_1)
=
(q_{i,M}^2+2x_M+2\gamma_Mq_{i,M})\ind(j=i)
-(x_M+\gamma_Mq_{i,M})\ind(j=i+1).
\end{equation}
Moreover, $X_{i,M}^0=q_{i,M}Z_{i,M}$, while $X_{i,M}$ depends only on $q_{i,M}$ and $Z_{j,M},G_{j,M}$ for $j\in\{i-1,i\}$, with $Z_{0,M}$ omitted when $i=1$.

\item If Assumption~\ref{ass:noise-bounded-coefficients} holds and $M\omega_M^2\to0$ as $M\to\infty$, then, for all sufficiently large $M$,
\(
\widetilde\E\{(X_{i,M}^0)^2\mid\mathcal G_1\},
\,
\widetilde\E\{X_{i,M}^2\mid\mathcal G_1\}
\in[(2m)^{-2},C]
\)
for all $i=1,\ldots,M$ almost surely.
\end{enumerate}
\end{lemma}

\begin{proof}
\textbf{(i).} Under Assumption~\ref{ass:microstructure-noise-null}, $\widetilde\sigma\equiv0$ and continuity remove the integral with respect to $W$ and the sum of volatility jumps, respectively, in \eqref{eq:volatility-decomposition}, so we can write $\sigma_t=\sigma_0+\int_0^t\widetilde\mu_s^\sigma\dd s+L_t^\sigma$.
Since $\sigma_0$ is $\mathcal G_0$-measurable, $\widetilde\mu^\sigma$ is $(\mathcal G_t)$-progressively measurable, and $L^\sigma$ is a $(\mathcal G_t)$-local martingale, $\sigma$ is $(\mathcal G_t)$-adapted and hence $\mathcal G_1$-measurable.
It follows that $\{\sigma_t\}_{t\in[0,1]}$ is $\mathcal G_1$-measurable.

Since the path of $\sigma$ is $\mathcal G_1$-measurable and $W$ is independent of $\mathcal G_1$, conditionally on $\mathcal G_1$, the volatility path is deterministic and $W$ remains a Brownian motion.
Therefore, the normality of Wiener integrals and It\^o's isometry imply that
\[
\left[
\left.
\int_{I_{i,M}}\sigma_s\dd W_s
\,\right|\,\mathcal G_1
\right]
\sim
\Normal\left(0,\int_{I_{i,M}}\sigma_s^2\dd s\right),
\qquad i=1,\ldots,M.
\]
These integrals are conditionally independent since $I_{i,M}$ are disjoint intervals.
It follows that
\(
\left.
(Z_{1,M},\ldots,Z_{M,M})
\,\right|\,\mathcal G_1
\sim\Normal_M(0,I_M)
\).
As the conditional distribution does not depend on $\mathcal G_1$, the $Z_{i,M}$ are mutually independent standard normal variables, jointly independent of $\mathcal G_1$.
Finally, recall that $G_{0,M},\ldots,G_{M,M}\simIID\Normal(0,1)$ are independent of $\mathcal F$, which contains $\mathcal G_1$ as its subset.
This proves~(i).

\textbf{(ii).} By Assumption~\ref{ass:microstructure-noise-null}, $X_{i,M}^0=q_{i,M}Z_{i,M}$.
Also, by definition of $\varepsilon_{i,M}$, we can represent $e_{i,M}^{\varepsilon}$ as a linear combination of $Z_{i-1,M}$, $Z_{i,M}$, $G_{i-1,M}$, and $G_{i,M}$ with deterministic coefficients.
For $i=1$, $\varepsilon_{0,M}=\omega_M G_{0,M}$ and $Z_{0,M}$ does not appear.
By~(i), conditionally on $\mathcal G_1$, the $Z_{j,M}$ and $G_{j,M}$ are mutually independent standard normal variables.
Since $q_{i,M}$ is $\mathcal G_1$-measurable and the coefficients defining $e_{i,M}^{\varepsilon}$ are deterministic, the variables $X_{i,M}^0$ and $e_{i,M}^{\varepsilon}$ for $i=1,\ldots,M$ are jointly centered Gaussian conditional on $\mathcal G_1$.

By~(i) and the definition of $\varepsilon_{i,M}$, $[\varepsilon_{0,M},\ldots,\varepsilon_{M,M}|\mathcal G_1]\simIID \Normal(0,\omega_M^2)$.
Recalling that
$e_{i,M}^{\varepsilon}
=M^{1/2}(\varepsilon_{i,M}-\varepsilon_{i-1,M})$
and $x_M=M\omega_M^2$, we have, for $1\le i,j\le M$,
\begin{equation}\label{eq: ee-identity}
\widetilde\E(e_{i,M}^{\varepsilon}e_{j,M}^{\varepsilon}
\mid\mathcal G_1)
=x_M\{2\ind(j=i)-\ind(|j-i|=1)\}.
\end{equation}
It follows from the definition of $\varepsilon_{j,M}$ that
$\widetilde\E(\varepsilon_{j,M}Z_{i,M}\mid\mathcal G_1)
=\rho_M\omega_M\ind(j=i)$
for $1\le i\le M$ and $0\le j\le M$.
Using $X_{i,M}^0=q_{i,M}Z_{i,M}$ and $M^{1/2}\rho_M\omega_M=\gamma_M$, we obtain, for $1\le i,j\le M$,
\begin{equation}\label{eq: Xe-identity}
\widetilde\E(X_{i,M}^0e_{j,M}^{\varepsilon}
\mid\mathcal G_1)
=\gamma_Mq_{i,M}\{\ind(j=i)-\ind(j=i+1)\}.    
\end{equation}
Finally, for $1\le i\le j\le M$, \eqref{eq: Xe-identity} gives
$\widetilde\E(e_{i,M}^{\varepsilon}X_{j,M}^0
\mid\mathcal G_1)
=\gamma_Mq_{i,M}\ind(j=i)$.
Expanding
$X_{i,M}X_{j,M}
=(X_{i,M}^0+e_{i,M}^{\varepsilon})
(X_{j,M}^0+e_{j,M}^{\varepsilon})$
and using
$\widetilde\E(X_{i,M}^0X_{j,M}^0\mid\mathcal G_1)
=q_{i,M}^2\ind(j=i)$
together with \eqref{eq: ee-identity} and \eqref{eq: Xe-identity} yield
\[
\widetilde\E(X_{i,M}X_{j,M}\mid\mathcal G_1)
=(q_{i,M}^2+2x_M+2\gamma_Mq_{i,M})\ind(j=i)
-(x_M+\gamma_Mq_{i,M})\ind(j=i+1).
\]

\textbf{(iii).} Since $\gamma_M^2=\rho_M^2x_M$, we can rewrite \eqref{eq:noise-covariance} as
\[\widetilde\E(X_{i,M}^2\mid\mathcal G_1)
=(q_{i,M}+\gamma_M)^2+(2-\rho_M^2)x_M.\] 
We then show that it lies in $[(2m)^{-2},C]$.
Under Assumption~\ref{ass:noise-bounded-coefficients}, $q_{i,M}^2=M\int_{I_{i,M}}\sigma_s^2\dd s\in[m^{-2},m^2]$.
When $x_M=M\omega_M^2\to0$, we have $|\gamma_M|\le x_M^{1/2}\le(2m)^{-1}$ for all sufficiently large $M$.
Therefore, $q_{i,M}+\gamma_M\ge m^{-1}-(2m)^{-1}=(2m)^{-1}$ and $x_M\to0$.
It follows that $\widetilde\E(X_{i,M}^2\mid\mathcal G_1)\geq (2m)^{-2}$ eventually.
Meanwhile, as $|q_{i,M}|\leq m,|\gamma_M|\leq (2m)^{-1}$ with $x_M\to0$, we have $\widetilde\E(X_{i,M}^2\mid\mathcal G_1)\leq C$ for some constant $C$ eventually.
\end{proof}

For $(H,d)$ in \eqref{eq:noise-sum-form} and $i\le M-d+1$, put $U_{i,M}^0=(X_{i,M}^0,\ldots,X_{i+d-1,M}^0)^\T$ and $U_{i,M}=(X_{i,M},\ldots,X_{i+d-1,M})^\T$.

\begin{lemma}[Negligibility of drift–noise interaction]
\label{lem:noise-drift}
Suppose Assumptions~\ref{ass:continuous-price}, \ref{ass:microstructure-noise-null}, and~\ref{ass:noise-bounded-coefficients} hold and $M\omega_M^2\to0$ as $M\to\infty$.
Then, as $M\to\infty$,
\[
M^{1/2}\Big\{
\Big(\hat\theta_{g_1}[\widetilde y]-\hat\theta_{g_1}[y]\Big)
-
\Big(\hat\theta_{g_1}[M^{-1/2}X]-\hat\theta_{g_1}[M^{-1/2}X^0]\Big)
\Big\}
\inP0,
\]
and the same holds with $\hat\theta_{g_1}$ replaced by $Q_k^{(n)}$, $k=1,\ldots,\ell$.
\end{lemma}

\begin{proof}
Since $\hat\theta_{g_1}=\hat{\QV}-\hat{\IV}_{g_1}$, it suffices to treat each pair $(H,d)$ in \eqref{eq:noise-sum-form}, with $\hat\theta_{g_1}$ in the lemma replaced by the corresponding statistic.
For $i\le M-d+1$, put
\[
u_i=M^{1/2}\left(\int_{I_{i,M}}\mu_s^y\dd s,\ldots,
\int_{I_{i+d-1,M}}\mu_s^y\dd s\right)^\T,
\]
so that $M^{1/2}(\widetilde y_i,\ldots,\widetilde y_{i+d-1})^\T=U_{i,M}+u_i$ and $M^{1/2}(y_i,\ldots,y_{i+d-1})^\T=U_{i,M}^0+u_i$.
By two-homogeneity of $H$, it suffices to prove the statement by showing
\begin{equation}\label{eq: H-bound}
    M^{-1/2}
\sum_{i=1}^{M-d+1}\Big\{
H(U_{i,M}+u_i)-H(U_{i,M})-H(U_{i,M}^0+u_i)+H(U_{i,M}^0)
\Big\}
\end{equation}
converges to 0 in probability for each pair $(H,d)$ in \eqref{eq:noise-sum-form}.
We next show that the conditional expectations given $\mathcal G_1$ of the absolute values of the summands in \eqref{eq: H-bound} can be bounded by a deterministic sequence that is $o(M^{-1/2})$ uniformly in $i$.

Since $H$ is continuously differentiable whenever all coordinates of its argument are nonzero, define
\begin{equation}\label{eq: taylor H}
    R(w,u)=H(w+u)-H(w)-\nabla H(w)^\T u
\end{equation}
for such $w$ and $u\in\mathbb R^d$.
By Lemma~\ref{lem:noise-conditional-law}(ii) and~(iii), for all sufficiently large $M$ and $V\in\{U_{i,M},U_{i,M}^0\}$, $V$ is conditionally Gaussian with strictly positive conditional variances.
Thus every coordinate of $V$ is nonzero almost surely, so $\nabla H(U_{i,M})$ and $\nabla H(U_{i,M}^0)$ are well defined almost everywhere.
We can then apply \eqref{eq: taylor H} to rewrite the $i$-th summand of \eqref{eq: H-bound} as
\[
\{\nabla H(U_{i,M})-\nabla H(U_{i,M}^0)\}^\T u_i
+R(U_{i,M},u_i)-R(U_{i,M}^0,u_i).
\]
The vector $u_i$ need not be $\mathcal G_1$-measurable.
We bypass this problem by using the bound $\|u_i\|_\infty\le mM^{-1/2}$ from Assumption~\ref{ass:noise-bounded-coefficients}.
It suffices to show both $\nabla H(U_{i,M})-\nabla H(U_{i,M}^0)$ and $R(U_{i,M},u_i)-R(U_{i,M}^0,u_i)$ vanish uniformly over $i$ at the rates established below.

We first derive a bound for the remainder term $R$.
We will show that for all sufficiently large $M$, all $i$, $V\in\{U_{i,M},U_{i,M}^0\}$, and $0<h\le1$, with $p=2/n$,
\begin{equation}
\label{eq:noise-drift-remainder}
\widetilde\E\left\{\sup_{\|u\|_\infty\le h}|R(V,u)|\middle|\mathcal G_1\right\}
\le Ch^{1+p}.
\end{equation}
For $H(v)=v^2$, direct algebra gives $R(w,u)=u^2$, so $\sup_{|u|\le h}|R(w,u)|=h^2\le h^{1+p}$.  

For $H=g_1$, define $\delta_j=|w_j+u_j|-|w_j|$ for $j=1,2$ and we can rewrite $\mu_1^2g_1(w+u)=(|w_1|+\delta_1)(|w_2|+\delta_2)$.
It follows from \eqref{eq: taylor H} that
\[
\mu_1^2R(w,u)=\delta_1\delta_2+|w_2|\{\delta_1-\operatorname{sgn}(w_1)u_1\}
+|w_1|\{\delta_2-\operatorname{sgn}(w_2)u_2\}.
\]
When $\|u\|_\infty\le h$, we have $|\delta_j|\le h$ and hence $|\delta_1\delta_2|\leq h^2$.
Notice that for $|w_j|>h$, we have $|\delta_j-\operatorname{sgn}(w_j)u_j|=0$ and thus $|w_{j'}||\delta_j-\operatorname{sgn}(w_j)u_j|\leq |w_{j'}|(h+h)\ind(|w_j|\leq h)$ for $j'\neq j$.
Therefore, 
\begin{equation}\label{eq: Rbound-g_1}
    \mu_1^2\sup_{\|u\|_\infty\le h}|R(w,u)|\le h^2+2h|w_2|\ind(|w_1|\le h)
+2h|w_1|\ind(|w_2|\le h).
\end{equation}
We then take $w=V$ and control this upper bound after taking conditional expectation given $\mathcal G_1$.
By Lemma~\ref{lem:noise-conditional-law}(ii) and~(iii), conditionally on $\mathcal G_1$, $(V_1,V_2)$ is centered Gaussian with $\Var(V_j\mid\mathcal G_1)\in[(2m)^{-2},C]$.
Moreover, the property of conditional Gaussian distribution gives
\[
\widetilde\E(V_2\mid\mathcal G_1,V_1)
=\frac{\widetilde\E(V_1V_2\mid\mathcal G_1)}
{\widetilde\E(V_1^2\mid\mathcal G_1)}V_1,
\qquad
\left|
\frac{\widetilde\E(V_1V_2\mid\mathcal G_1)}
{\widetilde\E(V_1^2\mid\mathcal G_1)}
\right|^2
\le
\frac{\widetilde\E(V_2^2\mid\mathcal G_1)}
{\widetilde\E(V_1^2\mid\mathcal G_1)}
\le C.
\]
Therefore, we have $|\widetilde\E(V_2\mid\mathcal G_1,V_1)|\le |C V_1|$ and $\Var(V_2\mid\mathcal G_1,V_1)\leq \Var(V_2\mid\mathcal G_1) \le C$.
Also, we have
\[
\begin{aligned}
\widetilde\E(|V_2|\mid\mathcal G_1,V_1)
&\le
\left\{\widetilde\E(V_2^2\mid\mathcal G_1,V_1)\right\}^{1/2}\\
&=
\left\{
\widetilde\E(V_2\mid\mathcal G_1,V_1)^2 
+
\Var(V_2\mid\mathcal G_1,V_1)
\right\}^{1/2}\\
&\le C(1+|V_1|).
\end{aligned}
\]
As $\Var(V_j\mid\mathcal G_1)\ge(2m)^{-2}>0$, the density of $[V_j\mid\mathcal G_1]$ is bounded above by
$\{2\pi\Var(V_j\mid\mathcal G_1)\}^{-1/2}
\le 2m(2\pi)^{-1/2}$.
It follows that
\begin{equation}\label{eq: cond bound density}
\pr(|V_1|\le h\mid\mathcal G_1)
=\int_{-h}^h f_{V_1\mid\mathcal G_1}(v)\dd v\le Ch.
\end{equation}
By the tower property,
\[
\begin{aligned}
&\widetilde\E\{|V_2|\ind(|V_1|\le h)\mid\mathcal G_1\}\\
&=\widetilde\E\!\left[
\widetilde\E(|V_2|\mid\mathcal G_1,V_1)
\ind(|V_1|\le h)
\mid\mathcal G_1\right]\\
&\le C(1+h)\pr(|V_1|\le h\mid\mathcal G_1)
\le C(1+h)h\leq Ch.
\end{aligned}
\]
Similarly, we have $\widetilde\E\{|V_1|\ind(|V_2|\le h)\mid\mathcal G_1\}\leq Ch$.
Substituting these into \eqref{eq: Rbound-g_1} gives
\[
\widetilde\E\left\{
\mu_1^2\sup_{\|u\|_\infty\le h}|R(V,u)|
\mid\mathcal G_1
\right\}
\le h^2+4Ch^2
\le Ch^{1+p}.
\]

For $H=F_{Q,k}$, let $p=2/n\in(0,1)$ and $f_p(v)=|v|^p$.
We can write $F_{Q,k}(w)
=\mu_{2/n}^{-n}\prod_{r\in\mathcal I_k}f_p(w_r)$.
By the finite product expansion and the product rule, we have
\begin{align*}
\mu_{2/n}^n F_{Q,k}(V+u)
&=\prod_{r\in\mathcal I_k}
\bigl\{f_p(V_r)+f_p(V_r+u_r)-f_p(V_r)\bigr\}\\
&=\prod_{r\in\mathcal I_k}f_p(V_r)
+\sum_{r\in\mathcal I_k}
\left[
\bigl(f_p(V_r+u_r)-f_p(V_r)\bigr)
\prod_{s\in\mathcal I_k\setminus\{r\}}f_p(V_s)
\right]\\
&\quad+\sum_{\substack{J\subseteq\mathcal I_k\\|J|\ge2}}
\left[
\prod_{r\in J}\bigl(f_p(V_r+u_r)-f_p(V_r)\bigr)
\prod_{s\in\mathcal I_k\setminus J}f_p(V_s)
\right],\\
\mu_{2/n}^n\nabla F_{Q,k}(V)^\T u
&=\sum_{r\in\mathcal I_k}
\left[
f_p'(V_r)u_r
\prod_{s\in\mathcal I_k\setminus\{r\}}f_p(V_s)
\right].
\end{align*}

It follows from \eqref{eq: taylor H} that
\[
\begin{aligned}
\mu_{2/n}^n R(V,u)
={}&\sum_{r\in\mathcal I_k}
\left[
\bigl(f_p(V_r+u_r)-f_p(V_r)-f_p'(V_r)u_r\bigr)
\prod_{s\in\mathcal I_k\setminus\{r\}}f_p(V_s)
\right]\\
&+\sum_{\substack{J\subseteq\mathcal I_k\\|J|\ge2}}
\left[
\prod_{r\in J}\bigl(f_p(V_r+u_r)-f_p(V_r)\bigr)
\prod_{s\in\mathcal I_k\setminus J}f_p(V_s)
\right].
\end{aligned}
\]
By Lemma~\ref{lem:noise-conditional-law}(ii), $\{V_r\}_{r\in\mathcal I_k}$ are conditionally independent, since distinct elements of $\mathcal I_k$ differ by at least two.
Therefore, we have
\begin{equation}\label{eq:noise-product-conditional-bound}
\begin{aligned}
&\mu_{2/n}^n\widetilde\E\left\{
\sup_{\|u\|_\infty\le h}|R(V,u)|
\middle|\mathcal G_1\right\}\\
&\quad\le
\sum_{r\in\mathcal I_k}
\left[
\widetilde\E\Bigl(
T_{r,h}
\Bigm|\mathcal G_1\Bigr)
\prod_{s\in\mathcal I_k\setminus\{r\}}
\widetilde\E\{f_p(V_s)\mid\mathcal G_1\}
\right]\\
&\qquad+
\sum_{\substack{J\subseteq\mathcal I_k\\|J|\ge2}}
\left[
\prod_{r\in J}
\widetilde\E\Bigl(
D_{r,h}
\Bigm|\mathcal G_1\Bigr)
\prod_{s\in\mathcal I_k\setminus J}
\widetilde\E\{f_p(V_s)\mid\mathcal G_1\}
\right],
\end{aligned}
\end{equation}
where
$D_{r,h}
=\sup_{|t|\le h}|f_p(V_r+t)-f_p(V_r)|$ and
$T_{r,h}
=\sup_{|t|\le h}
|f_p(V_r+t)-f_p(V_r)-f_p'(V_r)t|$.
By Lemma~\ref{lem:noise-conditional-law}(iii), we have
\begin{equation}\label{eq:noise-fp-moment}
\widetilde\E\{f_p(V_r)\mid\mathcal G_1\}
\le\{\widetilde\E(V_r^2\mid\mathcal G_1)\}^{p/2}
\le C.
\end{equation}
Therefore, it remains to derive the order of $\widetilde\E(D_{r,h}|\mathcal G_1)$ and $\widetilde\E(T_{r,h}|\mathcal G_1)$.

For $v\ne0$, as $|a^p-b^p|\le b^{p-1}|a-b|$ for $a\ge0$ and $b>0$, we have
\begin{equation}\label{eq: f_p tri}
\sup_{|t|\le h}|f_p(v+t)-f_p(v)|
\le h|v|^{p-1}.    
\end{equation}
By Lemma~\ref{lem:noise-conditional-law}(iii), the conditional density of $V_r$ is bounded by $C$.
Since $p-1>-1$ and $|v|^{p-1}\le1$ for $|v|>1$, we have
\begin{equation}\label{eq:noise-negative-moment}
\begin{aligned}
\widetilde\E(|V_r|^{p-1}\mid\mathcal G_1)
\le C\int_{|v|\le1}|v|^{p-1}\dd v
+\pr(|V_r|>1\mid\mathcal G_1)
\le \frac{2C}{p}+1\le C.
\end{aligned}
\end{equation}
Consequently,
\begin{equation}\label{eq: D bound}
\widetilde\E(D_{r,h}\mid\mathcal G_1)
\le h\,\widetilde\E(|V_r|^{p-1}\mid\mathcal G_1)
\le Ch.
\end{equation}
Next, since $|f_p'(v)|=p|v|^{p-1}$, \eqref{eq: f_p tri} implies that
\[\sup_{|t|\le h}
|f_p(v+t)-f_p(v)-f_p'(v)t|\leq
\sup_{|t|\le h}
|f_p(v+t)-f_p(v)|+\sup_{|t|\le h}|f_p'(v)t|\le 
Ch|v|^{p-1}\] when $|v|\le2h$.
For $|v|>2h$, with $|t|\le h$ the interval between $v$ and $v+t$ stays at least $|v|/2>0$ away from zero.
Thus Taylor's theorem and $|f_p''(v)|=p(1-p)|v|^{p-2}$ give
\[
\begin{aligned}
|f_p(v+t)-f_p(v)-f_p'(v)t|
\le \frac{t^2}{2}
\sup_{|z-v|\le h}|f_p''(z)|
\le Ch^2|v|^{p-2}.
\end{aligned}
\]
Notice that this upper bound is uniform over $|t|\le h$.
Therefore, we have
\begin{equation}\label{eq: T bound}
\widetilde\E(T_{r,h}\mid\mathcal G_1)
\le C\left\{
h\int_{|v|\le2h}|v|^{p-1}\dd v
+h^2\int_{|v|>2h}|v|^{p-2}\dd v
\right\} 
\le Ch^{1+p}.    
\end{equation}
Substituting \eqref{eq: D bound} and \eqref{eq: T bound} into \eqref{eq:noise-product-conditional-bound}, and using $|\mathcal I_k|=n$ and $0<h\le1$, we obtain
\[
\begin{aligned}
\widetilde\E\left\{
\sup_{\|u\|_\infty\le h}|R(V,u)|
\middle|\mathcal G_1\right\}
\le C\left[
\sum_{r\in\mathcal I_k}h^{1+p}
+\sum_{\substack{J\subseteq\mathcal I_k\\|J|\ge2}}h^{|J|}
\right]
\le C\left(h^{1+p}+\sum_{j=2}^n h^j\right)
\le Ch^{1+p}.
\end{aligned}
\]
This proves \eqref{eq:noise-drift-remainder} for $H=F_{Q,k}$.

Finally, $\|u_i\|_\infty\le mM^{-1/2}\le1$ for all sufficiently large $M$.
Applying \eqref{eq:noise-drift-remainder} with $h=mM^{-1/2}$ to $V=U_{i,M}$ and $V=U_{i,M}^0$ gives
\[
\begin{aligned}
&\widetilde\E\bigl\{
|R(U_{i,M},u_i)-R(U_{i,M}^0,u_i)|
\bigm|\mathcal G_1\bigr\}
\le 2Cm^{1+p}M^{-(1+p)/2}
=o(M^{-1/2}),
\end{aligned}
\]
uniformly over $i\le M-d+1$.

It remains to handle the gradient term $\nabla H(U_{i,M})-\nabla H(U_{i,M}^0)$.
We will show that, for all sufficiently large $M$ and all $i$,
\begin{equation}
\label{eq:noise-drift-gradient}
\widetilde\E\bigl\{\|\nabla H(U_{i,M})-\nabla H(U_{i,M}^0)\|_1
\bigm|\mathcal G_1\bigr\}\le Cx_M^{1/(16n)}.
\end{equation}
As $x_M\to 0$ as $M\to\infty$, without loss of generality, we can consider only $0\le x_M\le1$.
If $x_M=0$, then $U_{i,M}=U_{i,M}^0$ almost surely, so \eqref{eq:noise-drift-gradient} holds with both sides zero.
For the later argument, let $0<x_M\le1$.

For $H(v)=v^2$, since $X_{i,M}-X_{i,M}^0=e_{i,M}^{\varepsilon}$, we have
\[
\begin{aligned}
&\widetilde\E\bigl\{
\|\nabla H(U_{i,M})-\nabla H(U_{i,M}^0)\|_1
\mid\mathcal G_1\bigr\}
=2\widetilde\E\{
|e_{i,M}^{\varepsilon}|\mid\mathcal G_1\}
=2\mu_1(2x_M)^{1/2}
\le Cx_M^{1/(16n)}.
\end{aligned}
\]

For $H=g_1$ and $H=F_{Q,k}$, the derivatives involve $\operatorname{sgn}(v)$ and $f_p'(v)=p\operatorname{sgn}(v)|v|^{p-1}$, which are discontinuous at zero, with $f_p'$ also unbounded near zero.
We first bound the difference between the sign functions.
If $|w|>2t$ and $|v-w|\le t$, then $v$ and $w$ have the same sign.
Consequently,
\[
|\operatorname{sgn}(v)-\operatorname{sgn}(w)|
\le 2\left[
\ind(|w|\le2t)+\ind(|v-w|>t)
\right].
\]
Consider $v=X_{j,M}$ and $w=X_{j,M}^0$.
It follows from an argument similar to that for \eqref{eq: cond bound density} that
\begin{equation}\label{eq: X_0 bound}
\pr(|X_{j,M}^0|\le2t\mid\mathcal G_1)\le Ct.    
\end{equation}
By Chebyshev's inequality, we have
\begin{equation}\label{eq: X diff bound}
\pr(|X_{j,M}-X_{j,M}^0|>t\mid\mathcal G_1)
\le
\frac{\widetilde\E\{(e_{j,M}^{\varepsilon})^2
\mid\mathcal G_1\}}{t^2}
=\frac{2x_M}{t^2}.
\end{equation}
Letting $t=x_M^{1/8}$ yields
\[
\begin{aligned}
&\widetilde\E\{
|\operatorname{sgn}(X_{j,M})
-\operatorname{sgn}(X_{j,M}^0)|
\mid\mathcal G_1\}
\le C\left(t+\frac{x_M}{t^2}\right)
=C\left(x_M^{1/8}+x_M^{3/4}\right)
\le Cx_M^{1/8}.
\end{aligned}
\]

For $H=g_1$, since $\mu_1^2\partial_1g_1(v)=\operatorname{sgn}(v_1)|v_2|$, we have
\begin{align*}
&\mu_1^2|\partial_1g_1(U_{i,M})
-\partial_1g_1(U_{i,M}^0)|\\
&=\Bigl|
\{\operatorname{sgn}(X_{i,M})
-\operatorname{sgn}(X_{i,M}^0)\}|X_{i+1,M}|
+
\operatorname{sgn}(X_{i,M}^0)
\{|X_{i+1,M}|-|X_{i+1,M}^0|\}
\Bigr|\\
&\le
|\operatorname{sgn}(X_{i,M})
-\operatorname{sgn}(X_{i,M}^0)|\,|X_{i+1,M}|+
\bigl||X_{i+1,M}|-|X_{i+1,M}^0|\bigr|\\
&\le
|\operatorname{sgn}(X_{i,M})
-\operatorname{sgn}(X_{i,M}^0)|\,|X_{i+1,M}|
+|e_{i+1,M}^{\varepsilon}|.
\end{align*}
Since $|\operatorname{sgn}(X_{i,M})-\operatorname{sgn}(X_{i,M}^0)|^2
\le2|\operatorname{sgn}(X_{i,M})-\operatorname{sgn}(X_{i,M}^0)|$,
we have
\[
\begin{aligned}
&\widetilde\E\bigl\{
|\operatorname{sgn}(X_{i,M})-\operatorname{sgn}(X_{i,M}^0)|
|X_{i+1,M}|
\bigm|\mathcal G_1\bigr\}\\
&\quad\le
\left[
2\widetilde\E\bigl\{
|\operatorname{sgn}(X_{i,M})-\operatorname{sgn}(X_{i,M}^0)|
\bigm|\mathcal G_1\bigr\}
\right]^{1/2}
\left[
\widetilde\E\{X_{i+1,M}^2\mid\mathcal G_1\}
\right]^{1/2}
\le Cx_M^{1/16}.
\end{aligned}
\]
Also, $\widetilde\E\{|e_{i+1,M}^{\varepsilon}|\mid\mathcal G_1\}\le Cx_M^{1/2}$.
The same bound holds for $\partial_2g_1$ by symmetry.
Therefore, \eqref{eq:noise-drift-gradient} holds for $H=g_1$.

For $H=F_{Q,k}$, recall that
\[
\partial_rF_{Q,k}(w)
=\mu_{2/n}^{-n}f_p'(w_r)
\prod_{s\in\mathcal I_k\setminus\{r\}}f_p(w_s),
\qquad r\in\mathcal I_k,
\]
and $\partial_rF_{Q,k}=0$ for $r\notin\mathcal I_k$.
We first bound the difference between the factors $f_p'$.
For nonzero $v,w$ and $0<t\le1$, if $|w|>2t$ and $|v-w|\le t$, every point between $v$ and $w$ has absolute value greater than $t$.
Hence, by the mean value theorem and $|f_p''(z)|=p(1-p)|z|^{p-2}$,
\[
|f_p'(v)-f_p'(w)|
\le p(1-p)t^{p-2}|v-w|.
\]
Otherwise, we can simply consider $|f_p'(v)-f_p'(w)|\le |f_p'(v)|+|f_p'(w)|$.
Combining these two scenarios gives
\begin{equation}\label{eq: derivative bound}
|f_p'(v)-f_p'(w)|
\le p(1-p)t^{p-2}|v-w|
+p(|v|^{p-1}+|w|^{p-1})
\ind(|w|\le2t\text{ or }|v-w|>t).
\end{equation}
We again consider $v=X_{j,M}$, $w=X_{j,M}^0$, and $t=x_M^{1/8}$.
Since $p=2/n$ and $n>2$, we have
\[
(p-1)\frac{n-1}{n-2}
=-\frac{n-1}{n}>-1.
\]
By Lemma~\ref{lem:noise-conditional-law}(iii), the conditional densities of $v$ and $w$ given $\mathcal G_1$ are uniformly bounded by $C$.
Consequently, for $z\in\{v,w\}$,
\[
\begin{aligned}
\widetilde\E\left\{
|z|^{(p-1)(n-1)/(n-2)}
\Bigm|\mathcal G_1\right\}
&=
\int_{\mathbb R}|a|^{-(n-1)/n}
f_{z\mid\mathcal G_1}(a)\dd a\\
&\le
C\int_{|a|\le1}|a|^{-(n-1)/n}\dd a
+\pr(|z|>1\mid\mathcal G_1)
\le C,
\end{aligned}
\]
where the integral is finite since $(n-1)/n<1$.
Thus,
\[
\widetilde\E\left\{
|v|^{(p-1)(n-1)/(n-2)}
+|w|^{(p-1)(n-1)/(n-2)}
\Bigm|\mathcal G_1
\right\}\le C.
\]

Let $A=\{|w|\le2t\}\cup\{|v-w|>t\}$.
By conditional H\"older's inequality with exponents $(n-1)/(n-2)$ and $n-1$, together with \eqref{eq: X_0 bound} and \eqref{eq: X diff bound}, we have
\[
\begin{aligned}
&\widetilde\E\Bigl\{
(|v|^{p-1}+|w|^{p-1})\ind(A)
\Bigm|\mathcal G_1\Bigr\}\\
&\quad\le
\left[
\widetilde\E\Bigl\{
(|v|^{p-1}+|w|^{p-1})^{(n-1)/(n-2)}
\Bigm|\mathcal G_1\Bigr\}
\right]^{(n-2)/(n-1)}
\pr(A\mid\mathcal G_1)^{1/(n-1)}\\
&\quad\le
C\left\{
\pr(|w|\le2t\mid\mathcal G_1)
+\pr(|v-w|>t\mid\mathcal G_1)
\right\}^{1/(n-1)}\\
&\quad\le
C\left(t+\frac{x_M}{t^2}\right)^{1/(n-1)}.
\end{aligned}
\]
Moreover,
$\widetilde\E(|v-w|\mid\mathcal G_1)
\le(2x_M)^{1/2}$.
It then follows from \eqref{eq: derivative bound} that
\[
\begin{aligned}
\widetilde\E\{
|f_p'(X_{j,M})-f_p'(X_{j,M}^0)|
\mid\mathcal G_1\}
&\le C\left\{
t^{p-2}x_M^{1/2}
+\left(t+\frac{x_M}{t^2}\right)^{1/(n-1)}
\right\}\\
&\quad\le C\left\{
x_M^{(2+p)/8}
+x_M^{1/\{8(n-1)\}}
\right\}
\le Cx_M^{1/\{8(n-1)\}}.
\end{aligned}
\]

We next bound the difference between the gradients of $F_{Q,k}$.
Write $V=U_{i,M}$ and $V^0=U_{i,M}^0$.
For $r\in\mathcal I_k$, adding and subtracting $f_p'(V_r^0)\prod_{s\in\mathcal I_k\setminus\{r\}}f_p(V_s)$ and telescoping the remaining product give
\[
\begin{aligned}
&\mu_{2/n}^n
\{\partial_rF_{Q,k}(V)-\partial_rF_{Q,k}(V^0)\}\\
&=\{f_p'(V_r)-f_p'(V_r^0)\}
\prod_{s\in\mathcal I_k\setminus\{r\}}f_p(V_s)\\
&\quad+f_p'(V_r^0)
\sum_{j\in\mathcal I_k\setminus\{r\}}
\left[
\bigl(f_p(V_j)-f_p(V_j^0)\bigr)
\prod_{\substack{s\in\mathcal I_k\setminus\{r\}\\s<j}}
f_p(V_s)
\prod_{\substack{s\in\mathcal I_k\setminus\{r\}\\s>j}}
f_p(V_s^0)
\right].
\end{aligned}
\]
Furthermore, since $|f_p'(v)|=p|v|^{p-1}$, \eqref{eq:noise-negative-moment} gives
\(
\widetilde\E\{
|f_p'(V_r^0)|\mid\mathcal G_1\}
=p\widetilde\E\{
|V_r^0|^{p-1}\mid\mathcal G_1\}
\le C.
\)
By \eqref{eq:noise-fp-moment}, we also have
\(
\widetilde\E\{
f_p(V_s)+f_p(V_s^0)\mid\mathcal G_1\}
\le C.
\)
In addition, since $|f_p(v)-f_p(w)|\le|v-w|^p$ for $0<p<1$,
\[
\begin{aligned}
\widetilde\E\{
|f_p(V_j)-f_p(V_j^0)|
\mid\mathcal G_1\}
\le
\widetilde\E\{
|e_{i+j-1,M}^{\varepsilon}|^p
\mid\mathcal G_1\}
\le Cx_M^{p/2}
=Cx_M^{1/n}.
\end{aligned}
\]
By Lemma~\ref{lem:noise-conditional-law}(i) and~(ii), the pairs $(V_s,V_s^0)$ for $s\in\mathcal I_k$ are conditionally independent given $\mathcal G_1$.
We can thus conclude for $r\in\mathcal I_k$,
\[
\begin{aligned}
&\widetilde\E\{
|\partial_rF_{Q,k}(V)-\partial_rF_{Q,k}(V^0)|
\mid\mathcal G_1\}
\le
C\left\{
x_M^{1/\{8(n-1)\}}+x_M^{1/n}
\right\}.
\end{aligned}
\]
Since $\partial_rF_{Q,k}=0$ for $r\notin\mathcal I_k$ and $|\mathcal I_k|=n$,
\[
\begin{aligned}
\widetilde\E\{
\|\nabla F_{Q,k}(V)-\nabla F_{Q,k}(V^0)\|_1
\mid\mathcal G_1\}
\le
C\left\{
x_M^{1/\{8(n-1)\}}+x_M^{1/n}
\right\}
\le Cx_M^{1/(16n)}.
\end{aligned}
\]
Therefore, \eqref{eq:noise-drift-gradient} also holds for $H=F_{Q,k}$.

Since $\|u_i\|_\infty\le mM^{-1/2}$, \eqref{eq:noise-drift-gradient} implies
\[
\begin{aligned}
\widetilde\E\Bigl\{
\bigl|\{\nabla H(U_{i,M})
-\nabla H(U_{i,M}^0)\}^\T u_i\bigr|
\Bigm|\mathcal G_1\Bigr\}
&\le mM^{-1/2}
\widetilde\E\bigl\{
\|\nabla H(U_{i,M})-\nabla H(U_{i,M}^0)\|_1
\bigm|\mathcal G_1\bigr\}\\
&\quad\le CmM^{-1/2}x_M^{1/(16n)}
=o(M^{-1/2}),
\end{aligned}
\]
uniformly over $i\le M-d+1$.
Combining this with \eqref{eq:noise-drift-remainder} implies that for every $i\le M-d+1$,
\[
\begin{aligned}
&\widetilde\E\bigl\{
|H(U_{i,M}+u_i)-H(U_{i,M})
-H(U_{i,M}^0+u_i)+H(U_{i,M}^0)|
\bigm|\mathcal G_1\bigr\}\\
&\quad\le
CmM^{-1/2}x_M^{1/(16n)}
+2Cm^{1+p}M^{-(1+p)/2}.
\end{aligned}
\]
The right-hand side is deterministic and uniform in $i$.
Taking expectations and applying the triangle inequality, we bound the expected absolute value of \eqref{eq: H-bound} by
\[
\begin{aligned}
&M^{-1/2}(M-d+1)
\left\{
CmM^{-1/2}x_M^{1/(16n)}
+2Cm^{1+p}M^{-(1+p)/2}
\right\}\\
&\quad\le
Cmx_M^{1/(16n)}
+2Cm^{1+p}M^{-p/2}
\longrightarrow0,
\end{aligned}
\]
since $x_M\to0$ and $p>0$.
Thus \eqref{eq: H-bound} converges to zero in $L^1$, and hence in probability.
This completes the proof.
\end{proof}

\begin{lemma}[Leading-order effects of microstructure noise]
\label{lem:noise-effect}
Suppose Assumptions~\ref{ass:microstructure-noise-null} and~\ref{ass:noise-bounded-coefficients} hold, $M^{5/4}\omega_M^2\to c\in[0,\infty)$, and $M^{3/4}\rho_M\omega_M\to\iota\in\mathbb R$ as $M\to\infty$.
Then, as $M\to\infty$,
\begin{equation}\label{eq: theta shift}
M^{1/2}\bigl\{\hat\theta_{g_1}[M^{-1/2}X]-\hat\theta_{g_1}[M^{-1/2}X^0]\bigr\}
\inP-\frac12\int_0^1\Bigl(\frac c{\sigma_t}+\iota\Bigr)^2\dd t,
\end{equation}
\begin{equation}\label{eq: Q shift}
M^{1/2}D\{\mathbb Q_\ell^{(n)}(1;M^{-1/2}X)
-\mathbb Q_\ell^{(n)}(1;M^{-1/2}X^0)\}\inP0    
\end{equation}
for every fixed deterministic matrix $D$ with $\ell$ columns satisfying $DJ_\ell=0$.
\end{lemma}

\begin{proof}
By \eqref{eq:noise-sum-form}, it suffices to analyze
\begin{equation}\label{eq: diff decomp}
M^{-1/2}\sum_{i=1}^{M-d+1}
\widetilde\E\{H(U_{i,M})-H(U_{i,M}^0)\mid\mathcal G_1\}
+M^{-1/2}\sum_{i=1}^{M-d+1}\chi_{i,M}^H
\end{equation}
for each pair $(H,d)$ in \eqref{eq:noise-sum-form}, where
\[
\chi_{i,M}^H=H(U_{i,M})-H(U_{i,M}^0)
-\widetilde\E\{H(U_{i,M})-H(U_{i,M}^0)\mid\mathcal G_1\}.
\]
Since $\hat\theta_{g_1}=\hat{\QV}-\hat{\IV}_{g_1}$, the difference between \eqref{eq: diff decomp} for $(v\mapsto v^2,1)$ and $(g_1,2)$ gives the left-hand side of \eqref{eq: theta shift}.
Similarly, applying $D$ to the vector of \eqref{eq: diff decomp} for $(F_{Q,k},K)$, $k=1,\ldots,\ell$, gives the left-hand side of \eqref{eq: Q shift}.
We will show that $M^{-1/2}\sum_{i=1}^{M-d+1}\chi_{i,M}^H\inP0$ for each $(H,d)$ in \eqref{eq:noise-sum-form}.

Since $\widetilde\E(\chi_{i,M}^H\mid\mathcal G_1)=0$ and $\chi_{i,M}^H$ and $\chi_{i',M}^H$ are conditionally independent given $\mathcal G_1$ for $|i-i'|>d$, it suffices to show that
$\max_{1\le i\le M-d+1}
\widetilde\E\{(\chi_{i,M}^H)^2\mid\mathcal G_1\}\to0$
to yield $M^{-1/2}\sum_{i=1}^{M-d+1}\chi_{i,M}^H\inP0$.

Since $x_M=M\omega_M^2\to0$, it follows from Lemma~\ref{lem:noise-conditional-law}(ii)--(iii) that $e_{j,M}^{\varepsilon}$, $X_{j,M}$, and $X_{j,M}^0$ are centered Gaussian conditionally on $\mathcal G_1$ with $\Var(X_{j,M}\mid\mathcal G_1)
+\Var(X_{j,M}^0\mid\mathcal G_1)\le C$ and $\Var(e_{j,M}^{\varepsilon}\mid\mathcal G_1)=2x_M$.
Also, recall the identity $\E Y^4=3(\E Y^2)^2$ for a centered Gaussian random variable $Y$.
It follows that for all sufficiently large $M$ and all $j$, we have
\begin{equation}
\label{eq:noise-fourth-moments}
\widetilde\E\{(e_{j,M}^{\varepsilon})^4\mid\mathcal G_1\}=12x_M^2,
\qquad
\widetilde\E(X_{j,M}^4\mid\mathcal G_1)
+\widetilde\E\{(X_{j,M}^0)^4\mid\mathcal G_1\}\le C.
\end{equation}

For $H(v)=v^2$, since $X_{i,M}-X_{i,M}^0=e_{i,M}^{\varepsilon}$, it follows from \eqref{eq:noise-fourth-moments} that
\[
\begin{aligned}
&\widetilde\E\left\{\left[H(U_{i,M})-H(U_{i,M}^0)\right]^2\middle|\mathcal G_1\right\}\\
&=\widetilde\E\{(e_{i,M}^{\varepsilon})^2
(X_{i,M}+X_{i,M}^0)^2\mid\mathcal G_1\}\\
&\le
\{\widetilde\E[(e_{i,M}^{\varepsilon})^4\mid\mathcal G_1]\}^{1/2}
\{\widetilde\E[(X_{i,M}+X_{i,M}^0)^4\mid\mathcal G_1]\}^{1/2}\\
&\le
(12x_M^2)^{1/2}
\{8\widetilde\E[X_{i,M}^4+(X_{i,M}^0)^4
\mid\mathcal G_1]\}^{1/2}
\le Cx_M.
\end{aligned}
\]

For $H=g_1$, we first notice that $\bigl||a|-|b|\bigr|\le|a-b|$ gives
\[
\begin{aligned}
\mu_1^2|g_1(U_{i,M})-g_1(U_{i,M}^0)|
&\le |X_{i,M}X_{i+1,M}
-X_{i,M}^0X_{i+1,M}^0|\\
&=|(X_{i,M}-X_{i,M}^0)X_{i+1,M}
+X_{i,M}^0(X_{i+1,M}-X_{i+1,M}^0)|\\
&\le |e_{i,M}^{\varepsilon}||X_{i+1,M}|
+|X_{i,M}^0||e_{i+1,M}^{\varepsilon}|.
\end{aligned}
\]
Also, by \eqref{eq:noise-fourth-moments} and the elementary inequality $(a+b)^2\le2a^2+2b^2$, we have
\[
\begin{aligned}
&\mu_1^4\widetilde\E\left\{
\left[g_1(U_{i,M})-g_1(U_{i,M}^0)\right]^2\middle|\mathcal G_1\right\}\\
&\le
2\widetilde\E\{(e_{i,M}^{\varepsilon})^2X_{i+1,M}^2
\mid\mathcal G_1\}
+2\widetilde\E\{(X_{i,M}^0)^2(e_{i+1,M}^{\varepsilon})^2
\mid\mathcal G_1\}\\
&\le
2\{\widetilde\E[(e_{i,M}^{\varepsilon})^4\mid\mathcal G_1]\}^{1/2}
\{\widetilde\E[X_{i+1,M}^4\mid\mathcal G_1]\}^{1/2}
+
2\{\widetilde\E[(X_{i,M}^0)^4\mid\mathcal G_1]\}^{1/2}
\{\widetilde\E[(e_{i+1,M}^{\varepsilon})^4\mid\mathcal G_1]\}^{1/2}\\
&\le Cx_M,
\end{aligned}
\]
where the last inequality follows from \eqref{eq:noise-fourth-moments}.

For $H=F_{Q,k}$, let $p=2/n\in(0,1)$.
Also, assume the elements of $\mathcal I_k$ are ordered increasingly.
We have
\[
\begin{aligned}
&\mu_{2/n}^n\{F_{Q,k}(U_{i,M})-F_{Q,k}(U_{i,M}^0)\}\\
&=\sum_{j\in\mathcal I_k}
\bigl(|X_{i+j-1,M}|^p-|X_{i+j-1,M}^0|^p\bigr)
\prod_{\substack{r\in\mathcal I_k\\r<j}}|X_{i+r-1,M}|^p
\prod_{\substack{r\in\mathcal I_k\\r>j}}|X_{i+r-1,M}^0|^p.
\end{aligned}
\]
Since $\bigl||v|^p-|w|^p\bigr|\le|v-w|^p$ for $0<p\le1$ and $X_{i,M}-X_{i,M}^0=e_{i,M}^{\varepsilon}$,
\[
\begin{aligned}
\mu_{2/n}^n|F_{Q,k}(U_{i,M})-F_{Q,k}(U_{i,M}^0)|\le\sum_{j\in\mathcal I_k}|e_{i+j-1,M}^{\varepsilon}|^p
\prod_{\substack{r\in\mathcal I_k\\r<j}}|X_{i+r-1,M}|^p
\prod_{\substack{r\in\mathcal I_k\\r>j}}|X_{i+r-1,M}^0|^p.
\end{aligned}
\]
As $|\mathcal I_k|=n$ is finite and fixed with $2pn=4$, H\"older's inequality and \eqref{eq:noise-fourth-moments} imply that
\[
\begin{aligned}
&\widetilde\E\left\{
|e_{i+j-1,M}^{\varepsilon}|^{2p}
\prod_{\substack{r\in\mathcal I_k\\r<j}}|X_{i+r-1,M}|^{2p}
\prod_{\substack{r\in\mathcal I_k\\r>j}}|X_{i+r-1,M}^0|^{2p}
\middle|\mathcal G_1\right\}\\
&\le
\{\widetilde\E[(e_{i+j-1,M}^{\varepsilon})^4
\mid\mathcal G_1]\}^{1/n}
\prod_{\substack{r\in\mathcal I_k\\r<j}}
\{\widetilde\E[X_{i+r-1,M}^4\mid\mathcal G_1]\}^{1/n}\\
&\quad\times
\prod_{\substack{r\in\mathcal I_k\\r>j}}
\{\widetilde\E[(X_{i+r-1,M}^0)^4
\mid\mathcal G_1]\}^{1/n}
\le Cx_M^{2/n}.
\end{aligned}
\]
Using $(\sum_{j\in\mathcal I_k}a_j)^2
\le n\sum_{j\in\mathcal I_k}a_j^2$, we obtain
\[
\begin{aligned}
&\widetilde\E\left\{
\left[F_{Q,k}(U_{i,M})-F_{Q,k}(U_{i,M}^0)\right]^2
\middle|\mathcal G_1\right\}\\
&\le n\mu_{2/n}^{-2n}\sum_{j\in\mathcal I_k}
\widetilde\E\left\{
|e_{i+j-1,M}^{\varepsilon}|^{2p}
\prod_{\substack{r\in\mathcal I_k\\r<j}}|X_{i+r-1,M}|^{2p}
\prod_{\substack{r\in\mathcal I_k\\r>j}}|X_{i+r-1,M}^0|^{2p}
\middle|\mathcal G_1\right\}
\le Cx_M^{2/n}.
\end{aligned}
\]
Combining the bounds for $H(v)=v^2$, $H=g_1$, and $H=F_{Q,k}$, we have, for each $(H,d)$ in \eqref{eq:noise-sum-form}, all sufficiently large $M$, and $i=1,\ldots,M-d+1$,
\begin{equation}
\label{eq:noise-difference-second-moment}
\widetilde\E\left\{\left[H(U_{i,M})-H(U_{i,M}^0)\right]^2
\middle|\mathcal G_1\right\}
\le C(x_M+x_M^{2/n}).
\end{equation}
Furthermore,
\(
\widetilde\E\{(\chi_{i,M}^H)^2\mid\mathcal G_1\}
\le\widetilde\E[
\{H(U_{i,M})-H(U_{i,M}^0)\}^2
\mid\mathcal G_1]
\le C(x_M+x_M^{2/n}).
\)

By Lemma~\ref{lem:noise-conditional-law}(i)--(ii), conditionally on $\mathcal G_1$, $\chi_{i,M}^H$ depends only on $Z_{j,M},G_{j,M}$ for $i-1\le j\le i+d-1$, with $Z_{0,M}$ omitted when $i=1$.
Since these index sets are disjoint when $|i-i'|>d$, we have
\(
\widetilde\E(\chi_{i,M}^H\chi_{i',M}^H\mid\mathcal G_1)=0,
\)
whenever $|i-i'|>d$.
For $|i-i'|\le d$, consider
\[
\begin{aligned}
|\widetilde\E(\chi_{i,M}^H\chi_{i',M}^H\mid\mathcal G_1)|
\le
\{\widetilde\E[(\chi_{i,M}^H)^2\mid\mathcal G_1]\}^{1/2}
\{\widetilde\E[(\chi_{i',M}^H)^2\mid\mathcal G_1]\}^{1/2}
\le
\max_{1\le j\le M-d+1}
\widetilde\E[(\chi_{j,M}^H)^2\mid\mathcal G_1].
\end{aligned}
\]
It then follows from \eqref{eq:noise-difference-second-moment} that
\begin{equation}
\label{eq:noise-centered-perturbation}
\begin{aligned}
\widetilde\E\left\{
\left[M^{-1/2}\sum_{i=1}^{M-d+1}\chi_{i,M}^H\right]^2
\middle|\mathcal G_1\right\}
&=\frac1M\sum_{i=1}^{M-d+1}
\sum_{\substack{1\le i'\le M-d+1\\|i-i'|\le d}}
\widetilde\E(\chi_{i,M}^H\chi_{i',M}^H\mid\mathcal G_1)\\
&\le\frac{(2d+1)(M-d+1)}{M}
\max_{1\le j\le M-d+1}
\widetilde\E[(\chi_{j,M}^H)^2\mid\mathcal G_1]\\
&\le C(x_M+x_M^{2/n})\to0.
\end{aligned}
\end{equation}
Therefore, we have $M^{-1/2}\sum_{i=1}^{M-d+1}\chi_{i,M}^H\inP0$.
This completes the first part of the proof.

We then evaluate the first sum in \eqref{eq: diff decomp}.
First, as $M^{1/4}x_M=M^{5/4}\omega_M^2\to c$ and $M^{1/4}\gamma_M=M^{3/4}\rho_M\omega_M\to\iota$, we have $x_M+|\gamma_M|=O(M^{-1/4})$.
Also, \eqref{eq:noise-covariance} and Lemma~\ref{lem:noise-conditional-law}(iii) together imply that
\[
\widetilde q_{i,M}:= \{\Var(X_{i,M}\mid\mathcal G_1)\}^{1/2}
=(q_{i,M}^2+2x_M+2\gamma_Mq_{i,M})^{1/2}
\ge(2m)^{-1}
\]
for all sufficiently large $M$.
Throughout the following arguments, all constants in the $O\{\cdot\}$ bounds are independent of $M$, $i$, and the sample path.

For \eqref{eq: theta shift}, the difference between the first sums in \eqref{eq: diff decomp} for $(v\mapsto v^2,1)$ and $(g_1,2)$ is
$M^{1/2}\widetilde\E\{\hat\theta_{g_1}[M^{-1/2}X]
-\hat\theta_{g_1}[M^{-1/2}X^0]\mid\mathcal G_1\}$.
Since $\widetilde\E(X_{i,M}^2\mid\mathcal G_1)=\widetilde q_{i,M}^2$ and $\widetilde\E\{(X_{i,M}^0)^2\mid\mathcal G_1\}=q_{i,M}^2$, it follows from \eqref{eq:noise-sum-form} that
\[
\begin{aligned}
&\widetilde\E\{\hat\theta_{g_1}[M^{-1/2}X]-\hat\theta_{g_1}[M^{-1/2}X^0]
\mid\mathcal G_1\}\\
&\quad=\frac1M\sum_{i=1}^M(\widetilde q_{i,M}^2-q_{i,M}^2)\\
&\qquad-\frac1M\sum_{i=1}^{M-1}
\bigl[\widetilde\E\{g_1(X_{i,M},X_{i+1,M})\mid\mathcal G_1\}
-\widetilde\E\{g_1(X_{i,M}^0,X_{i+1,M}^0)\mid\mathcal G_1\}\bigr].
\end{aligned}
\]
To evaluate $\widetilde\E\{g_1(X_{i,M},X_{i+1,M})\mid\mathcal G_1\}$ and $\widetilde\E\{g_1(X_{i,M}^0,X_{i+1,M}^0)\mid\mathcal G_1\}$, we use the following result from \citet[p.~3]{Nabeya1951}: if $(V_1,V_2)$ is a centered Gaussian pair with standard deviations $s_1,s_2>0$ and correlation $r$, then
\begin{equation}
\label{eq:noise-abs-product-moment}
\E|V_1V_2|
=\mu_1^2s_1s_2\{(1-r^2)^{1/2}+r\arcsin r\}.
\end{equation}
By Lemma~\ref{lem:noise-conditional-law}(ii), $(X_{i,M},X_{i+1,M})$ is a centered Gaussian pair conditionally on $\mathcal G_1$.
Since $g_1(x_1,x_2)=\mu_1^{-2}|x_1x_2|$, applying \eqref{eq:noise-abs-product-moment} conditionally on $\mathcal G_1$ with $s_1=\widetilde q_{i,M}$ and $s_2=\widetilde q_{i+1,M}$ gives
\[
\widetilde\E\{g_1(X_{i,M},X_{i+1,M})\mid\mathcal G_1\}
=\widetilde q_{i,M}\widetilde q_{i+1,M}
\{(1-r^2)^{1/2}+r\arcsin r\},
\]
where, by \eqref{eq:noise-covariance},
\[
r=\operatorname{Corr}(X_{i,M},X_{i+1,M}\mid\mathcal G_1)
=-\frac{x_M+\gamma_Mq_{i,M}}
{\widetilde q_{i,M}\widetilde q_{i+1,M}}.
\]
Since $q_{i,M}\le m$ and $\widetilde q_{j,M}\ge(2m)^{-1}$,
\[
|r|\le(2m)^2(x_M+m|\gamma_M|)
=O(x_M+|\gamma_M|)\to0.
\]
Hence $|r|\le1/2$ for all sufficiently large $M$ and all $i$, so $(1-r^2)^{1/2}+r\arcsin r$ can be expanded about $r=0$.
Since its derivative is $\arcsin r$, and $\arcsin r=r+O(r^3)$ for $|r|\le1/2$,
\begin{equation}
\label{eq:noise-abs-product-expansion}
(1-r^2)^{1/2}+r\arcsin r
=1+\int_0^r\arcsin z\dd z
=1+\frac{r^2}{2}+O(r^4).
\end{equation}
Substituting \eqref{eq:noise-abs-product-expansion} and using $\widetilde q_{i,M}\widetilde q_{i+1,M}\le C$, which follows from Lemma~\ref{lem:noise-conditional-law}(iii), give the first equality in
\begin{equation}
\label{eq:noise-g1-mean}
\begin{aligned}
&\widetilde\E\{g_1(X_{i,M},X_{i+1,M})\mid\mathcal G_1\}\\
&=\widetilde q_{i,M}\widetilde q_{i+1,M}
+\frac{(x_M+\gamma_Mq_{i,M})^2}
{2\widetilde q_{i,M}\widetilde q_{i+1,M}}
+O\{(x_M+|\gamma_M|)^4\}\\
&=\widetilde q_{i,M}\widetilde q_{i+1,M}
+\frac{(x_M+\gamma_Mq_{i,M})^2}
{2q_{i,M}q_{i+1,M}}
+O\{(x_M+|\gamma_M|)^3\}.
\end{aligned}
\end{equation}
The second equality replaces the denominator $\widetilde q_{i,M}\widetilde q_{i+1,M}$ of the middle term by $q_{i,M}q_{i+1,M}$.
To justify this replacement, note that, as $q_{j,M}\ge m^{-1}$,
\[
\widetilde q_{j,M}-q_{j,M}
=\frac{2x_M+2\gamma_Mq_{j,M}}
{\widetilde q_{j,M}+q_{j,M}}
=O(x_M+|\gamma_M|).
\]
Since, in addition, $(2m)^{-1}\le\widetilde q_{j,M}\le C$ and $q_{j,M}\le m$,
\[
\begin{aligned}
\left|
\frac1{\widetilde q_{i,M}\widetilde q_{i+1,M}}
-\frac1{q_{i,M}q_{i+1,M}}
\right|
&=\frac{|\widetilde q_{i,M}\widetilde q_{i+1,M}
-q_{i,M}q_{i+1,M}|}
{\widetilde q_{i,M}\widetilde q_{i+1,M}
q_{i,M}q_{i+1,M}}\\
&\le4m^4\bigl(|\widetilde q_{i,M}-q_{i,M}|\,
\widetilde q_{i+1,M}
+q_{i,M}|\widetilde q_{i+1,M}-q_{i+1,M}|\bigr)\\
&\le C(x_M+|\gamma_M|).
\end{aligned}
\]
As $(x_M+\gamma_Mq_{i,M})^2\le(x_M+m|\gamma_M|)^2$, the replacement changes the middle term by $O\{(x_M+|\gamma_M|)^3\}$.

For $(X_{i,M}^0,X_{i+1,M}^0)$, the conditional correlation is zero by Lemma~\ref{lem:noise-conditional-law}(ii), so \eqref{eq:noise-abs-product-moment} gives
$\widetilde\E\{g_1(X_{i,M}^0,X_{i+1,M}^0)\mid\mathcal G_1\}
=q_{i,M}q_{i+1,M}$.
Substituting this identity and \eqref{eq:noise-g1-mean} into the expression for
$\widetilde\E\{\hat\theta_{g_1}[M^{-1/2}X]-\hat\theta_{g_1}[M^{-1/2}X^0]
\mid\mathcal G_1\}$ gives the first equality below, where the remainder is $O\{(x_M+|\gamma_M|)^3\}$ because the remainder in \eqref{eq:noise-g1-mean} is uniform in $i$.
The first two sums after the first equality need not be $o_P(M^{-1/2})$ separately.
The identity
\[
\sum_{i=1}^Mw_i^2-\sum_{i=1}^{M-1}w_iw_{i+1}
=\frac12\sum_{i=1}^{M-1}(w_{i+1}-w_i)^2+\frac{w_1^2+w_M^2}2
\]
follows by expanding the squares on the right.
Applied with $w_i=\widetilde q_{i,M}$ and with $w_i=q_{i,M}$, it rewrites their difference as squared increments and boundary terms and gives the second equality:
\begin{equation}
\label{eq:noise-theta-mean-decomposition}
\begin{aligned}
&\widetilde\E\{\hat\theta_{g_1}[M^{-1/2}X]-\hat\theta_{g_1}[M^{-1/2}X^0]
\mid\mathcal G_1\}\\
&\quad=\frac1M\sum_{i=1}^M(\widetilde q_{i,M}^2-q_{i,M}^2)
-\frac1M\sum_{i=1}^{M-1}(\widetilde q_{i,M}\widetilde q_{i+1,M}
-q_{i,M}q_{i+1,M})\\
&\qquad-\frac1{2M}\sum_{i=1}^{M-1}
\frac{(x_M+\gamma_Mq_{i,M})^2}{q_{i,M}q_{i+1,M}}
+O\{(x_M+|\gamma_M|)^3\}\\
&\quad=\frac1{2M}\sum_{i=1}^{M-1}\{(\widetilde q_{i+1,M}-\widetilde q_{i,M})^2
-(q_{i+1,M}-q_{i,M})^2\}\\
&\qquad+\frac{\widetilde q_{1,M}^2-q_{1,M}^2+\widetilde q_{M,M}^2-q_{M,M}^2}{2M}\\
&\qquad-\frac1{2M}\sum_{i=1}^{M-1}
\frac{(x_M+\gamma_Mq_{i,M})^2}{q_{i,M}q_{i+1,M}}
+O\{(x_M+|\gamma_M|)^3\}.
\end{aligned}
\end{equation}
After multiplication by $M^{1/2}$,
\[
-\frac1{2M}\sum_{i=1}^{M-1}
\frac{(x_M+\gamma_Mq_{i,M})^2}{q_{i,M}q_{i+1,M}}
\]
is the only term that can have a non-trivial probability limit.
We will show that all other terms vanish as $M\to\infty$.

We first handle the boundary terms.
Since $\widetilde q_{j,M}^2-q_{j,M}^2=2x_M+2\gamma_Mq_{j,M}$ and $x_M+|\gamma_M|=O(M^{-1/4})$, the boundary term and the remainder satisfy
\begin{align*}
\frac{\widetilde q_{1,M}^2-q_{1,M}^2+\widetilde q_{M,M}^2-q_{M,M}^2}{2M^{1/2}}
&=O\{M^{-1/2}(x_M+|\gamma_M|)\}=O(M^{-3/4}),\\
M^{1/2}O\{(x_M+|\gamma_M|)^3\}&=O(M^{-1/4}).
\end{align*}

We next bound $(2M)^{-1}\sum_{i=1}^{M-1}\{(\widetilde q_{i+1,M}-\widetilde q_{i,M})^2
-(q_{i+1,M}-q_{i,M})^2\}$.
Since
$\widetilde q_{i+1,M}^2-\widetilde q_{i,M}^2
=(q_{i+1,M}-q_{i,M})(q_{i+1,M}+q_{i,M}+2\gamma_M)$,
\[
\begin{aligned}
&(\widetilde q_{i+1,M}-\widetilde q_{i,M})-(q_{i+1,M}-q_{i,M})\\
&\quad=(q_{i+1,M}-q_{i,M})\,
\frac{(q_{i+1,M}+\gamma_M-\widetilde q_{i+1,M})
+(q_{i,M}+\gamma_M-\widetilde q_{i,M})}
{\widetilde q_{i+1,M}+\widetilde q_{i,M}}.
\end{aligned}
\]
In order to bound the factor multiplying $(q_{i+1,M}-q_{i,M})$, note that $\gamma_M^2=\rho_M^2x_M$ gives $\widetilde q_{j,M}^2=(q_{j,M}+\gamma_M)^2+(2-\rho_M^2)x_M$, and $q_{j,M}+\gamma_M\ge m^{-1}-|\gamma_M|\ge(2m)^{-1}$.
Hence
$0\le\widetilde q_{j,M}-q_{j,M}-\gamma_M
=(2-\rho_M^2)x_M/(\widetilde q_{j,M}+q_{j,M}+\gamma_M)\le2mx_M$.
As $\widetilde q_{i+1,M}+\widetilde q_{i,M}\ge m^{-1}$, the fraction is at most $4m^2x_M$ in absolute value.
It follows that
\[
\begin{aligned}
|(\widetilde q_{i+1,M}-\widetilde q_{i,M})
-(q_{i+1,M}-q_{i,M})|
\le 4m^2x_M|q_{i+1,M}-q_{i,M}|.
\end{aligned}
\]
Using $|a^2-b^2|\le|a-b|(|a-b|+2|b|)$, we obtain
\[
\begin{aligned}
|(\widetilde q_{i+1,M}-\widetilde q_{i,M})^2
-(q_{i+1,M}-q_{i,M})^2|
&\le 4m^2x_M(4m^2x_M+2)
(q_{i+1,M}-q_{i,M})^2\\
&\le Cx_M(q_{i+1,M}-q_{i,M})^2,
\end{aligned}
\]
where the last inequality uses $x_M\to0$.

Therefore, it remains to bound $\sum_{i=1}^{M-1}(q_{i+1,M}-q_{i,M})^2$.
We proceed by controlling the increments of $\sigma$.
By the proof of Lemma~\ref{lem:noise-conditional-law}(i), $\sigma_t-\sigma_s=\int_s^t\widetilde\mu_u^\sigma\dd u+L_t^\sigma-L_s^\sigma$ for $0\le s\le t\le1$.
Under Assumption~\ref{ass:noise-bounded-coefficients}, $|a_t|\le m$ and hence $\langle L^\sigma,L^\sigma\rangle_1=\int_0^1a_u\dd u\le m$, so $L^\sigma$ is a square-integrable martingale on $[0,1]$.
The martingale isometry gives
$$\widetilde\E(L_t^\sigma-L_s^\sigma)^2=\widetilde\E\int_s^ta_u\dd u\le m(t-s).$$  Hence
\begin{equation}\label{eq: sig lip}
\widetilde\E(\sigma_t-\sigma_s)^2
\le2\widetilde\E\left(\int_s^t\widetilde\mu_u^\sigma\dd u\right)^2
+2\widetilde\E\int_s^ta_u\dd u
\leq 2m^2(t-s)^2+2m(t-s)
\le4m^2(t-s).
\end{equation}
The last inequality holds as $m\ge 2$.

For $d\ge1$ and $i\le M-d$, since $I_{i+d,M}=I_{i,M}+d/M$, the definition of $q_{i,M}$ and the reverse triangle inequality give
\[
\begin{aligned}
|q_{i+d,M}-q_{i,M}|
&=\left|
\left(M\int_{I_{i,M}}\sigma_{s+d/M}^2\dd s\right)^{1/2}
-\left(M\int_{I_{i,M}}\sigma_s^2\dd s\right)^{1/2}
\right|\\
&\le\left(
M\int_{I_{i,M}}
(\sigma_{s+d/M}-\sigma_s)^2\dd s
\right)^{1/2}.
\end{aligned}
\]
It then follows from \eqref{eq: sig lip} that $\widetilde\E(\sigma_t-\sigma_s)^2\le4m^2(t-s)$.
Consequently,
\[
\begin{aligned}
\widetilde\E\sum_{i=1}^{M-d}
(q_{i+d,M}-q_{i,M})^2
&\le\sum_{i=1}^{M-d}M\int_{I_{i,M}}
\widetilde\E(\sigma_{s+d/M}-\sigma_s)^2\dd s\\
&\le\sum_{i=1}^{M-d}M\int_{I_{i,M}}
\frac{4m^2d}{M}\dd s
\le4m^2d.
\end{aligned}
\]
By Markov's inequality, we have
\begin{equation}
\label{eq:noise-local-variation}
\sum_{i=1}^{M-d}(q_{i+d,M}-q_{i,M})^2=O_P(1).
\end{equation}
With $d=1$, \eqref{eq:noise-local-variation} and $x_M=O(M^{-1/4})$ give
\[
\frac1{2M^{1/2}}\sum_{i=1}^{M-1}
\{(\widetilde q_{i+1,M}-\widetilde q_{i,M})^2-(q_{i+1,M}-q_{i,M})^2\}
=O_P(M^{-1/2}x_M)=O_P(M^{-3/4}).
\]

Recall that $M^{1/4}x_M\to c$ and $M^{1/4}\gamma_M\to\iota$.
Since $q_{i,M}^2$ is the average of $\sigma^2$ over $I_{i,M}$, $q_{i,M}$ lies between the minimum and the maximum of $\sigma$ on the closure of $I_{i,M}$.
By the continuity of $\sigma$, $\max_i\sup_{t\in I_{i,M}}|q_{i,M}-\sigma_t|\to0$ and $\max_{i<M}|q_{i,M}/q_{i+1,M}-1|\to0$ almost surely, so the following Riemann sums converge almost surely:
\begin{align*}
-\frac{M^{1/2}}{2M}\sum_{i=1}^{M-1}
\frac{(x_M+\gamma_Mq_{i,M})^2}{q_{i,M}q_{i+1,M}}
&=-\frac1{2M}\sum_{i=1}^{M-1}
\Bigl(\frac{M^{1/4}x_M}{q_{i,M}}+M^{1/4}\gamma_M\Bigr)^2
\frac{q_{i,M}}{q_{i+1,M}}\\
&\to-\frac12\int_0^1\Bigl(\frac c{\sigma_t}+\iota\Bigr)^2\dd t.
\end{align*}
It then follows from \eqref{eq:noise-theta-mean-decomposition} that
\begin{equation}
\label{eq:noise-theta-mean-shift}
M^{1/2}\widetilde\E\{\hat\theta_{g_1}[M^{-1/2}X]
-\hat\theta_{g_1}[M^{-1/2}X^0]\mid\mathcal G_1\}
\inP-\frac12\int_0^1\Bigl(\frac c{\sigma_t}+\iota\Bigr)^2\dd t.
\end{equation}
Combining \eqref{eq:noise-theta-mean-shift} with \eqref{eq:noise-centered-perturbation}, applied to $(H,d)=(v\mapsto v^2,1)$ and $(g_1,2)$, proves \eqref{eq: theta shift}.

We show \eqref{eq: Q shift} following a similar strategy.
For $(F_{Q,k},K)$, the first sum in the decomposition \eqref{eq: diff decomp} is given by
$M^{1/2}\widetilde\E\{Q_k^{(n)}[M^{-1/2}X]-Q_k^{(n)}[M^{-1/2}X^0]
\mid\mathcal G_1\}$, $k=1,\ldots,\ell$.  
We will show in \eqref{eq:noise-Q-mean-shift} that it can be written as a term independent of $k$ and a remainder term of order $o_P(M^{-1/2})$. 

We first recall from the definition that
\[
\widetilde\E\{Q_k^{(n)}[M^{-1/2}X]\mid\mathcal G_1\}
=\frac{\mu_{2/n}^{-n}}M\sum_{i=1}^{M-K+1}
\widetilde\E\left\{\prod_{r\in\mathcal I_k}|X_{i+r-1,M}|^{2/n}
\middle|\mathcal G_1\right\},
\]
and similarly for $X^0$.
By Lemma~\ref{lem:noise-conditional-law}(ii) and \eqref{eq:noise-covariance}, we know that $X_{j,M}$ and $X_{j',M}$ are conditionally uncorrelated when $|j-j'|\ge2$.
Hence $X_{i+r-1,M}$, $r\in\mathcal I_k$, are conditionally independent given $\mathcal G_1$, and so are $X_{i+r-1,M}^0$, $r\in\mathcal I_k$.
We can write
\[
\begin{aligned}
&\widetilde\E\{Q_k^{(n)}[M^{-1/2}X]-Q_k^{(n)}[M^{-1/2}X^0]
\mid\mathcal G_1\}\\
&\quad=\frac1M\sum_{i=1}^{M-K+1}\left\{\prod_{r\in\mathcal I_k}
\widetilde q_{i+r-1,M}^{2/n}
-\prod_{r\in\mathcal I_k}q_{i+r-1,M}^{2/n}\right\}.
\end{aligned}
\]
Since $\widetilde q_{j,M}^2=q_{j,M}^2+2x_M+2\gamma_Mq_{j,M}$, we have, for $i\le M-K+1$,
\begin{equation}
\label{eq:noise-Q-product-difference}
\begin{aligned}
&\prod_{r\in\mathcal I_k}\widetilde q_{i+r-1,M}^{2/n}
-\prod_{r\in\mathcal I_k}q_{i+r-1,M}^{2/n}\\
&\quad=\psi\bigl((q_{i+r-1,M})_{r\in\mathcal I_k};x_M,\gamma_M\bigr)
+2x_M+\frac{2\gamma_M}n\sum_{r\in\mathcal I_k}q_{i+r-1,M},
\end{aligned}
\end{equation}
where
\[
\psi(u;x,\gamma)
=\left\{\prod_{j=1}^n(u_j^2+2x+2\gamma u_j)\right\}^{1/n}
-\left\{\prod_{j=1}^nu_j^2\right\}^{1/n}
-2x-\frac{2\gamma}n\sum_{j=1}^nu_j.
\]
Recall that $q_{j,M}\in[m^{-1},m]$ and $x_M+|\gamma_M|\to0$.
Then for all sufficiently large $M$,
\[
\widetilde q_{j,M}^2
=q_{j,M}^2+2x_M+2\gamma_Mq_{j,M}
\ge m^{-2}-2m(x_M+|\gamma_M|)
\ge(2m^2)^{-1}.
\]
We then handle each term in \eqref{eq:noise-Q-product-difference} one by one.
After summation over $i\le M-K+1$ and division by $M$, the term $2x_M$ in \eqref{eq:noise-Q-product-difference} contributes $2x_M(M-K+1)/M$, which is independent of $k$.
For the term $2\gamma_Mn^{-1}\sum_{r\in\mathcal I_k}q_{i+r-1,M}$, shifting the summation index by $r-1$ changes only the boundary terms.
Indeed, for each $r\in\mathcal I_k$, since $q_{j,M}\in[m^{-1},m]$ and $r\le K$,
\[
\begin{aligned}
\left|\sum_{i=1}^{M-K+1}q_{i+r-1,M}
-\sum_{i=1}^{M-K+1}q_{i,M}\right|
=\left|\sum_{i=M-K+2}^{M-K+r}q_{i,M}
-\sum_{i=1}^{r-1}q_{i,M}\right|
\le m(r-1)\le mK.
\end{aligned}
\]
As $|\mathcal I_k|=n$, we have
\begin{equation}
\label{eq:noise-Q-index-shift}
\frac{2\gamma_M}{nM}
\sum_{i=1}^{M-K+1}\sum_{r\in\mathcal I_k}q_{i+r-1,M}
=\frac{2\gamma_M}M\sum_{i=1}^{M-K+1}q_{i,M}
+O\Bigl(\frac{|\gamma_M|}M\Bigr).
\end{equation}

It remains to handle the term involving $\psi$ in \eqref{eq:noise-Q-product-difference}.
We first show that, for $u\in[m^{-1},m]^n$ and $|x|+|\gamma|\le(4m^3)^{-1}$, the following bound holds:
\begin{equation}
\label{eq:noise-psi-bound}
|\psi(u;x,\gamma)|
\le C(|x|+|\gamma|)\sum_{j=2}^n(u_j-u_1)^2.
\end{equation}

To prove \eqref{eq:noise-psi-bound}, first observe that for every $\bar u\in[m^{-1},m]$,
\[
\begin{aligned}
\psi(\bar uJ_n;x,\gamma)&=0,\\
\frac{\partial\psi}{\partial u_j}
(\bar uJ_n;x,\gamma)
&=\frac{2(\bar u+\gamma)}n
-\frac{2\bar u}n-\frac{2\gamma}n=0,
\qquad j=1,\ldots,n.
\end{aligned}
\]
Moreover, $\psi(w;0,0)=0$ for every $w$, so $\nabla_u^2\psi(w;0,0)=0$.
Since $w_j^2+2x+2\gamma w_j\ge(2m^2)^{-1}$ for $w\in[m^{-1},m]^n$ and $|x|+|\gamma|\le(4m^3)^{-1}$, the partial derivatives of $\nabla_u^2\psi$ with respect to $x$ and $\gamma$ are uniformly bounded.
Thus, applying the fundamental theorem of calculus along $t\mapsto(tx,t\gamma)$ gives
\begin{equation*}
\begin{aligned}
\nabla_u^2\psi(w;x,\gamma)
&=\int_0^1\bigl[
x\,\partial_x\nabla_u^2\psi(w;tx,t\gamma)
+\gamma\,\partial_\gamma\nabla_u^2\psi(w;tx,t\gamma)
\bigr]\dd t,\\
\|\nabla_u^2\psi(w;x,\gamma)\|
&\le\int_0^1(C|x|+C|\gamma|)\dd t
=C(|x|+|\gamma|),
\end{aligned}
\end{equation*}
where the bound is uniform over $w\in[m^{-1},m]^n$.
Since both $\psi$ and its gradient vanish at $u_1J_n$, the second-order Taylor theorem about $u_1J_n$ yields
\[
\begin{aligned}
|\psi(u;x,\gamma)|
&\le\frac12\|u-u_1J_n\|^2
\sup_{w\in[m^{-1},m]^n}
\|\nabla_u^2\psi(w;x,\gamma)\|\\
&\le C(|x|+|\gamma|)
\sum_{j=2}^n(u_j-u_1)^2,
\end{aligned}
\]
which proves \eqref{eq:noise-psi-bound}.

Recall that $(q_{i+r-1,M})_{r\in\mathcal I_k}$ lies in $[m^{-1},m]^n$.
Also, all $r\in\mathcal I_k$ satisfy $r-1<K$ and hence $M-K+1\le M-(r-1)$.
Therefore, \eqref{eq:noise-local-variation} with $d=r-1$ for $r\in\mathcal I_k\setminus\{1\}$ implies \eqref{eq:q-L2 bound}, while \eqref{eq:noise-psi-bound} implies \eqref{eq:noise-psi-sum-bound}:
\begin{align}
\sum_{i=1}^{M-K+1}\sum_{r\in\mathcal I_k}
(q_{i+r-1,M}-q_{i,M})^2
&=O_P(1),\label{eq:q-L2 bound}\\
\frac1M\sum_{i=1}^{M-K+1}
\bigl|\psi\bigl((q_{i+r-1,M})_{r\in\mathcal I_k};x_M,\gamma_M\bigr)\bigr|
&=O_P\Bigl(\frac{x_M+|\gamma_M|}M\Bigr).
\label{eq:noise-psi-sum-bound}
\end{align}
Summing \eqref{eq:noise-Q-product-difference} over $i\le M-K+1$, dividing by $M$, and using \eqref{eq:noise-Q-index-shift} and \eqref{eq:noise-psi-sum-bound} give
\begin{equation}
\label{eq:noise-Q-mean-shift}
\begin{aligned}
&\widetilde\E\{Q_k^{(n)}[M^{-1/2}X]
-Q_k^{(n)}[M^{-1/2}X^0]\mid\mathcal G_1\}\\
&\qquad=\frac{2x_M(M-K+1)}M+\frac{2\gamma_M}M\sum_{i=1}^{M-K+1}q_{i,M}
+O_P\!\left(\frac{x_M+|\gamma_M|}M\right).
\end{aligned}
\end{equation}
Because the two leading terms in \eqref{eq:noise-Q-mean-shift} do not depend on $k$, we have
\begin{align*}
&M^{1/2}D\,\widetilde\E\{\mathbb Q_\ell^{(n)}(1;M^{-1/2}X)
-\mathbb Q_\ell^{(n)}(1;M^{-1/2}X^0)\mid\mathcal G_1\}\\
&\qquad=M^{1/2}\left\{\frac{2x_M(M-K+1)}M
+\frac{2\gamma_M}M\sum_{i=1}^{M-K+1}q_{i,M}\right\}DJ_\ell
+O_P\left(\frac{x_M+|\gamma_M|}{M^{1/2}}\right),
\end{align*}
which is $O_P(M^{-3/4})$ because $DJ_\ell=0$ and $x_M+|\gamma_M|=O(M^{-1/4})$.
Since $\ell$ and $D$ are fixed, \eqref{eq:noise-centered-perturbation}, applied with $(H,d)=(F_{Q,k},K)$ for $k=1,\ldots,\ell$, gives
\[
D\left(M^{-1/2}\sum_{i=1}^{M-K+1}\chi_{i,M}^{F_{Q,k}}\right)_{k=1}^{\ell}
\inP0.
\]
This proves \eqref{eq: Q shift}.
\end{proof}

\begin{proof}[Proof of Theorem~\ref{thm:microstructure-noise}]
\textbf{\textit{(Validity)}} Suppose first that Assumption~\ref{ass:continuous-price} holds, and let $D$ satisfy \eqref{eq:D-conditions}.
We aim to prove that, as $M\to\infty$,
\begin{equation}
\label{eq:noise-statistic-transfer}
\begin{aligned}
M^{1/2}\bigl\{\hat\theta_{g_1}^{\mathrm{tr}}[\widetilde y]
-\hat\theta_{g_1}[y]\bigr\}
&\inP-\frac12\int_0^1\Bigl(\frac c{\sigma_t}+\iota\Bigr)^2\dd t,\\
M^{1/2}D\bigl\{\mathbb Q_\ell^{(n),\mathrm{tr}}(1;\widetilde y)
-\mathbb Q_\ell^{(n)}(1;y)\bigr\}&\inP0,
\end{aligned}
\end{equation}
and then combine it with the joint stable limit of the noise-free statistics.
Lemmas~\ref{lem:noise-drift} and~\ref{lem:noise-effect} justify the above convergence for the unthresholded version under Assumption~\ref{ass:noise-bounded-coefficients}, which Theorem~\ref{thm:microstructure-noise} does not assume.
Instead, we apply them to a localized model and use Lemma~\ref{lem:noise-conditional-law} to show that under $H_0$ no squared localized return exceeds its threshold with probability tending to one, and then remove the localization.

We localize $\sigma$, $\widetilde\mu^\sigma$, $a$, and $\mu^y$ at a level $m$ so that Assumption~\ref{ass:noise-bounded-coefficients} is satisfied.
For an integer $m\ge2$, set $\Omega_m^0=\{m^{-1}\le\sigma_0\le m\}\in\mathcal G_0$ and
\[
\begin{aligned}
\mathcal E_m^\sigma
&=\bigl\{(t,\omega)\in[0,1]\times\Omega:
\sigma_t(\omega)\notin(m^{-1},m)
\text{ or }
|\widetilde\mu_t^\sigma(\omega)|+|a_t(\omega)|>m\bigr\},\\
\tau_m^\sigma(\omega)
&=\inf\{t\in[0,1]:(t,\omega)\in\mathcal E_m^\sigma\},
\end{aligned}
\]
with the convention $\inf\varnothing=\infty$.
Since $\mathcal E_m^\sigma$ is $(\mathcal G_t)$-progressively measurable and $(\mathcal G_t)$ is right-continuous and contains all $\pr$-null sets, the d\'ebut theorem implies that $\tau_m^\sigma$ is a $(\mathcal G_t)$-stopping time.

Put $\sigma_t^{(m)}=\ind(\Omega_m^0)\sigma_{t\wedge\tau_m^\sigma}+\ind((\Omega_m^0)^c)$ and $\mu_t^{y,(m)}=(-m)\vee(\mu_t^y\wedge m)$.
Then $\sigma^{(m)}$ admits the form \eqref{eq:volatility-decomposition} with the $\mathcal G_0$-measurable $\sigma_0^{(m)}$, $\widetilde\sigma\equiv0$, and with $\widetilde\mu_t^\sigma$, $a_t$, and $L_t^\sigma$ replaced by $\ind(\Omega_m^0)\ind(t<\tau_m^\sigma)\widetilde\mu_t^\sigma$, $\ind(\Omega_m^0)\ind(t<\tau_m^\sigma)a_t$, and $\ind(\Omega_m^0)L_{t\wedge\tau_m^\sigma}^\sigma$, respectively.
Moreover, $\sigma^{(m)}$ is continuous and $(\mathcal G_t)$-adapted, and $\mu^{y,(m)}$ is adapted and c\`adl\`ag.
The variables $Z_{i,M}^{(m)}$, $\varepsilon_{i,M}^{(m)}$, $y_i^{(m)}$, and $\widetilde y_i^{(m)}$ are constructed as in Section~\ref{sec:microstructure-noise} and Assumption~\ref{ass:microstructure-noise-null} using $\sigma^{(m)}$, $\mu^{y,(m)}$, $W$ and $G_{0,M},\ldots,G_{M,M}$.
We then denote the corresponding drift-free returns by $X_{i,M}^{0,(m)}$ and $X_{i,M}^{(m)}$.
Such a localized model satisfies Assumptions~\ref{ass:continuous-price} and~\ref{ass:microstructure-noise-null}.
It also satisfies Assumption~\ref{ass:noise-bounded-coefficients} as we have
\begin{equation}\label{eq: local bound}
    m^{-1}\le\sigma_t^{(m)}\le m,\qquad
|\mu_t^{y,(m)}|\le m,\qquad
\ind(\Omega_m^0)\ind(t<\tau_m^\sigma)
\bigl(|\widetilde\mu_t^\sigma|+|a_t|\bigr)\le m,
\end{equation}
for every $t\in[0,1]$.

We then work under such a localized model for some fixed $m$ unless otherwise specified.
Since $M^{5/4}\omega_M^2\to c$ implies $M\omega_M^2\to0$, it follows from Lemma~\ref{lem:noise-drift} that
\[
M^{1/2}\bigl\{\hat\theta_{g_1}[\widetilde y^{(m)}]
-\hat\theta_{g_1}[y^{(m)}]\bigr\}
=M^{1/2}\bigl\{\hat\theta_{g_1}[M^{-1/2}X^{(m)}]
-\hat\theta_{g_1}[M^{-1/2}X^{0,(m)}]\bigr\}+o_P(1),
\]
and the same relation holds with $\hat\theta_{g_1}$ replaced by $Q_k^{(n)}$ for $k=1,\ldots,\ell$.
Then, Lemma~\ref{lem:noise-effect} implies that
\begin{equation}
\label{eq:noise-localized-comparison}
\begin{aligned}
M^{1/2}\bigl\{\hat\theta_{g_1}[\widetilde y^{(m)}]
-\hat\theta_{g_1}[y^{(m)}]\bigr\}
&\inP-\frac12\int_0^1\Bigl(\frac c{\sigma_t^{(m)}}+\iota\Bigr)^2\dd t,\\
M^{1/2}D\bigl\{\mathbb Q_\ell^{(n)}(1;\widetilde y^{(m)})
-\mathbb Q_\ell^{(n)}(1;y^{(m)})\bigr\}&\inP0.
\end{aligned}
\end{equation}
We want to show the same holds for the thresholded statistics.
It suffices to show that no squared return exceeds its threshold with probability tending to one because every thresholding indicator equals 1 in this case.
Let $\underline v_M=\underline\xi^{1/2}c_\vartheta M^{-\varpi}/2$.
By \eqref{eq:Theta-power}--\eqref{eq:xi-bounded}, $\vartheta_{i,M}=\xi_{i,M}c_\vartheta^2M^{-2\varpi}\ge4\underline v_M^2$ for all $i$ almost surely.
Therefore, it remains to justify that
\[\max_{i\le M}|\widetilde y_i^{(m)}|\le 2\underline v_M \le \min_{i\le M}(\vartheta_{i,M})^{1/2}\] 
with probability tending to one.  

By \eqref{eq: local bound}, we obtain $|\int_{I_{i,M}}\mu_s^{y,(m)}\dd s|\le m/M\le\underline v_M$ for all sufficiently large $M$.
Therefore, the inclusion $\{|\widetilde y_i^{(m)}|>2\underline v_M\}\subseteq \{|X_{i,M}^{(m)}|>M^{1/2}\underline v_M\}$ holds for large $M$.
We then recall from Lemma~\ref{lem:noise-conditional-law}(ii) and~(iii) that $[X_{i,M}^{(m)}\mid \mathcal G_1]\sim\Normal(0,\Var(X_{i,M}^{(m)}\mid \mathcal G_1))$ with $\Var(X_{i,M}^{(m)}\mid \mathcal G_1)\le C$.
We can thus apply the tail bound for a Gaussian random variable to yield
\begin{equation}
\label{eq:noise-localized-threshold-probability}
\pr\Bigl\{\max_{i\le M}|\widetilde y_i^{(m)}|>2\underline v_M\Bigr\}
\le\sum_{i=1}^M\pr\bigl(|X_{i,M}^{(m)}|>M^{1/2}\underline v_M\bigr)
\le2M\exp\{-M\underline v_M^2/(2C)\}\to0,
\end{equation}
where the upper bound vanishes as $M\underline v_M^2=\underline\xi c_\vartheta^2M^{1-2\varpi}/4$ with $\varpi<1/2$.
Therefore, \eqref{eq:noise-localized-comparison} also holds with the thresholded analogue.

We then remove the localization.
First, observe that the localized and original models agree on the event
\[
\Omega_m=\Omega_m^0\cap\{\tau_m^\sigma>1\}
\cap\left\{\sup_{t\in[0,1]}|\mu_t^y|\le m\right\}.
\]
On $\Omega_m$, we have $\sigma_t^{(m)}=\sigma_t$ and $\mu_t^{y,(m)}=\mu_t^y$ for every $t\in[0,1]$.
Since $\sigma^{(m)}$ and $\sigma$ coincide up to $\tau_m^\sigma$ on $\Omega_m^0\in\mathcal G_0$, we have
\[
\int_{I_{i,M}}\sigma_s^{(m)}\dd W_s
=\int_{I_{i,M}}\sigma_s\dd W_s,
\qquad
\int_{I_{i,M}}(\sigma_s^{(m)})^2\dd s
=\int_{I_{i,M}}\sigma_s^2\dd s,
\]
for $i=1,\ldots,M$ almost surely on $\Omega_m$.
Similarly, we have
\[
X_{i,M}^{0,(m)}=X_{i,M}^0,\quad
Z_{i,M}^{(m)}=Z_{i,M},\quad
\varepsilon_{i,M}^{(m)}=\varepsilon_{i,M},\quad
y_i^{(m)}=y_i,\quad
\widetilde y_i^{(m)}=\widetilde y_i,
\qquad i=1,\ldots,M,
\]
almost surely on $\Omega_m$.
Under Assumption~\ref{ass:microstructure-noise-null}, $\sigma$ has continuous and strictly positive paths on $[0,1]$.
Together with the local boundedness of $\widetilde\mu^\sigma$, $a$, and $\mu^y$, we have
\[
0<\inf_{t\in[0,1]}\sigma_t
\le\sup_{t\in[0,1]}\sigma_t<\infty,
\qquad
\sup_{t\in[0,1]}
\bigl(|\widetilde\mu_t^\sigma|+|a_t|+|\mu_t^y|\bigr)
<\infty
\quad\text{a.s.}
\]
Consequently, almost surely, there exists an integer $m_0\ge2$ such that $\Omega_m$ occurs for every $m\ge m_0$.
By construction, $\Omega_m\subseteq\Omega_{m+1}$, and hence
\[
\lim_{m\to\infty}\pr(\Omega_m^c)
=\lim_{m\to\infty}\{1-\pr(\Omega_m)\}
=
1-\pr\left(\bigcup_{m\ge2}\Omega_m\right)
=0.
\]

We then remove the localization restriction from \eqref{eq:noise-localized-comparison} and \eqref{eq:noise-localized-threshold-probability} simultaneously.
Fix $\eta>0$ and define
\[
\begin{aligned}
\mathcal C_{M,1}(\eta)
&=\left\{
\left\vert M^{1/2}\{\hat\theta_{g_1}[\widetilde y]
-\hat\theta_{g_1}[y]\}
+\frac12\int_0^1
\left(\frac c{\sigma_t}+\iota\right)^2\dd t\right\vert>\eta
\right\},\\
\mathcal C_{M,2}(\eta)
&=\Bigl\{
\bigl\|M^{1/2}D\{
\mathbb Q_\ell^{(n)}(1;\widetilde y)
-\mathbb Q_\ell^{(n)}(1;y)\}\bigr\|_2>\eta
\Bigr\},\\
\mathcal C_{M,3}
&=\Bigl\{
\max_{i\le M}\ind(\widetilde y_i^2>\vartheta_{i,M})=1
\Bigr\}.
\end{aligned}
\]
Set
$\mathcal C_M(\eta)=
\mathcal C_{M,1}(\eta)\cup
\mathcal C_{M,2}(\eta)\cup
\mathcal C_{M,3}$.
Let $\mathcal C_{M,1}^{(m)}(\eta)$, $\mathcal C_{M,2}^{(m)}(\eta)$, and $\mathcal C_{M,3}^{(m)}$ be the corresponding events with $\widetilde y$, $y$, and $\sigma$ replaced by $\widetilde y^{(m)}$, $y^{(m)}$, and $\sigma^{(m)}$, respectively, while retaining the same thresholds $\vartheta_{i,M}$.
Define
\(
\mathcal C_M^{(m)}(\eta)
=\mathcal C_{M,1}^{(m)}(\eta)
\cup\mathcal C_{M,2}^{(m)}(\eta)
\cup\mathcal C_{M,3}^{(m)}.
\)
As $\mathcal C_M(\eta)=\mathcal C_M^{(m)}(\eta)$ almost surely on $\Omega_m$, we have
\[
\pr\{\mathcal C_M(\eta)\}
\le\pr\{\mathcal C_M^{(m)}(\eta)\}
+\pr(\Omega_m^c).
\]
For each fixed $m$, \eqref{eq:noise-localized-comparison} gives
$
\pr\{\mathcal C_{M,1}^{(m)}(\eta)\}\to0
$
and
$
\pr\{\mathcal C_{M,2}^{(m)}(\eta)\}\to0.
$
Since $\vartheta_{i,M}\ge4\underline v_M^2$ almost surely for every $i$,
\[
\mathcal C_{M,3}^{(m)}
\subseteq
\Bigl\{
\max_{i\le M}|\widetilde y_i^{(m)}|
>2\underline v_M
\Bigr\}
\quad\text{a.s.},
\]
and hence $\pr\{\mathcal C_{M,3}^{(m)}\}\to0$ by \eqref{eq:noise-localized-threshold-probability}.
Consequently, we have $\pr\{\mathcal C_M^{(m)}(\eta)\}\to0$ for each fixed $m$.
It follows that
\[
\limsup_{M\to\infty}\pr\{\mathcal C_M(\eta)\}
\le\lim_{M\to\infty}
\pr\{\mathcal C_M^{(m)}(\eta)\}
+\pr(\Omega_m^c)
=\pr(\Omega_m^c),
\qquad m\ge2.
\]
Letting $m\to\infty$ gives $\pr\{\mathcal C_M(\eta)\}\to0$ as $M\to\infty$. 

On the complement of $\mathcal C_M(\eta)$, we have $\widetilde y_i^2\le\vartheta_{i,M}$ for every $i$, so every threshold indicator in \eqref{eq:ivg-tr} and \eqref{eq:Qk-tr} evaluated at $\widetilde y$ equals one, $\hat\theta_{g_1}^{\mathrm{tr}}[\widetilde y]=\hat\theta_{g_1}[\widetilde y]$, and $\mathbb Q_\ell^{(n),\mathrm{tr}}(1;\widetilde y)
=\mathbb Q_\ell^{(n)}(1;\widetilde y)$.
Since $\eta>0$ is arbitrary, this proves \eqref{eq:noise-statistic-transfer}.

Under Assumption~\ref{ass:continuous-price}, $\Omega_0=\Omega$, and Theorem~\ref{thm:joint-stable-CLT-voljump} applies to $g_1$ by Remark~\ref{rem:concrete-g-verification}.
Since the noise-free statistics are $\mathcal F$-measurable, their stable limit remains valid on the noise-extended probability space.
Hence, \eqref{eq:noise-statistic-transfer} and stable Slutsky give, with $Z_0=\mathcal Z_0/\sigma_{g_1}$ and $Z=D\mathcal Z_{\mathbb Q}=(Z_1,\ldots,Z_{\ell-1})^\T$,
\[
M^{1/2}
\begin{pmatrix}
\hat\theta_{g_1}^{\mathrm{tr}}[\widetilde y]\\
D\mathbb Q_\ell^{(n),\mathrm{tr}}(1;\widetilde y)
\end{pmatrix}
\inST 
\begin{pmatrix}
(v^{[4]})^{1/2}\sigma_{g_1}Z_0
-\dfrac12\displaystyle\int_0^1
\Bigl(\frac c{\sigma_t}+\iota\Bigr)^2\dd t\\
(v^{[4]})^{1/2}Z
\end{pmatrix}.
\]
The argument following \eqref{eq:local-projected-stable-limit} in the proof of Theorem~\ref{thm:trunc-local-power} then gives the stable limit and conditional Student-$t$ law in item~(i), with $\lambda_J$ replaced by $\lambda_{c,\iota}$.

Recall from the beginning of the proof that the argument so far works under Assumption~\ref{ass:continuous-price}.
It remains to extend the result to $\Omega_0$ under Assumption~\ref{ass:microstructure-noise-null}.
Let $y^{(q)}$ and $E_q$ be as in Lemma~\ref{lem:no-jump-reduction}.
For each $q$, evaluate $T_{\ell,*}^{(n),\mathrm{tr}}(g_1)$ using the returns of $y^{(q)}(i/M)+\varepsilon_{i,M}$, with the same thresholds $\vartheta_{i,M}$.
Since $y^{(q)}$ has the same $\sigma$ and $W$ as $y^*$, the variables $Z_{i,M}$ and $\varepsilon_{i,M}$ are unchanged.
Thus, the stable convergence established above applies to each $y^{(q)}$, with the same limit for all $q$.
Moreover, on $E_q$,
\[
y^{(q)}(i/M)+\varepsilon_{i,M}
=y^*(i/M)+\varepsilon_{i,M},
\qquad i=0,\ldots,M.
\]
Therefore, the stable limit in item~(i) in restriction to $\Omega_0$ follows from Lemma~\ref{lem:no-jump-reduction}(ii) with
$\zeta_M=(\varepsilon_{0,M},\ldots,\varepsilon_{M,M},
\vartheta_{1,M},\ldots,\vartheta_{M,M})$.

Since the conditional law $t_{\ell-1}[\lambda_{c,\iota}]$ given $\mathcal F$ is continuous and $\Omega_0\in\mathcal F$, the stable convergence in restriction to $\Omega_0$ and \eqref{eq:cross-multiplied-rejection} imply that
\begin{align*}
\pr\{\Psi_{\ell,*}^{(n),\mathrm{tr}}(t_{\ell-1,\alpha_0})=1,\ \Omega_0\}
&=\pr\{T_{\ell,*}^{(n),\mathrm{tr}}(g_1)>t_{\ell-1,\alpha_0},\ \Omega_0\}\\
&\to\E\{\ind(\Omega_0)\beta_{\ell-1,\alpha_0}(\lambda_{c,\iota})\}
\end{align*}
for every $\alpha_0\in(0,1)$ as $M\to\infty$.
Dividing both sides by $\pr(\Omega_0)>0$ proves item~(i).
As $\lambda_{c,\iota}\le0$ and $\beta_{\ell-1,\alpha_0}$ is a monotonically increasing function, we have $\E\{\beta_{\ell-1,\alpha_0}(\lambda_{c,\iota})\mid\Omega_0\}\le\alpha_0$.

\textbf{\textit{(Consistency)}} Suppose Assumption~\ref{ass:price-model} holds and $\varepsilon_{i,M}\sim\Normal(0,\omega_M^2)$ for $i=0,\ldots,M$.
We do not impose independence among the $(\varepsilon_{i,M})$ or between $\varepsilon$ and $y^*$ throughout the following argument.
By the same argument in the proof of Theorem~\ref{thm:thresholded-consistency}, it suffices to show
\[
\hat\theta_{g_1}^{\mathrm{tr}}[\widetilde y]\inP\theta,
\qquad
\hat V_{\ell,*}^{(n),\mathrm{tr}}(1)[\widetilde y]\inP0.
\]

First, recall from Proposition~\ref{prop:tr-iv-robust} that $\hat{\IV}_{g_1}^{\mathrm{tr}}[y]\inP v^{[2]}$, $\hat\theta_{g_1}^{\mathrm{tr}}[y]\inP\theta$ and $\hat{\QV}[y]=\sum_iy_i^2\inP v^{[2]}+\theta$.
Since $\widetilde y_i-y_i=\varepsilon_{i,M}-\varepsilon_{i-1,M}$, we have
\(
\widetilde\E\left\{\sum_{i=1}^M(\widetilde y_i-y_i)^2\right\}\le
4M\omega_M^2\to0
\)
and hence by Markov's inequality,
\begin{equation}
\label{eq:noise-H1-return-perturbation}
\sum_{i=1}^M(\widetilde y_i-y_i)^2\inP 0.
\end{equation}
Notice that $\widetilde y_i^2-y_i^2=2y_i(\widetilde y_i-y_i)+(\widetilde y_i-y_i)^2$.
It then follows from \eqref{eq:noise-H1-return-perturbation} that
\begin{equation}
\label{eq:noise-H1-QV-perturbation}
\bigl|\hat{\QV}[\widetilde y]-\hat{\QV}[y]\bigr|
\le2\left(\sum_{i=1}^M y_i^2\right)^{1/2}\left\{\sum_{i=1}^M(\widetilde y_i-y_i)^2\right\}^{1/2}
+\sum_{i=1}^M(\widetilde y_i-y_i)^2=o_P(1).
\end{equation}
Therefore, we have $\hat{\QV}[\widetilde y]\inP v^{[2]}+\theta$.

For $(H,d)=(g_1,2)$ or $(F_{Q,k},K)$, write $H(w)=C_H\prod_{r\in\mathcal J}|w_r|^p$, where $C_H>0$ and $(\mathcal J,p)=(\{1,2\},1)$ or $(\mathcal I_k,2/n)$, respectively.
In both cases, $0<p\le1$ and $|\mathcal J|p=2$.
By \eqref{eq:noise-sum-form}, \eqref{eq:ivg-tr}, and \eqref{eq:Qk-tr}, both $\hat{\IV}_{g_1}^{\mathrm{tr}}[\widetilde y]$ and $Q_k^{(n),\mathrm{tr}}[\widetilde y]$ can be decomposed as
\begin{equation}
\label{eq:noise-H1-threshold-decomposition}
\begin{aligned}
&\sum_{i=1}^{M-d+1}H(\widetilde y_i,\ldots,\widetilde y_{i+d-1})
\prod_{r\in\mathcal J}
\ind(|\widetilde y_{i+r-1}|\le\vartheta_{i+r-1,M}^{1/2})\\
&\quad=\sum_{i=1}^{M-d+1}H(y_i,\ldots,y_{i+d-1})
\prod_{r\in\mathcal J}
\ind(|\widetilde y_{i+r-1}|\le\vartheta_{i+r-1,M}^{1/2})\\
&\qquad+\sum_{i=1}^{M-d+1}
\{H(\widetilde y_i,\ldots,\widetilde y_{i+d-1})
-H(y_i,\ldots,y_{i+d-1})\}
\prod_{r\in\mathcal J}
\ind(|\widetilde y_{i+r-1}|\le\vartheta_{i+r-1,M}^{1/2}).
\end{aligned}
\end{equation}
We first show that the second sum converges to zero in probability.

Since $0<p\le1$, we have
\begin{align*}
&|H(\widetilde y_i,\ldots,\widetilde y_{i+d-1})
-H(y_i,\ldots,y_{i+d-1})|\\
&\quad=C_H\left\vert
\sum_{r\in\mathcal J}
\bigl(|\widetilde y_{i+r-1}|^p-|y_{i+r-1}|^p\bigr)
\prod_{\substack{s\in\mathcal J\\s<r}}|y_{i+s-1}|^p
\prod_{\substack{s\in\mathcal J\\s>r}}|\widetilde y_{i+s-1}|^p
\right\vert\\
&\quad\le C_H\sum_{r\in\mathcal J}
|\widetilde y_{i+r-1}-y_{i+r-1}|^p
\prod_{s\in\mathcal J\setminus\{r\}}
(|y_{i+s-1}|+
|\widetilde y_{i+s-1}-y_{i+s-1}|)^p,
\end{align*}
where the inequality follows from the fact that $\big||v+w|^p-|v|^p\big|\le|w|^p$.
As each summand of the summation in the upper bound is a product of $|\mathcal J|=2/p$ factors, we can apply H\"older's inequality with exponent $2/p$ for every factor to yield
\[
\begin{aligned}
&\sum_{i=1}^{M-d+1}
|\widetilde y_{i+r-1}-y_{i+r-1}|^p
\prod_{s\in\mathcal J\setminus\{r\}}
(|y_{i+s-1}|+|\widetilde y_{i+s-1}-y_{i+s-1}|)^p\\
&\quad\le
\left\{\sum_{i=1}^M(\widetilde y_i-y_i)^2\right\}^{p/2}
\prod_{s\in\mathcal J\setminus\{r\}}
\left\{\sum_{i=1}^M
(|y_i|+|\widetilde y_i-y_i|)^2\right\}^{p/2}.
\end{aligned}
\]

Notice that $(|a|+|b|)^2\le2(a^2+b^2)$ and $(|\mathcal J|-1)p/2=1-p/2$.
We obtain
\begin{equation}
\label{eq:noise-H1-product-perturbation}
\begin{aligned}
&\sum_{i=1}^{M-d+1}
|H(\widetilde y_i,\ldots,\widetilde y_{i+d-1})
-H(y_i,\ldots,y_{i+d-1})|\\
&\quad\le
C_H|\mathcal J|2^{1-p/2}\left\{\sum_{i=1}^M(\widetilde y_i-y_i)^2\right\}^{p/2}
\left\{\sum_{i=1}^My_i^2+
\sum_{i=1}^M(\widetilde y_i-y_i)^2\right\}^{1-p/2}
\inP 0,
\end{aligned}
\end{equation}
where the convergence follows from \eqref{eq:noise-H1-return-perturbation} and $\hat{\QV}[y]\inP v^{[2]}+\theta$.
Since every threshold indicator is at most one, the second sum on the right of \eqref{eq:noise-H1-threshold-decomposition} is bounded in absolute value by the left side of \eqref{eq:noise-H1-product-perturbation} and is therefore $o_P(1)$.

It remains to show that 
\begin{equation}\label{eq: H noisy threshold main}
\sum_{i=1}^{M-d+1}H(y_i,\ldots,y_{i+d-1})
\prod_{r\in\mathcal J}
\ind(|\widetilde y_{i+r-1}|\le\vartheta_{i+r-1,M}^{1/2})
\inP
v^{[2]}.    
\end{equation}
We first recall that $\underline v_M=\underline\xi^{1/2}c_\vartheta M^{-\varpi}/2$ and it satisfies $\vartheta_{i,M}^{1/2}\ge2\underline v_M$ almost surely.
Similarly, let $\overline v_M=2\overline\xi^{1/2}c_\vartheta M^{-\varpi}$.
Then $\vartheta_{i,M}^{1/2}\le\overline v_M/2$ almost surely.
On the event $\{\max_{i\le M}|\widetilde y_i-y_i|\le\underline v_M\}$, we have $|y_i|\le\underline v_M$ implies $|\widetilde y_i|\le2\underline v_M\le\vartheta_{i,M}^{1/2}$, while $|\widetilde y_i|\le\vartheta_{i,M}^{1/2}$ implies $|y_i|\le\overline v_M/2+\underline v_M\le\overline v_M$.
Thus, the following holds over all $i$ almost surely on this event
\[
\ind(|y_i|\le\underline v_M)\le\ind(|\widetilde y_i|\le\vartheta_{i,M}^{1/2})
\le\ind(|y_i|\le\overline v_M).
\]
It follows that, almost surely on this event,
\begin{align*}
&\sum_{i=1}^{M-d+1}H(y_i,\ldots,y_{i+d-1})
\prod_{r\in\mathcal J}\ind(|y_{i+r-1}|\le\underline v_M)\\
&\qquad\le\sum_{i=1}^{M-d+1}H(y_i,\ldots,y_{i+d-1})
\prod_{r\in\mathcal J}\ind(|\widetilde y_{i+r-1}|\le\vartheta_{i+r-1,M}^{1/2})\\
&\qquad\le\sum_{i=1}^{M-d+1}H(y_i,\ldots,y_{i+d-1})
\prod_{r\in\mathcal J}\ind(|y_{i+r-1}|\le\overline v_M).
\end{align*}
By Propositions~\ref{prop:tr-iv-robust} and~\ref{prop:tr-Q-robust}, the first and last sums in the above inequality converge in probability to $v^{[2]}$.  

It remains to show that $\{\max_{i\le M}|\widetilde y_i-y_i|\le\underline v_M\}$ has probability tending to one.
For $\omega_M>0$, $|\widetilde y_i-y_i|>\underline v_M$ implies $|\varepsilon_{i,M}|>\underline v_M/2$ or $|\varepsilon_{i-1,M}|>\underline v_M/2$.
Therefore, we have
\begin{equation*}
    \pr\left\{\max_{i\le M}|\widetilde y_i-y_i|>\underline v_M\right\}
\le\sum_{j=0}^M\pr(|\varepsilon_{j,M}|>\underline v_M/2)
\le2(M+1)\exp\{-\underline v_M^2/(8\omega_M^2)\}.
\end{equation*}
As $\omega_M^2=O(M^{-5/4})$ and $\underline v_M^2=\underline\xi c_\vartheta^2M^{-2\varpi}/4$, we have $\underline v_M^2/\omega_M^2\ge C^{-1}M^{5/4-2\varpi}$ with $5/4-2\varpi>1/4$.
Therefore, we have $\pr\{\max_{i\le M}|\widetilde y_i-y_i|>\underline v_M\}\to0.$
If $\omega_M=0$, this probability is zero.

Therefore, by the squeeze theorem, we have proven \eqref{eq: H noisy threshold main}.
It follows immediately that $\hat{\IV}_{g_1}^{\mathrm{tr}}[\widetilde y]\inP v^{[2]}$ and $\max_{1\le k\le\ell}|Q_k^{(n),\mathrm{tr}}[\widetilde y]-v^{[2]}|\inP0$.
By \eqref{eq:noise-H1-QV-perturbation}, this gives
$\hat\theta_{g_1}^{\mathrm{tr}}[\widetilde y]
=\hat{\QV}[\widetilde y]-\hat{\IV}_{g_1}^{\mathrm{tr}}[\widetilde y]
\inP\theta$.
Also, $\hat V_{\ell,*}^{(n),\mathrm{tr}}(1)[\widetilde y]\inP0$ due to \eqref{eq:VD-tr}.  

Applied with $a=t_{\ell-1,\alpha_0}$ to the statistics evaluated at $\widetilde y_i$, the argument in the proof of Theorem~\ref{thm:thresholded-consistency} yields
$\pr\{\Psi_{\ell,*}^{(n),\mathrm{tr}}(t_{\ell-1,\alpha_0})=1,\ \Omega_1\}
\to\pr(\Omega_1)$.
Dividing by $\pr(\Omega_1)>0$ proves item~(ii).
\end{proof}